\documentclass[aps,prd,preprintnumbers,showpacs,nofootinbib,amssymb]{revtex4}
\usepackage{graphicx}
\usepackage{amsmath}
\usepackage{amssymb}
\usepackage{amsfonts}
\usepackage{bm}
\usepackage{color}

\def\be{\begin{equation}}
\def\ee{\end{equation}}
\def\bea{\begin{eqnarray}}
\def\eea{\end{eqnarray}}

\begin{document}

\title{Kinetic theory of inhomogeneous systems with long-range
interactions and fluctuation-dissipation theorem}
\author{Pierre-Henri Chavanis}
\affiliation{Laboratoire de
Physique Th\'eorique, Universit\'e de Toulouse, CNRS, UPS, France}

\begin{abstract}

We complete the  kinetic theory  of inhomogeneous systems with long-range
interactions initiated in previous works. We use a simpler
and more physical
formalism. We
consider a system of particles submitted to a small external stochastic
perturbation and determine the response of the system to the perturbation. We
derive the diffusion tensor and the friction  by polarization of a test
particle.
We introduce a general Fokker-Planck equation involving a diffusion term and a
friction term. When the friction by polarization can be neglected, we obtain a
secular dressed diffusion (SDD) equation sourced by the external noise. When the
external
perturbation is created by a discrete collection of $N$ field particles, we
obtain the inhomogeneous Lenard-Balescu kinetic equation reducing to the
inhomogeneous Landau kinetic
equation when collective effects are neglected. We consider a multi-species
system of particles.  When the field particles are at statistical
equilibrium (thermal bath), we establish the proper expression of the
fluctuation-dissipation theorem for systems with long-range interactions
relating the power
spectrum of the fluctuations to the response
function of the system.  In
that case, the friction and diffusion coefficients satisfy the
Einstein relation and the Fokker-Planck equation reduces to the inhomogeneous
Kramers equation.  We also
consider a gas of Brownian particles with long-range interactions  described by
$N$ coupled stochastic Langevin
equations and determine its mean and mesoscopic evolution. We discuss
the notion
of stochastic kinetic equations and the role of fluctuations possibly 
triggering random
transitions from one equilibrium state to the other. Our
presentation parallels the one given for two-dimensional point vortices in a
previous paper [P.H. Chavanis, Eur. Phys. J. Plus {\bf 138}, 136 (2023)].

\end{abstract}

\pacs{95.30.Sf, 95.35.+d, 95.36.+x, 98.62.Gq, 98.80.-k}

\maketitle

\section{Introduction}
\label{sec_intro}

The dynamics and thermodynamics  of systems with long-range interactions is a
topic of active research \cite{campabook}. Systems
with long-range interactions generically experience two successive types
of relaxation \cite{houches}. In a first regime, the
encounters (correlations) between the particles can be neglected -- provided
that the
number of particles $N$ is sufficiently large -- and the evolution of the system
is collisionless. In that case, a mean field approximation can be implemented.
The evolution of the
distribution function (DF) of the system $f({\bf r},{\bf v},t)$ is governed
by the collisionless Boltzmann equation \cite{jeans}, also known as the Vlasov
equation \cite{vlasov1,vlasov2},\footnote{See H\'enon \cite{henonvlasov} for a
discussion
about the name that should be given to this equation.} which reads
\begin{equation}
\frac{\partial f}{\partial t}+{\bf v}\cdot\frac{\partial f}{\partial {\bf
r}}-\nabla\Phi\cdot \frac{\partial f}{\partial {\bf v}}=0,
\label{intro1}
\end{equation}
where
\begin{equation}
\Phi({\bf r},t)=\int u(|{\bf r}-{\bf r}'|)f({\bf r}',{\bf v}',t)\,
d{\bf r}'d{\bf v}'
\label{intro2}
\end{equation}
is the mean field potential produced by the particles and $u(|{\bf r}-{\bf
r}'|)$ is the potential of interaction. The Vlasov equation can
develop a complicated process of violent collisionless relaxation towards a
quasistationary state (QSS) which is a stable steady state of the Vlasov
equation (virialized state) \cite{kingvrnewref,henonvr,lb}. This process takes
place on a few dynamical times $t_D$. If the system mixes efficiently, the QSS
can be predicted by the statistical theory of
Lynden-Bell \cite{lb} which assumes that the system has reached the most mixed
state consistent with the collisionless evolution,
i.e., the one that maximizes a mixing entropy while conserving all the
constraints of the Vlasov equation: the energy and all the Casimirs (see
\cite{kinvr,ewart} for recent studies on the Lynden-Bell theory
of violent relaxation). This maximum entropy principle relies on an assumption
of ergodicity which is not always fulfilled in practice. Indeed, the system may
relax towards a QSS which is different from the Lynden-Bell prediction because
of the phenomenon of incomplete relaxation \cite{lb,incomplete}. In that case,
the QSS is a dynamically stable steady state of the Vlasov equation which is not
the most mixed state.  It is difficult to know in advance if the system will
reach the
most mixed state or not because the outcome depends on the efficiency of the
collisionless relaxation which is a self-sustained process. The fluctuations of
the potential, which are the engine of the relaxation, weaken as the system
approaches the QSS and the 
collisionless evolution may stop before the system has reached the most mixed
state. Sometimes, the Lynden-Bell distribution provides a good prediction of the
equilibium
state and sometimes the prediction fails.\footnote{As discussed below the
Lynden-Bell prediction always fails for self-gravitating systems because there
is no maximum entropy state even in theory \cite{lb,incomplete}.
However, the Lynden-Bell theory has been applied to other systems with
long-range interactions such as 2D
vortices, the HMF model and Coulombian
plasmas -- for which a maximum entropy state exists -- with different degrees of
success depending on the type of initial
condition considered (see Sec. 3 of \cite{hb3}, Sec. 6 of
\cite{assise}, Appendix B of \cite{closer}, and Ref. \cite{fit} for a
detailed discussion).} In
the latter case, the
prediction of the QSS effectively reached by the system is
complicated or even impossible. Then, in a second regime which takes place on a
secular timescale increasing algebraically with the number $N$ of particles, the
system evolves under the effect of encounters (correlations). This corresponds
to intrinsic internal noise (Poisson shot noise) arising from the graininess of
the system (finite $N$ effects, granularities,...).  For $t\rightarrow +\infty$
the DF $f({\bf r},{\bf v},t)$ relaxes towards the Boltzmann
distribution which is the most probable state (maximum entropy state at fixed
mass and energy) predicted
by the statistical mechanics of $N$ particles.

The case of self-gravitating systems is special because there is no maximum
entropy state (i.e. no statistical equilibrium state) in a
strict sense \cite{btnew}. When coupled to gravity, the Lynden-Bell distribution
and the Boltzmann distribution have an infinite mass (the spatial density
decreases as $r^{-2}$ at large distances). The entropy
can be made infinitely large while conserving the mass and the energy by
forming a core-halo structure and extending the halo to infinity. Concerning the
collisionless regime (violent relaxation) we have seen that the relaxation is
incomplete \cite{lb,incomplete}. The system reaches a partially mixed QSS which
is a stable steady state of the Vlasov equation (virialized state). It remains
blocked in that state until collisions come into play. Concerning the
collisional regime (secular relaxation) the evolution of the system is marked
by the evaporation of stars \cite{amba,spitzer1940} and the
gravothermal catastrophe \cite{antonov,lbw}. Due to star encounters and tidal
effects, the system reaches a truncated Boltzmann distribution
called the Michie-King model \cite{michie,king}. As the stellar system
slowly evaporates, it increases its central density (following the King
sequence) until it becomes thermodynamically unstable and
undergoes core collapse \cite{lbe,cohn}. Core collapse leads to the
formation of binary stars \cite{henonbinary}.
The energy released by the binaries can stop the collapse and induce a
re-expansion of the halo in a post-collapse regime \cite{inagakilb}. Then, a
series of
gravothermal oscillations follows \cite{sugimoto}. During this secular
evolution, the Boltzmann entropy increases without bound. In
that respect, the statistical mechanics of stellar systems is essentially an
out-of-equilibrium problem. Therefore, in order to understand the evolution of
self-gravitating systems, and systems
with long-range interactions in general, it is necessary to develop a kinetic
theory.
Below, we briefly review the kinetic theories that have been developed in the
past (see Refs. \cite{epjp2,aa} for complementary reviews on this subject).

The first kinetic equation applicable to systems with long-range
interactions was derived by Landau \cite{landau} for a spatially homogeneous
neutral Coulombian plasma. He considered a gas of electrons of mass $m$
and charge $-e$ moving in a uniform, continuous, positively charged background
of ions, which insures the electroneutrality of the system. He emphasized the
importance of collisions between particles at large distances which are
scattered through small angles. Starting from the Boltzmann equation
\cite{boltzmann}, expanding this
equation in terms of a
weak deflexion parameter, and making a  linear trajectory approximation, he
obtained a kinetic equation governing the evolution
of the velocity distribution $f({\bf v},t)$ of the electrons. The original form
of the
equation given by Landau \cite{landau} reads 
\begin{equation}
\label{intro3} 
\frac{\partial f}{\partial t}=A\frac{\partial}{\partial
v_i}\int \frac{u^2\delta_{ij}-u_iu_j}{u^3} \left (f'\frac{\partial
f}{\partial v_j}-f\frac{\partial
f'}{\partial v'_j}\right)\, d{\bf v}',
\end{equation}
where  ${\bf u}={\bf v}-{\bf v}'$ is the relative velocity of the particles
(charges) engaged in a collision, $f$ and $f'$ stands for $f({\bf v},t)$
and $f({\bf v}',t)$, and $A=(2\pi e^4/m^3)\ln\Lambda$ is a prefactor
where $\ln \Lambda=\int_0^{+\infty}db/b$ is the so-called
Coulomb
logarithm, defined in terms of the impact parameter $b$, that has to be
regularized with
appropriate cut-offs. The approach
of Landau ignores collective effects and exhibits a logarithmic
divergence at large impact parameters arising from the long-range nature of the
Coulomb force that he cured heuristically by
introducing an
upper cut-off
at the Debye screening length $\lambda_D\sim [k_BT/(ne^2)]^{1/2}$ \cite{dh}.
There is
also
a logarithmic divergence at small impact parameters, due
to the neglect of strong collisions, that Landau cured heuristically by
introducing a lower cut-off at the distance $b_{90}\sim e^2/(m v^2)\sim
e^2/(k_B T)\sim 1/(n\lambda_D^2)$ at which the particles are
deflected by
about  $90^{\rm o}$. This leads to a Coulomb logarithm of the form
$\ln\Lambda\sim \ln
(\lambda_D/b_{90})\sim \ln (n\lambda_D^3)$ where
$\Lambda=n\lambda_D^3$ is the plasma parameter. It gives the typical number of
electrons in the Debye sphere and usually satisfies $\Lambda\gg 1$.

The kinetic theory of stellar systems was pioneered by
Jeans \cite{jeansk1,jeansk2,jeansbook}, Eddington
\cite{eddingtonkin}, Charlier \cite{charlier}, Schwarzschild
\cite{schwarzschild1924}, Rosseland \cite{rosseland}, Smart
\cite{smart}, Ambarzumian \cite{amba}, Spitzer
\cite{spitzer1940} and
Chandrasekhar \cite{cvn0,cvn1,chandra,chandrabrown,cvn2,chandra1,
chandra2,chandra3,chandrany,cvn3,cvn4,cvn5,nice}. When the
particles interact
according to inverse-square forces, Jeans \cite{jeansk1,jeansk2,jeansbook}
showed that the cumulative
effect of the weak deflections resulting from the relatively distant encounters
is more important than the effect of occasional large deflections (relatively
close encounters). As Eddington \cite{eddingtonkin} writes: ``Close approaches
which produce an appreciable deflection are exceedingly rare. It is of greater
importance to determine what is the cumulative effect of the large number of
infinitesimal encounters experienced by a star in the course of a long period of
time.'' Using this argument, Chandrasekhar
\cite{cvn0,cvn1,chandra,chandrabrown,cvn2,chandra1,
chandra2,chandra3,chandrany,cvn3,cvn4,cvn5,nice}
developed an analogy with Brownian motion where the evolution of the velocity
distribution of the particles is due to a large number of small deflections. He
argued that the
force experienced by a  star can be decomposed in two terms.
There is
a mean field force due to the system as a whole and a
force arising from
discrete collisions. The
first force is smooth and the second accounts for the granularity of
the system. Because of
the fluctuations of the gravitational force
created by the field stars, a test star experiences a diffusion in
velocity space.\footnote{The notion of ``test'' and ``field''
stars was introduced by Chandrasekhar \cite{chandra} (see also \cite{cohen}).}
However,
diffusion alone would cause the energy of the test star to diverge and would not
result in the Maxwell-Boltzmann equilibrium. Therefore, Chandrasekhar 
\cite{chandra1,chandra2,chandra3} concluded that the test star must experience,
in addition to
diffusion, a dynamical friction. Starting from a
phenomenological Langevin \cite{langevin} equation of the form $d{\bf
v}/{dt}=-\nabla\Phi-\xi{\bf v}+{\bf A}(t)$ incorporating a friction force
$-\xi {\bf v}$ and a
stochastic force ${\bf A}(t)$ (Gaussian white noise)
in addition to the mean field force $-\nabla\Phi$ he 
derived the corresponding Fokker-Planck \cite{fokker0,fokker,planck,fokkerF}
equation
\begin{eqnarray}
\label{intro4}
\frac{\partial f}{\partial t}+{\bf v}\cdot
\frac{\partial f}{\partial
{\bf r}}-\nabla\Phi\cdot \frac{\partial f}{\partial
{\bf v}}=\frac{\partial}{\partial {\bf v}}\cdot\left (
D\frac{\partial f}{\partial {\bf v}}+\xi f{\bf v}\right
),
\end{eqnarray}
where $D$ is the diffusion coefficient of the stars. This is the so-called
Kramers-Chandrasekhar equation.\footnote{This equation was
previously introduced by Klein \cite{klein} and Kramers \cite{kramers} in the
theory of Brownian motion. Actually, this
equation, without the advection
term, was first introduced by Lord Rayleigh \cite{lr} as early as 1891 before
all the classical works on Brownian motion. He considered
the kinetic theory of massive particles bombarded by numerous small
projectiles. The heuristic argument used by
Chandrasekhar \cite{chandra1} to justify the dynamical friction was borrowed to
Lord Rayleigh
\cite{chandrabrown}.} It involves a
diffusion term and
a friction term. The condition that this equation relaxes towards the
Maxwell-Boltzmann distribution $f_{\rm eq}=Ae^{-\beta m v^2/2}$ of
statistical equilibrium with $\beta=1/k_B T$ yields the Einstein relation
$\xi=D\beta
m$ between the diffusion and the friction
coefficients \cite{chandra1,chandra2,chandra3}. In order to determine
the expression of the diffusion coefficient, Chandrasekhar and von Neumann
\cite{cvn0,cvn1,cvn2,cvn3,cvn4,cvn5} developed a completely
stochastic formalism of gravitational fluctuations and showed that the
fluctuations of the gravitational force are described by the
Holtsmark \cite{holtsmark}
distribution (a particular L\'evy law) in which the nearest neighbor
plays a prevalent role. Their calculation of the diffusion coefficient was
completed by Kandrup \cite{kandruprep}.

In parallel, Chandrasekhar \cite{chandra,chandra1} developed a more
accurate kinetic theory of stellar systems from a
binary collision picture. He
explicitly computed the first and
second
moments of the velocity increment of a test star (diffusion and friction
coefficients)
assuming that the test star experiences a succession of two-body
encounters with field stars. His calculations were completed by
Spitzer and collaborators \cite{cohen,ss,sh,sh2} and Rosenbluth {\it et al.}
\cite{rosenbluth}.
These authors proceeded as if the system were spatially homogeneous (by making
a local approximation) and neglected collective effects.
They obtained
a Fokker-Planck equation of the form  
\begin{eqnarray}
\label{intro5}
\frac{\partial f}{\partial t}=\frac{\partial^2}{\partial
v_i\partial v_j}( D_{ij} f)-\frac{\partial}{\partial v_i}(f
F_i^{\rm tot}),
\end{eqnarray}
with a diffusion tensor and a friction force given
by
\begin{eqnarray}
\label{intro6}
D_{ij}=A\int  f' \frac{u^2\delta_{ij}-u_iu_j}{u^3}\, d{\bf v}'
,\qquad {\bf F}_{\rm tot}=-4A \int f'
\frac{{\bf u}}{u^3} \, d{\bf v}'.
\end{eqnarray}
In the dominant approximation, the prefactor is
given by $A=2\pi G^2
m\ln\Lambda$ where  $\ln\Lambda$ is the gravitational Coulomb
logarithm.\footnote{The expressions from Eq. (\ref{intro6}) were obtained by
Rosenbluth {\it et al.}
\cite{rosenbluth} for an arbitrary DF. They generalize the results of
Chandrasekhar \cite{chandra,chandra1} who computed the
diffusion coefficient and the friction force under the assumption that $f(v)$
is isotropic in velocity space. In his book, Chandrasekhar
\cite{chandra} only computes
the diffusion coefficient  in
order to obtain
an estimate of the relaxation time of stellar systems (following previous works
by \cite{jeansk1,jeansk2,jeansbook,eddingtonkin,charlier,schwarzschild1924,
rosseland,smart,amba,spitzer1940}). He
realized later \cite{chandra1,chandra2,chandra3} while developing the
analogy with Brownian motion \cite{chandrabrown}, and possibly influenced
by the work
of Lord Rayleigh \cite{lr}, that a fundamental friction term  was missing in
the theory exposed in his monograph. For an isotropic DF the diffusion
coefficient is given by Eqs. (2.352) and (5.723) of \cite{chandra} and the
friction force by Eq. (32) of
\cite{chandra1}. When these expressions are substituted into the Fokker-Planck 
equation (a substitution that Chandrasekhar did not carry out explicitly) one
obtains after a few calculations a
self-consistent integrodifferential equation of the form
\begin{eqnarray}
\label{intro7}
\frac{\partial f}{\partial t}=8\pi A \frac{1}{v^{2}}\frac{\partial}{\partial
v}\biggl\lbrack \frac{1}{3}\frac{\partial f}{\partial v}\biggl (
\frac{1}{v}\int_{0}^{v}{v_{1}^{4}} f(v_{1},t)dv_{1}
+v^{2}\int_{v}^{+\infty}v_{1}f(v_{1},t)dv_{1}\biggr
)+f\int_{0}^{v}f(v_{1},t)v_{1}^{2}dv_{1}\biggr\rbrack,
\end{eqnarray}
which governs the evolution
of the system as a whole. This self-consistent equation is implicit in the
paper of
Rosenbluth {\it et al.} \cite{rosenbluth} (it can be deduced from Eqs.
(\ref{intro5}) and (\ref{intro6}) for an isotropic DF) and was first written
explicitly by King \cite{kingkin} and H\'enon \cite{henonbinary}. Equivalent
equations
were written by McDonald
{\it
et al.} \cite{chuck} and Ipser \cite{ipser} in a slightly
different form. 
If we make a thermal bath approximation, replacing $f(v_1,t)$ by a Maxwellian
as first done by Chandrasekhar
\cite{chandra,chandra1,chandra2,chandra3} (see also Spitzer and Schwarzschild
\cite{ss} and Spitzer and H\"arm
\cite{sh2}), Eq.
(\ref{intro7}) returns the
Kramers-Chandrasekhar equation (\ref{intro4}) and allows an explicit calculation
of the velocity-dependent diffusion coefficient $D(v)$ (see,
e.g., \cite{aa}).}
The
calculations of Chandrasekhar \cite{chandra,chandra1} involve an
integral over the
impact parameter.  There is no divergence
at small impact
parameters because strong deflections are taken into account
in his kinetic theory.
Therefore, the classical distance of closest approach
$b_{90}\sim Gm/v^2\sim Gm^2/(k_B T)\sim 1/(n\lambda_J^2)$ at which the particles
are deflected by about  $90^{\rm o}$  appears naturally in $\ln\Lambda$.
However, there
is
a logarithmic divergence at large impact parameters because his treatment
does not take into account the spatial inhomogeneity of the system and its
finite extent.\footnote{This logarithmic divergence can be related to the
fact that, in an infinite medium, the autocorrelation of the gravitational force
$\langle {\bf F}(0)\cdot {\bf F}(t)\rangle$ decreases as $t^{-1}$
\cite{cvn4,cohen} (see also
\cite{ps66,lee,severne72,lcohen,cohenamad,hb2,aa}) and that
$D=\frac{1}{3}\int_0^{+\infty} \langle {\bf F}(0)\cdot {\bf F}(t)\rangle\, dt$.
This implies that the variance of the velocity increases as $\langle
v^2\rangle\sim t\ln t$ \cite{henon58,ps66,od,lee,lcohen,cohenamad,hb2,aa}.}
Initially,
Chandrasekhar
\cite{chandra,chandra1}  argued that the Coulombian factor
$\ln\Lambda$
should be cut-off at the interparticle distance. This claim was
corroborated by his stochastic approach \cite{cvn0,cvn1,cvn2,cvn3,cvn4,cvn5}. 
However, this choice was later criticized by Spitzer \cite{cohen,sh,sh2}
who
argued
that the Coulombian factor has to be cut-off at the the Jeans length
$\lambda_J\sim [k_BT/(Gm^2 n)]^{1/2}$ \cite{jeans1902}
which is the gravitational analogue of the Debye length \cite{dh} in
plasma physics.\footnote{The logarithmic divergence of the
diffusion coefficient for systems interacting via a $1/r$
potential was originally discussed
by Gruner \cite{gruner}, Eddington \cite{eddingtonkin}, Jeans \cite{jeansk2},
Chapman \cite{chapmanplasma}, Schwarzschild \cite{schwarzschild1924},
Persico \cite{persico}, Jeans \cite{jeansbook}, Landau
\cite{landau}, Ambartsumian
\cite{amba}, Spitzer \cite{spitzer1940}, Chandrasekhar \cite{chandra},
Cowling \cite{cowling}, Bohm and Aller \cite{aller}, Cahn
\cite{cahn}, Landshoff \cite{landshoff},  Cohen {\it et al.} 
\cite{cohen}, Pines and Bohm \cite{pines}, Spitzer and H\"arm \cite{sh},
Gasiorowicz {\it et al.}
\cite{gasiorowicz}, Rosenbluth
{\it et al.} \cite{rosenbluth}, and  Spitzer and H\"arm \cite{sh2}.
Some authors cured the divergence at large scales by 
introducing  an upper cut-off at the interparticle distance $d$
\cite{chapmanplasma,schwarzschild1924,jeansbook,spitzer1940,chandra,
cowling,landshoff}. Other authors introduced an upper cut-off at the Debye
length $\lambda_{D}$ for Coulombian
plasmas
\cite{persico,landau,aller,cahn,cohen,pines,sh,gasiorowicz,rosenbluth} or at
the 
Jeans $\lambda_{J}$ length (i.e. at the system's size) for stellar systems
\cite{eddingtonkin,jeansk2,amba,sh2}.
Landau \cite{landau} argued that
in the absence of screening the
upper cut-off should be the system's size.} The Jeans length is of
the order of the system size $R$ (using a virial type argument $v^2\sim GM/R$
one gets $R^2\sim {v^2}/{Gnm}\sim
{k_BT}/{Gnm^2}\sim \lambda_J^2$).
This leads
to a gravitational Coulomb logarithm of the form
$\ln\Lambda\sim \ln
({\lambda_J}/b_{90})\sim \ln (n\lambda_J^3)\sim \ln N$ where
$N\sim n\lambda_J^3\sim nR^3$ is the number of stars in the
cluster. In a globular cluster and in a galaxy, we have $N\gg 1$
\cite{btnew}. The relaxation time scales like $t_R\sim (N/\ln N)\, t_D$, where
$t_D\sim R/v$ is the dynamical time.\footnote{This $N/\ln N$
scaling is explicitly given in Eq. (5.227) of \cite{chandra}.}
Chandrasekhar \cite{chandra,chandra1}, Spitzer and collaborators
\cite{cohen,sh,sh2,ss} and Rosenbluth {\it et al.} \cite{rosenbluth} apparently
did not know the Landau \cite{landau} kinetic equation of plasma physics.
However, the
Fokker-Planck equation (\ref{intro5}) with (\ref{intro6}) is equivalent to
the Landau equation (\ref{intro3}) albeit written in a different form (see Ref.
\cite{aa} for a
detailed comparison of the two approaches).\footnote{The Landau
equation was not known (or misunderstood) by the astrophysical
community
for a long
time. Landau's paper
was wrongly criticized by Allis \cite{allis,landauREP,enoch}. There is also a
strange
comment in footnote 5 of Cohen {\it et al.} \cite{cohen}.  Rosenbluth {\it et
al.} \cite{rosenbluth} mention ``the often used method in which one expands the
collision integrand of the Boltzmann equation in powers of the momentum change
per collision'' but they do not give any reference. The Landau equation
was first explicitly mentioned in astrophysics by Kandrup \cite{kandrup1} in
1981.} Obviously, the Landau
form [Eq. (\ref{intro3})] is much more symmetric and elegant than the 
 Fokker-Planck form
[Eqs. (\ref{intro5}) and (\ref{intro6})].

By using the general theory of weakly coupled systems developed by
Brout and Prigogine \cite{brout}, and being unaware of the former work of
Landau \cite{landau}, Prigogine and Balescu \cite{pb} derived a kinetic
equation for homogeneous Coulombian plasmas of the form 
\begin{eqnarray}
\frac{\partial f}{\partial t}=\pi (2\pi)^d m
\frac{\partial}{\partial {\bf v}}\cdot  \int\, d{\bf k}d{\bf v}'\,
{\bf k} \, \hat{u}({\bf k})^2 \delta \lbrack
{\bf k}\cdot ({\bf v}-{\bf v}')\rbrack {\bf k}\cdot
\left (\frac{\partial}{\partial {\bf v}}-\frac{\partial}{\partial
{\bf v'}}\right ) f({\bf v},t)f({\bf v}',t).
\label{intro8}
\end{eqnarray}
Their equation involves the Fourier transform of the potential of interaction
${\hat u}({\bf k})$ and exhibits a condition of resonance 
encapsulated in the delta function $\delta[{\bf k}\cdot ({\bf v}-{\bf v}')]$.
The kinetic equation (\ref{intro8}), which ignores collective effects, is
equivalent to the Landau
equation (\ref{intro3}). Indeed, if we perform the integral over the wavenumber 
${\bf k}$
(see, e.g., Appendix C of \cite{aa}) we recover the original form of the Landau
equation.\footnote{Prigogine and Balescu \cite{pb} (see also \cite{balescugrav})
cured the logarithmic
divergence at large scales by introducing an upper cut-off at the 
the Debye length $\lambda_D$ for Coulombian plasmas and at the interparticle
distance $d$ for stellar systems. Curiously, they argued that the gravitational
analogue
of the Debye length $\lambda_D$ is the interparticle distance $d$.}

Several authors attempted to take into account collective effects in order to
solve the large scale divergence appearing in the Landau equation of Coulombian
plasmas.
Starting from the Liouville equation governing the evolution of the $N$-body
DF and neglecting three-body correlations,
Bogoliubov \cite{bogoliubov} obtained two coupled
equations for the DF and the two-body
correlation function (they are now called
the first two equations of the BBGKY hierarchy). By assuming a
timescale separation according to which the two-body correlation
function relaxes much faster than the one-body DF (Bogoliubov
ansatz), he reduced the second equation
to an integral equation and showed that collective
effects regularize the logarithmic divergence at large scales. Lenard
\cite{lenard} managed to solve the Bogoliubov integral equation explicitly and
obtained a single closed kinetic equation for
the DF. The same kinetic equation was derived independently
by
Balescu
\cite{balescu} from the general theory of irreversible processes based on
diagram techniques developed by Prigogine and Balescu \cite{pb1,pb2}. The
Lenard-Balescu equation reads\footnote{This kinetic equation was
also derived by Guernsey in his thesis \cite{guernseyphd} so it is sometimes
called the Lenard-Balescu-Guernsey equation. The kinetic theory
of Guernsey, and other kinetic theories, are presented in the book of Wu
\cite{wubook}.}
\begin{eqnarray}
\frac{\partial f}{\partial t}=\pi (2\pi)^d m
\frac{\partial}{\partial {\bf v}}\cdot  \int\, d{\bf k}d{\bf v}'\,
{\bf k}  \frac{\hat{u}({\bf k})^2}{|\epsilon({\bf k},{\bf k}\cdot {\bf
v})|^{2}} \delta \lbrack {\bf k}\cdot ({\bf v}-{\bf v}')\rbrack {\bf k}\cdot
\left (\frac{\partial}{\partial {\bf v}}-\frac{\partial}{\partial
{\bf v'}}\right ) f({\bf v},t)f({\bf v}',t).
\label{intro9}
\end{eqnarray}
It takes into account collective effects (dynamical Debye shielding)
through the dielectric function
\begin{eqnarray}
\epsilon({\bf k},\omega)=1+(2\pi)^{d}\hat{u}({\bf k})\int \frac{{\bf k}\cdot
\frac{\partial f}{\partial {\bf v}}}{\omega-{\bf k}\cdot {\bf v}}\, d{\bf v}.
\label{intro10}
\end{eqnarray}
This amounts to replacing the bare potential of interaction $\hat{u}({\bf k})$
in the form  of the Landau equation  given by Prigogine and
Balescu \cite{pb} [see Eq. (\ref{intro8})] by a
dressed potential of interaction $\hat{u}^d({\bf
k})=\hat{u}({\bf k})/|\epsilon({\bf k},{\bf k}\cdot {\bf
v})|$ taking into account the effect of the dynamical polarization cloud.
Collective effects remove the divergence at large scales present in the
Landau equation which only accounts for two-body collisions. As a result, the
Debye length appears naturally in the
Lenard-Balescu equation. Prior to Lenard \cite{lenard} and
Balescu
\cite{balescu}, several
authors \cite{temko,kadomtsev,liboff,tchen,ichikawa,willis} 
attempted to derive a
kinetic
equation taking into account collective effects by using more or less heuristic
arguments. These early kinetic theories are reviewed in \cite{epjp2}. The
Lenard-Balescu equation can also be derived in
a simpler manner from
the Klimontovich formalism by using a quasilinear approximation
\cite{klimontovich}.

A
kinetic theory of spatially homogeneous plasmas was also developed by
Hubbard \cite{th,hubbard1,hubbard2} in parallel to Lenard \cite{lenard} and
Balescu \cite{balescu}. The
kinetic theory of Hubbard is less famous  than the approaches of Lenard and
Balescu, but it appears to be much simpler (on a technical point of view) and
more physical. Hubbard started from the Fokker-Planck
equation and 
derived the diffusion coefficient and the friction force by a direct
calculation, taking
collective effects into
account. In a preliminary work, Thompson and Hubbard \cite{th} calculated
the
diffusion tensor by assuming that the field particles are at statistical
equilibrium and by using the fluctuation-dissipation theorem to obtain
the power spectrum of the potential fluctuations from the dielectric
function. Then, by using the notion of dressed particles,\footnote{The notion
of particles dressed by their polarization cloud made of 
particles with opposite charge was introduced and developed in
plasma physics by Debye and H\"uckel \cite{dh} in the static case and by
Pines and Bohm \cite{pines}, Gasiorowicz {\it et al.}
\cite{gasiorowicz}, Rostoker and Rosenbluth \cite{rr60}, Balescu \cite{balescu},
Thompson and Hubbard \cite{th}, Hubbard \cite{hubbard1,hubbard2}, Rosenbluth
and Rostoker \cite{rr62}, and Rostoker \cite{rostoker64,rostoker64b} in the
dynamical case. They have been called ``effective free particles'' (electron
plus its associated cloud) \cite{pines}, ``quasiparticles'' (electrons
or ions ``dressed'' by their polarization clouds)  \cite{balescu} or ``dressed
test particles'' \cite{rr62,rostoker64}. The notion of wake
corresponding to an asymmetrical polarization cloud trailing behind
a moving particle was introduced by Pines and Bohm \cite{pines}.}
Hubbard
\cite{hubbard1,hubbard2} obtained the power spectrum,
the diffusion
coefficient and the friction force by a direct calculation which is valid
for field particles that are not necessarily at statistical equilibrium
(see
also \cite{gasiorowicz,temko,tchen,helfand}). In
the thermal bath approximation, this returns the
results of Thompson and Hubbard \cite{th} and
Rostoker and Rosenbluth \cite{rr60}. When
collective effects are neglected, this returns the results of
Chandrasekhar \cite{chandra,chandra1}, Spitzer and collaborators
\cite{cohen,sh,sh2,ss} and Rosenbluth {\it et al.}
\cite{rosenbluth}. 
When the diffusion tensor and the friction force are substituted into the
Fokker-Planck equation and a simple integration
by parts is carried out (a substitution that
Hubbard did not make explicitly), one obtains an equation identical to the
Lenard-Balescu equation (see, e.g., \cite{epjp}). Therefore, the Fokker-Planck
equation derived by
Hubbard is equivalent to the  Lenard-Balescu
equation. The first paper of Hubbard \cite{hubbard1} corresponds to the
{\it wave theory} that is appropriate to describe collective effects. In his
second paper, Hubbard \cite{hubbard2} connected the wave theory to the
{\it binary encounter theory} that is appropriate to describe strong collisions
like in the standard approach of Boltzmann described by Chapman and Cowling
\cite{chapman}.
As a result, he obtained Fokker-Planck coefficients which include
both collective effects and the contribution of close binary encounters.
In his  theory, no {\it ad hoc} cut-off procedures of any kind are needed
so the exact expression of the Coulombian logarithm is obtained.
In the dominant approximation, he recovered
the results of Landau but now without any divergence. The classical
distance of closest approach and the Debye length
appear naturally in his treatment. Other attempts
\cite{baldwin,fb63,weinstock,guernsey,ka,gould} to obtain a kinetic equation
that takes strong collisions and collective effects into account and exhibits no
divergence are reviewed in \cite{epjp2}.

The Landau and the Lenard-Balescu equations
can be
applied to a large class of systems with long-range interactions in
different dimensions of space provided that they are spatially homogeneous
\cite{epjp,epjp2,epjp3}. These kinetic equations are valid at the order $1/N$
(since they take into account two-body correlations and neglect three-body and
higher correlations) so they describe the evolution of the system of a timescale
of order $N\, t_D$, where $t_D$ is the dynamical time.\footnote{For plasmas,
the number of particles $N$ is replaced by
$\Lambda$, the number of
particles in the Debye sphere.} In $d\ge 2$ dimensions, the Landau and the
Lenard-Balescu equations relax towards the
Boltzmann distribution of statistical equilibrium. However, when the kinetic
theory is specifically applied to one-dimensional systems, such as
one-dimensional plasmas \cite{ef,dawson,kp} and the Hamiltonian Mean Field (HMF)
model
\cite{cvb,bd}, it is found that the Landau and the Lenard-Balescu collision
operators vanish identically for spatially homogeneous systems:
$\partial f/\partial t=0$.\footnote{By
contrast, the Fokker-Planck operator governing the relaxation of a test particle
in a bath of field particles with a prescribed DF $f(v)$ does not vanish
\cite{epjp}:
\begin{eqnarray}
\frac{\partial P}{\partial t}=\frac{\partial}{\partial v}\left\lbrack
m f(v)\int_0^{+\infty}dk\, \frac{4\pi^2{\hat u}(k)^2 k}{|\epsilon(k,kv)|^2}\left
(\frac{\partial P}{\partial v}-P\frac{d\ln f}{dv}\right )\right\rbrack.
\end{eqnarray}
The distribution $P(v,t)$ of the
test particle relaxes towards the
distribution $f(v)$ of the field particles (whatever its shape) on a timescale
of order $N\, t_D$. The fact that a ``distinguished''
population of particles relaxes towards the overall
DF in a time of order $N\, t_D$ while the
overall DF does not change on this timescale has been
illustrated numerically in \cite{rf}.} This is a situation of kinetic
blocking.\footnote{A similar situation of kinetic
blocking occurs for an
axisymmetric distribution of 2D point vortices with a monotonic profile of
angular velocity \cite{cl}. The term ``kinetic blocking'' was introduced
in that paper.} It implies that the
relaxation towards statistical equilibrium  takes more time that naively
expected (i.e. larger than $N\, t_D$), being sourced by three-body or higher
order correlations \cite{epjp}. Recently,  a new kinetic equation applying to
spatially
homogeneous one-dimensional systems with long-range interactions (with
collective effects neglected) has been
obtained by Fouvry {\it et al.} \cite{fbcn2,fcpn2}.
This equation, which takes
three-body correlations into account, is valid at the order $1/N^2$. It
satisfies an $H$-theorem and relaxes towards the Boltzmann distribution of
statistical equilibrium on a timescale of order $N^2\, t_D$.

The Landau equation (\ref{intro3}) [or Eqs. (\ref{intro5}) and (\ref{intro6})]
can be applied to stellar systems, which are
intrinsically spatially inhomogeneous, if we make a {\it local approximation}.
In that case, we must introduce an advection (Vlasov) operator on the
l.h.s. of this equation like in Eq. (\ref{intro1}). The resulting
kinetic equation must be
self-consistently coupled to the Poisson equation (\ref{intro2}). This leads to
the Vlasov-Landau-Poisson equation. Since the collisions have a weak effect on
the evolution of the system, one can  simplify this equation further by
averaging over the orbits. When the DF $f=f(\epsilon,t)$ depends only on the
individual energy of the stars we get the orbit-averaged-Fokker-Planck
equation \cite{kuzmin,henonbinary} 
\begin{eqnarray}
\label{oak14}
\frac{\partial q}{\partial \epsilon}\frac{\partial f}{\partial t}-\frac{\partial
q}{\partial t}\frac{\partial f}{\partial \epsilon}=8\pi A
\frac{\partial}{\partial\epsilon}\left\lbrace
\frac{\partial f}{\partial \epsilon}\left (
\int_{-\infty}^{\epsilon} f_{1}q_1\, d\epsilon_{1}
+q\int_{\epsilon}^{+\infty}f_1\,
d\epsilon_{1}\right
)+f\int_{-\infty}^{\epsilon}f_1 \frac{\partial q_1}{\partial\epsilon_1}\,
d\epsilon_{1}\right\rbrace,
\end{eqnarray}  
\begin{eqnarray}
\label{oak15} q(\epsilon,t)={16\pi^{2}\over
3}\int_{0}^{r_{max}(\epsilon,t)}\lbrack
2(\epsilon-\Phi(r,t))\rbrack^{3/2}r^{2}dr,
\end{eqnarray}
\begin{eqnarray}
\label{oak16} {1\over r^{2}}{\partial\over\partial r}\left
(r^{2}{\partial\Phi\over\partial r}\right
)=16\pi^{2}G\int_{\Phi(r,t)}^{+\infty}f(\epsilon,t)\lbrack
2(\epsilon-\Phi(r,t))\rbrack^{1/2}\, d\epsilon.
\end{eqnarray}
This equation gives relatively good results in the case of globular
clusters. However, there remains the problem of the logarithmic
divergence of the diffusion and friction coefficients which imposes one to
``tune'' the value of $\ln\Lambda$ in a rather {\it ad hoc} manner.
On the other hand, if one tries to account for collective effects in
self-gravitating systems by
making a local approximation and by using the
homogeneous Lenard-Balescu equation, one gets a strong (algebraic) divergence at
large scales associated with the Jeans instability of a uniform
self-gravitating medium \cite{jeans1902}.\footnote{Self-gravitating
systems are never spatially homogeneous even in theory.
Indeed, an infinite
homogeneous distribution of stars is not a steady state of the Vlasov-Poisson
equations (except in an expanding background). Even if we advocate the ``Jeans
swindle'' \cite{btnew}, or take into account the expansion of the Universe
\cite{peebles}, we
find that an infinite homogeneous distribution of stars is linearly unstable to
perturbations whose wavelength exceeds the Jeans length
$\lambda_J$.} If we
truncate the integration over the 
wavelength at a maximum wavelength $\lambda_{\rm max}$ smaller than the Jeans
length $\lambda_J$, we find that the diffusion and friction coefficients
diverge
algebraically when $\lambda_{\rm max}\rightarrow \lambda_J$ (see
Ref. \cite{weinberglb} and Appendix E of Ref. \cite{aa}). This is because the
fluctuations become very large as the system's size approaches the Jeans
length of marginal stability. This suggests that, if we were
able to account for spatial inhomogeneity rigorously,
collective effects (Jeans anti-shielding) would tend to decrease the relaxation
time of stellar systems.\footnote{As discussed below, this naive argument is
not always correct.}

It is therefore necessary to develop an appropriate kinetic theory of
spatially inhomogeneous stellar systems. This can be done most
conveniently by using angle-action variables because, in these
variables, the trajectories (orbits)  of the particles are simple
\cite{goldstein,btnew}. For a collisionless equilibrium, the actions ${\bf
J}$ are
constant and the angles ${\bf w}$ advance at a constant rate ${\bf\Omega}({\bf
J})$. The DF then depends on the actions alone, $f=f({\bf J})$. The
angle-action
variables were first used by Kalnajs \cite{kalnajs71,kalnajs,kalnajs77} in
astrophysics to investigate the dynamical
stability of collisionless stellar systems like flat galaxies (stellar discs)
and calculate their small-amplitude oscillations. He introduced  a biorthogonal
basis of density-potential pairs (see also Clutton-Brock \cite{clutton}) and
a ``dielectric'' matrix which
is the proper generalization of the dielectric function in plasma physics.
With this formalism he could derive the appropriate dispersion relation of
inhomogeneous stellar systems (see also the Appendix of Polyachenko and
Shukhman
\cite{ps80}).\footnote{In plasma physics, a formalism of angle-action
variables was independently developed by Kaufman
\cite{kaufman1,kaufman2,kaufman3} to study
the interaction of waves and particles.}

Using the angle-action formalism, Lynden-Bell and Kalnajs \cite{lbk}
 calculated the angular momentum exchange
between the stars and the wave in a disk galaxy (see also Kato \cite{kato})
and showed that all the
contribution to the integral comes from the resonances. These resonances are
encapsulated in a delta function that reflects the fact that
only orbits with nearly commensurate frequencies give rise to
secular changes.
Their calculation was extended by Tremaine and Weinberg
\cite{tw84} and Weinberg \cite{w85}  who investigated dynamical friction on a
test object
(such as a bar or a satellite) which rotates or revolves through an
inhomogeneous spherical stellar system. 
Dynamical friction on a test mass moving through
an infinite homogeneous medium
may be interpreted as the gravitational force due to the enhanced density wake
formed behind the moving test object. It can be obtained by computing the
density response of the background stellar system to a moving point source by
solving the linearized Vlasov-Poisson equations. The density perturbation
(polarization cloud) leads to a back reaction on the perturber giving rise to a
drag force
\cite{pines,as,gasiorowicz,rr60,th,hubbard1,marochnik,kalnajsF,mulder,
kandrupfriction,bz,cp,hb4,aa,magorrian}. 
Using angle and action variables, Tremaine and Weinberg
\cite{tw84} and Weinberg \cite{w85}  developed a
formalism which generalizes dynamical friction to spherical systems. The halo
exerts a drag on the bar or on the satellite just as a homogeneous medium exerts
a drag on a test mass in the standard picture of dynamical
friction. They  derived an analog to Chandrasekhar's
dynamical friction formula
which applies to spherical systems. They showed that the dynamical friction
torque
on a test object or on a stellar bar in galaxies  is given by a simple
generalization of the formula derived by
Lynden-Bell and Kalnajs \cite{lbk} to describe angular momentum
transport in disc galaxies. The dominant contribution to the
torque  occurs at
commensurabilities or resonances between the orbital frequencies of the field
particles and the angular velocity of the test particle (satellite,
bar,...).
Palmer and Papaloizou \cite{palmer} considered this problem using a
different formalism and derived an equivalent expression for
the torque. Actually, they derived an expression for the energy
exchange rate $dE/dt$ (see also Polyachenko and Shukhman \cite{ps80}) which is
similar to the expression for
the angular momentum exchange $dL/dt$ obtained by
Lynden-Bell and Kalnajs \cite{lbk} for disc galaxies.
Weinberg \cite{w86} explicitly computed the response density of an
inhomogeneous singular isothermal sphere to a satellite (i.e., the
wake of the satellite) and compared the result with the wake in the
infinite homogeneous case. He used this theory to compute the drag and
the rate of orbital decay of a sinking satellite moving in a spherical galaxy.
In this theory applicable to inhomogeneous systems, the orbits are
determined exactly in a given spherical potential $\Phi(r)$. This differs from
the Chandrasekhar theory of dynamical friction \cite{chandra1} where the
background stellar
system  is isotropic, infinite and homogeneous and the trajectories are
straight lines. The drag force in Chandrasekhar's theory diverges
logarithmically with impact parameter and
must be truncated {\it ad hoc} at a maximum impact parameter $b_{\rm max}$ at
the assumed boundary of the system. In the spherical theory, there is no
divergence at large
scales since the spatial inhomogeneity and the finite extent of the system have
been taken into account (the large impact parameter cut-off is intrinsic to the
equilibrium model) and this allows a determination of the logarithmic factor
$\ln\Lambda$ appearing in Chandrasekhar's theory.\footnote{There remains,
however, a logarithmic divergence $\sim \ln k$ for large harmonic
order $k$ corresponding to a  logarithmic divergence at small impact parameters
when one makes the linear trajectory approximation.} 
The previous studies neglect the self-gravity of the stellar background, i.e.,
they consider only the interaction of the stars with the test object, not with
one another. In other words they neglect the self-gravity of
the wake induced by the gravitational focusing of the test particle. However,
self-gravity is important for the features at the
largest scales. Weinberg \cite{w89} presented a method for computing the
response of a spherical stellar system (galaxy) to the perturbation of a
satellite by taking into account collective effects. As an application, he
derived an analytic expression for the orbital decay of the satellite due to the
friction force (torque) raised by the wake. He showed that
collective effects (self-gravity) increase the orbital decay time of the sinking
satellite.

Binney and Lacey \cite{bl} considered a collisionless stellar system submitted
to an externally imposed stochastic field. In the absence of external
perturbation the system is assumed to have reached a virialized state through
the process of violent collisionless relaxation and the phase-space density of
stars is a function $f({\bf J})$  of the actions
only (Jeans theorem \cite{jeans}). In the presence of an external perturbation
the DF 
changes due to irregularities in the system's potential. Making an adiabatic
approximation, they assumed that the evolution is slow so the phase-space
density
of stars $f$ remains at any given time a function of the actions only, though a
function $f=f({\bf J},t)$ that changes slowly in time. They studied the
diffusion of stars through three-dimensional orbit space due to the fluctuating
gravitational field and obtained a nonlinear diffusion equation with a diffusion
coefficient determined by the external forcing. However, their
approach neglects collective effects (self-gravity). Weinberg \cite{weinberg}
independently developed a theoretical framework
for studying the evolution of a near-equilibrium galaxy caused by scattering of
orbits by fluctuations in the gravitational potential. All the information about
the stochastic
perturbation (noise) is encapsulated in
the
Fourier transform of the density correlation function. As stochastic events he
considered both uncorrelated point mass perturbers (such as orbiting
super-massive
black holes) and transient perturbers (such as dwarf galaxy mergers, 
orbiting substructure decaying as a result of dynamical friction,  transient
spiral
structure, mixing tidal
debris...). He took
into account collective effects (self-gravity) because his previous work
\cite{weinberglb}   had
shown that fluctuations in stellar systems on the largest scales can
be strongly amplified by their own self-gravity. He obtained a
nonlinear Fokker-Planck equation in which the diffusion
coefficients depend on the properties of the external forcing and on the DF of
the system.\footnote{The
secular dressed diffusion
(SDD) equation
\cite{weinberg,mab,pa,nardini,nardini2,epjp,epjp2,fpp,fp15,fbp,fb} is
reviewed in \cite{sdduniverse}. When collective effects are
neglected it reduces to the secular bare diffusion (SBD) equation
\cite{bl,ba,bf}.}
In a companion paper \cite{weinberg2} he used this equation to investigate the
evolution of haloes in noisy environments. Such environments may be found in the
epoch just after galaxy formation or in clusters and groups at present
times. He argued that the repetitive stochastic response of a halo
drives the equilibrium toward an asymptotically universal
profile, independent of its initial conditions.

The previous works focus either on the process of dynamical friction or on
the process of diffusion. However, for an isolated self-gravitating system
evolving under the effect of Poisson shot noise (granularities), both processes
are in action simultaneously as a result of the fluctuation-dissipation
theorem. Therefore, a self-consistent kinetic equation should include both
diffusion and friction terms. It should also be written in angle-action
variables in order to account for spatial inhomogeneity. Finally, it should take
collective effects into account. This master equation, called the inhomogeneous
Lenard-Balescu equation, was derived by Heyvaerts \cite{heyvaerts} from the
BBGKY hierarchy and by Chavanis \cite{physicaA} from the Klimontovich approach.
It can also be derived from the Fokker-Planck equation
\cite{physicaA,hfcp,fb,hamilton} by calculating
the first two moments (diffusion and friction) of the increment in action  of
the test particle. When
collective effects are neglected, it reduces to
the inhomogeneous Landau equation
\cite{aa}.\footnote{Previous works on kinetic theories for inhomogeneous
systems include Prigogine and Severne
\cite{ps68}, Gilbert
\cite{gilbert,gilbert71}, Lerche
\cite{lerche}, Severne and Haggerty \cite{severne,hs76},
Parisot and Severne
\cite{ps}, Kandrup \cite{kandrup1}, Polyachenko and Shukhman \cite{ps82}, 
Bernstein and Molvig \cite{bm83}, Cohen {\it et al.} \cite{chmb},  Heggie and
Retterer \cite{hr}, Luciani and Pellat \cite{lp}, Mynick \cite{mynick}, Valageas
\cite{valageas} and Chavanis
\cite{hb3,hb4,kin2007,unified,aa}.}

The inhomogeneous Landau and Lenard-Balescu equations have  been
studied recently in relation to stellar discs \cite{fpc,fpmc,fpcm}, globular
clusters \cite{hfbp,fhrp} and galactic nuclei containing a
supermassive black hole
\cite{st1,st2,st3,fpm,fpcBH,bf,tep}.\footnote{The
inhomogeneous Landau and Lenard-Balescu equations are structurally very
different from the
orbit-averaged-Fokker-Planck equation \cite{kuzmin,henonbinary} used to
describe globular clusters.}  The inhomogeneous Landau and Lenard-Balescu
equations have also
been
applied to other systems with long-range interactions such as the inhomogeneous
phase of the HMF model \cite{bm} and one-dimensional
self-gravitating systems \cite{roule}. In
Ref. \cite{fbc} they have been used to study the
dynamics of spins with long-range interactions
moving on a sphere (recovering and generalizing previous works on the
subject \cite{gm2,gm,bg}) in relation to the process of vector resonant
relaxation
(VRR) in galactic nuclei \cite{rt,kt,rkt}. In some cases, collective
effects can
considerably accelerate the relaxation. For example,
the
work \cite{fpmc} demonstrated that collective effects
decrease the relaxation time of stellar discs by three orders of
magnitude. A similar acceleration is obtained for the periodic
stellar cube   as its mass
approaches the Jeans mass \cite{weinberglb,aa,magorrian} and for the
homogeneous HMF
model close to the critical temperature $T_c$ \cite{epjp3}. In other cases,
such as a spherical galaxy modeled by an
$n=3$
polytrope \cite{w89}, the inhomogeneous phase of the  HMF model
\cite{bm}, globular clusters represented by a spherical isotropic
isochrone DF \cite{fhrp} and 1D
self-gravitating systems
\cite{roule},
collective effects unexpectedly reduce the diffusion and the friction and
increase the
relaxation time. Therefore, in certain situations, collective
effets delay the relaxation
contrary to the intuition formed in \cite{weinberglb,aa,magorrian,epjp3} (see
footnote 17). The inhomogeneous Landau and
Lenard-Balescu equations also share many analogies with the
kinetic equations obtained in the context of 2D point vortices in
hydrodynamics (see \cite{Kvortex2023} and
references therein). The numerous analogies between stellar systems
and 2D vortices are discussed in Refs. \cite{houches,tcfd,unified,csr}.

In the present paper, we complement the kinetic theory of inhomogeneous systems
with long-range interactions in the following manner:

(i) We provide a simpler and more physical derivation of the inhomogeneous
Lenard-Balescu equation. We compute the diffusion tensor $D_{ij}$ and the
friction by polarization ${\bf F}_{\rm pol}$ by a direct approach and substitute
these expressions into the Fokker-Planck equation written in a suitable form in
which the diffusion tensor is ``sandwiched'' between the two action derivative
so that the friction by polarization appears naturally.

(ii) We derive the proper expression of the fluctuation-dissipation 
theorem for inhomogeneous systems with long-range interactions.

(iii) We consider a multi-species system of particles  while 
previous works were mostly restricted to particles with the same mass.

(iv) We consider a collisionless system of particles with long-range
interactions submitted to a small external stochastic
perturbation of arbitrary origin and derive a SDD
equation sourced by the external noise. 

(v) The Landau and Lenard-Balescu equations are associated with the
microcanonical ensemble where the system of particles is isolated. In that
case, the particles are fundamentally described by $N$-body Hamiltonian
equations \cite{btnew}. We compare these results with those obtained for a gas
of  Brownian particles with long-range interactions \cite{hb5} described by 
$N$-body stochastic Langevin equations. We
establish a deterministic Kramers equation governing their mean evolution as
well
as a stochastic Kramers equation governing their
mesoscopic evolution.

The paper is organized as follows. In Sec. \ref{sec_inhos}, we
present the basic equations describing a system of particles with long-range
interactions submitted to a small external stochastic perturbation and introduce
the quasilinear approximation. In Sec. \ref{sec_bog},
we explain how the
linearized equation for the perturbation can be analytically solved with Fourier
transforms by making the Bogoliubov ansatz. In Sec. \ref{sec_diel}, we
present the matrix method. In Sec. \ref{sec_inhosrf}, we
determine the linear response of the system to a small external perturbation. In
Sec. \ref{sec_cf}, we relate the dressed power spectrum of the total
fluctuating potential to the correlation function
of the external perturbation and specifically consider
the case where the external perturbation  is due to a random distribution of
$N$ particles. In Sec. \ref{sec_fdi}, we derive the fluctuation-dissipation
theorem satisfied by an isolated inhomogeneous system of particles with
long-range
interactions at statistical equilibrium. In
Sec. \ref{sec_fp}, we introduce the general Fokker-Planck equation in action
space. In Sec. \ref{sec_diffco}, we derive the diffusion tensor
of a test particle experiencing an external stochastic perturbation and
specifically consider
the case where the external perturbation  is due to a random distribution of
$N$ field particles. In Sec. \ref{sec_ifpol}, we derive the friction by
polarization experienced by a test particle traveling in a background medium
created by a smooth distribution of particles. In
Sec. \ref{sec_ein}, we consider the evolution of a test particle in a bath of
field particles at statistical equilibrium and establish the appropriate form of
Einstein relation between the friction and the diffusion coefficients. In Sec.
\ref{sec_avp}, we derive the kinetic equation (inhomogeneous Lenard-Balescu
equation) of particles with long-range
interactions. In Sec.
\ref{sec_mono}, we show how this kinetic
equation
simplifies itself for 1D systems with a monotonic frequency profile.  In Sec.
\ref{sec_tdbv}, we
contrast the kinetic theory of an isolated Hamiltonian system of particles
to the
kinetic theory of a gas of Brownian particles with long-range interactions.
In Sec. \ref{sec_sdd}, we  derive the SDD equation
describing the mean evolution of a system of particles with long-range
interactions submitted to
a small external stochastic perturbation of arbitrary origin. In Sec.
\ref{sec_fd}, we study
the mean evolution and the mesoscopic evolution of the DF governed  by the
stochastic damped Vlasov equation. We recover by
this approach the
power spectrum and the diffusion tensor of a system
of particles with long-range interactions.   In Sec.
\ref{sec_spv}, we apply the same approach to the case of stochastically forced
Brownian particles with long-range interactions. In Sec. \ref{sec_diff},
we summarize our results and compare the
different kinetic equations obtained in this paper. We present a 
kinetic equation that generalizes the Lenard-Balescu equation and the SDD
equation. This equation includes a diffusion
term of arbitrary origin (instead of being produced by a collection of $N$ field
particles) and a friction by polarization induced by the wake produced by the
moving particle. The Appendices provide useful complements to the results
established in the main text.

\section{Basic equations}
\label{sec_inhos}

We consider a system of material particles of individual mass
$m$ interacting
via a long-range binary potential $u(|{\bf r}-{\bf r}'|)$ decreasing more slowly
than $r^{-\gamma}$ with $\gamma\le d$ in a
space of
dimension
$d$. We assume that the particles are subjected to an external
stochastic
force  (exterior perturbation) of zero mean arising from a
fluctuating
potential $\Phi_e({\bf r},t)$. The equations of motion of the
particles are
\begin{eqnarray}
\frac{d{\bf r}_{i}}{dt}={\bf v}_i,\qquad \frac{d{\bf
v}_{i}}{dt}=-\nabla\Phi_d({\bf r}_i)-\nabla\Phi_e({\bf r}_i,t),
\label{n1zero}
\end{eqnarray}
where $\Phi_d({\bf r})=m\sum_j u(|{\bf r}-{\bf r}_j|)$ is the exact potential
produced by the particles.  They can be written in
Hamiltonian form as $m d{\bf r}_i/dt=\partial{(H_d+H_e)}/\partial {\bf v}_i$ and
$m d{\bf v}_i/dt=-\partial{(H_d+H_e)}/\partial {\bf r}_i$, where
$H_d=(1/2)\sum_i mv_i^2+\sum_{i<j} m^2 u(|{\bf r}_i-{\bf r}_j|)$ is the
Hamiltonian of the particles and $H_e=\sum_i m\Phi_e({\bf r}_i,t)$ is
the Hamiltonian associated with the external force. The discrete (or singular
exact) DF of the
particles,
$f_d({\bf r},{\bf
v},t)=m\sum_i \delta({\bf r}-{\bf r}_i(t))\delta({\bf v}-{\bf v}_i(t))$,
satisfies the Klimontovich equation
\begin{eqnarray}
\frac{\partial f_d}{\partial t}+{\bf v}\cdot \frac{\partial f_d}{\partial {\bf
r}}-\nabla(\Phi_d+\Phi_e)\cdot \frac{\partial f_d}{\partial {\bf v}}=0,
\label{n1}
\end{eqnarray}
where
\begin{eqnarray}
\Phi_d({\bf r},t)=\int u(|{\bf r}-{\bf r}'|)\rho_d({\bf r}',t)\, d{\bf
r}'
\label{n2}
\end{eqnarray}
is the potential produced by the discrete density of particles
$\rho_d({\bf
r},t)=\int f_d({\bf r},{\bf
v},t)\, d{\bf v}=m\sum_i \delta({\bf r}-{\bf r}_i(t))$. The
Klimontovich equation (\ref{n1}) can be written under the form
\begin{eqnarray}
\frac{\partial f_d}{\partial t}+\lbrace f_d,H_d+H_e\rbrace=0,
\label{n2y}
\end{eqnarray}
where $H_d+H_e=p^2/2m+m(\Phi_d+\Phi_e)$ denotes here the total Hamiltonian of a
particle and $\lbrace f,g\rbrace=\partial_{\bf r}f\partial_{\bf
p}g-\partial_{\bf p}f\partial_{\bf r}g$ is the usual Poisson bracket (${\bf
p}=m{\bf v}$ is the impulse). This Poisson structure is valid for all canonical
coordinates and is particularly useful to treat spatially inhomogeneous systems
\cite{physicaA,weinberg,fpp}.

We introduce the mean DF $f({\bf r},{\bf v},t)=\langle
f_{d}({\bf r},{\bf v},t)\rangle$ corresponding to an ensemble
average of
$f_{d}({\bf r},{\bf v},t)$. We then write
$f_d({\bf r},{\bf v},t)=f({\bf r},{\bf v},t)+\delta f({\bf r},{\bf v},t)$, where
$\delta f({\bf r},{\bf v},t)$ denotes the fluctuations about the mean DF.
Similarly, we write $\Phi_d({\bf r},t)=\Phi({\bf
r},t)+\delta\Phi({\bf r},t)$, where $\delta\Phi({\bf
r},t)$ denotes the fluctuations about the mean potential $\Phi({\bf
r},t)=\langle \Phi_d({\bf
r},t)\rangle $. Substituting these
decompositions into Eq. (\ref{n1}), we obtain
\begin{equation}
\frac{\partial f}{\partial t}+\frac{\partial\delta f}{\partial t}+{\bf
v}\cdot\frac{\partial f}{\partial {\bf r}}+{\bf v}\cdot\frac{\partial \delta
f}{\partial {\bf r}}-\nabla\Phi\cdot \frac{\partial f}{\partial {\bf
v}}-\nabla\Phi\cdot \frac{\partial \delta f}{\partial {\bf v}}-\nabla
(\delta\Phi+\Phi_e)\cdot\frac{\partial f}{\partial {\bf
v}}-\nabla(\delta\Phi+\Phi_e)\cdot
\frac{\partial \delta f}{\partial {\bf v}}=0.
\label{n3}
\end{equation}
Taking the ensemble  average of this equation, we get
\begin{equation}
\frac{\partial f}{\partial t}+{\bf v}\cdot\frac{\partial f}{\partial {\bf
r}}-\nabla\Phi\cdot \frac{\partial f}{\partial {\bf v}}=\frac{\partial}{\partial
{\bf v}}\cdot \left\langle \delta f\nabla(\delta\Phi+\Phi_e)\right\rangle.
\label{n4}
\end{equation}
This equation governs the evolution of the mean DF. Its right hand side
can be interpreted as a ``collision''
term arising from the
granularity of the system (finite $N$
effects) and the  external
stochastic force.\footnote{We generically 
call it the ``collision'' term although it may have a more general meaning due
to the external perturbation. A more
proper name could be the ``correlational'' term.}
Subtracting
this expression from Eq. (\ref{n3}), we obtain the equation for the
DF fluctuations
\begin{equation}
\frac{\partial \delta f}{\partial t}+{\bf v}\cdot\frac{\partial \delta
f}{\partial {\bf r}}-\nabla\Phi\cdot \frac{\partial \delta f}{\partial {\bf
v}}-\nabla(\delta\Phi+\Phi_e)\cdot \frac{\partial f}{\partial {\bf
v}}=\frac{\partial}{\partial {\bf v}}\cdot  (\delta
f\nabla(\delta\Phi+\Phi_e))-\frac{\partial}{\partial {\bf v}}\cdot \left\langle
\delta
f\nabla(\delta\Phi+\Phi_e)\right\rangle.
\label{n5}
\end{equation}
The foregoing equations are exact since no
approximation has been made for the
moment.

We now assume that the
external force
is weak and treat the stochastic potential
$\Phi_e({\bf r},t)$ as a small perturbation to the mean field dynamics.
We also assume that  the
fluctuations $\delta\Phi({\bf r},t)$ of the long-range potential  created by
the particles are weak.
Since the mass of the particles scales as $m\sim 1/N$ this approximation is
valid when $N\gg
1$ \cite{campabook}.
If we ignore the external stochastic perturbation and the fluctuations of the
potential
due to finite $N$ effects altogether, the collision
term vanishes and Eq. (\ref{n4}) reduces to
the Vlasov
equation
\begin{equation}
\frac{\partial f}{\partial t}+{\bf v}\cdot\frac{\partial f}{\partial {\bf
r}}-\nabla\Phi\cdot \frac{\partial f}{\partial {\bf v}}=0,
\label{n4b}
\end{equation}
where
\begin{eqnarray}
\Phi({\bf r},t)=\int u(|{\bf r}-{\bf r}'|)\rho({\bf r}',t)\, d{\bf
r}'
\label{n2bzero}
\end{eqnarray}
is the potential produced by the mean density of particles $\rho({\bf
r},t)=\int f({\bf r},{\bf v},t)\, d{\bf v}=\langle \rho_{d}({\bf r},t)\rangle$. 
The Vlasov equation describes a collisionless dynamics driven only by
self-consistent mean
field effects. It is valid in the limit $\Phi_e\rightarrow 0$ and in a
proper thermodynamic limit
$N\rightarrow +\infty$ with $m\sim 1/N$. It is also valid for
sufficiently short times.

We now take into account a small correction to the Vlasov equation obtained by
keeping the 
collision term on the right hand side of Eq. (\ref{n4}) but neglecting the
quadratic terms on the right hand side of Eq. (\ref{n5}). We therefore obtain
a set of two coupled equations
\begin{equation}
\frac{\partial f}{\partial t}+{\bf v}\cdot \frac{\partial f}{\partial {\bf
r}}-\nabla\Phi\cdot \frac{\partial f}{\partial {\bf v}}=\frac{\partial}{\partial
{\bf v}}\cdot \left\langle \delta f \nabla(\delta\Phi+\Phi_e)\right\rangle,
\label{n6}
\end{equation}
\begin{equation}
\frac{\partial\delta f}{\partial t}+{\bf v}\cdot \frac{\partial \delta
f}{\partial {\bf r}}-\nabla\Phi\cdot \frac{\partial \delta f}{\partial {\bf
v}}-\nabla(\delta\Phi+\Phi_e)\cdot \frac{\partial f}{\partial {\bf
v}}=0.
\label{n7}
\end{equation}
These equations form the starting point of the quasilinear theory which is
valid in a weak
coupling  approximation ($m\sim 1/N\ll 1$) and for a weak external stochastic
perturbation ($\Phi_e\ll \Phi$). Eq. (\ref{n6})
describes the evolution of the mean DF sourced by
the correlations of the fluctuations and Eq. (\ref{n7}) describes the evolution
of the fluctuations due to the
granularities of the system (finite $N$ effects) and the external noise.
These equations are valid at the order $1/N$ and to leading
order in  
$\Phi_e$.

For spatially inhomogeneous systems, it is convenient to work with angle-action
variables $({\bf w},{\bf J})$ \cite{btnew}. The initial
stage of the evolution of the system is governed by the Vlasov equation
involving only mean
field effects. During the collisionless regime, the system
reaches a QSS as a result of a process of violent relaxation
\cite{lb}.
This is a stable steady state of the Vlasov equation. According to the Jeans
theorem \cite{jeans}, the DF is a
function of the actions alone: $f=f({\bf J})$. Then, on a longer (secular)
timescale, the DF evolves under the effect of collisions between the
particles (granularities) and the effect of the external potential, but the
DF
remains a function of the actions alone, $f=f({\bf J},t)$, that changes
slowly with time (adiabatic approximation).\footnote{We assume
that the background 
potential of the system is stationary and integrable so that we may
always remap
the usual phase space
coordinates $({\bf r}, {\bf v})$ to the angle-action coordinates $({\bf w}, {\bf
J})$. This is consistent with the Bogoliubov ansatz discussed in Sec.
\ref{sec_bog}. We also assume
that the system remains Vlasov stable during the whole evolution. This
may not always be the case. Even if we start from a Vlasov stable DF 
$f_0({\bf J})$, the ``collision'' term (r.h.s. in Eq. (\ref{i56})) will change
it and induce a temporal evolution of $f({\bf J},t)$. The system may
become dynamically (Vlasov) unstable and undergo a dynamical
phase transition from one state to the other \cite{ccgm}. We assume
here
that this transition does not take place or we consider a period of time
preceding this transition.} By construction, the mean field Hamiltonian
$H$ of a
particle in angle and action variables depends only on the actions ${\bf J}$
that are constants of the motion: $H=H({\bf J})$. The conjugate
coordinates ${\bf w}$ are the angles. The Hamilton equations
become $\dot{\bf
J}=-\partial H/\partial {\bf w}={\bf 0}$ and $\dot{\bf
w}=\partial H/\partial {\bf J}={\bf \Omega}({\bf J})$, where ${\bf
\Omega}({\bf J})=\partial H/\partial {\bf J}$ is the angular
frequency (pulsation) of the orbit with action ${\bf J}$, and the unperturbed
equations of motion of the
particles (at leading order) are simply straight lines in angle-action space
traveled at constant
action: ${\bf
J}(t)={\bf J}$ and ${\bf w}(t)={\bf w}+{\bf \Omega}({\bf J})t$. As
a result, the fundamental
equations of the
quasilinear theory written in terms of angle-action variables are
\cite{physicaA,weinberg,fpp} 
\begin{eqnarray}
\label{i56}
\frac{\partial f}{\partial t}=\frac{\partial}{\partial
{\bf J}}\cdot \left\langle \delta
f \frac{\partial}{\partial {\bf w}} (\delta\Phi+\Phi_e)\right\rangle,
\end{eqnarray}
\begin{eqnarray}
\label{i57}
\frac{\partial \delta f}{\partial t}+{\bf \Omega}\cdot \frac{\partial\delta
f}{\partial {\bf w}}- \frac{\partial}{\partial {\bf w}}
(\delta\Phi+\Phi_e)\cdot \frac{\partial
f}{\partial {\bf J}}=0.
\end{eqnarray}

{\it Remark:} For the simplicity of the presentation, we have assumed that
the external perturbation $\Phi_e$ is of zero mean. If it has a mean field
component, it can be included in $\Phi$. Then, ${\bf
\Omega}$ represents the total pulsation produced by the system and by the
external perturbation.

\section{Bogoliubov ansatz}
\label{sec_bog}

In order to solve Eq. (\ref{i57}) for the fluctuations, we resort
to the Bogoliubov
ansatz. We  assume that there exists a timescale separation between a slow and a
fast dynamics and we regard ${\bf \Omega}({\bf J})$ and $f({\bf J})$ in Eq.
(\ref{i57}) as
``frozen'' (independent of time) at the scale of the fast
dynamics. This amounts to neglecting the temporal
variations
of the mean field when we consider the evolution of the fluctuations. This is
possible when the mean DF evolves on a secular timescale that is
long compared to the correlation time (the time over which the
correlations of the fluctuations have their essential
support). We can
then introduce Fourier
transforms in ${\bf w}$ and $t$ for the
fluctuations, writing
\begin{equation}
\delta \hat f({\bf k},{\bf J},\omega)=\int \frac{d{\bf
w}}{(2\pi)^d}\int_{-\infty}^{+\infty}dt\, e^{-i({\bf k}\cdot{\bf w}-\omega
t)}\delta
f({\bf w},{\bf J},t),
\label{i58}
\end{equation}
\begin{equation}
\delta f({\bf w},{\bf J},t)=\sum_{{\bf k}}\int\frac{d\omega}{2\pi}\,
e^{i({\bf k}\cdot{\bf w}-\omega t)}\delta\hat f({\bf k},{\bf J},\omega).
\label{i59}
\end{equation}
Similar expressions hold for the fluctuating potential $\delta\Phi({\bf w},{\bf
J},t)$ and for the external potential  $\Phi_e({\bf w},{\bf
J},t)$. Since the angles ${\bf w}$ are
$2\pi$-periodic,
we have introduced discrete Fourier expansions with respect to these
variables. For future reference, we recall the Fourier
representation of the Dirac $\delta$-function
\begin{equation}
\delta(t)=\int_{-\infty}^{+\infty} e^{-i \omega t}\, \frac{d\omega}{2\pi},\qquad
\delta_{{\bf k},{\bf 0}}=\int e^{-i {\bf k}\cdot {\bf w}}\, \frac{d{\bf
w}}{(2\pi)^d}.
\label{deltac}
\end{equation} 

Before going further, some comments about our procedure of derivation are
required.
In order to derive the
Lenard-Balescu equation describing the mean evolution of a system of particles
under discreteness effects (``collisions'') at the order
$1/N$  we usually take
$\Phi_e=0$ in Eq. (\ref{n1}) and consider an initial value problem as
described in Sec. 2 of \cite{physicaA} (see also Appendix \ref{sec_j}). In that
case, Eq. (\ref{i57}) has
to be solved by
introducing
a Fourier transform in angle and a Laplace transform in time.
This brings a term $\delta{\hat f}({\bf k},{\bf J},0)$ related to the initial
condition
in the equation for the
fluctuations [see Eq. (\ref{j3})]. This is how discreteness
effects (granularities) are taken into account in this approach. Calculating the
correlation
function and
substituting the result into Eq.
(\ref{i56}), one obtains the inhomogeneous Lenard-Balescu equation in which the 
diffusion
and the friction terms appear simultaneously.   This derivation involves,
however,
rather technical calculations. In the
present paper, we shall
derive the inhomogeneous Lenard-Balescu equation differently by using a simpler
and more
physical approach based on the
Fokker-Planck
equation (see Sec. \ref{sec_fp}). In this approach, 
discreteness effects are taken into account  in the  external perturbation
$\Phi_e$.  Indeed, we can regard $\Phi_e$ either as having an arbitrary 
origin
(see
Sec. \ref{sec_inhosgr}) or as being generated by a random distribution 
of $N$ particles, the so-called field particles (see Sec.
\ref{sec_inhoscf}). In the presence of an
external perturbation, Eq.
(\ref{i57}) can be solved
by introducing
Fourier transforms in space and time.\footnote{In the
absence of external perturbation, we need to introduce a
Laplace transform in time in order to take into account the initial
condition as
explained above.
In our approach, the initial condition is rejected to the infinite past but we
have to add a small imaginary term $i0^+$ in the (real)
pulsation $\omega$ of the Fourier transform in order to make the fluctuations
vanish for
$t\rightarrow -\infty$. In a sense, this procedure amounts to
using a Laplace
transform in time but ignoring the term $\delta{\hat f}({\bf k},{\bf
J},0)$ related to the initial
condition. Our approach does not take into account the
interaction of the particles with the damped modes. It therefore implicitly
assumes that these
modes are very damped and that the system is strongly stable.} We
can then derive the 
dressed power spectrum and the diffusion coefficient of a test particle due to
the external perturbation by
taking
collective effects
into
account (see Sec. \ref{sec_diffco}). On the other hand, the friction by
polarization of a test particle can be
obtained by determining the response of the system to the perturbation
that it has caused (see Sec. \ref{sec_ifpol}). In the first
case (diffusion) the external perturbation is stochastic: it is due to the
fluctuations (noise) of the $N$ field particles. In the
second case (friction by polarization)
the external perturbation is deterministic: it is caused by the
wake created by the test particle.
Substituting these
coefficients
into
the Fokker-Planck equation, we obtain the inhomogeneous Lenard-Balescu equation
(see
Secs. \ref{sec_avp} and \ref{sec_mono}). This formalism also
allows us to treat situations in which the external perturbation $\Phi_e$ is
not necessarily due to a discrete distribution of particles. In that case,
we can derive more
general kinetic equations. When $N\rightarrow +\infty$, i.e. when the collisions
between the particles are negligible, we obtain the SDD equation 
involving a diffusion term due to the external perturbation (see
Sec. \ref{sec_sdd}). For 
finite $N$, we obtain a mixed kinetic equation involving a SDD term and a
Lenard-Balescu term (see
Sec. \ref{sec_diff}).

\section{Dielectric matrix}
\label{sec_diel}

Let us determine the linear response of the system to a
small external perturbation
$\Phi_e({\bf w},{\bf J},t)$. 
In the present case, the perturbation is not necessarily a stochastic
perturbation. Since the perturbation is small, we can use the linearized Vlasov
equation (\ref{i57}). Taking the Fourier transform of this equation in
${\bf w}$ and $t$, we obtain
\begin{equation}
\delta\hat f ({\bf k},{\bf J},\omega)=\frac{{\bf k}\cdot \frac{\partial
f}{\partial {\bf J}}}{{\bf k}\cdot {\bf
\Omega}-\omega}\left\lbrack \delta\hat\Phi({\bf
k},{\bf J},\omega)+\hat\Phi_e({\bf
k},{\bf J},\omega)\right\rbrack.
\label{i60}
\end{equation}
The fluctuations of the potential are
related to the fluctuations of the density
by
\begin{eqnarray}
\delta\Phi({\bf r},t)=\int u(|{\bf r}-{\bf r}'|)\delta\rho({\bf r}',t)\, d{\bf
r}'.
\label{n2b}
\end{eqnarray}
The coupled equations
(\ref{i57}) and (\ref{n2b}) are difficult to
solve because their natural coordinates are different. The natural
coordinates for solving the linearized Vlasov equation (\ref{i57}) are
angle-action variables which make the solution by Fourier transform easy (see
Eq. (\ref{i60})).
On the other hand, the natural coordinates for solving Eq.  (\ref{n2b})
 are
the position coordinates ${\bf r}$. If we  transform the fluctuating potential
to angle-action variables, we get (see Appendix \ref{sec_ts})
\begin{equation}
\delta{\hat\Phi}({\bf k},{\bf J},\omega)=(2\pi)^d\sum_{{\bf k}'}\int d{\bf J}'
A_{{\bf k},{\bf k}'}({\bf J},{\bf J}')\delta {\hat f}({\bf k}',{\bf J}',\omega),
\label{i60g}
\end{equation}
where $A_{{\bf k},{\bf k}'}({\bf J},{\bf J}')$ is the Fourier transform of
the bare potential of interaction in angle-action variables
\cite{physicaA} (see Appendix \ref{sec_bdpi}). If we
substitute Eq. (\ref{i60}) into Eq. (\ref{i60g}) we obtain a Fredholm integral
equation
\begin{equation}
\delta{\hat\Phi}({\bf k},{\bf J},\omega)=(2\pi)^d\sum_{{\bf k}'}\int d{\bf J}'
A_{{\bf k},{\bf k}'}({\bf J},{\bf J}')\frac{{\bf k}'\cdot \frac{\partial
f'}{\partial {\bf J}'}}{{\bf k}'\cdot {\bf
\Omega}'-\omega}\left\lbrack \delta\hat\Phi({\bf
k}',{\bf J}',\omega)+\hat\Phi_e({\bf
k}',{\bf J}',\omega)\right\rbrack,
\end{equation}
which determines $\delta{\hat \Phi}({\bf k},{\bf
J},\omega)$. However, this
equation is complicated to solve explicitly because of the summation on ${\bf
k}'$ and the integration over ${\bf J}'$. Proceeding like in
Appendix \ref{sec_dnwout} its formal solution is
\begin{equation}
\label{myle}
\delta{\hat\Phi}({\bf k},{\bf J},\omega)=(2\pi)^d\sum_{{\bf k}'}\int d{\bf J}'
A^d_{{\bf k},{\bf k}'}({\bf J},{\bf J}',\omega)\frac{{\bf k}'\cdot
\frac{\partial
f'}{\partial {\bf J}'}}{{\bf k}'\cdot {\bf
\Omega}'-\omega}\hat\Phi_e({\bf
k}',{\bf J}',\omega),
\end{equation}
where $A^d_{{\bf k},{\bf k}'}({\bf J},{\bf J}',\omega)$ is the Fourier transform
of the dressed potential of interaction in angle-action variables
\cite{physicaA} (see Appendix \ref{sec_bdpi}).  A more explicit manner to solve
Eq. (\ref{i60}) with Eq. (\ref{n2b}) is to follow Kalnajs' matrix method
\cite{kalnajs}. We
assume that the fluctuating density $\delta\rho({\bf r},t)$, the fluctuating
potential $\delta\Phi({\bf r},t)$ and the external potential $\Phi_e({\bf r},t)$
can be expanded
on a complete biorthogonal basis such
that
\begin{equation}
\delta\rho({\bf r},t)=\sum_{\alpha}A_{\alpha}(t)\rho_{\alpha}({\bf r}),\qquad
\delta\Phi({\bf r},t)=\sum_{\alpha}A_{\alpha}(t)\Phi_{\alpha}({\bf r}),\qquad
\Phi_e({\bf r},t)=\sum_{\alpha}A^e_{\alpha}(t)\Phi_{\alpha}({\bf r}),
\label{i62}
\end{equation}
where the density-potential pairs $\rho_{\alpha}({\bf r})$ and
$\Phi_{\alpha}({\bf r})$
satisfy\footnote{We consider an attractive self-interaction, like gravity, hence
the sign $-$
in the second term of Eq. (\ref{i63}).} 
\begin{equation}
\Phi_{\alpha}({\bf r})=\int u(|{\bf r}-{\bf
r}'|) \rho_{\alpha}({\bf
r}')\,
d{\bf r}',\qquad \int \rho_{\alpha}({\bf r})\Phi_{\alpha'}({\bf r})^*\, d{\bf
r}=-\delta_{\alpha\alpha'}.
\label{i63}
\end{equation}
Multiplying the first relation in  Eq. (\ref{i62}) by $\Phi_{\alpha}^*({\bf
r})$, integrating
over ${\bf r}$, and using Eq. (\ref{i63}), we get
\begin{equation}
A_{\alpha}(t)=-\int\delta\rho({\bf
r},t) \Phi_{\alpha}^*({\bf
r})\, d{\bf r}.
\label{i64}
\end{equation}
Substituting the relation
\begin{equation}
\delta\rho({\bf
r},t)=\int \delta f({\bf
r},{\bf v},t)\, d{\bf v}=\int d{\bf v}\sum_{{\bf
k}}\,
e^{i{\bf k}\cdot{\bf w}}\delta\hat f({\bf k},{\bf J},t)
\label{i65}
\end{equation}
into Eq. (\ref{i64}), we obtain
\begin{eqnarray}
A_{\alpha}(t)&=&-\int d{\bf r} d{\bf v}\sum_{{\bf
k}}\,
e^{i{\bf k}\cdot{\bf w}}\delta\hat f({\bf k},{\bf J},t)  \Phi_{\alpha}^*({\bf
r})\nonumber\\
&=&-\int d{\bf w} d{\bf J}\sum_{{\bf
k}}\,
e^{i{\bf k}\cdot{\bf w}}\delta\hat f({\bf k},{\bf J},t)  \Phi_{\alpha}^*({\bf
w},{\bf J})\nonumber\\
&=&-(2\pi)^d\int d{\bf J}\sum_{{\bf
k}}\, \delta\hat f({\bf k},{\bf J},t)  \hat\Phi_{\alpha}^*({\bf
k},{\bf J}),
\label{i66}
\end{eqnarray}
where we have use the fact that the transformation $({\bf r},{\bf
v})\rightarrow ({\bf w},{\bf J})$ is canonical so
that $d{\bf r}d{\bf v}=d{\bf w}d{\bf J}$. Taking the temporal Fourier transform
of Eq. (\ref{i66}), we get
\begin{eqnarray}
\hat A_{\alpha}(\omega)
=-(2\pi)^d\int d{\bf J}\sum_{{\bf
k}}\, \delta\hat f({\bf k},{\bf J},\omega)  \hat\Phi_{\alpha}^*({\bf
k},{\bf J}).
\label{i67}
\end{eqnarray}
Using Eqs. (\ref{i60}) and (\ref{i62}) in conjuction with Eq. (\ref{i67}), we
obtain
\begin{eqnarray}
\hat A_{\alpha}(\omega)
&=&-(2\pi)^d\int d{\bf J}\sum_{{\bf
k}}\, \frac{{\bf k}\cdot \frac{\partial
f}{\partial {\bf J}}}{{\bf k}\cdot {\bf
\Omega}-\omega}\left\lbrack \delta\hat\Phi({\bf
k},{\bf J},\omega)+\hat\Phi_e({\bf
k},{\bf J},\omega)\right\rbrack \hat\Phi_{\alpha}^*({\bf
k},{\bf J})\nonumber\\
&=&-(2\pi)^d\int d{\bf J}\sum_{{\bf
k}}\, \frac{{\bf k}\cdot \frac{\partial
f}{\partial {\bf J}}}{{\bf k}\cdot {\bf
\Omega}-\omega}\sum_{\alpha'} \left\lbrack {\hat
A}_{\alpha'}(\omega)+{\hat
A^e}_{\alpha'}(\omega)\right\rbrack \hat\Phi_{\alpha'}({\bf
k},{\bf J})\hat\Phi_{\alpha}^*({\bf
k},{\bf J}).
\label{i68}
\end{eqnarray}
Introducing the ``dielectric'' matrix
\begin{eqnarray}
\epsilon_{\alpha\alpha'}(\omega)=\delta_{\alpha\alpha'}+(2\pi)^d\int d{\bf
J}\sum_{{\bf
k}}\, \frac{{\bf k}\cdot \frac{\partial
f}{\partial {\bf J}}}{{\bf k}\cdot {\bf
\Omega}-\omega} \hat\Phi_{\alpha'}({\bf
k},{\bf J})\hat\Phi_{\alpha}^*({\bf
k},{\bf J}),
\label{i70}
\end{eqnarray}
the foregoing
equation can be rewritten as
\begin{eqnarray}
\hat A_{\alpha}(\omega)
=\sum_{\alpha'} \left\lbrack {\hat
A}_{\alpha'}(\omega)+{\hat
A^e}_{\alpha'}(\omega)\right\rbrack \left
(\delta_{\alpha\alpha'}-\epsilon_{\alpha\alpha'}(\omega)\right
).
\label{i71}
\end{eqnarray}
Solving this equation for ${\hat A}_{\alpha}(\omega)$ we get
\begin{eqnarray}
\hat A_{\alpha}(\omega)
=\sum_{\alpha'} \left\lbrack
(\epsilon^{-1})_{\alpha\alpha'}(\omega)-\delta_{\alpha\alpha'}\right\rbrack
{\hat A^e}_{\alpha'}(\omega).
\label{fdi6}
\end{eqnarray}
Without the external perturbation, the system would be described by
the DF $f({\bf J})$. The external perturbation $\Phi_{e}({\bf w},{\bf
J},t)$  polarizes the system and creates a small
change in the DF $\delta f({\bf w},{\bf
J},t)$ [see Eq. (\ref{i57})] producing in turn a weak potential $\delta
\Phi({\bf w},{\bf
J},t)$ [see Eq. (\ref{n2b})]. As a result, the
total potential acting on a particle, which is
sometimes called the dressed or effective potential, is $\delta\Phi_{\rm
tot}({\bf
w},{\bf
J},t)=\Phi_{e}({\bf w},{\bf
J},t)+\delta \Phi({\bf w},{\bf
J},t)$.  This  is the sum of the external potential plus
the potential fluctuation induced by
the system itself (i.e. the system's own
response). Eqs. (\ref{i57}) and (\ref{n2b}) are
coupled together and written in terms of different variables. The manner to
solve this loop is to use a biorthogonal basis and Fourier
transforms as
we have done above. Using Eq. (\ref{fdi6}), we find that the Fourier amplitudes
$\hat A_{\alpha}^{\rm
tot}(\omega)=\hat A_{\alpha}(\omega)+{\hat
A^e}_{\alpha}(\omega)$ of the
total fluctuating potential $\delta\Phi_{\rm tot}=\delta\Phi+\Phi_e$ acting on
a particle are related to the Fourier amplitudes of the
external stochastic potential $\Phi_e$ by\footnote{In matrix
form, Eq. (\ref{i71}) can be written as ${\hat
A}=(1-\epsilon)({\hat A}+{\hat
A}_e)$ yielding $\epsilon{\hat A}=(1-\epsilon){\hat A}_e$, then ${\hat
A}=\epsilon^{-1}(1-\epsilon){\hat A}_e=(\epsilon^{-1}-1){\hat A}_e$, and finally
${\hat A}_{\rm tot}={\hat A}+{\hat A}_e=\epsilon^{-1}{\hat A}_e$. If we neglect
collective effects, we have ${\hat A}_{\rm tot}={\hat A}_e$ corresponding to
$\epsilon=1$ and  ${\hat
A}=(1-\epsilon){\hat
A}_e\simeq 0$.}
\begin{eqnarray}
\hat A_{\alpha}^{\rm tot}(\omega)=\sum_{\alpha'}
(\epsilon^{-1})_{\alpha\alpha'}(\omega){\hat
A^e}_{\alpha'}(\omega).
\label{i72}
\end{eqnarray}
Although not explicitly written, we must use the Landau
prescription $\omega\rightarrow \omega+i0^{+}$ in Eq. (\ref{i70}) (see footnote
24). As
a result, $(\epsilon)^{-1}_{\alpha\alpha'}(\omega)$ is a complex
matrix which plays the role of the response function in plasma physics (see
Sec. \ref{sec_inhosrf}).\footnote{We note that
$\epsilon_{\alpha\alpha'}(-\omega)=\epsilon_{\alpha\alpha'}(\omega)^*$.}
It determines the
response of the system
$\hat A_{\alpha}^{\rm tot}(\omega)$ to an external perturbation ${\hat
A^e}_{\alpha'}(\omega)$ through Eq. (\ref{i72}). 
The dielectric matrix  takes into account the polarization of the medium caused
by the self-interaction of the particles. This corresponds to the so-called
``collective
effects''. Depending on the form of the self-interaction and the geometry of
the system, the polarization cloud may amplify
or shield the action of the imposed external perturbation.
Therefore, the action of the
external potential is modified by collective effects. This amounts to
replacing the 
bare potential $\hat\Phi_e({\bf k},{\bf J},\omega)$
by the dressed potential  $\delta\hat\Phi_{\rm
tot}({\bf k},{\bf J},\omega)$ which
is the potential ${\hat\Phi_e({\bf k},{\bf J},\omega)}$ dressed by
the polarization cloud.  Without
self-interaction, or if we neglect collective effects, we just have
$\delta\hat\Phi_{\rm tot}({\bf k},{\bf J},\omega)=\hat\Phi_e({\bf
k},{\bf J},\omega)$,
corresponding to $\epsilon=1$ and $\delta\Phi=0$.

{\it Remark:} The dielectric matrix (\ref{i70}) can be written as
$\epsilon=1-M$ where $M$ is the polarization matrix such that
${\hat A}=M{\hat A}_{\rm tot}$. On the other
hand,
$\chi=\epsilon^{-1}-1$ is the response matrix such that ${\hat A}=\chi{\hat
A_e}$. Eq.
(\ref{i72}) may be written in matrix form as $\epsilon {\hat A}_{\rm
tot}={\hat A}_e$. When ${\hat A}_e=0$ this equation has nontrivial
solutions only if
\begin{eqnarray}
{\rm det}[\epsilon(\omega)]=0.
\label{disrel}
\end{eqnarray}
This is the dispersion
equation
which determines the proper complex pulsations $\omega$ of the system associated
with the DF $f({\bf J})$.\footnote{In that case, it is necessary
to solve the linearized Vlasov equation by using a  Laplace transform in time
and compute the integral in the dielectric
matrix (\ref{i70})  along the Landau contour (analytical continuation).} It
can be used to study the
(Vlasov) linear dynamical
stability of the system. A
zero of the dispersion relation ${\rm det}[\epsilon(\omega)]=0$ for
$\omega_i>0$ signifies an unstable growing mode. The condition of marginal
stability corresponds to $\omega_i=0$. In the following, we
assume that the system is stable so that $\omega_i<0$ for all modes. This
implies that $\epsilon(\omega)$ does not vanish for any real $\omega$. This
assumption will be necessary to make the inhomogeneous Lenard-Balescu equation
derived below well-defined.

\section{Response function}
\label{sec_inhosrf}

In this section, we consider the response of an inhomogeneous system with
long-range interactions to a small external perturbation $\Phi_e({\bf w},{\bf
J},t)$. We define the response matrix
$\chi_{\alpha\alpha'}(\omega)$ by
\begin{eqnarray}
\hat A_{\alpha}(\omega)
=\sum_{\alpha'} \chi_{\alpha\alpha'}(\omega)
{\hat A^e}_{\alpha'}(\omega).
\label{fdi7}
\end{eqnarray}
According to Eq. (\ref{fdi6}) we have
\begin{eqnarray}
\chi_{\alpha\alpha'}(\omega)=
(\epsilon^{-1})_{\alpha\alpha'}(\omega)-\delta_{\alpha\alpha'}.
\label{fdi8}
\end{eqnarray}
Therefore, the response matrix is equal to the inverse of the
dielectric matrix minus the identity. We then introduce the potential response
function
\begin{eqnarray}
\chi_{{\bf k},{\bf k}'}({\bf J},{\bf
J'},\omega)=-\sum_{\alpha\alpha'}{\hat\Phi}_{\alpha}({\bf k},{\bf
J})\chi_{\alpha\alpha'}(\omega){\hat\Phi}^*_{\alpha'}({\bf k}',{\bf
J}').
\label{fdi10}
\end{eqnarray}
Using the expressions (\ref{pi37}) and (\ref{can1b}) of the bare and dressed
potentials of interaction it can be written as 
\begin{eqnarray}
\chi_{{\bf k},{\bf k}'}({\bf J},{\bf
J'},\omega)=A^d_{{\bf k},{\bf k}'}({\bf J},{\bf
J'},\omega)-A_{{\bf k},{\bf k}'}({\bf J},{\bf
J'}).
\label{fdi9}
\end{eqnarray}
Taking ${\bf
J}'={\bf J}$ and ${\bf k}'={\bf k}$, we obtain
\begin{eqnarray}
\chi_{{\bf k},{\bf k}}({\bf J},{\bf
J},\omega)=A^d_{{\bf k},{\bf k}}({\bf J},{\bf
J},\omega)-A_{{\bf k},{\bf k}}({\bf J},{\bf
J}).
\label{fdi9b}
\end{eqnarray}

Setting $\omega=0$ we obtain the static response matrix
\begin{eqnarray}
\chi_{\alpha\alpha'}=
(\epsilon^{-1})_{\alpha\alpha'}(0)-\delta_{\alpha\alpha'}
\label{fdi8j}
\end{eqnarray}
with
\begin{eqnarray}
\epsilon_{\alpha\alpha'}(0)=\delta_{\alpha\alpha'}+(2\pi)^d\int d{\bf
J}\sum_{{\bf
k}}\, \frac{{\bf k}\cdot \frac{\partial
f}{\partial {\bf J}}}{{\bf k}\cdot {\bf
\Omega}} \hat\Phi_{\alpha'}({\bf
k},{\bf J})\hat\Phi_{\alpha}^*({\bf
k},{\bf J}),
\end{eqnarray}
and the static response function
\begin{eqnarray}
\chi_{{\bf k},{\bf k}'}({\bf J},{\bf
J}')=\chi_{{\bf k},{\bf k}'}({\bf J},{\bf
J}',\omega=0).
\label{sps3v}
\end{eqnarray}

\section{Correlation function}
\label{sec_cf}

In this section we assume that the external perturbation $\Phi_e({\bf
w},{\bf J},t)$ is a stochastic
process and we determine the auto-correlation function (power spectrum) of
the total fluctuating potential that it induces. We first give general results
and then consider
the case where the
external perturbation is produced by a random distribution of $N$ field
particles.

\subsection{General results}
\label{sec_inhosgr}

We assume that the time evolution of the perturbing potential is a
stationary stochastic  process and write the auto-correlation function
of its different components as
\begin{equation}
\langle
A^e_{\alpha}(t) A^e_{\alpha'}(t')^*\rangle=C_{\alpha\alpha'}(t-t').
\label{i80}
\end{equation}
The function $C_{\alpha\alpha'}(t-t')$ describes a possibly colored noise. The
spectral auto-correlation function of the components of the external 
stochastic potential is
\begin{equation}
\langle {\hat
A^e}_{\alpha}(\omega){\hat
A^e}_{\alpha'}(\omega')^*\rangle=2\pi\delta(\omega-\omega'){\hat
C}_{\alpha\alpha'}(\omega),
\label{i81}
\end{equation}
where ${\hat C}_{\alpha\alpha'}(\omega)$ is the temporal Fourier transform of
$C_{\alpha\alpha'}(t-t')$  (Wiener-Khinchin theorem \cite{wiener,khinchin}). For
a white noise ${\hat C}_{\alpha\alpha'}(\omega)$ is independent of
$\omega$. The auto-correlation function of the
external
potential is
\begin{equation}
\label{mar1}
\langle \hat\Phi_{e}({\bf
k},{\bf J},t) \hat\Phi_{e}({\bf
k},{\bf J},t')^*\rangle={C}({\bf
k},{\bf J},t-t'),
\end{equation}
and its temporal  Fourier transform is
\begin{equation}
\label{mar1b}
\langle \hat\Phi_{e}({\bf
k},{\bf J},\omega) \hat\Phi_{e}({\bf
k},{\bf J},\omega')^*\rangle=2\pi \delta(\omega-\omega'){\hat C}({\bf
k},{\bf J},\omega).
\end{equation}
According to Eq. (\ref{i62}) we have
\begin{equation}
{\hat \Phi}_e({\bf k},{\bf J},\omega)=\sum_{\alpha}{\hat
A}^e_{\alpha}(\omega){\hat \Phi}_{\alpha}({\bf k},{\bf J}),
\label{ciw8}
\end{equation}
hence
\begin{equation}
\langle {\hat \Phi}_e({\bf k},{\bf J},\omega){\hat \Phi}_e({\bf k},{\bf
J},\omega')^*\rangle=\sum_{\alpha\alpha'}\langle {\hat
A}^e_{\alpha}(\omega){\hat A}^e_{\alpha'}(\omega')^*\rangle {\hat
\Phi}_{\alpha}({\bf k},{\bf J}) {\hat \Phi}_{\alpha'}({\bf k},{\bf J})^*.
\label{mar3}
\end{equation}
Combining Eqs. (\ref{i81}), (\ref{mar1b}) and (\ref{mar3}) we find that
\begin{equation}
{\hat C}({\bf k},{\bf J},\omega)=\sum_{\alpha\alpha'}  {\hat \Phi}_{\alpha}({\bf
k},{\bf J}) {\hat
C}_{\alpha\alpha'}(\omega) {\hat
\Phi}_{\alpha'}({\bf k},{\bf J})^*.
\label{mar5}
\end{equation}

Similarly, we write the spectral auto-correlation function of
the
components of the total fluctuating potential as
\begin{equation}
\langle {\hat
A^{\rm tot}}_{\alpha}(\omega){\hat
A^{\rm
tot}}_{\alpha'}(\omega')^*\rangle=2\pi\delta(\omega-\omega')P_{\alpha\alpha'
}(\omega),
\label{cf1}
\end{equation}
where $P_{\alpha\alpha'}(\omega)$ is the power spectrum tensor. The
auto-correlation function of the total fluctuating potential in Fourier
space is given by 
\begin{equation}
\label{psaab}
\langle \delta\hat\Phi_{\rm tot}({\bf
k},{\bf J},\omega) \delta\hat\Phi_{\rm tot}({\bf
k},{\bf J},\omega')^*\rangle=2\pi \delta(\omega-\omega')P({\bf
k},{\bf J},\omega),
\end{equation}
where  $P({\bf k},{\bf J},\omega)$  is the power
spectrum.  According to Eq. (\ref{i62}) we have
\begin{equation}
\delta{\hat \Phi}_{\rm tot}({\bf k},{\bf J},\omega)=\sum_{\alpha}{\hat
A}^{\rm tot}_{\alpha}(\omega){\hat \Phi}_{\alpha}({\bf k},{\bf J}),
\label{ciw8y}
\end{equation}
hence
\begin{equation}
\label{kl1}
\langle \delta\hat\Phi_{\rm tot}({\bf
k},{\bf J},\omega) \delta\hat\Phi_{\rm tot}({\bf
k},{\bf J},\omega')^*\rangle=\sum_{\alpha\alpha'}\langle  {\hat
A}_{\alpha}^{\rm tot}(\omega)
{\hat
A}^{\rm tot}_{\alpha'}(\omega')^*\rangle\hat\Phi_{\alpha}({\bf
k},{\bf J})\hat\Phi_{\alpha'}({\bf
k},{\bf J})^*.
\end{equation}
Comparing Eqs. (\ref{cf1}), (\ref{psaab}) and (\ref{kl1})
we find that
\begin{equation}
P({\bf k},{\bf J},\omega)=\sum_{\alpha\alpha'}  {\hat \Phi}_{\alpha}({\bf
k},{\bf J}) {P}_{\alpha\alpha'}(\omega) {\hat
\Phi}_{\alpha'}({\bf k},{\bf J})^*.
\label{cf2}
\end{equation}
On the other hand, using Eq. (\ref{i72}), we get
\begin{equation}
\label{kl2}
\langle  {\hat
A}_{\alpha}^{\rm tot}(\omega)
{\hat
A}^{\rm tot}_{\alpha'}(\omega')^*\rangle=\sum_{\alpha''\alpha'''}
\epsilon^{-1}_{\alpha\alpha''}(\omega)
\langle {\hat
A^e}_{\alpha''}(\omega){\hat
A^e}_{\alpha'''}(\omega')^*\rangle\epsilon^{-1}_{\alpha'\alpha'''}(\omega')^*.
\end{equation}
Comparing Eqs. (\ref{i81}), (\ref{cf1}) and
(\ref{kl2}), we find that
\begin{equation}
 {P}_{\alpha\alpha'}(\omega) = \left \lbrack\epsilon^{-1} {\hat
C} (\epsilon^{-1})^{\dagger}\right
\rbrack_{\alpha\alpha'}(\omega).
\label{cf3}
\end{equation}
Substituting Eq. (\ref{cf3}) into Eq. (\ref{cf2}), we finally obtain
\begin{eqnarray}
P({\bf
k},{\bf J},\omega)=\sum_{
\alpha\alpha'} \, \hat\Phi_{\alpha}({\bf
k},{\bf J})
 \left \lbrack\epsilon^{-1} {\hat
C} (\epsilon^{-1})^{\dagger}\right
\rbrack_{\alpha\alpha'}(\omega)\, \hat\Phi_{\alpha'}({\bf
k},{\bf J})^*,
\label{i83}
\end{eqnarray}
where we have used
$(\epsilon^{-1})^{\dagger}_{\alpha\alpha'}=(\epsilon^{-1})^{*}_{
\alpha'\alpha}$. This returns the result of
\cite{weinberg,pa,fb,sdduniverse}. This equation relates the power
spectrum $P({\bf
k},{\bf J},\omega)$ of
the total fluctuating potential acting on the particles to the auto-correlation
function ${\hat
C}_{\alpha\alpha'}(\omega)$ of the external stochastic
potential. We note that ${\hat
C}_{\alpha\alpha'}(\omega)$, ${\hat C}({\bf
k},{\bf J},\omega)$, ${P}_{\alpha\alpha'}(\omega)$ and $P({\bf k},{\bf
J},\omega)$
are real and positive. The power spectrum $P({\bf
k},{\bf J},\omega)$  takes into account collective
effects through the dielectric matrix $\epsilon_{\alpha\alpha'}(\omega)$. It can
be seen
as a dressed correlation function.
If we neglect collective
effects ($\epsilon=1$) we get the bare power spectrum $P_{\rm bare}({\bf k},{\bf
J},\omega)={\hat C}({\bf k},{\bf J},\omega)$ coinciding with the correlation
function (\ref{mar5}) of the external stochastic potential.
It can be directly obtained from Eqs. (\ref{mar1b}) and
(\ref{psaab}) with $\delta\hat\Phi_{\rm tot}({\bf
k},{\bf J},\omega)={\hat \Phi}_e({\bf k},{\bf J},\omega)$. 

{\it Remark:} From Eq. (\ref{psaab}) we can write 
\begin{eqnarray}
\left\langle
\delta{\hat \Phi}_{\rm tot}({\bf k},{\bf J},t)\delta{\hat \Phi}_{\rm tot}({\bf
k},{\bf J},t')^*\right\rangle={\cal P}({\bf k},{\bf J},t-t'),
\label{an11bh}
\end{eqnarray}
where ${\cal P}({\bf k},{\bf J},t)$ is the inverse  Fourier transform
in time of $P({\bf k},{\bf J},\omega)$. We note
that
${\cal P}({\bf k},{\bf J},t)$ is complex while $P({\bf
k},{\bf J},\omega)$ is real. They satisfy the identities ${\cal P}({\bf -k},{\bf
J},t)={\cal P}({\bf k},{\bf J},t)^*={\cal P}({\bf
k},{\bf J},-t)$ and $P({\bf k},{\bf J},\omega)=P({\bf k},{\bf
J},\omega)^*=P(-{\bf k},{\bf
J},-\omega)$. The static power spectrum
$P({\bf k},{\bf J})={\cal P}({\bf k},{\bf J},0)$ is
\begin{eqnarray}
P({\bf k},{\bf J})=\int P({\bf k},{\bf
J},\omega)\,
\frac{d\omega}{2\pi}.
\label{sps2}
\end{eqnarray}

\subsection{Correlation function of the external potential created by a random
distribution of $N$ field particles}
\label{sec_inhoscf}

We now assume
that the external noise is produced by a random
distribution of $N$
field (or background) particles. We allow for different
species of particles with
masses $\lbrace m_b\rbrace$. Developing the
external potential $\Phi_e({\bf r},t)$ and the corresponding density
$\rho_e({\bf r},t)$ [related by $\Phi_e({\bf r},t)=\int u(|{\bf r}-{\bf
r}'|)\rho_e({\bf r}',t)\, d{\bf
r}'$] on a biorthogonal basis as
in Eq.
(\ref{i62}), i.e. writing $\Phi_e({\bf
r},t)=\sum_{\alpha}A_{\alpha}^e(t)\Phi_{\alpha}({\bf r})$  and $\rho_e({\bf
r},t)=\sum_{\alpha}A_{\alpha}^e(t)\rho_{\alpha}({\bf r})$, we have [see Eq.
(\ref{i64})] 
\begin{eqnarray}
A^e_{\alpha}(t)&=&-\int \rho_e({\bf
r},t) \Phi_{\alpha}^*({\bf
r})\, d{\bf r}\nonumber\\
&=&-\int f_e({\bf
r},{\bf v},t) \Phi_{\alpha}^*({\bf
r})\, d{\bf r}d{\bf v}\nonumber\\
&=&-\int f_e({\bf
w},{\bf J},t) \Phi_{\alpha}^*({\bf
w},{\bf J})\, d{\bf w}d{\bf J}.
\label{ci1}
\end{eqnarray}
To get the third line, we have use the fact that the transformation $({\bf
r},{\bf
v})\rightarrow ({\bf w},{\bf J})$ is canonical. The
discrete DF of the field particles is
\begin{eqnarray}
f_e({\bf w},{\bf J},t)&=&\sum_i m_i \delta({\bf w}-{\bf w}_i(t))\delta({\bf
J}-{\bf J}_i(t)).
\label{ci2z}
\end{eqnarray}
The initial angles and actions $({\bf w}_i,{\bf
J}_i)$ of the
field particles are assumed to be uncorrelated and
randomly distributed. The probability density of finding a field particle of
species $b$ with an action ${\bf J}$ is $P_1^{(b)}({\bf J})$.
The mean DF  of particles of species $b$ is therefore $f_b({\bf J})=N_b
m_b P_1^{(b)}({\bf J})$,
where $N_b$ is the total number of particles of species $b$. When
$N\rightarrow +\infty$ with $m\sim 1/N$, the particles follow mean field
trajectories with pulsation  ${\bf\Omega}({\bf J})$ produced by the total
DF $f=\sum_b f_b({\bf J})$. These
trajectories are straight lines in angle-action variables at constant ${\bf J}$:
\begin{eqnarray}
{\bf J}_i(t)={\bf
J}_i,\qquad {\bf w}_i(t)={\bf
w}_i+{\bf\Omega}({\bf J}_i) t.
\label{eqmot}
\end{eqnarray}
As a result, the discrete DF of the field particles can be written as
\begin{eqnarray}
f_e({\bf w},{\bf J},t)=\sum_i m_i \delta({\bf w}-{\bf
w}_i-{\bf \Omega}_i t)\delta({\bf
J}-{\bf J}_i).
\label{ci2}
\end{eqnarray}
Taking the Fourier
transform of Eq. (\ref{ci2}) with
respect to the angle ${\bf w}$ and the time $t$, we get
\begin{eqnarray}
{\hat f}_e({\bf k},{\bf J},\omega)=\frac{1}{(2\pi)^{d-1}}\sum_i m_i e^{-i{\bf
k}\cdot {\bf w}_i}\delta(\omega-{\bf k}\cdot {\bf\Omega}_i)\delta({\bf J}-{\bf
J}_i).
\label{ci3}
\end{eqnarray}
Taking the inverse Fourier transform of this expression with respect to the
wavenumber ${\bf k}$, we obtain 
\begin{eqnarray}
{\hat f}_e({\bf w},{\bf J},\omega)=\frac{1}{(2\pi)^{d-1}}\sum_i \sum_{\bf k} m_i
e^{i{\bf
k}\cdot ({\bf w}-{\bf w}_i)}\delta(\omega-{\bf k}\cdot {\bf\Omega}_i)\delta({\bf
J}-{\bf
J}_i).
\label{ci4}
\end{eqnarray}
The Fourier transform of Eq. (\ref{ci1})
reads
\begin{eqnarray}
{\hat A}^e_{\alpha}(\omega)=-\int {\hat f}_e({\bf
w},{\bf J},\omega) \Phi_{\alpha}^*({\bf
w},{\bf J})\, d{\bf w}d{\bf J}.
\label{ci1b}
\end{eqnarray}
Using Eqs. (\ref{ci4}) and (\ref{ci1b}), we find that the autocorrelation function of the Fourier components of the
external
potential is 
\begin{eqnarray}
\langle {\hat A}^e_{\alpha}(\omega)
{\hat A}^e_{\alpha'}(\omega')^*\rangle&=&\frac{1}{(2\pi)^{2d-2}}\biggl\langle
\int
\sum_{ij}
\sum_{{\bf k},{\bf k}'} m_i m_j
e^{i{\bf
k}\cdot ({\bf w}-{\bf w}_i)}e^{-i{\bf
k}'\cdot ({\bf w}'-{\bf w}_j)}\nonumber\\
&\times&\delta(\omega-{\bf k}\cdot {\bf\Omega}_i)
\delta(\omega'-{\bf k}'\cdot {\bf\Omega}_j)
\delta({\bf J}-{\bf J}_i)
\delta({\bf J}'-{\bf J}_j) \Phi_{\alpha}^*({\bf
w},{\bf J})   \Phi_{\alpha'}({\bf
w}',{\bf J}')    \, d{\bf w}d{\bf J}d{\bf w}'d{\bf J}'\biggr\rangle.
\label{ci5}
\end{eqnarray}
Integrating over ${\bf w}$, ${\bf w}'$, ${\bf J}$ and ${\bf J}'$, this expression reduces to
\begin{eqnarray}
\langle {\hat A}^e_{\alpha}(\omega)
{\hat A}^e_{\alpha'}(\omega')^*\rangle= (2\pi)^{2}\biggl\langle \sum_{ij}
\sum_{{\bf k},{\bf k}'} m_i m_j
e^{-i{\bf
k}\cdot{\bf w}_i}e^{i{\bf
k}'\cdot {\bf w}_j}\delta(\omega-{\bf k}\cdot {\bf\Omega}_i)
\delta(\omega'-{\bf k}'\cdot {\bf\Omega}_j) {\hat \Phi}_{\alpha}^*({\bf
k},{\bf J}_i)   {\hat \Phi}_{\alpha'}({\bf
k}',{\bf J}_j) \biggr\rangle.
\label{ci6pa}
\end{eqnarray}
We then get
\begin{eqnarray}
\langle {\hat A}^e_{\alpha}(\omega)
{\hat A}^e_{\alpha'}(\omega')^*\rangle
&=& (2\pi)^{2}\biggl\langle  \sum_{i}
\sum_{{\bf k},{\bf k}'} m_i^2
e^{-i({\bf
k}-{\bf k}')\cdot{\bf w}_i}
\delta(\omega-{\bf k}\cdot {\bf\Omega}_i)
\delta(\omega'-{\bf k}'\cdot {\bf\Omega}_i) {\hat \Phi}_{\alpha}^*({\bf
k},{\bf J}_i)   {\hat \Phi}_{\alpha'}({\bf
k}',{\bf J}_i)\biggr\rangle\nonumber\\
&=& (2\pi)^{2}\sum_b N_b m_b^2
\sum_{{\bf k},{\bf k}'}\int
e^{-i({\bf
k}-{\bf k}')\cdot{\bf w}}\delta(\omega-{\bf k}\cdot {\bf\Omega})
\delta(\omega'-{\bf k}'\cdot {\bf\Omega}) {\hat \Phi}_{\alpha}^*({\bf
k},{\bf J})   {\hat \Phi}_{\alpha'}({\bf
k}',{\bf J}) P_1^{(b)}({\bf J})\, d{\bf w}d{\bf J},\nonumber\\
&=& (2\pi)^{2}\sum_b
\sum_{{\bf k},{\bf k}'}\int m_b
e^{-i({\bf
k}-{\bf k}')\cdot{\bf w}}\delta(\omega-{\bf k}\cdot {\bf\Omega})
\delta(\omega'-{\bf k}'\cdot {\bf\Omega}) {\hat \Phi}_{\alpha}^*({\bf
k},{\bf J})   {\hat \Phi}_{\alpha'}({\bf
k}',{\bf J}) f_b({\bf J})\, d{\bf w}d{\bf J}.
\label{ci6}
\end{eqnarray}
To obtain the first line, we have used the 
decomposition $\sum_{ij}=\sum_i+\sum_{i\neq j}$ and the fact that the terms
involving different particles  ($i\neq j$) vanish in average since the particles
are initially uncorrelated. To obtain the second line, we have
used the fact that the particles of the same species are
identical. Then, integrating over ${\bf w}$
and summing over ${\bf k}'$, we get
\begin{eqnarray}
\langle {\hat A}^e_{\alpha}(\omega)
{\hat A}^e_{\alpha'}(\omega')^*\rangle
&=& (2\pi)^{d+2}
\sum_b\sum_{{\bf k}} \int m_b \delta(\omega-{\bf k}\cdot {\bf\Omega})
\delta(\omega'-{\bf k}\cdot {\bf\Omega}) {\hat \Phi}_{\alpha}^*({\bf
k},{\bf J})   {\hat \Phi}_{\alpha'}({\bf
k},{\bf J}) f_b({\bf J})\, d{\bf J}\nonumber\\
&=& (2\pi)^{d+2} \delta(\omega-\omega')
\sum_b\sum_{{\bf k}} \int m_b \delta(\omega-{\bf k}\cdot {\bf\Omega})
{\hat \Phi}_{\alpha}^*({\bf
k},{\bf J})   {\hat \Phi}_{\alpha'}({\bf
k},{\bf J}) f_b({\bf J})\, d{\bf J}.
\label{ci9}
\end{eqnarray}
Comparing Eqs.  (\ref{i81}) and (\ref{ci9}), we find that the temporal
Fourier transform of
the auto-correlation function of the external stochastic potential is
\begin{eqnarray}
{\hat C}_{\alpha\alpha'}(\omega)
= (2\pi)^{d+1}\sum_b m_b
\sum_{{\bf k}}\int \delta(\omega-{\bf k}\cdot {\bf\Omega})
{\hat \Phi}_{\alpha}^*({\bf
k},{\bf J})   {\hat \Phi}_{\alpha'}({\bf
k},{\bf J}) f_b({\bf J})\, d{\bf J}.
\label{ci10}
\end{eqnarray}
Substituting Eq. (\ref{ci10}) into  Eq. (\ref{mar5}) and
using
Eq. (\ref{pi37}) we obtain
\begin{eqnarray}
{\hat C}({\bf k},{\bf J},\omega)=(2\pi)^{d+1}\sum_b
m_b\sum_{{\bf k}'}\int d{\bf
J}'\, |A_{{\bf k},{\bf k}'}({\bf J},{\bf
J'})|^2 \delta(\omega-{\bf k}'\cdot {\bf\Omega}')f_b({\bf J}'),
\label{pi5myl}
\end{eqnarray}
where ${\bf\Omega}'$ stands for ${\bf\Omega}({\bf J}')$.
This is the correlation function of the external potential created by
the
background particles. Integrating over $\omega$, we obtain the static
correlation function of the external potential 
\begin{eqnarray}
{\hat C}_{\alpha\alpha'}
= (2\pi)^{d}\sum_b m_b
\sum_{{\bf k}}\int 
{\hat \Phi}_{\alpha}^*({\bf
k},{\bf J})   {\hat \Phi}_{\alpha'}({\bf
k},{\bf J}) f_b({\bf J})\, d{\bf J},
\end{eqnarray}
\begin{eqnarray}
{\hat C}({\bf k},{\bf J})=(2\pi)^{d}\sum_b
m_b\sum_{{\bf k}'}\int d{\bf
J}'\, |A_{{\bf k},{\bf k}'}({\bf J},{\bf
J'})|^2 f_b({\bf J}').
\label{tw}
\end{eqnarray}

The power spectrum from Eq. (\ref{cf3}) can be written in
components
form as
\begin{eqnarray}
{P}_{\alpha\alpha'}(\omega)=\sum_{\alpha''\alpha'''} 
 (\epsilon^{-1})_{\alpha\alpha''}(\omega) {\hat
C}_{\alpha''\alpha'''}(\omega)
(\epsilon^{\dagger})_{\alpha'''\alpha'}^{-1}(\omega).
\label{pi1n}
\end{eqnarray}
Substituting Eq. (\ref{ci10}) into Eq. (\ref{pi1n}) we obtain
the
dressed power
spectrum of the total
fluctuating potential created by a random distribution  of $N$ field particles
\begin{eqnarray}
{P}_{\alpha\alpha'}(\omega)&=&(2\pi)^{d+1}\sum_b m_b
\sum_{\alpha''\alpha'''}\sum_{{\bf k}}\int \,
 (\epsilon^{-1})_{\alpha\alpha''}(\omega)
{\hat \Phi}_{\alpha''}^*({\bf
k},{\bf J})  {\hat \Phi}_{\alpha'''}({\bf
k},{\bf J})
(\epsilon^{\dagger})_{\alpha'''\alpha'}^{-1}(\omega)\,
\delta(\omega-{\bf k}\cdot {\bf\Omega})f_b({\bf J})\, d{\bf
J}\nonumber\\
&=&(2\pi)^{d+1}\sum_b m_b \sum_{\alpha''\alpha'''}\sum_{{\bf k}}\int \,
 (\epsilon^{-1})_{\alpha\alpha''}(\omega)
{\hat \Phi}_{\alpha''}^*({\bf
k},{\bf J}) (\epsilon^{-1})^*_{\alpha'\alpha'''}(\omega)\,
 {\hat \Phi}_{\alpha'''}({\bf
k},{\bf J})
 \delta(\omega-{\bf k}\cdot {\bf\Omega})f_b({\bf J})\, d{\bf
J}\nonumber\\
&=& (2\pi)^{d+1}\sum_b m_b
\sum_{{\bf k}}\int \delta(\omega-{\bf k}\cdot {\bf\Omega})
\left\lbrack \epsilon^{-1}(\omega){\hat \Phi}^*({\bf
k},{\bf J})\right\rbrack_{\alpha}  \left\lbrack \epsilon^{-1}(\omega)^*{\hat
\Phi}({\bf
k},{\bf J})\right\rbrack_{\alpha'}f_b({\bf J})\,
d{\bf J}.
\label{pi2n}
\end{eqnarray}
Substituting Eq. (\ref{pi2n}) into Eq. (\ref{cf2}) and using Eq. (\ref{can1b})
we obtain
\begin{eqnarray}
P({\bf
k},{\bf J},\omega)=(2\pi)^{d+1}\sum_b m_b\sum_{{\bf k}'}\int d{\bf
J}'\,
|A^d_{{\bf k},{\bf k}'}({\bf J},{\bf
J'},\omega)|^2 \delta(\omega-{\bf k}'\cdot {\bf\Omega}')f_b({\bf J}').
\label{pi4}
\end{eqnarray}
This returns the expression  (43) of the power spectrum
given in Ref. \cite{physicaA} which was obtained from the Klimontovich
formalism (see also the Remark at the end of Appendix
\ref{sec_jbl}). The
present
approach provides an alternative, more physical,
manner to derive this result. Integrating over $\omega$, we
obtain the static power spectrum
\begin{eqnarray}
{P}_{\alpha\alpha'}
= (2\pi)^{d}\sum_b m_b
\sum_{{\bf k}}\int
\left\lbrack \epsilon^{-1}({\bf k}\cdot {\bf\Omega}){\hat \Phi}^*({\bf
k},{\bf J})\right\rbrack_{\alpha}  \left\lbrack \epsilon^{-1}({\bf
k}\cdot {\bf\Omega})^*{\hat
\Phi}({\bf
k},{\bf J})\right\rbrack_{\alpha'}f_b({\bf J})\,
d{\bf J},
\label{paas}
\end{eqnarray}
\begin{eqnarray}
P({\bf
k},{\bf J})=(2\pi)^{d}\sum_b m_b\sum_{{\bf k}'}\int d{\bf
J}'\,
|A^d_{{\bf k},{\bf k}'}({\bf J},{\bf
J'},{\bf k}'\cdot {\bf\Omega}')|^2 f_b({\bf J}').
\label{pi4stat}
\end{eqnarray}

If we neglect the collective effects, we find that the bare correlation function
of the total
fluctuating potential produced by a random
distribution of field particles is given by $P_{\rm bare}({\bf
k},{\bf J},\omega)={\hat C}({\bf k},{\bf J},\omega)$ with Eq. (\ref{pi5myl}).
It can be obtained from Eq. (\ref{pi4}) by replacing the dressed potential of
interaction (\ref{can1b}) by the bare potential of interaction (\ref{pi37}).
Similarly, the bare static power spectrum is given by $P_{\rm bare}({\bf
k},{\bf J})={\hat C}({\bf k},{\bf J})$ with Eq. (\ref{tw}).

\subsection{Correlation function of the external potential created by a
random distribution of $N$ field particles without using the
biorthogonal basis}
\label{sec_ciw}

The correlation function of the external
potential created by a
random distribution of field particles can be calculated without using the
biorthogonal basis. According to Eq.
(\ref{bof3b}),
the
autocorrelation function of the Fourier
components of the
external
potential is 
\begin{eqnarray}
\langle {\hat \Phi}_{e}({\bf k},{\bf J},\omega) {\hat \Phi}_{e}({\bf k},{\bf
J},\omega')^*\rangle=(2\pi)^{2d}\sum_{{\bf k}'{\bf k}''}\int d{\bf J}' d{\bf
J}''\,  A_{{\bf k}{\bf k}'}({\bf J},{\bf J}') A_{{\bf k}{\bf k}''}({\bf J},{\bf
J}'')^* \langle {\hat f}_{e}({\bf k}',{\bf J}',\omega) {\hat f}_{e}({\bf
k}'',{\bf J}'',\omega')^*\rangle.
\label{ciw4}
\end{eqnarray}
If the external noise is produced by a random distribution of field particles,
using Eq. (\ref{ci3}), we find
\begin{eqnarray}
\langle {\hat f}_e({\bf k},{\bf J},\omega) {\hat f}_e({\bf k}',{\bf
J}',\omega')^*\rangle=\frac{1}{(2\pi)^{2(d-1)}}\left\langle \sum_{ij} m_i m_j
e^{-i{\bf
k}\cdot {\bf w}_i}e^{i{\bf
k}'\cdot {\bf w}_j}\delta(\omega-{\bf k}\cdot {\bf\Omega}_i)\delta(\omega'-{\bf
k}'\cdot {\bf\Omega}_j)\delta({\bf J}-{\bf
J}_i)\delta({\bf J}'-{\bf
J}_j)\right\rangle\nonumber\\
=\frac{1}{(2\pi)^{2(d-1)}}\sum_b  m_b \int d{\bf w}_1d{\bf J}_1\, f_b({\bf
J}_1)
e^{-i({\bf
k}-{\bf k}')\cdot {\bf w}_1}\delta(\omega-{\bf k}\cdot
{\bf\Omega}_1)\delta(\omega'-{\bf k}'\cdot {\bf\Omega}_1)\delta({\bf J}-{\bf
J}_1)\delta({\bf J}'-{\bf
J}_1)\nonumber\\
=\frac{1}{(2\pi)^{d-2}}\sum_b  m_b  f_b({\bf J}) \delta_{{\bf
k},{\bf k}'}\delta(\omega-\omega')\delta(\omega-{\bf k}\cdot
{\bf\Omega})\delta({\bf J}-{\bf
J}'),\quad
\label{ciw5}
\end{eqnarray}
where we have used the fact that the particles are initially 
uncorrelated and that the particles of the same species are identical.
Substituting Eq. (\ref{ciw5}) into Eq. (\ref{ciw4}) we obtain
\begin{eqnarray}
\langle {\hat \Phi}_{e}({\bf k},{\bf J},\omega) {\hat \Phi}_{e}({\bf k},{\bf
J},\omega')^*\rangle=(2\pi)^{d+2}\sum_b m_b\sum_{{\bf k}'}\int d{\bf J}' \, 
A_{{\bf
k}{\bf k}'}({\bf J},{\bf J}') A_{{\bf k}{\bf k}'}({\bf J},{\bf J}')^*f_b({\bf
J}')\delta(\omega-\omega')\delta(\omega-{\bf k}'\cdot {\bf\Omega}').
\label{ciw6}
\end{eqnarray}
Comparing this expression with Eq. (\ref{mar1b}) we get
\begin{eqnarray}
{\hat C}({\bf k},{\bf J},\omega)=(2\pi)^{d+1}\sum_b m_b\sum_{{\bf k}'}\int
d{\bf
J}' \,  |A_{{\bf k}{\bf k}'}({\bf J},{\bf J}')|^2 \delta(\omega-{\bf
k}'\cdot {\bf\Omega}')f_b({\bf
J}'),
\label{ciw7}
\end{eqnarray}
returning Eq. (\ref{pi5myl}). If we neglect collective effects, we have $P_{\rm
bare}({\bf
k},{\bf
J},\omega)={\hat C}({\bf k},{\bf J},\omega)$.

{\it Remark:} If we start from Eq. (\ref{ciw7}) and decompose $A_{{\bf k}{\bf
k}'}({\bf J},{\bf
J}')$ on the biorthogonal basis using Eq.  (\ref{pi37}) we obtain
Eq. (\ref{mar5}) with Eq. (\ref{ci10}).

\subsection{Power spectrum of the total fluctuating potential created by a
random distribution of $N$ field particles without using the biorthogonal basis}
\label{sec_ciwtay}

The power spectrum of the total fluctuating potential created by a
random distribution of field particles can be calculated similarly without using
the biorthogonal basis. According to Eq. (\ref{dn6b}), we have
\begin{equation}
\langle \delta{\hat \Phi}_{\rm tot}({\bf k},{\bf J},\omega) \delta{\hat
\Phi}_{\rm tot}({\bf k},{\bf
J},\omega')^*\rangle=(2\pi)^{2d}\sum_{{\bf k}'{\bf k}''}\int d{\bf J}' d{\bf
J}''\,  A^d_{{\bf k}{\bf k}'}({\bf J},{\bf J}',\omega) A^d_{{\bf k}{\bf
k}''}({\bf J},{\bf
J}'',\omega')^* \langle {\hat f}_{e}({\bf k}',{\bf J}',\omega) {\hat f}_{e}({\bf
k}'',{\bf J}'',\omega')^*\rangle.
\label{ciw4tay}
\end{equation}
If the external noise is produced by a random distribution of field
particles,
using Eq. (\ref{ciw5}), we obtain
\begin{equation}
\langle \delta{\hat \Phi}_{\rm tot}({\bf k},{\bf J},\omega) \delta{\hat
\Phi}_{\rm tot}({\bf k},{\bf
J},\omega')^*\rangle=(2\pi)^{d+2}\sum_b m_b\sum_{{\bf k}'}\int d{\bf J}' \, 
A^d_{{\bf
k}{\bf k}'}({\bf J},{\bf J}',\omega) A^d_{{\bf k}{\bf k}'}({\bf J},{\bf
J}',\omega)^*f_b({\bf
J}')\delta(\omega-\omega')\delta(\omega-{\bf k}'\cdot {\bf\Omega}').
\label{ciw6tay}
\end{equation}
Comparing this expression with Eq. (\ref{psaab}) we get
\begin{eqnarray}
P({\bf k},{\bf J},\omega)=(2\pi)^{d+1}\sum_b m_b\sum_{{\bf
k}'}\int
d{\bf
J}' \,  |A^d_{{\bf k}{\bf k}'}({\bf J},{\bf J}',\omega)|^2 \delta(\omega-{\bf
k}'\cdot {\bf\Omega}')f_b({\bf
J}'),
\label{ciw7tay}
\end{eqnarray}
returning Eq. (\ref{pi4}).

{\it Remark:} If we start from Eq. (\ref{ciw7tay}) and decompose $A^d_{{\bf
k}{\bf
k}'}({\bf J},{\bf
J}',\omega)$ on the biorthogonal basis using Eq.  (\ref{can1b}) we obtain
Eq. (\ref{cf2}) with Eq. (\ref{pi2n}).

\subsection{Energy of fluctuations}
\label{sec_ieof}

The energy of fluctuations
\begin{equation}
{\cal E}=\frac{1}{2}\int \langle \delta\rho_{\rm tot}\delta\Phi_{\rm
tot}\rangle\, d{\bf r},
\label{ieof1}
\end{equation}
where $\delta\rho_{\rm tot}=\delta\rho+\rho_e$ and $\delta\Phi_{\rm
tot}=\delta\Phi+\Phi_e$ are the total fluctuations of density and potential, can
be calculated as follows.
Decompositing the fluctuations of density and
potential on the biorthogonal basis, using Eq. (\ref{i62}), we get
\begin{equation}
{\cal E}=\frac{1}{2}\sum_{\alpha\alpha'} \langle
A_{\alpha}^{\rm tot}(t)A_{\alpha'}^{\rm tot}(t)^*\rangle \int
\rho_{\alpha}({\bf r})\Phi_{\alpha'}^*({\bf r})\, d{\bf r}.
\label{ieof2}
\end{equation}
Using the orthonormalization condition from Eq. (\ref{i63}), we find that
\begin{equation}
{\cal E}=-\frac{1}{2}\sum_{\alpha} \langle
|A_{\alpha}^{\rm tot}(t)|^2\rangle.
\label{ieof3}
\end{equation}
Decomposing the amplitudes of the total potential in temporal Fourier modes, we
can
rewrite the
foregoing equation as
\begin{equation}
{\cal E}=-\frac{1}{2}\sum_{\alpha}
\int\frac{d\omega}{2\pi}\frac{d\omega'}{2\pi} e^{-i\omega t}
e^{i\omega't} \langle
{\hat A}_{\alpha}^{\rm tot}(\omega){\hat A}_{\alpha}^{\rm
tot}(\omega')^*\rangle.
\label{ieof3b}
\end{equation}
Introducing the power spectrum tensor from Eq. (\ref{cf1}) and integrating over
$\omega'$, we obtain 
\begin{equation}
{\cal E}=-\frac{1}{2}\sum_{\alpha}\int
\frac{d\omega}{2\pi}P_{\alpha\alpha}(\omega)=-\frac{1}{2}\sum_{\alpha}P_{
\alpha\alpha}.
\label{ieof4}
\end{equation}
Finally, using Eq. (\ref{cf3}), we can rewrite the energy of fluctuations as 
\begin{equation}
{\cal E}=-\frac{1}{2}\sum_{\alpha}\int
\frac{d\omega}{2\pi}\left\lbrack \epsilon^{-1}{\hat
C}(\epsilon^{-1})^{\dagger}\right\rbrack_{\alpha\alpha}(\omega).
\label{ieof5}
\end{equation}
When the perturbation is due to a random distribution of $N$ particles, using
Eqs. (\ref{paas}) and (\ref{ieof4}), we obtain
\begin{eqnarray}
{\cal E}=-\frac{1}{2}(2\pi)^d\sum_b m_b \sum_{\bf k}\int
|\epsilon^{-1}({\bf k}\cdot {\bf\Omega}){\hat\Phi}^*({\bf k},{\bf J})|^2
f_b({\bf J})\, d{\bf J},
\end{eqnarray}
where $|\epsilon^{-1}({\bf k}\cdot {\bf\Omega}){\hat\Phi}^*({\bf
k},{\bf J})|^2=\sum_{\alpha}\lbrack \epsilon^{-1}({\bf k}\cdot
{\bf\Omega}){\hat \Phi}^*({\bf
k},{\bf J})\rbrack_{\alpha}  \lbrack \epsilon^{-1}({\bf
k}\cdot {\bf\Omega})^*{\hat
\Phi}({\bf
k},{\bf J})\rbrack_{\alpha}$. We note that ${\cal E}$ is constant in time. If we
neglect collective effects, using Eq. (\ref{fdi13}), the
energy of fluctuations 
can be written as 
\begin{eqnarray}
{\cal E}=\frac{1}{2}(2\pi)^d\sum_b m_b \sum_{\bf k}\int
A_{{\bf k},{\bf k}}({\bf J},{\bf J})
f_b({\bf J})\, d{\bf J}.
\end{eqnarray}

\section{Fluctuation-dissipation theorem for an isolated system of particles at
statistical equilibrium}
\label{sec_fdi}

\subsection{Fluctuation-dissipation theorem}
\label{sec_fdia}

The fluctuation-dissipation theorem for an inhomogeneous system of particles
with long-range interactions can be written as
\begin{eqnarray}
 P({\bf
k},{\bf J},\omega)=-\frac{2k_B T}{\omega} {\rm Im}\left\lbrack\chi_{{\bf k},{\bf
k}}({\bf J},{\bf
J},\omega)\right\rbrack.
\label{fdi15}
\end{eqnarray}
It relates the power spectrum $P({\bf
k},{\bf J},\omega)$ of the fluctuations  to 
the response function $\chi_{{\bf k},{\bf
k}}({\bf J},{\bf
J},\omega)$  of a system of particles at statistical
equilibrium with the temperature $T$.  Using Eq. (\ref{fdi9b}) and noting
that $A_{{\bf k},{\bf k}}({\bf J},{\bf J})$ is real (see Appendix
\ref{sec_bdpi}), 
the fluctuation-dissipation theorem  can be rewritten
as\footnote{The left hand side of Eq. (\ref{fdi5}) measures the
correlations of the total fluctuating  potential. The fact that the right hand
side of
Eq. (\ref{fdi5}) is related to the dissipation will be clear from the expression
of
the
friction by polarization established below [see Eq. (\ref{ci1bnh})].}
\begin{eqnarray}
 P({\bf
k},{\bf J},\omega)=-\frac{2k_B T}{\omega} {\rm Im}\left\lbrack A^d_{{\bf
k},{\bf k}}({\bf J},{\bf
J},\omega)\right\rbrack.
\label{fdi5}
\end{eqnarray}
Although the
fluctuation-dissipation theorem can be established from very general
arguments
\cite{nyquist,welton,green,greene1,greene2,green1,green2,kubo0,kubo2,kubo},
we shall derive Eq. (\ref{fdi5}) directly from the results obtained in the
preceding sections.

We start from the
general identity (see Eq. (54) in \cite{physicaA}) 
\begin{eqnarray}
{\rm Im}\left\lbrack A^d_{{\bf k},{\bf k}}({\bf J},{\bf
J},\omega)\right\rbrack=\pi (2\pi)^{d}\sum_{{\bf k}'}\int d{\bf
J}'\, |A^d_{{\bf k},{\bf k}'}({\bf J},{\bf
J'},\omega)|^2 \delta(\omega-{\bf k}'\cdot {\bf\Omega}')\left ({\bf
k}'\cdot \frac{\partial f'}{\partial {\bf J}'}\right ).
\label{fdi2a}
\end{eqnarray}
For a multi-species system of particles, it can be rewritten as
\begin{eqnarray}
{\rm Im}\left\lbrack A^d_{{\bf k},{\bf k}}({\bf J},{\bf
J},\omega)\right\rbrack
=\pi
(2\pi)^{d}\sum_b\sum_{{\bf k}'}\int d{\bf
J}'\, |A^d_{{\bf k},{\bf k}'}({\bf J},{\bf
J'},\omega)|^2 \delta(\omega-{\bf k}'\cdot {\bf\Omega}')\left ({\bf
k}'\cdot \frac{\partial f'_b}{\partial {\bf J}'}\right ).
\label{fdi2}
\end{eqnarray}
We assume that the particles are at statistical equilibrium with the Boltzmann
distribution
\begin{eqnarray}
f_b({\bf J})=A_b\, e^{-\beta m_b H({\bf J})},
\label{fdi3voit}
\end{eqnarray}
where $H({\bf J})$ is the individual energy of the particles
by unit of mass. In that case, we have the
identity
\begin{eqnarray}
\frac{\partial f_b}{\partial {\bf J}}=-\beta m_b f_b({\bf J}){\bf\Omega}({\bf
J}),
\label{fdi3}
\end{eqnarray}
where we have used  $\partial H/\partial {\bf J}={\bf\Omega}({\bf
J})$. Substituting Eq. (\ref{fdi3}) into Eq. (\ref{fdi2}), we obtain
\begin{eqnarray}
{\rm Im}\left\lbrack A^d_{{\bf k},{\bf k}}({\bf J},{\bf
J},\omega)\right\rbrack&=&-\beta\pi (2\pi)^{d} \sum_b m_b\sum_{{\bf k}'}\int
d{\bf
J}'\,
|A^d_{{\bf k},{\bf k}'}({\bf J},{\bf
J'},\omega)|^2 \delta(\omega-{\bf k}'\cdot {\bf\Omega}')\left ({\bf
k}'\cdot {\bf\Omega}'\right )f_b({\bf J}')\nonumber\\
&=&-\beta \omega\pi (2\pi)^{d}\sum_b m_b \sum_{{\bf k}'}\int d{\bf
J}'\, |A^d_{{\bf k},{\bf k}'}({\bf J},{\bf
J'},\omega)|^2 \delta(\omega-{\bf k}'\cdot {\bf\Omega}')f_b({\bf
J}')\nonumber\\
&=&-\frac{1}{2}\beta \omega P({\bf
k},{\bf J},\omega).
\label{fdi4}
\end{eqnarray}
To get the second line we have used the property of the $\delta$-function and to
get the third line we have used the expression (\ref{pi4}) of the power
spectrum. This establishes Eq. (\ref{fdi5}), which
is only valid for the Boltzmann distribution. We refer to
Nelson and Tremaine \cite{nt99} for an alternative derivation of the
fluctuation-dissipation theorem for inhomogeneous self-gravitating systems and
for further discussions.

\subsection{Static power spectrum}
\label{sec_sps}

At statistical equilibrium, we have the following relation
\begin{eqnarray}
P({\bf k},{\bf J})=-k_B T   \chi_{{\bf k},{\bf k}}({\bf J},{\bf
J})
\label{sps6}
\end{eqnarray}
between the static  power spectrum $P({\bf k},{\bf J})=\frac{1}{2\pi}\int P({\bf
k},{\bf J},\omega)\, d\omega$ of
the fluctuations (see Sec. \ref{sec_cf})  and
the static response
function $\chi_{{\bf k},{\bf k}}({\bf J},{\bf
J})=\chi_{{\bf k},{\bf k}}({\bf J},{\bf
J},\omega=0)$ (see Sec. \ref{sec_inhosrf}). This general  relation can be
derived
directly 
from the microcanonical distribution of the particles at statistical
equilibrium. It can also be recovered from the fluctuation-dissipation theorem
(\ref{fdi15}) as follows.

Integrating Eq.
(\ref{fdi15}) between $-\infty$ and $+\infty$, we get
\begin{eqnarray}
P({\bf k},{\bf J})=-k_B T\int_{-\infty}^{+\infty}
\frac{{\rm Im}\left\lbrack\chi_{{\bf k},{\bf
k}}({\bf J},{\bf
J},\omega)\right\rbrack}{\pi\omega}\, d\omega.
\label{sps3}
\end{eqnarray}
On the other hand, applying at $\omega=0$ the Kramers-Kronig relation
\cite{kram,kron}
\begin{eqnarray}
\label{sps4}
\chi_{{\bf k},{\bf k}}({\bf J},{\bf
J},\omega)={\rm P} \frac{1}{\pi}\int_{-\infty}^{+\infty}
\frac{{\rm
Im}\left\lbrack\chi_{{\bf k},{\bf k}}({\bf J},{\bf J},\omega')
\right\rbrack}{\omega'-\omega}\, d\omega',
\end{eqnarray}
where P is the principal value, we get
\begin{eqnarray}
\label{sps5}
\chi_{{\bf k},{\bf k}}({\bf J},{\bf
J})=\frac{1}{\pi}\int_{-\infty}^{+\infty}
\frac{{\rm
Im}\left\lbrack\chi_{{\bf k},{\bf k}}({\bf J},{\bf J},\omega)
\right\rbrack}{\omega}\, d\omega.
\end{eqnarray}
Combining Eqs. (\ref{sps3}) and (\ref{sps5}) we obtain Eq. (\ref{sps6}).

\section{Fokker-Planck equation}
\label{sec_fp}

Let us consider the evolution of a test particle of mass $m$ moving in a
spatially
inhomogeneous medium and
experiencing a stochastic perturbation $\Phi_e({\bf r},t)$. The
equations of motion of the test particle, written with angle-action variables,
are 
\begin{equation}
\label{fp1b}
\frac{d{\bf w}}{dt}={\bf\Omega}({\bf
J})+\frac{\partial \delta\Phi_{\rm
tot}}{\partial
{\bf J}}({\bf w},{\bf J},t),\qquad \frac{d{\bf
J}}{dt}=-\frac{\partial\delta\Phi_{\rm
tot}}{\partial {\bf
w}}({\bf w},{\bf J},t),
\end{equation}
where $\delta\Phi_{\rm tot}({\bf w},{\bf J},t)$ is the total
fluctuating
potential acting on the
particle. They can be written in Hamiltonian form as $\dot{\bf
w}=\partial (H+\delta H_{\rm tot})/\partial {\bf J}$ and $\dot{\bf
J}=-\partial (H+\delta H_{\rm tot})/\partial {\bf w}$, where $H$
is the mean
Hamiltonian and $\delta H_{\rm tot}$ is the total fluctuating
Hamiltonian. At leading order, the test particle moves on an orbit characterized
by a constant action ${\bf J}$ and a pulsation ${\bf \Omega}({\bf
J})={\partial H}/{\partial {\bf J}}$ (mean field motion) but it also
experiences
a small stochastic perturbation $\delta\Phi_{\rm tot}=\Phi_e+\delta\Phi$  which
is equal
to the external potential  $\Phi_e$ plus the fluctuating potential $\delta\Phi$
produced by the
system itself
(collective effects).  Eq. (\ref{fp1b}) can be formally integrated
into 
\begin{equation}
\label{fp2b}
{\bf w}(t)={\bf w}+\int_{0}^{t}{\bf\Omega}({\bf J}(t'))\,
dt'+\int_{0}^{t}\frac{\partial\delta\Phi_{\rm tot}}{\partial {\bf J}}({\bf
w}(t'),{\bf
J}(t'),t')\, dt',
\end{equation}
\begin{equation}
\label{fp3}
{\bf J}(t)={\bf J}-\int_{0}^{t}\frac{\partial\delta\Phi_{\rm tot}}{\partial {\bf
w}}({\bf
w}(t'),{\bf J}(t'),t')\, dt',
\end{equation}
where we have assumed that, initially, the test particle has an angle  ${\bf w}$
and an action ${\bf
J}$. Since the fluctuations
$\delta\Phi_{\rm tot}$ of the potential are small, the changes in the action of
the test particle are also small. On the other hand, the fluctuation time
is short with
respect to the evolution time of the DF. As a result, the
dynamics of the test
particle can
be represented by a stochastic process governed by a Fokker-Planck equation
\cite{chandrabrown,risken}.   The
Fokker-Planck equation  can be derived from the Master
equation by using the
Kramers-Moyal \cite{kramers,moyal} expansion truncated at the level of the
second moments of the
increment in action. If we denote by $P({\bf J},t)$ the  probability density
that the
test particle has an action ${\bf J}$ at time $t$, the general form
of this
equation is
\begin{equation}
\label{fp5}
\frac{\partial P}{\partial t}=\frac{\partial^{2}}{\partial J_{i}\partial
J_{j}}\left (D_{ij}P\right )-\frac{\partial}{\partial J_{i}}\left
(PF^{\rm tot}_{i}\right
).
\end{equation}
The diffusion tensor and the friction force are defined by
\begin{equation}
\label{fp6}
D_{ij}({\bf J})=\lim_{t\rightarrow +\infty}\frac{1}{2t} \langle
(J_{i}(t)-J_{i})(J_{j}(t)-J_{j})\rangle=\frac{\langle\Delta J_i\Delta J_j\rangle}{2\Delta t},
\end{equation}
\begin{equation}
\label{fp7}
F^{\rm tot}_{i}({\bf J})=\lim_{t\rightarrow +\infty}\frac{1}{t} \langle
J_{i}(t)-J_{i}\rangle=\frac{\langle\Delta J_i\rangle}{\Delta t}.
\end{equation}
In writing these limits, we have implicitly assumed that the time $t$ is long
compared  to the fluctuation time but short compared to the evolution
time of the DF.

As shown in our previous paper
\cite{physicaA}, it is relevant to rewrite the Fokker-Planck
equation in the alternative form
\begin{equation}
\label{fp8}
\frac{\partial P}{\partial t}=\frac{\partial}{\partial J_{i}} \left
(D_{ij}\frac{\partial P}{\partial
J_{j}}- P F_i^{\rm pol}\right ).
\end{equation}
The total friction can be written as
\begin{equation}
\label{fp9}
F_{i}^{\rm tot}=F_{i}^{\rm pol}+\frac{\partial D_{ij}}{\partial J_{j}},
\end{equation}
where ${\bf F}_{\rm pol}$ is the friction by polarization (see Sec. 3.3 of
\cite{physicaA} and Sec.
\ref{sec_ifpol} of the present paper), while
the second term
is due to the
variations of the diffusion tensor with ${\bf J}$ (see Sec. 3.4
of
\cite{physicaA}). The friction by
polarization ${\bf F}_{\rm pol}$ arises from the retroaction (response) of the
system to the perturbation (wake)
caused by the test particle, just like in a polarization process. It
represents, however, only one
component of the dynamical friction ${\bf F}_{\rm tot}$ experienced by the test
particle, the other
component being $\partial D_{ij}/{\partial J_{j}}$.

{\it Remark:} The two
expressions (\ref{fp5}) and (\ref{fp8}) of the
Fokker-Planck equation  have their own interest. The
expression (\ref{fp5}) where the diffusion tensor is placed after the
second derivative $\partial^2(DP)$ involves the total friction ${\bf
F}_{\rm tot}$ and the expression (\ref{fp8}) where the diffusion tensor is
``sandwiched'' between the derivatives $\partial D\partial P$ isolates the
friction by polarization ${\bf F}_{\rm pol}$.  We shall see that this second
form is directly related to the
Lenard-Balescu equation and the SDD equation. It has
therefore a clear physical meaning.

\section{Diffusion tensor}
\label{sec_diffco}

\subsection{General expression of the diffusion tensor}
\label{sec_gei}

We now calculate the diffusion tensor 
from Eq. (\ref{fp6}) following the approach developed in Sec.
3.2 of \cite{physicaA}. According to Eq. (\ref{fp1b}) the
increment in action of the test particle is
\begin{equation}
\Delta {\bf J}=-\int_0^t\frac{\partial\delta\Phi_{\rm tot}}{\partial
{\bf w}}({\bf w}(t'),{\bf J}(t'),t')\, dt'.
\label{delty}
\end{equation}
Substituting Eq. (\ref{delty}) into Eq. (\ref{fp6}) and assuming that the
correlations of the fluctuating force persist for a time less than the
time for the trajectory of the test particle to be much altered, we can make a
linear trajectory approximation in angle-action space (i.e. the
test particle follows an unperturbed orbit)
\begin{equation}
{\bf w}(t')={\bf w}+{\bf\Omega} t',\qquad  {\bf J}(t')={\bf J},
\label{unp}
\end{equation}
and write
\begin{equation}
D_{ij}=\lim_{t\rightarrow +\infty}\frac{1}{2t}\int_{0}^{t}dt'\int_{0}^{t}dt'' \,
\left\langle
\frac{\partial\delta\Phi_{\rm tot}}{\partial w_i}({\bf w}+{\bf\Omega}t', {\bf J},
t')\frac{\partial\delta\Phi_{\rm tot}}{\partial w_j}({\bf w}+{\bf\Omega}t'', {\bf J},
t'')\right\rangle.
\label{diff1}
\end{equation}
Introducing
the Fourier transform of the total fluctuating potential, we
obtain
\begin{eqnarray}
\left\langle \frac{\partial\delta\Phi_{\rm tot}}{\partial w_i}({\bf w}+{\bf\Omega}t', {\bf
J}, t')\frac{\partial\delta\Phi_{\rm tot}}{\partial w_j}({\bf w}+{\bf\Omega}t'', {\bf J},
t'')\right\rangle=\sum_{\bf k}\int_{\cal C}\frac{d\omega}{2\pi}\sum_{{\bf
k}'}\int_{\cal C}\frac{d\omega'}{2\pi} \, k_i k'_j e^{i{\bf k}\cdot ({\bf
w}+{\bf \Omega} t')}e^{-i\omega t'}e^{-i {\bf k}'\cdot ({\bf w}+{\bf \Omega}
t'')}e^{i\omega' t''} \nonumber\\
\times\langle \delta{\hat \Phi}_{\rm tot}({\bf k},{\bf J},\omega)\delta{\hat\Phi}_{\rm tot}({\bf
k}',{\bf J},\omega')^*\rangle.\qquad
\label{diff2}
\end{eqnarray}
Since the correlation function depends only on the
action ${\bf J}$, we can average the diffusion tensor over the angle ${\bf
w}$ without loss of
information. This
brings a Kronecker factor
$\delta_{{\bf k},{\bf k}'}$ which amounts to taking ${\bf
k}'={\bf k}$.
Then,
substituting the power spectrum from Eq. (\ref{psaab}) into Eq. (\ref{diff2}),
and carrying out
the
integrals over $\omega'$, we end up with the result
\begin{eqnarray}
\left\langle \frac{\partial\delta\Phi_{\rm tot}}{\partial w_i}({\bf w}+{\bf\Omega}t', {\bf
J}, t')\frac{\partial\delta\Phi_{\rm tot}}{\partial w_j}({\bf w}+{\bf\Omega}t'', {\bf J},
t'')\right\rangle=\sum_{{\bf k}}\int\frac{d\omega}{2\pi} \, 
k_i k_j
e^{i({\bf k}\cdot {\bf \Omega}-\omega)(t'-t'')}P({\bf k},{\bf J},\omega).
\label{diff3}
\end{eqnarray}
This expression shows that the correlation function appearing in Eq.
(\ref{diff1})  depends only on the difference of times $s=t'-t''$. Using the
identity  
\begin{eqnarray}
\int_{0}^{t}dt'\int_{0}^{t}dt''\, f(t'-t'')=2\int_{0}^{t}dt'\int_{t'}^{t}dt''\,
f(t'-t'')=2\int_{0}^{t}dt'\int_0^{t-t'} ds\, f(s)\nonumber\\
=2\int_{0}^{t}ds\int_0^{t-s} dt'\, f(s)=  2\int_{0}^{t}ds\, (t-s)f(s),
\label{diff4}
\end{eqnarray}
and assuming that the autocorrelation function of the total fluctuating
force $f(s)$
decreases more rapidly
than $s^{-1}$, we find for $t\rightarrow +\infty$
that\footnote{This formula can also be obtained by using the
identity 
\begin{eqnarray}
D_{ij}=\frac{1}{2}\frac{d}{dt}\langle \Delta J_i\Delta
J_j\rangle=\frac{1}{2}\langle \dot J_i\Delta J_j+ \dot J_j\Delta J_i
\rangle=\frac{1}{2}\left \langle \int_0^t dt'\, \left\lbrack {\cal F}_i({\bf
w},{\bf J},0){\cal F}_j({\bf w}(t'),{\bf J}(t'),t')+{\cal F}_j({\bf w},{\bf
J},0){\cal F}_i({\bf
w}(t'),{\bf J}(t'),t')\right\rbrack
\right\rangle,
\label{th1b}
\end{eqnarray}
where ${\cal F}=-\partial\delta\Phi_{\rm tot}/\partial {\bf w}$ is the total
fluctuating force (by
unit of mass) experienced by the test particle. Making the approximations
discussed
above and taking
the limit $t\rightarrow +\infty$, we obtain
\begin{eqnarray}
D_{ij}= \int_0^{+\infty} \left\langle {\cal F}_i({\bf w},{\bf J},0){\cal
F}_j({\bf w}+{\bf
\Omega}({\bf J})s,s)
\right\rangle\, ds,
\label{th2b}
\end{eqnarray}
in agreement with Eq. (\ref{gei1}).}
\begin{eqnarray}
D_{ij}=\int_0^{+\infty}\left\langle \frac{\partial\delta\Phi_{\rm tot}}{\partial
w_i}({\bf w},{\bf J},0)\frac{\partial\delta\Phi_{\rm tot}}{\partial
w_j}({\bf w}+{\bf\Omega}s,{\bf J},s)\right\rangle\, ds.
\label{gei1}
\end{eqnarray}
Therefore, as in the theory of
Brownian motion
\cite{uo,chandrabrown,green1,lhp,kubo0,kubo2,helfand60,helfand,
kubo,rubin,zwanzig,risken}, turbulent
fluids
\cite{taylor,osterbrock}, plasma
physics \cite{cohen,gabor,gasiorowicz,th,hubbard1,hubbard2,mynick} and stellar
dynamics
\cite{cvn0,lee,cohenamad,lcohen,kandruprep,kandrup1,kandrupfriction,bm92,maoz,
nt99,cp,hb4,aa}, the
diffusion tensor of the test particle is equal to the integral of
the temporal auto-correlation
function $\langle {\cal F}_i(0){\cal F}_j(t)\rangle$ of the fluctuating force
acting on it:
\begin{eqnarray}
D_{ij}=\int_0^{+\infty}\langle {\cal F}_i(0){\cal F}_j(t)\rangle\, dt.
\label{dcorr}
\end{eqnarray}
Replacing
the auto-correlation function by
its expression from Eq. (\ref{diff3}), which can be written as
\begin{equation}
\langle {\cal F}_i(0){\cal F}_j(t)\rangle=\sum_{{\bf
k}}\int\frac{d\omega}{2\pi}k_i k_j
e^{-i({\bf
k}\cdot {\bf\Omega}-\omega)t} P({\bf k},{\bf J},\omega),
\end{equation}
we obtain
\begin{eqnarray}
D_{ij}=\int_0^{+\infty}dt\, \sum_{{\bf k}}\int\frac{d\omega}{2\pi}k_i k_j
e^{-i({\bf
k}\cdot {\bf\Omega}-\omega)t} P({\bf k},{\bf J},\omega).
\label{gei2}
\end{eqnarray}
Making the change of variables $t\rightarrow -t$, ${\bf k}\rightarrow -{\bf k}$
and $\omega\rightarrow -\omega$, and using the fact
that $P(-{\bf k},{\bf J},-\omega)=P({\bf k},{\bf J},\omega)$, we see that we can
replace
$\int_{0}^{+\infty}dt$ by $(1/2)\int_{-\infty}^{+\infty}dt$ in Eq. (\ref{gei2}).
Therefore, we get
\begin{eqnarray}
D_{ij}=\frac{1}{2}\int_{-\infty}^{+\infty}dt\, \sum_{{\bf
k}}\int\frac{d\omega}{2\pi}k_i k_j
e^{-i({\bf
k}\cdot {\bf\Omega}-\omega)t} P({\bf k},{\bf J},\omega).
\label{gei3b}
\end{eqnarray}
Using the identity (\ref{deltac}), we find that
\begin{eqnarray}
D_{ij}
=\pi \sum_{{\bf
k}}\int\frac{d\omega}{2\pi}k_i k_j
\delta({\bf
k}\cdot {\bf\Omega}-\omega) P({\bf k},{\bf J},\omega).
\label{gei3c}
\end{eqnarray}
The time integration has given a $\delta$-function which creates a
resonance
condition for interaction. Integrating over the $\delta$-function
(resonance), we arrive at the
following equation
\begin{eqnarray}
D_{ij}=\frac{1}{2} \sum_{{\bf
k}}k_i k_j  P({\bf k},{\bf J},{\bf
k}\cdot {\bf\Omega}({\bf J})).
\label{gei3}
\end{eqnarray}
This equation
expresses the diffusion tensor of the test particle in terms of the
power spectrum of the fluctuations at the resonances $\omega={\bf k}\cdot
{\bf \Omega}({\bf J})$. This is the general
expression of the
diffusion
tensor of a test particle submitted to a stochastic perturbation in an
inhomogeneous system with long-range interactions. When
collective effects are neglected,
we get
\begin{eqnarray}
D^{\rm bare}_{ij}=\frac{1}{2} \sum_{{\bf
k}}k_i k_j  P_{\rm bare}({\bf k},{\bf J},{\bf
k}\cdot {\bf\Omega}({\bf J})).
\label{gei3bare}
\end{eqnarray}

Using the relation between the power spectrum and the correlation function of the external perturbation [see Eq. (\ref{i83})],  we get
\begin{equation}
D_{ij}[f,{\bf
J}]=\frac{1}{2}\sum_{\bf k}\sum_{
\alpha\alpha'} \, k_i k_j \hat\Phi_{\alpha}({\bf
k},{\bf J})\left \lbrack\epsilon^{-1} {\hat
C} (\epsilon^{\dagger})^{-1}\right
\rbrack_{\alpha\alpha'}({\bf k}\cdot {\bf\Omega}({\bf J}))
\hat\Phi_{\alpha'}({\bf
k},{\bf J})^*.
\label{i88b}
\end{equation}
This returns the result of \cite{weinberg,pa,fpp,fp15,fbp,fb,sdduniverse}. This
expression
shows that the diffusion tensor of the test particle depends on the spectral
auto-correlation tensor
of the
external perturbation ${\hat C}_{\alpha\alpha'}(\omega)$ and on
the inverse dielectric matrix (response function)  ${\epsilon^{-1}}(\omega)$
both evaluated
at the
resonance
frequencies $\omega={\bf k}\cdot {\bf\Omega}({\bf
J})$.\footnote{We have assumed that the system is stable so the
dielectric matrix
is invertible on the real axis.}   As a result, the
diffusion tensor $D_{ij}[f,{\bf
J}]$ depends not only  on the action ${\bf J}$ but is also a functional of the
DF $f({\bf J},t)$ itself through the
dielectric matrix $\epsilon_{\alpha\alpha'}({\bf k}\cdot {\bf \Omega}({\bf
J}))$ defined by Eq. (\ref{i70}). It also depends
implicitly on $f$ through the orbital frequencies ${\bf \Omega}({\bf J})$.

When collective
effects
are neglected, i.e., if we make $\epsilon=1$ in Eq. (\ref{i88b}) and use Eq.
(\ref{mar5}), or if we use Eq. (\ref{gei3bare}) with $P_{\rm bare}({\bf k},{\bf
J},\omega)={\hat C}({\bf k},{\bf J},\omega)$, the
diffusion tensor reduces to
\begin{eqnarray}
D^{\rm bare}_{ij}=\frac{1}{2} \sum_{{\bf
k}}k_i k_j  {\hat C}({\bf k},{\bf J},{\bf
k}\cdot {\bf\Omega}({\bf J})).
\label{gei3barej}
\end{eqnarray}
This returns the result of \cite{bl,ba,bf}. In that case, it is
not necessary to
introduce a biorthogonal basis.

{\it Remark:} Alternative derivations of the general expression of the 
diffusion tensor of a test particle are given in Appendices \ref{sec_eug}
and \ref{sec_ele}  (see also
Appendix \ref{sec_geic} where we
compute the force auto-correlation function of the test particle).

\subsection{Expression of the diffusion tensor due to $N$ particles}
\label{sec_geij}

We now assume that the external noise is due to a  discrete collection 
of  $N$ particles. In that case,  $P({\bf k},{\bf J},\omega)$ is given by Eq.
(\ref{pi4}) and we obtain 
\begin{eqnarray}
D_{ij}=\pi(2\pi)^{d}\sum_b m_b\sum_{{\bf k},{\bf k}'}\int d{\bf
J}'\, k_ik_j
|A^d_{{\bf k},{\bf k}'}({\bf J},{\bf
J'},{\bf k}\cdot {\bf\Omega})|^2 \delta({\bf
k}\cdot {\bf\Omega}-{\bf k}'\cdot
{\bf\Omega}')f_b({\bf J}').
\label{di1}
\end{eqnarray}
This  returns Eq. (79) of
\cite{physicaA}. This is the general
expression of  the diffusion tensor of a test particle of mass $m$ produced by
$N$ field particles of masses
$\lbrace m_b\rbrace$. The diffusion of the test particle is due to the
fluctuations
of the field particles induced by finite
$N$ effects (granularities). This is why it depends on $\lbrace m_b\rbrace$
but not on
$m$.\footnote{Note, however, that some field particles may have the mass
$m_b=m$, i.e., they belong to the
same species as the test particle.} We note that in order to derive the
diffusion tensor (\ref{di1}) we do not need to introduce a biorthogonal basis if
we use the results of Sec. \ref{sec_ciwtay}.

If we neglect collective effects, the previous results remain valid provided
that $A^d_{{\bf k},{\bf k}'}({\bf J},{\bf
J'},{\bf k}\cdot {\bf\Omega})$ is 
replaced by $A_{{\bf k},{\bf k}'}({\bf J},{\bf
J'})$. The bare diffusion coefficient of a test
particle is
\begin{eqnarray}
D^{\rm bare}_{ij}=\pi(2\pi)^{d}\sum_b m_b\sum_{{\bf k},{\bf k}'}\int d{\bf
J}'\, k_ik_j
|A_{{\bf k},{\bf k}'}({\bf J},{\bf
J'})|^2 \delta({\bf
k}\cdot {\bf\Omega}-{\bf k}'\cdot
{\bf\Omega}')f_b({\bf J}').
\label{gei5}
\end{eqnarray}
This returns Eq. (117) of \cite{aa}.

{\it Remark:} An alternative derivation of the 
diffusion tensor of a test particle produced by $N$ field particles is given
in Appendix \ref{sec_dn} (see also
Appendix \ref{sec_geic}
where we
compute the force auto-correlation function of the test particle).

\section{Friction by polarization}
\label{sec_ifpol}

Let us consider a system of particles 
with long-range interactions with a DF $f({\bf J})$ and let us introduce a
test
particle with a small mass $m$ in this system. The DF $f({\bf J})$
may be due to a collection
of field particles with masses $\lbrace m_b\rbrace$, in which case it
represent
their mean DF in the limit $N_b\rightarrow +\infty$ with $m_b\sim
1/N_b$, but it can also have a more general origin. 
We want to determine the
friction
by polarization experienced by the test particle
due to the perturbation that it causes to the system. The friction by
polarization arise from the density wake
created by the moving particle. We
use the formalism of
linear response theory developed in the previous sections and treat the
perturbation induced by the test particle as a
small external perturbation $f_e({\bf
r},{\bf v},t)$ to the system.

\subsection{Friction by polarization with collective effects}

The temporal Fourier transform of the DF of the test particle is [see Eq.
(\ref{ci4})]
\begin{eqnarray}
{\hat f}_e({\bf w},{\bf J},\omega)=\frac{1}{(2\pi)^{d-1}}m\sum_{{\bf k}'} 
e^{i{\bf
k}'\cdot ({\bf w}-{\bf w}_0)}\delta(\omega-{\bf k}'\cdot {\bf\Omega}_0)\delta({\bf
J}-{\bf
J}_0),
\label{ci4n}
\end{eqnarray}
where ${\bf w}_0$ and ${\bf J}_0$ denote the initial angle and action of
the test particle. The Fourier transform of the amplitude of the potential
created by the test particle is given by Eq. (\ref{ci1b}) with Eq. (\ref{ci4n})
yielding
\begin{eqnarray}
{\hat A}^e_{\alpha}(\omega)&=&-\frac{1}{(2\pi)^{d-1}}m\sum_{{\bf k}'}\int  d{\bf w}d{\bf J}\,  
e^{i{\bf
k}'\cdot ({\bf w}-{\bf w}_0)}\delta(\omega-{\bf k}'\cdot {\bf\Omega}_0)\delta({\bf
J}-{\bf
J}_0)
 \Phi_{\alpha}^*({\bf
w},{\bf J})\nonumber\\
&=&-2\pi m\sum_{{\bf k}'}
e^{-i{\bf k}'\cdot {\bf w}_0}\delta(\omega-{\bf k}'\cdot {\bf\Omega}_0)
 {\hat \Phi}_{\alpha}^*({\bf
k}',{\bf J}_0).
\label{ci1bnb}
\end{eqnarray}
According to Eqs. (\ref{fdi6}) and (\ref{ci1bnb}), the Fourier transform of the
amplitude
of the 
perturbation
created by the test particle is
\begin{eqnarray}
{\hat A}_{\alpha}(\omega)
= -2\pi m\sum_{\alpha'} \left\lbrack
(\epsilon^{-1})_{\alpha\alpha'}(\omega)-\delta_{\alpha\alpha'}\right\rbrack\sum_{{\bf k}'}
e^{-i{\bf k}'\cdot {\bf w}_0}\delta(\omega-{\bf k}'\cdot {\bf\Omega}_0)
 {\hat \Phi}_{\alpha'}^*({\bf
k}',{\bf J}_0).
\label{ci1bnc}
\end{eqnarray}
Returning to physical space, we get
\begin{eqnarray}
{A}_{\alpha}(t)&=&-2\pi m \int \frac{d\omega}{2\pi} e^{-i\omega t}\sum_{{\bf k}'}\sum_{\alpha'} \left\lbrack
(\epsilon^{-1})_{\alpha\alpha'}(\omega)-\delta_{\alpha\alpha'}\right\rbrack
e^{-i{\bf k}'\cdot {\bf w}_0}\delta(\omega-{\bf k}'\cdot {\bf\Omega}_0)
 {\hat \Phi}_{\alpha'}^*({\bf
k}',{\bf J}_0)\nonumber\\
&=& -m \sum_{{\bf k}'}\sum_{\alpha'} \left\lbrack
(\epsilon^{-1})_{\alpha\alpha'}({\bf k}'\cdot {\bf\Omega}_0)-\delta_{\alpha\alpha'}\right\rbrack
 {\hat \Phi}_{\alpha'}^*({\bf
k}',{\bf J}_0)e^{-i{\bf k}'\cdot ({\bf\Omega}_0 t+{\bf w}_0)}.
\label{ci1bnd}
\end{eqnarray}
Using Eqs. (\ref{i62}) and (\ref{ci1bnd}), we find that the perturbed potential
created by the test particle is
\begin{eqnarray}
\delta\Phi({\bf r}({\bf w},{\bf J}),t)&=& \sum_{\alpha}A_{\alpha}(t)\Phi_{\alpha}({\bf r}({\bf w},{\bf J}))\nonumber\\
&=& -m \sum_{{\bf k}'}\sum_{\alpha\alpha'} {\Phi}_{\alpha}({\bf r}({\bf
w},{\bf J}))\left\lbrack (\epsilon^{-1})_{\alpha\alpha'}({\bf k}'\cdot {\bf\Omega}_0)-\delta_{\alpha\alpha'}\right\rbrack
 {\hat \Phi}_{\alpha'}^*({\bf
k}',{\bf J}_0)e^{-i{\bf k}'\cdot ({\bf\Omega}_0 t+{\bf w}_0)}
\nonumber\\
&=& -m \sum_{{\bf k}{\bf k}'}\sum_{\alpha\alpha'} e^{i{\bf k}\cdot {\bf w}} {\hat\Phi}_{\alpha}({\bf
k},{\bf J})\left\lbrack (\epsilon^{-1})_{\alpha\alpha'}({\bf k}'\cdot {\bf\Omega}_0)-\delta_{\alpha\alpha'}\right\rbrack
 {\hat \Phi}_{\alpha'}^*({\bf
k}',{\bf J}_0)e^{-i{\bf k}'\cdot ({\bf\Omega}_0 t+{\bf w}_0)}.
\label{ci1bne}
\end{eqnarray}
The corresponding field is
\begin{eqnarray}
-\frac{\partial \delta\Phi}{\partial {\bf w}}= m \sum_{{\bf k}{\bf k}'}\sum_{\alpha\alpha'} i{\bf k} e^{i{\bf k}\cdot {\bf w}} {\hat\Phi}_{\alpha}({\bf
k},{\bf J})\left\lbrack (\epsilon^{-1})_{\alpha\alpha'}({\bf k}'\cdot {\bf\Omega}_0)-\delta_{\alpha\alpha'}\right\rbrack
 {\hat \Phi}_{\alpha'}^*({\bf
k}',{\bf J}_0)e^{-i{\bf k}'\cdot ({\bf\Omega}_0 t+{\bf w}_0)}.
\label{ci1bnf}
\end{eqnarray}
The test particle is submitted to the force resulting from the perturbation that
it has caused and, as a
result, it experiences a dynamical
friction.
Applying Eq. (\ref{ci1bnf}) at the position of the test particle at time $t$
(${\bf w}={\bf
\Omega}_0 t+{\bf w}_0$,  ${\bf
J}={\bf J}_0$), we obtain
the friction by polarization 
\begin{eqnarray}
{\bf F}_{\rm pol}&=& m \sum_{{\bf k}{\bf k}'}\sum_{\alpha\alpha'} i{\bf k}
e^{i({\bf k}-{\bf k}')\cdot ({\bf\Omega}_0 t+{\bf w}_0)}
{\hat\Phi}_{\alpha}({\bf
k},{\bf J})\left\lbrack (\epsilon^{-1})_{\alpha\alpha'}({\bf k}\cdot
{\bf\Omega}_0)-\delta_{\alpha\alpha'}\right\rbrack
 {\hat \Phi}_{\alpha'}^*({\bf
k},{\bf J}_0)\nonumber\\
&=& i m \sum_{{\bf k}}{\bf k}\, \sum_{\alpha\alpha'}  
{\hat\Phi}_{\alpha}({\bf
k},{\bf J})\left\lbrack (\epsilon^{-1})_{\alpha\alpha'}({\bf k}\cdot
{\bf\Omega})-\delta_{\alpha\alpha'}\right\rbrack
 {\hat \Phi}_{\alpha'}^*({\bf
k},{\bf J}),
\label{ci1bnha}
\end{eqnarray}
where we used the fact that the result should not depend on ${\bf w}_0$.
Since ${\bf F}_{\rm pol}$ is real, we can write
\begin{eqnarray}
{\bf F}_{\rm pol}
= - m \sum_{{\bf k}}{\bf k}\, {\rm Im}  \left\lbrack \sum_{\alpha\alpha'}
{\hat\Phi}_{\alpha}({\bf
k},{\bf J}) (\epsilon^{-1})_{\alpha\alpha'}({\bf k}\cdot {\bf\Omega})
 {\hat \Phi}_{\alpha'}^*({\bf
k},{\bf J})\right\rbrack.
\label{ci1bnhb}
\end{eqnarray}
Introducing the dressed potential of interaction from Eq. (\ref{can1b}),
we get 
\begin{eqnarray}
{\bf F}_{\rm pol}
= m  \sum_{{\bf k}}{\bf k}\,  {\rm Im}\left\lbrack A^d_{{\bf k}{\bf
k}}({\bf J},{\bf J},{\bf k}\cdot {\bf
\Omega}) \right\rbrack.
\label{ci1bnh}
\end{eqnarray}
Finally, using the identity from Eq. (\ref{fdi2a}), we can rewrite
the foregoing
equation as
\begin{eqnarray}
{\bf F}_{\rm pol}=\pi(2\pi)^{d}m\sum_{{\bf k},{\bf k}'}\int d{\bf
J}'\, {\bf k}
|A^d_{{\bf k},{\bf k}'}({\bf J},{\bf
J'},{\bf k}\cdot {\bf\Omega})|^2 \delta({\bf
k}\cdot {\bf\Omega}-{\bf k}'\cdot
{\bf\Omega}')\left ({\bf k}'\cdot \frac{\partial f'}{\partial {\bf J}'}\right ).
\label{gei4zz}
\end{eqnarray}
This  returns Eqs. (91) and (92) of \cite{physicaA} (an expression equivalent to
Eq. (\ref{ci1bnh}) was
previously established in \cite{w89}). This is the
general expression of the
friction by polarization of the test particle.\footnote{The fact
that the friction calculated in this section corresponds to ${\bf F}_{\rm
pol}$ in Eq. (\ref{fp9}) is
justified in Ref. \cite{physicaA} by calculating ${\bf F}_{\rm tot}$ directly
from Eq.
(\ref{fp7}).}  The friction by polarization of
the test particle is due to the retroaction (response) of the perturbation that
it has
caused to the system. This is why it
is proportional to $m$. The calculation of the polarization
cloud (wake) created by the test particle is detailed in
Appendix \ref{sec_pc}.

If the DF is due to a collection of $N$ field particles with masses
$\lbrace m_b\rbrace$, the
friction by polarization takes the form
\begin{eqnarray}
{\bf F}_{\rm pol}=\pi(2\pi)^{d}m\sum_b\sum_{{\bf k},{\bf k}'}\int d{\bf
J}'\, {\bf k} |A^d_{{\bf k},{\bf k}'}({\bf J},{\bf
J'},{\bf k}\cdot {\bf\Omega})|^2 \delta({\bf
k}\cdot {\bf\Omega}-{\bf k}'\cdot
{\bf\Omega}')\left ({\bf k}'\cdot \frac{\partial f'_b}{\partial {\bf J}'}\right ).
\label{gei4zzb}
\end{eqnarray}
We note the similarity between Eq. (\ref{gei4zzb}) and the expression
(\ref{di1}) of the diffusion tensor created by a collection of field
particles. The main difference is that the friction by polarization involves the
gradient of the DF instead of the DF itself. In addition,
the friction by polarization is proportional to the mass $m$ of the
test particle while the diffusion tensor involves the masses $\lbrace
m_b\rbrace$ of the  
field particles.

{\it Remark:} The friction by polarization ${\bf
F}_{\rm pol}$
is just one component of
the total friction  ${\bf F}_{\rm tot}$ of the test particle which is given by
Eq. (\ref{fp9}).
Substituting Eqs. (\ref{di1}) and (\ref{gei4zzb}) into Eq.
(\ref{fp9}) and
making an integrating by parts,
we find that the total friction is
\begin{eqnarray}
{\bf F}_{\rm tot}=\pi(2\pi)^{d}\sum_b\sum_{{\bf k},{\bf k}'}\int d{\bf
J}'\, f_b({\bf J}') {\bf k} \left (m_b{\bf k}\cdot \frac{\partial}{\partial {\bf
J}}-m{\bf k}'\cdot \frac{\partial}{\partial {\bf
J}'}\right )|A^d_{{\bf k},{\bf k}'}({\bf J},{\bf
J'},{\bf k}\cdot {\bf\Omega})|^2 \delta({\bf
k}\cdot {\bf\Omega}-{\bf k}'\cdot
{\bf\Omega}').
\label{totfric}
\end{eqnarray}
This  returns Eq. (114) of \cite{physicaA}.

\subsection{Friction by polarization without collective effects}

It is instructive to redo the 
calculation of the friction by polarization by neglecting collective effects
from
the start.\footnote{We note that we cannot simply
replace $A^d$ by $A$ in
Eq. (\ref{ci1bnh}) otherwise we would find ${\bf F}_{\rm pol}={\bf 0}$ since $A$
is real. We first have to use Eq. (\ref{fdi2a}), then replace $A^d$ by
$A$ in
Eq. (\ref{gei4zz}).}
In that case, the change of the DF caused by the external
perturbation is determined by the equation
\begin{eqnarray}
\label{gh1}
\frac{\partial \delta f}{\partial t}+{\bf \Omega}\cdot \frac{\partial\delta
f}{\partial {\bf w}}- \frac{\partial\Phi_e}{\partial {\bf w}}\cdot \frac{\partial
f}{\partial {\bf J}}=0,
\end{eqnarray}
where we have neglected the term $\delta\Phi$ in Eq. (\ref{i57}).
Written in
Fourier space, we get
\begin{equation}
\delta\hat f ({\bf k},{\bf J},\omega)=\frac{{\bf k}\cdot \frac{\partial
f}{\partial {\bf J}}}{{\bf k}\cdot {\bf
\Omega}-\omega}\hat\Phi_e({\bf
k},{\bf J},\omega).
\label{gh2}
\end{equation}
The Fourier transform of the external potential created by a DF
$f_e$ is [see Eq. (\ref{gh4h})]
\begin{eqnarray}
\hat\Phi_{e}({\bf k},{\bf J},\omega)=(2\pi)^d\sum_{{\bf k}'} \int
d{\bf
J}'\,  A_{{\bf k}{\bf k}'}({\bf J},{\bf J}')
 {\hat f}_e({\bf k}',{\bf J}',\omega).
\label{lg1}
\end{eqnarray}
The Fourier transform of the DF of the test particle is [see 
Eq. (\ref{ci3})]
\begin{eqnarray}
{\hat f}_e({\bf k},{\bf J},\omega)=\frac{1}{(2\pi)^{d-1}}m
e^{-i{\bf
k}\cdot {\bf w}_0}\delta(\omega-{\bf k}\cdot
{\bf\Omega}_0)\delta({\bf
J}-{\bf
J}_0),
\label{lg2}
\end{eqnarray}
where $({\bf w}_0,{\bf J}_0)$ denotes the initial position in angle-action
variables of the test particle. According to
Eqs. (\ref{lg1}) and (\ref{lg2}), the Fourier transform of the 
bare potential (neglecting collective effects) created by the test
particle is given by [see also Eq. (\ref{ciw3})]
\begin{eqnarray}
{\hat \Phi_e}({\bf k},{\bf J},\omega)=2\pi m \sum_{{\bf k}'} A_{{\bf
k}{\bf k}'}({\bf J},{\bf J}_0) e^{-i{\bf k}'\cdot {\bf w}_0}\delta(\omega-{\bf
k}'\cdot{\bf \Omega}_0).
\label{ciw3rep}
\end{eqnarray}
Therefore, according to Eq. (\ref{gh2}), the perturbed DF
is\footnote{This corresponds to the polarization cloud without collective
effects (see Appendix \ref{sec_pc}).} 
\begin{equation}
\delta\hat f ({\bf k},{\bf J},\omega)=2\pi m\frac{{\bf k}\cdot \frac{\partial
f}{\partial {\bf J}}}{{\bf k}\cdot {\bf
\Omega}-\omega} \sum_{{\bf k}'} A_{{\bf
k}{\bf k}'}({\bf J},{\bf J}_0) e^{-i{\bf k}'\cdot {\bf w}_0}\delta(\omega-{\bf
k}'\cdot{\bf \Omega}_0).
\label{gh2nj}
\end{equation}
The Fourier transform of the potential produced by the perturbed DF is  [see Eq.
(\ref{edi})] 
\begin{eqnarray}
\delta{\hat \Phi}({\bf k},{\bf J},\omega)=(2\pi)^d\sum_{{\bf k}'} \int d{\bf J}'\,  A_{{\bf k}{\bf k}'}({\bf J},{\bf J}')
 \delta{\hat f}({\bf k}',{\bf J}',\omega).
\label{gh4}
\end{eqnarray}
Combining
Eqs. (\ref{gh2nj}) and (\ref{gh4}), we get
\begin{eqnarray}
\delta{\hat \Phi}({\bf k},{\bf J},\omega)
=(2\pi)^{d+1} m \sum_{{\bf k}'{\bf k}''} \int d{\bf J}'\,  A_{{\bf k}{\bf
k}'}({\bf J},{\bf J}')A_{{\bf k}'{\bf k}''}({\bf J}',{\bf J}_0)\frac{{\bf
k}'\cdot \frac{\partial
f'}{\partial {\bf J}'}}{{\bf k}'\cdot {\bf
\Omega}'-\omega}   
 e^{-i{\bf
k}''\cdot {\bf w}_0}\delta(\omega-{\bf k}''\cdot {\bf\Omega}_0).
\label{gh5}
\end{eqnarray}
Returning to physical space, we obtain
\begin{eqnarray}
\delta{\Phi}({\bf w},{\bf J},t)=(2\pi)^{d+1} m \sum_{\bf k} \int
\frac{d\omega}{2\pi}\, e^{i({\bf k}\cdot {\bf w}-\omega t)}\sum_{{\bf k}'{\bf
k}''} \int d{\bf J}'\,  A_{{\bf k}{\bf k}'}({\bf J},{\bf J}')A_{{\bf k}'{\bf
k}''}({\bf J}',{\bf J}_0)\frac{{\bf k}'\cdot \frac{\partial
f'}{\partial {\bf J}'}}{{\bf k}'\cdot {\bf
\Omega}'-\omega}   
 e^{-i{\bf
k}''\cdot {\bf w}_0}\delta(\omega-{\bf k}''\cdot {\bf\Omega}_0)\nonumber\\
=i\pi (2\pi)^{d+1} m \sum_{\bf k} \int
\frac{d\omega}{2\pi}\, e^{i({\bf k}\cdot {\bf w}-\omega t)}\sum_{{\bf k}'{\bf
k}''} \int d{\bf J}'\,  A_{{\bf k}{\bf k}'}({\bf J},{\bf J}')A_{{\bf k}'{\bf
k}''}({\bf J}',{\bf J}_0)\left ({\bf k}'\cdot \frac{\partial
f'}{\partial {\bf J}'}\right ) \delta({\bf k}'\cdot {\bf
\Omega}'-\omega)   
 e^{-i{\bf
k}''\cdot {\bf w}_0}\delta(\omega-{\bf k}''\cdot {\bf\Omega}_0)\nonumber\\
=i\pi (2\pi)^{d} m \sum_{{\bf k}{\bf k}'{\bf
k}''} \int d{\bf J}'\,   A_{{\bf k}{\bf k}'}({\bf J},{\bf J}')A_{{\bf k}'{\bf
k}''}({\bf J}',{\bf J}_0)\left ({\bf k}'\cdot \frac{\partial
f'}{\partial {\bf J}'}\right ) \delta({\bf k}'\cdot {\bf
\Omega}'-{\bf k}''\cdot {\bf\Omega}_0) e^{-i{\bf
k}''\cdot ({\bf w}_0+ {\bf {\Omega}}_0 t)}e^{i{\bf k}\cdot {\bf
w}}. \qquad
\label{gh6}
\end{eqnarray}
To get the second line we have used the Landau prescription
$\omega\rightarrow \omega+i0^{+}$ 
and the Sokhotski-Plemelj \cite{sokhotski,plemelj} formula
\begin{equation}
\frac{1}{x\pm i 0^+}={\rm P}\left (\frac{1}{x}\right )\mp i\pi\delta(x),
\label{n25}
\end{equation}
where ${\rm P}$ denotes the principal value. The
corresponding field is 
\begin{eqnarray}
-\frac{\partial\delta{\Phi}}{\partial {\bf w}}=-i\pi (2\pi)^{d} m
\sum_{{\bf k}{\bf k}'{\bf
k}''} i {\bf k} \int d{\bf J}'\,   A_{{\bf k}{\bf k}'}({\bf J},{\bf J}')A_{{\bf
k}'{\bf
k}''}({\bf J}',{\bf J}_0)\left ({\bf k}'\cdot \frac{\partial
f'}{\partial {\bf J}'}\right ) \delta({\bf k}'\cdot {\bf
\Omega}'-{\bf k}''\cdot {\bf\Omega}_0) e^{-i{\bf
k}''\cdot ({\bf w}_0+ {\bf {\Omega}}_0 t)}e^{i{\bf k}\cdot {\bf
w}}. \quad\label{gh7}
\end{eqnarray}
Applying Eq. (\ref{gh7}) at the position of the 
test particle at time $t$ (${\bf w}={\bf \Omega}_0 t+{\bf w}_0$, 
${\bf J}={\bf J}_0$), making
${\bf k}''={\bf k}$ since the result should be independent of ${\bf w}_0$,  and
using the
identity
$A_{{\bf k}'{\bf k}}({\bf
J}',{\bf J})=A_{{\bf k}{\bf k}'}({\bf J},{\bf J}')^*$  (see Appendix
\ref{sec_bdpia}) we obtain the
friction by polarization
\begin{eqnarray}
{\bf F}_{\rm pol}=\pi(2\pi)^{d}m\sum_{{\bf k},{\bf k}'}\int d{\bf
J}'\, {\bf k} |A_{{\bf k},{\bf k}'}({\bf J},{\bf
J'})|^2 \delta({\bf
k}\cdot {\bf\Omega}-{\bf k}'\cdot
{\bf\Omega}')\left ({\bf k}'\cdot \frac{\partial f'}{\partial {\bf J}'}\right ).
\label{gh8}
\end{eqnarray}
This returns Eq. (118) of \cite{aa} (an equivalent expression was
previously established in \cite{lbk,ps80,tw84,w85,palmer,rt}). This is
the general expression of the friction
by polarization
when collective effects are neglected.  It can be directly obtained from
Eq.
(\ref{gei4zz}) by
replacing the dressed potential of interaction $A^d_{{\bf k},{\bf k}'}({\bf
J},{\bf
J'},{\bf k}\cdot {\bf\Omega})$ by the bare potential of
interaction $A_{{\bf k},{\bf k}'}({\bf J},{\bf
J'})$ (see footnote 24). If the DF $f$ is due to a collection of field
particles, the friction by
polarization takes the form
\begin{eqnarray}
{\bf F}_{\rm pol}=\pi(2\pi)^{d} m
\sum_b\sum_{{\bf
k},{\bf k}'}  \int d{\bf J}'\, {\bf k} |A_{{\bf k},{\bf k}'}({\bf J},{\bf
J}')|^2 
\delta({{\bf k}\cdot {\bf
\Omega}-{\bf k}'\cdot {\bf\Omega}'})\left ({\bf k}'\cdot \frac{\partial
f'_b}{\partial {\bf J}'}\right ).
\label{gh10g}
\end{eqnarray}

{\it Remark:} We can also do the calculations as follows. Taking the inverse
Fourier transform in time of Eq. (\ref{ciw3rep}) we get
\begin{eqnarray}
{\hat \Phi_e}({\bf k},{\bf J},t)=m \sum_{{\bf k}'} A_{{\bf
k}{\bf k}'}({\bf J},{\bf J}_0) e^{-i{\bf k}'\cdot {\bf
w}_0}e^{-i {\bf
k}'\cdot{\bf \Omega}_0 t}.
\label{pan1}
\end{eqnarray}
Taking the Fourier transform in angle of Eq. (\ref{gh1}) and solving the
resulting differential equation in time like in Appendix \ref{sec_j} we obtain
\begin{eqnarray}
\label{pan2}
\delta{\hat f}({\bf k},{\bf J},t)=i{\bf k}\cdot
\frac{\partial f}{\partial {\bf J}}\int_0^t dt'\, {\hat \Phi}_e({\bf
k},{\bf J},t')e^{i{\bf k}\cdot{\bf\Omega}(t'-t)},
\end{eqnarray}
where we have assumed that the initial perturbation vanishes, $\delta{\hat
f}({\bf
k},{\bf J},0)=0$. Substituting Eq. (\ref{pan1}) into Eq. (\ref{pan2}) we find
that
\begin{eqnarray}
\label{pan3}
\delta{\hat f}({\bf k},{\bf J},t)=i\left ({\bf k}\cdot
\frac{\partial f}{\partial {\bf J}}\right ) e^{-i{\bf
k}\cdot{\bf\Omega}t} m \sum_{{\bf k}'} A_{{\bf
k}{\bf k}'}({\bf J},{\bf J}_0) e^{-i{\bf k}'\cdot {\bf
w}_0}\int_0^t dt'\, e^{i({\bf
k}\cdot{\bf \Omega}-{\bf
k}'\cdot{\bf \Omega}_0) t'}.
\end{eqnarray}
Taking the inverse Fourier transform of Eq. (\ref{gh4}) and combining the
resulting expression with Eq. (\ref{pan3}) we get
\begin{eqnarray}
\delta{\hat \Phi}({\bf w},{\bf J},t)=(2\pi)^d\sum_{{\bf k}}e^{i{\bf
k}\cdot{\bf w}}\sum_{{\bf k}'} \int d{\bf J}'\,  A_{{\bf k}{\bf k}'}({\bf
J},{\bf J}')i\left ({\bf k}'\cdot
\frac{\partial f'}{\partial {\bf J}'}\right ) e^{-i{\bf
k}'\cdot{\bf\Omega}'t}\nonumber\\
\times m \sum_{{\bf k}''} A_{{\bf
k}'{\bf k}''}({\bf J}',{\bf J}_0) e^{-i{\bf k}''\cdot {\bf
w}_0}\int_0^t dt'\, e^{i({\bf
k}'\cdot{\bf \Omega}'-{\bf
k}''\cdot{\bf \Omega}_0) t'}.
\label{pan4}
\end{eqnarray}
Integrating over time $t'$ and discarding the irrelevant terms we obtain
\begin{eqnarray}
\delta{\hat \Phi}({\bf w},{\bf J},t)=(2\pi)^d m\sum_{{\bf k}{\bf k}'{\bf
k}''}\int d{\bf J}'\,  A_{{\bf k}{\bf k}'}({\bf
J},{\bf J}')A_{{\bf
k}'{\bf k}''}({\bf J}',{\bf J}_0)\left ({\bf k}'\cdot
\frac{\partial f'}{\partial {\bf J}'} \right ) \frac{e^{-i{\bf
k}''\cdot({\bf w}_0+{\bf \Omega}_0 t)}}{{\bf
k}'\cdot{\bf \Omega}'-{\bf
k}''\cdot{\bf \Omega}_0}e^{i{\bf
k}\cdot{\bf w}}.
\label{pan5}
\end{eqnarray}
Using the Sokhotski-Plemelj
\cite{sokhotski,plemelj}  formula (\ref{n25}), we can replace $1/({\bf
k}'\cdot{\bf \Omega}'-{\bf
k}''\cdot{\bf \Omega}_0)$ by $i\pi\delta({\bf
k}'\cdot{\bf \Omega}'-{\bf
k}''\cdot{\bf \Omega}_0)$ returning Eq. (\ref{gh6}) and, finally, Eq.
(\ref{gh8}).

\subsection{Friction by polarization without using the biorthogonal basis}

We can also determine the friction by polarization without using the
biorthogonal basis. The Fourier transform of the
total fluctuating potential (including collective
effects) created by an external system with DF $f_e$ is [see Eq. (\ref{dn6b})] 
\begin{eqnarray}
\delta\hat\Phi_{\rm tot}({\bf k},{\bf J},\omega)=(2\pi)^d\sum_{{\bf k}'} \int
d{\bf
J}'\,  A^d_{{\bf k}{\bf k}'}({\bf J},{\bf J}',\omega)
 {\hat f}_e({\bf k}',{\bf J}',\omega).
\label{lg3}
\end{eqnarray}
The Fourier transform
of the DF of the test particle is given by Eq. (\ref{lg2}). According to
Eqs. (\ref{lg2}) and (\ref{lg3}), the Fourier transform of the total
fluctuating potential (including collective effects) created by the test
particle is given by [see also Eq. (\ref{dn7})] 
\begin{eqnarray}
\delta{\hat \Phi}_{\rm tot}({\bf k},{\bf J},\omega)=2\pi m \sum_{{\bf k}'} 
A_{{\bf
k}{\bf k}'}^d({\bf J},{\bf J}_0,\omega) e^{-i{\bf k}'\cdot {\bf
w}_0}\delta(\omega-{\bf
k}'\cdot{\bf \Omega}_0).
\label{arm1}
\end{eqnarray}
It arises from the polarization cloud created by the test particles (see
Appendix \ref{sec_pc}). Returning to
physical space, we obtain
\begin{eqnarray}
\delta{\Phi}_{\rm tot}({\bf w},{\bf J},t)&=&  2\pi m \sum_{\bf k} \int
\frac{d\omega}{2\pi}\, e^{i({\bf k}\cdot {\bf w}-\omega t)}\sum_{{\bf k}'} 
A^d_{{\bf k}{\bf k}'}({\bf J},{\bf J}_0,\omega)  
e^{-i{\bf k}'\cdot {\bf w}_0}\delta(\omega-{\bf k}'\cdot
{\bf\Omega}_0)\nonumber\\
&=& m \sum_{{\bf k}{\bf k}'} A^d_{{\bf k}{\bf k}'}({\bf J},{\bf
J}_0,{\bf k}'\cdot {\bf \Omega}_0)   e^{i({\bf
k}\cdot {\bf w}- {\bf
k}'\cdot {\bf w}_0-{\bf
k}'\cdot {\bf {\Omega}}_0 t)}.
\label{arm2}
\end{eqnarray}
The corresponding  field is 
\begin{eqnarray}
-\frac{\partial\delta{\Phi}_{\rm tot}}{\partial {\bf w}}=-m \sum_{{\bf k}{\bf
k}'} i{\bf k} A^d_{{\bf k}{\bf k}'}({\bf J},{\bf
J}_0,{\bf k}'\cdot {\bf \Omega}_0)   e^{i({\bf
k}\cdot {\bf w}- {\bf
k}'\cdot {\bf w}_0-{\bf
k}'\cdot {\bf {\Omega}}_0 t)}.
\label{arm3}
\end{eqnarray}
The test particle is submitted to the force resulting from the
perturbation that
it has caused and, as
a result,  it experiences a dynamical friction. Applying Eq.
(\ref{arm3}) at the position of the 
test particle at time $t$ (${\bf w}={\bf \Omega}_0 t+{\bf w}_0$, 
${\bf J}={\bf J}_0$), and averaging over  ${\bf w}$ (which amounts to making
${\bf k}'={\bf k}$),  we obtain the
friction by polarization 
\begin{eqnarray}
{\bf F}_{\rm pol}=-m \sum_{{\bf k}} i{\bf k} A^d_{{\bf k}{\bf k}}({\bf
J},{\bf
J},{\bf k}\cdot {\bf \Omega}). 
\label{arm4}
\end{eqnarray}
Since ${\bf F}_{\rm pol}$ is real, we can write
\begin{eqnarray}
{\bf F}_{\rm pol}
= m  \sum_{{\bf k}}{\bf k}\,  {\rm Im}\left\lbrack A^d_{{\bf k}{\bf
k}}({\bf J},{\bf J},{\bf k}\cdot {\bf
\Omega}) \right\rbrack,
\label{arm5}
\end{eqnarray}
which returns Eq. (\ref{ci1bnh}) then Eq. (\ref{gei4zz}).

\section{Einstein relation}
\label{sec_ein}

We consider here the evolution of a test particle  is a bath of field
particles.  If the field particles are at statistical equilibrium with the
Boltzmann
distribution (\ref{fdi3voit}), the friction by polarization from Eq.
(\ref{gei4zzb}) can
be rewritten as
\begin{eqnarray}
{\bf F}_{\rm pol}&=&-\beta\pi(2\pi)^{d}m\sum_b m_b \sum_{{\bf k},{\bf k}'}\int
d{\bf
J}'\, {\bf k} |A^d_{{\bf k},{\bf k}'}({\bf J},{\bf
J'},{\bf k}\cdot {\bf\Omega})|^2 \delta({\bf
k}\cdot {\bf\Omega}-{\bf k}'\cdot
{\bf\Omega}')\left ({\bf k}'\cdot {\bf \Omega}'\right) f_b({\bf J}')\nonumber\\
&=&-\beta\pi(2\pi)^{d}m\sum_b m_b \sum_{{\bf k},{\bf k}'}\int
d{\bf
J}'\, {\bf k}
|A^d_{{\bf k},{\bf k}'}({\bf J},{\bf
J'},{\bf k}\cdot {\bf\Omega})|^2 \delta({\bf
k}\cdot {\bf\Omega}-{\bf k}'\cdot
{\bf\Omega}')\left ({\bf k}\cdot {\bf \Omega}\right) f_b({\bf J}')\nonumber\\
&=&-\beta m \Omega_j\pi(2\pi)^{d}\sum_b m_b \sum_{{\bf k},{\bf k}'}\int
d{\bf
J}'\, k_i k_j |A^d_{{\bf k},{\bf k}'}({\bf J},{\bf
J'},{\bf k}\cdot {\bf\Omega})|^2 \delta({\bf
k}\cdot {\bf\Omega}-{\bf k}'\cdot
{\bf\Omega}')   f_b({\bf J}').
\label{gei4zzd}
\end{eqnarray}
To get the second line, we have used the properties of the $\delta$-function.
Recalling the expression (\ref{di1})  of the diffusion tensor, we
obtain\footnote{Since the Fokker-Planck current
${\cal J}_i=-D_{ij}\partial_jP+PF_i^{\rm
pol}$ [see Eq. (\ref{fp8})] must vanish at equilibrium for the Boltzmann
distribution
$P_{\rm eq}=Ae^{-\beta m H}$, we must
have ${\bf F}_{\rm pol}=-\beta m D_{ij}\Omega_j$. This relation was given by
Binney and Lacey \cite{bl}. It is valid at equilibrium. However, the relation
from Eq.
(\ref{ifpol3}) initially obtained in \cite{physicaA}  is more general because it
only assumes that the field particles are at statistical
equilibrium while the test particles are not necessarily at
statistical equilibrium.}
\begin{eqnarray}
{\bf F}_{\rm pol}=-\beta m D_{ij}\Omega_j.
\label{ifpol3}
\end{eqnarray}
We see that the
friction by polarization
is equal to a sum of terms which are proportional and opposite to the
components of the pulsation ${\bf\Omega}$ (i.e. the energy gradient
$\partial H/\partial {\bf J}$) and that the friction
tensor is given by a form of
Einstein relation \cite{physicaA}
\begin{eqnarray}
\xi_{ij}=D_{ij}\beta m,
\label{ham88}
\end{eqnarray}
like in the theory of Brownian motion \cite{chandrabrown}. The Einstein relation
connecting the friction to the diffusion tensor  is a
manifestation of the fluctuation-dissipation theorem (see the Remark below). We
note that
the Einstein relation is valid for the
friction by polarization  ${\bf
F}_{\rm pol}$, not for the total friction which has
a more complicated
expression
\begin{equation}
\label{ag}
F_{i}^{\rm tot}=-\beta m D_{ij}\Omega_j+\frac{\partial D_{ij}}{\partial J_{j}}
\end{equation}
due to the term $\partial_j D_{ij}$ [see Eqs. (\ref{fp9}) and
(\ref{totfric})]. 
We do not
have this subtlety
for the usual Brownian motion where the diffusion coefficient is constant. The
necessity  of the friction force and the Einstein relation
were discussed  by Chandrasekhar  \cite{chandra1,nice} in the case of
spatially homogeneous stellar
systems by using very general
arguments. Eq.
(\ref{ifpol3}) generalizes the expression of the Chandrasekhar dynamical
friction
experienced by a star in a spatially homogeneous stellar system at
statistical equilibrium (thermal bath) to the inhomogeneous case.

{\it Remark:} These results can also be obtained by substituting the 
fluctuation-dissipation theorem (\ref{fdi5})  valid at statistical equilibrium
into the expression (\ref{ci1bnh}) of the friction by polarization. This yields
\begin{eqnarray}
{\bf F}_{\rm pol}&=& - m \sum_{\bf k}\, {\bf k} \frac{{\bf k}\cdot {\bf
\Omega}}{2k_B T} P({\bf k},{\bf J},{\bf k}\cdot {\bf \Omega})\nonumber\\
&=& - \frac{1}{2} \beta m \Omega_j \sum_{\bf k}\, k_ik_j
P({\bf k},{\bf J},{\bf k}\cdot {\bf \Omega}).
\label{ifpol2}
\end{eqnarray}
Recalling the expression (\ref{gei3}) of the diffusion tensor, we recover
Eq. (\ref{ifpol3}). In this sense, the Einstein relation is another formulation
of the fluctuation-dissipation theorem. On the other hand, combining Eqs.
(\ref{dcorr}) and
(\ref{ham88}) we obtain the following relation
\begin{eqnarray}
\label{xi}
\xi_{ij}=\beta m \int_0^{+\infty} \langle {\cal F}_i(0){\cal
F}_j(t)\rangle\, dt
\end{eqnarray}
between the friction coefficient to the force auto-correlation function.
This is a form of
Green-Kubo \cite{green,green1,green2,kubo0,kubo2,kubo} formula expressing the
fluctuation-dissipation theorem.\footnote{Actually, this formula
was first derived by Kirkwood \cite{kirkwood} by a direct calculation of the
friction force. In the approach of Kirkwood \cite{kirkwood}, the
friction coefficient is  determined  via Eq. (\ref{xi}) by the
auto-correlation function of the inter-molecular forces. In this
sense, Eq. (\ref{xi}) may be viewed as a generalization of the Stokes formula.
Alternatively, in the usual theory of Brownian
motion \cite{einstein,chandrabrown}, the friction force is assumed to be given
by the Stokes formula and the Einstein relation relates the diffusion
coefficient to the friction coefficient.} This formula appeared in the theory of
Brownian motion, fluids,
plasmas and stellar systems
\cite{kirkwood,green,green1,green2,ross,mori56,kubo0,kubo2,mori2,mori,rubin,
zwanzig,
resibois,helfand60,helfand,kubo,kandruprep,kandrup1,kandrupfriction,bm92,maoz,
cp,nt99,hb4,aa}.

\section{Inhomogeneous Lenard-Balescu equation}
\label{sec_avp}

The kinetic equation of inhomogeneous systems with long-range interactions can
be obtained by substituting the expressions of the diffusion tensor (\ref{di1})
and
friction by polarization (\ref{gei4zzb}) into the Fokker-Planck equation (\ref{fp8}). This provides
an alternative  derivation of the inhomogeneous Lenard-Balescu kinetic equation
as compared to the one given
in \cite{physicaA}  which is based on the Klimontovich formalism.

\subsection{Multi-species case}

Substituting Eqs. (\ref{di1})
and (\ref{gei4zzb}) into the Fokker-Planck equation
(\ref{fp8}) and introducing the DF
$f_a=N_a m_a P_1^{(a)}$ of each species of particles we
obtain the integrodifferential equation
\begin{equation}
\frac{\partial f_a}{\partial t}=\pi (2\pi)^d  \frac{\partial}{\partial {\bf
J}}\cdot \sum_b\sum_{{\bf k},{\bf k}'}\int d{\bf J}' \, {\bf k}\,
|A^d_{{\bf k},{\bf k}'}({\bf J},{\bf J}',{\bf k}\cdot {\bf
\Omega})|^2\delta({\bf k}\cdot {\bf \Omega}-{\bf k}'\cdot {\bf \Omega}') \left
(m_b f'_b {\bf k}\cdot \frac{\partial f_a}{\partial {\bf J}}-m_a f_a {\bf
k}'\cdot \frac{\partial f'_b}{\partial {\bf J}'}\right ),
\label{ilb3}
\end{equation}
where $f_a$ stands for $f_a({\bf J},t)$ and ${\bf\Omega}$ stands for
${\bf\Omega}({\bf J},t)$. The pulsation ${\bf\Omega}({\bf J},t)$ is
determined by the total DF $f({\bf
J},t)=\sum_a f_a({\bf J},t)$. Equation (\ref{ilb3}) is the multi-species
inhomogeneous Lenard-Balescu
equation \cite{heyvaerts,physicaA}. It
describes the collisional evolution
of the system due to finite $N$ effects. When collective effects are
neglected,
i.e.,
when $A^d_{{\bf k},{\bf
k}'}({\bf J},{\bf J}',{\bf k}\cdot {\bf
\Omega})$ is replaced by
$A_{{\bf k},{\bf k}'}({\bf J},{\bf J}')$, it reduces to the
multi-species inhomogeneous Landau equation \cite{aa}. It can be
shown \cite{hfcp} that the kinetic equation (\ref{ilb3}) conserves the total
energy $E=\int f H\, d{\bf J}$ and the particle number $N_a$ (or the mass
$M_a=N_a
m_a=\int f_a\, d{\bf J}$) of each species of particles, and
that the Boltzmann entropy $S=-\sum_a\int
\frac{f_a}{m_a}\ln \frac{f_a}{m_a}\, d{\bf J}$ increases monotonically:  $\dot
S\ge 0$
($H$-theorem). Furthermore, it relaxes towards the
multi-species Boltzmann
distribution
\begin{eqnarray}
f_a^{\rm eq}({\bf J})=A_a e^{-\beta m_a H({\bf J})},
\label{ilb4}
\end{eqnarray}
where the inverse temperature $\beta$ is the same for all species of particles.
The Lenard-Balescu
equation is valid at the order
$1/N$ so it describes the evolution of the system on a timescale of order
$Nt_D$, where $t_D$ is the dynamical time. Therefore, in the inhomogeneous case,
the relaxation time of the system as a whole scales as\footnote{For 3D stellar
systems, the relaxation time scales as $t_R\sim (N/\ln\Lambda) t_D$. There is a
logarithmic correction due to
the effect of strong collisions at small scales (the divergence at large scales
is regularized by the inhomogeneity of the system and by its finite extension).}
\begin{equation}
t_{\rm R}\sim {N}t_D.
\label{rwhole}
\end{equation}
There is a case where these results are not valid because of a
situation of kinetic blocking at the order $1/N$. This situation is
specifically discussed in Sec. \ref{sec_mono}. We note that
the Boltzmann distributions of the different species of particles satisfy the
relation
\begin{eqnarray}
f_a^{\rm eq}({\bf J})=C_{ab} [f_b^{\rm eq}({\bf J})]^{m_a/m_b},
\label{ilb5}
\end{eqnarray}
where $C_{ab}$ is a constant (independent of ${\bf J}$). The properties of the
multi-species
Lenard-Balescu equation are detailed in Appendix C of \cite{hfcp} .

{\it Remark:} 
Substituting Eqs. (\ref{gei3}) and (\ref{ci1bnh})  into the Fokker-Planck
equation (\ref{fp8}) we obtain the
kinetic  equation
\begin{equation}
\label{ilb9}
\frac{\partial f_a}{\partial t}=\frac{\partial}{\partial J_{i}} \sum_{{\bf
k}}\left
\lbrace\frac{1}{2} k_i k_j  P({\bf k},{\bf J},{\bf
k}\cdot {\bf\Omega}({\bf J}))\frac{\partial f_a}{\partial
J_{j}}- m_a f_a   k_i\,  {\rm
Im}\left\lbrack A^d_{{\bf k},{\bf k}}({\bf J},{\bf J},{\bf k}\cdot {\bf
\Omega}) \right\rbrack\right \rbrace.
\end{equation}
When the external potential is due to  a
discrete collection of $N$ field particles, using Eqs. (\ref{pi4}) and
(\ref{fdi2a}), we can check that Eq. (\ref{ilb9}) returns Eq. (\ref{ilb3}).
However, Eq. (\ref{ilb9}) is more general
than  Eq. (\ref{ilb3}) because it is valid for an
arbitrary external forcing (see Sec. \ref{sec_gke}).

\subsection{Moment equations}

The kinetic equation (\ref{ilb3}) can be written under the form of a
Fokker-Planck equation
\begin{equation}
\label{ilb10}
\frac{\partial f_a}{\partial t}=\frac{\partial}{\partial J_{i}} \left
(D_{ij}\frac{\partial f_a}{\partial
J_{j}}- f_a F_{\rm pol,i}^{(a)}\right )
\end{equation} 
with a diffusion tensor [see Eq. (\ref{di1})]
\begin{eqnarray}
D_{ij}&=&\pi(2\pi)^{d}\sum_b \sum_{{\bf k},{\bf k}'}\int d{\bf
J}'\, k_ik_j
|A^d_{{\bf k},{\bf k}'}({\bf J},{\bf
J'},{\bf k}\cdot {\bf\Omega})|^2 \delta({\bf
k}\cdot {\bf\Omega}-{\bf k}'\cdot
{\bf\Omega}')m_bf_b({\bf J}')\nonumber\\
&=&\pi(2\pi)^{d}\sum_{{\bf k},{\bf k}'}\int d{\bf
J}'\, k_ik_j
|A^d_{{\bf k},{\bf k}'}({\bf J},{\bf
J'},{\bf k}\cdot {\bf\Omega})|^2 \delta({\bf
k}\cdot {\bf\Omega}-{\bf k}'\cdot
{\bf\Omega}')f_2({\bf J}')
\label{ilb11}
\end{eqnarray}
and a friction by polarization [see Eq. (\ref{gei4zzb})] 
\begin{eqnarray}
{\bf F}_{\rm pol}^{(a)}=\pi(2\pi)^{d}m_a\sum_b\sum_{{\bf k},{\bf k}'}\int d{\bf
J}'\, {\bf k}
|A^d_{{\bf k},{\bf k}'}({\bf J},{\bf
J'},{\bf k}\cdot {\bf\Omega})|^2 \delta({\bf
k}\cdot {\bf\Omega}-{\bf k}'\cdot
{\bf\Omega}')\left ({\bf k}'\cdot \frac{\partial f'_b}{\partial {\bf J}'}\right
)\nonumber\\
=\pi(2\pi)^{d}m_a\sum_{{\bf k},{\bf k}'}\int d{\bf
J}'\, {\bf k}
|A^d_{{\bf k},{\bf k}'}({\bf J},{\bf
J'},{\bf k}\cdot {\bf\Omega})|^2 \delta({\bf
k}\cdot {\bf\Omega}-{\bf k}'\cdot
{\bf\Omega}')\left ({\bf k}'\cdot \frac{\partial f'}{\partial {\bf J}'}\right
).
\label{ilb12}
\end{eqnarray}
In the second equalities of Eqs. (\ref{ilb11}) and
(\ref{ilb12}) we have introduced  the total DF $f=\sum_a f_a=\sum_a N_a m_a
P_1^{(a)}$ and the second
moment $f_2=\sum_a m_a f_a=\sum_a N_a m_a^2
P_1^{(a)}$ of the local mass distribution. As explained
previously, the diffusion tensor $D_{ij}$ of a test particle is due to the
fluctuations
of the field particles so it depends on $\lbrace
m_b\rbrace$ through
$f_2$. By
contrast, the friction by polarization acting on a test particle of species $a$
is due to the
retroaction (response) of the perturbation that this particle causes on the
systems' DF
$f$. As a
result, ${\bf F}_{\rm pol}^{(a)}$ is proportional to $m_a$.

The equation for the total DF is
\begin{equation}
\frac{\partial f}{\partial t}=\pi (2\pi)^d  \frac{\partial}{\partial {\bf
J}}\cdot \sum_{{\bf k},{\bf k}'}\int d{\bf J}' \, {\bf k}\,
|A^d_{{\bf k},{\bf k}'}({\bf J},{\bf J}',{\bf k}\cdot {\bf
\Omega})|^2\delta({\bf k}\cdot {\bf \Omega}-{\bf k}'\cdot {\bf \Omega}') \left
(f'_2 {\bf k}\cdot \frac{\partial f}{\partial {\bf J}}-f_2 {\bf
k}'\cdot \frac{\partial f'}{\partial {\bf J}'}\right ).
\label{ilb13z}
\end{equation}
This equation depends
on the second moment $f_2$. We
can write down a hierarchy of equations for the moments $f_n=\sum_a
m_a^{n-1}f_a=\sum_a N_a m_a^n
P_1^{(a)}$. The
generic term of this hierarchy is
\begin{equation}
\frac{\partial f_n}{\partial t}=\pi (2\pi)^d  \frac{\partial}{\partial {\bf
J}}\cdot \sum_{{\bf k},{\bf k}'}\int d{\bf J}' \, {\bf k}\,
|A^d_{{\bf k},{\bf k}'}({\bf J},{\bf J}',{\bf k}\cdot {\bf
\Omega})|^2\delta({\bf k}\cdot {\bf \Omega}-{\bf k}'\cdot {\bf \Omega}') \left
(f'_2 {\bf k}\cdot \frac{\partial f_n}{\partial {\bf J}}-f_{n+1} {\bf
k}'\cdot \frac{\partial f'}{\partial {\bf J}'}\right ).
\label{ilb13}
\end{equation}
This hierarchy is not closed (except for a single species system) since the
equation for $f_n$ depends on
$f_{n+1}$.  For a single species
system  the Lenard-Balescu equation  (\ref{ilb3})
reduces to
\begin{equation}
\frac{\partial f}{\partial t}=\pi (2\pi)^d m \frac{\partial}{\partial {\bf
J}}\cdot \sum_{{\bf k},{\bf k}'}\int d{\bf J}' \, {\bf k}\,
|A^d_{{\bf k},{\bf k}'}({\bf J},{\bf J}',{\bf k}\cdot {\bf
\Omega})|^2\delta({\bf k}\cdot {\bf \Omega}-{\bf k}'\cdot {\bf \Omega}') \left
(f' {\bf k}\cdot \frac{\partial f}{\partial {\bf J}}-f {\bf
k}'\cdot \frac{\partial f'}{\partial {\bf J}'}\right ).
\label{ilb6}
\end{equation}
In that case, all the equations (\ref{ilb13}) are equivalent to Eq.
(\ref{ilb6}) since $f_n=m^{n-1} f$.

\subsection{Thermal bath approximation}
\label{sec_tbaw}

We consider a test particle\footnote{This can be a single particle or an
ensemble of noninteracting particles of the same species. In the second case,
we take into
account the collisions
between the test particles of mass $m$ and the field particles
of masses $\lbrace m_b\rbrace$,
but we ignore the collisions between the test particles themselves.} of mass $m$
experiencing ``collisions'' with field particles  of masses $\lbrace
m_b\rbrace$. We assume
that the DF $f_b$ of the field particles 
is prescribed (the validity of this approximation is discussed below). The
pulsation ${\bf \Omega}({\bf J})$, which is determined
by $f=\sum_b f_b$, is also prescribed. This situation
corresponds to the bath approximation in its general form. Under these
conditions, the Lenard-Balescu equation (\ref{ilb3}) reduces  to
\begin{equation}
\frac{\partial P}{\partial t}=\pi (2\pi)^d  \frac{\partial}{\partial {\bf
J}}\cdot \sum_b\sum_{{\bf k},{\bf k}'}\int d{\bf J}' \, {\bf k}\,
|A^d_{{\bf k},{\bf k}'}({\bf J},{\bf J}',{\bf k}\cdot {\bf
\Omega})|^2\delta({\bf k}\cdot {\bf \Omega}-{\bf k}'\cdot {\bf \Omega}') \left
(m_b f'_b {\bf k}\cdot \frac{\partial P}{\partial {\bf J}}-m P {\bf
k}'\cdot \frac{\partial f'_b}{\partial {\bf J}'}\right ).
\label{ilb14}
\end{equation}
Equation (\ref{ilb14}) governs the evolution of the probability
density $P({\bf J},t)$ that the test particle of mass $m$ has an action 
${\bf J}$ at
time $t$.  It can be written under
the form of a Fokker-Planck equation
\begin{equation}
\label{ilb15}
\frac{\partial P}{\partial t}=\frac{\partial}{\partial J_{i}} \left
(D_{ij}\frac{\partial P}{\partial
J_{j}}- P F_{i}^{\rm pol}\right )
\end{equation}
with a diffusion tensor
\begin{eqnarray}
D_{ij}&=&\pi(2\pi)^{d}\sum_b \sum_{{\bf k},{\bf k}'}\int d{\bf
J}'\, k_ik_j
|A^d_{{\bf k},{\bf k}'}({\bf J},{\bf
J'},{\bf k}\cdot {\bf\Omega})|^2 \delta({\bf
k}\cdot {\bf\Omega}-{\bf k}'\cdot
{\bf\Omega}')m_bf_b({\bf J}')\nonumber\\
&=&\pi(2\pi)^{d}\sum_{{\bf k},{\bf k}'}\int d{\bf
J}'\, k_ik_j
|A^d_{{\bf k},{\bf k}'}({\bf J},{\bf
J'},{\bf k}\cdot {\bf\Omega})|^2 \delta({\bf
k}\cdot {\bf\Omega}-{\bf k}'\cdot
{\bf\Omega}')f_2({\bf J}')
\label{ilb16}
\end{eqnarray}
and a friction by polarization
\begin{eqnarray}
{\bf F}_{\rm pol}&=&\pi(2\pi)^{d}m\sum_b\sum_{{\bf k},{\bf k}'}\int d{\bf
J}'\, {\bf k}
|A^d_{{\bf k},{\bf k}'}({\bf J},{\bf
J'},{\bf k}\cdot {\bf\Omega})|^2 \delta({\bf
k}\cdot {\bf\Omega}-{\bf k}'\cdot
{\bf\Omega}')\left ({\bf k}'\cdot \frac{\partial f'_b}{\partial {\bf J}'}\right
)\nonumber\\
&=&\pi(2\pi)^{d}m\sum_{{\bf k},{\bf k}'}\int d{\bf
J}'\, {\bf k}
|A^d_{{\bf k},{\bf k}'}({\bf J},{\bf
J'},{\bf k}\cdot {\bf\Omega})|^2 \delta({\bf
k}\cdot {\bf\Omega}-{\bf k}'\cdot
{\bf\Omega}')\left ({\bf k}'\cdot \frac{\partial f'}{\partial {\bf J}'}\right
).
\label{ilb17}
\end{eqnarray}
In this manner, we have transformed
an integrodifferential equation of the Lenard-Balescu or Landau form [see Eq.
(\ref{ilb3})] into a  differential equation  of the Fokker-Planck form [see Eq.
(\ref{ilb14})].

This bath approach is self-consistent in the general case only if the field
particles are at statistical equilibrium, otherwise their distribution
$f_b$ evolves in time due to discreteness effects. If we assume that
the field particles are at statistical equilibrium with the Boltzmann
distribution (\ref{fdi3voit}), corresponding to the thermal bath
approximation, the Fokker-Planck
equation (\ref{ilb14}) becomes
\begin{eqnarray}
\frac{\partial P}{\partial t}&=&\pi (2\pi)^d  \frac{\partial}{\partial {\bf
J}}\cdot \sum_b\sum_{{\bf k},{\bf k}'}\int d{\bf J}' \, {\bf k}\,
|A^d_{{\bf k},{\bf k}'}({\bf J},{\bf J}',{\bf k}\cdot {\bf
\Omega})|^2\delta({\bf k}\cdot {\bf \Omega}-{\bf k}'\cdot {\bf \Omega}') \left
(m_b f'_b {\bf k}\cdot \frac{\partial P}{\partial {\bf J}}+\beta m_b m P {\bf
k}'\cdot {\bf \Omega}' f'_b\right )\nonumber\\
&=&\pi (2\pi)^d  \frac{\partial}{\partial {\bf
J}}\cdot \sum_b\sum_{{\bf k},{\bf k}'}\int d{\bf J}' \, {\bf k}\,
|A^d_{{\bf k},{\bf k}'}({\bf J},{\bf J}',{\bf k}\cdot {\bf
\Omega})|^2\delta({\bf k}\cdot {\bf \Omega}-{\bf k}'\cdot {\bf \Omega}')m_b
f'_b {\bf k}\cdot \left
 ( \frac{\partial P}{\partial {\bf J}}+\beta  m P {\bf \Omega}\right ),
\label{ilb18}
\end{eqnarray}
where we have used Eq. (\ref{fdi3}) to get the first equality and the property of the
$\delta$-function to get the second equality. We can then write
the Fokker-Planck equation under the form \cite{physicaA}
\begin{eqnarray}
\frac{\partial P}{\partial t}=\frac{\partial}{\partial J_i}\left
\lbrack D_{ij}\left
(\frac{\partial P}{\partial J_j}+\beta m P \Omega_j\right )\right\rbrack,
\label{ilb19}
\end{eqnarray}
where $D_{ij}$ is given by Eq. (\ref{ilb16}) with the Boltzmann
distribution (\ref{fdi3voit}). Equation (\ref{ilb19}) can also be directly obtained from Eq.
(\ref{ilb15}) by using the expression (\ref{ifpol3}) of the friction by polarization valid
for a
thermal bath (Einstein relation). The Fokker-Planck
equation
(\ref{ilb19}) conserves the normalization condition $\int P\, d{\bf J}=1$ and
decreases the free energy $F=E-TS$ with $E=\int PmH({\bf J})\, d{\bf J}$ and
$S=-\int
P\ln P\, d{\bf J}$ monotonically: $\dot F\le 0$ (canonical $H$-theorem). It
relaxes
towards the Boltzmann distribution
\begin{eqnarray}
P_{\rm eq}({\bf J})=A\, e^{-\beta m H({\bf J})}.
\label{ilb20}
\end{eqnarray}
Since the Fokker-Planck is valid at the
order $1/N$, the
relaxation time of a test particle in a thermal bath scales as
\begin{equation}
t_{\rm R}^{\rm bath}\sim {N}t_D,
\label{rbath1}
\end{equation}
where $t_D$ is the dynamical time (for self-gravitating systems the relaxation
time scales likes $t_{\rm R}^{\rm bath}\sim (N/\ln N)\, t_D$ as explained in
footnote 40). The Fokker-Planck equation (\ref{ilb19}) is similar to the
Kramers equation in angle-action variables describing the
evolution of an underdamped  Brownian particle evolving in an external potential
$\Phi$ \cite{kramers}. The
relaxation of the test particle towards statistical equilibrium is due to a
competition
between
diffusion and friction. For inhomogeneous systems, the friction force is
proportional and opposite to the components of the pulsation ${\bf \Omega}({\bf
J})$ while for homogeneous systems it  is
proportional and opposite to the velocity ${\bf v}$. Recalling that
${\bf\Omega}({\bf
J})=\partial H/\partial {\bf J}$,  the Fokker-Planck equation
(\ref{ilb19})  for a
test particle is also similar to the
 Smoluchowski (drift-diffusion) equation describing the evolution of the
DF $P({\bf J},t)$ of an overdamped Brownian particle in an ``effective
potential''
$U_{\rm eff}({\bf J})=H({\bf J})$. For homogeneous systems, the potential is
quadratic: $U_{\rm eff}({\bf v})=v^2/2$ (harmonic oscillator). For
inhomogeneous systems it has a more general expression determined by the DF of
the field
particles.

{\it Remark:} In the thermal bath
approximation, substituting the
fluctuation-dissipation theorem (\ref{fdi5}) into the kinetic equation
(\ref{ilb9}), or combining Eqs. (\ref{gei3}), (\ref{ifpol2}) and (\ref{ilb15}),
we
obtain
\begin{equation}
\label{ilb21}
\frac{\partial f}{\partial t}=\frac{1}{2}\frac{\partial}{\partial J_{i}}
\sum_{{\bf
k}} k_i k_j  P({\bf k},{\bf J},{\bf
k}\cdot {\bf\Omega}({\bf J}))\left (\frac{\partial f}{\partial
J_{j}}+\beta m f \Omega_j\right ).
\end{equation}
This equation  is equivalent to the Fokker-Planck equation  (\ref{ilb19}). The
diffusion tensor
$D_{ij}$ is given by Eq. (\ref{gei3}) with Eqs. (\ref{pi4}) and (\ref{fdi3voit})
which return Eq.
(\ref{ilb16}).

\subsection{Pure diffusion}
\label{sec_purediff}

When the test particle is much lighter than the field
particles ($m\ll m_b$), the friction by polarization can be neglected
(${\bf F}_{\rm
pol}={\bf 0}$) and the test particle has
a purely diffusive evolution. In that case, the Fokker-Planck equation
(\ref{ilb15}) reduces
to 
\begin{eqnarray}
\frac{\partial P}{\partial t}=\frac{\partial}{\partial J_i}\left
( D_{ij}\frac{\partial P}{\partial J_j}\right ),
\label{ilb22}
\end{eqnarray}
where $D_{ij}$ is given by Eq. (\ref{ilb16}). 
We note that the diffusion tensor $D_{ij}$ depends on the masses $\lbrace
m_b\rbrace$ of the
field particles through
the second moment $f_2$. This  reflects  the discrete
nature of the field particles. On the other hand, the diffusion tensor does
not
depend on the mass $m$ of the test particle.

{\it Remark:} Since the diffusion coefficient depends on ${\bf J}$ and since it
is placed between the two derivatives $\partial/\partial {\bf J}$, Eq.
(\ref{ilb22}) is
not exactly a diffusion equation. It can be rewritten as Eq. (\ref{fp5})  showing that
the test particle experiences a friction [see Eq.
(\ref{fp9})] 
\begin{eqnarray}
{\bf F}_{\rm tot}=\frac{\partial D_{ij}}{\partial J_{j}}.
\label{pd2}
\end{eqnarray}

{\it Remark:} This type of equations  was first introduced and studied by
Spitzer and Schwarzschild \cite{ss} in an
astrophysical context. They considered
spatially homogeneous systems with different populations of objects and
neglected collective effects. They
showed that the collisions of test particles (stars) with massive perturbers
like molecular clouds can considerably increase their diffusion
and heating.\footnote{A similar result was obtained by Osterbrock
\cite{osterbrock} who studied the diffusion of stars due to the fluctuations
in the density field of the interstellar medium by using a theory of
turbulence.} They studied the velocity distribution resulting from
star-cloud encounters as a function of
time numerically.\footnote{A recent discussion of this problem
has been given
in Appendix F of Ref.
\cite{kinquant} where a self-similar solution of the Spitzer-Schwarzschild
equation is obtained. The velocity distribution is non-Maxwellian and
the velocity dispersion increases with time as $\sigma^2\sim t^{2/5}$ instead of
$\sigma^2\sim t$.}
The extension of the Spitzer-Schwarzschild problem to razor-thin planar
discs was considered by Binney and Lacey \cite{bl} using
angle-action variables.

\subsection{Pure friction}

When the test particle is much heavier than the field
particles ($m\gg m_b$), the diffusion can be neglected ($D_{ij}=0$) and the test
particle has a purely
deterministic evolution. In that case, the Fokker-Planck equation (\ref{ilb15})
reduces
to 
\begin{equation}
\label{ilb24}
\frac{\partial P}{\partial t}=\frac{\partial}{\partial {\bf J}}\cdot \left
(- P {\bf F}_{\rm pol}\right )
\end{equation}
where ${\bf F}_{\rm pol}$ is given by Eq. (\ref{ilb17}). 
We note that the friction by polarization ${\bf F}_{\rm pol}$ is proportional to
the mass $m$ of the test
particle. On the other hand, it depends only on the total
DF $f$ of the background medium, not on the individual masses $\lbrace
m_b\rbrace$ of the
field particles reflecting their discrete nature.

{\it Remark:} The deterministic motion of the test particle induced by the
friction force can be written as
\begin{eqnarray}
\frac{d{\bf J}}{dt}={\bf F}_{\rm pol}({\bf J},t),
\label{ham96}
\end{eqnarray}
and we have ${\bf F}_{\rm drift}={\bf F}_{\rm pol}$ [see Eq. (\ref{fp9})].
This equation describes the frictional decay of a particle in a
collisionless medium. This corresponds to the sinking satellite problem in
stellar dynamics \cite{btnew}. The change of
energy of the test particle is
$\dot \epsilon=m {\bf F}_{\rm pol}\cdot {\bf\Omega}$. This relation can be
obtained by taking the
time variation of $E=m\int PH \, d{\bf J}$, using Eq. (\ref{ilb24}), and
integrating by parts. Using the general expression  (\ref{gei4zz}) of the
friction term, we obtain
\begin{eqnarray}
\dot\epsilon=\pi(2\pi)^{d}m^2\sum_{{\bf k},{\bf k}'}\int d{\bf
J}'\, ({\bf k}'\cdot{\bf\Omega}')
|A^d_{{\bf k},{\bf k}'}({\bf J},{\bf
J'},{\bf k}\cdot {\bf\Omega})|^2 \delta({\bf
k}\cdot {\bf\Omega}-{\bf k}'\cdot
{\bf\Omega}')\left ({\bf k}'\cdot \frac{\partial f'}{\partial {\bf J}'}\right
).
\label{cat1}
\end{eqnarray}
Only resonant orbits with ${\bf
k}\cdot {\bf\Omega}={\bf k}'\cdot {\bf\Omega}'$ contribute to the energy
transfer. The damping or growth of the perturbation depends on the components
of ${\bf \Omega}({\bf J})$ and on the gradient of $f({\bf J})$ along the
resonance line. For a stationary DF of the form $f=f(H)$, that depends on the
energy alone, we find that
\begin{eqnarray}
\dot\epsilon=\pi(2\pi)^{d}m^2\sum_{{\bf k},{\bf k}'}\int d{\bf
J}'\, ({\bf k}'\cdot{\bf\Omega}')^2
|A^d_{{\bf k},{\bf k}'}({\bf J},{\bf
J'},{\bf k}\cdot {\bf\Omega})|^2 \delta({\bf
k}\cdot {\bf\Omega}-{\bf k}'\cdot
{\bf\Omega}') \frac{df'}{dH'}.
\label{cat2}
\end{eqnarray}
Therefore $\dot\epsilon$ is negative when $df/dH<0$ and positive when
$df/dH>0$. When the DF  is a decreasing function of energy ($df/dH<0$), which is
the case in most model galaxies with $f=f(H)$, the test particle
loses energy to the embedding system (e.g. the halo). For the Boltzmann
distribution (\ref{fdi3voit}), we have
$df/dH=-\beta f_2<0$. These results (without
collective effects) were originally derived in Refs.
\cite{lbk,ps80,palmer,nt95,nt99}.

\section{Kinetic equation for one dimensional systems with a monotonic frequency
profile}
\label{sec_mono}

In this section, we consider an inhomogeneous 1D system with long-range
interactions and assume that the binary potential of interaction is of the form
$u=u(|w-w'|,J,J')$. This is the case for spins with long-range interactions
moving on a sphere in relation to the process of vector resonant
relaxation (VRR) in galactic nuclei \cite{fbc}. In that case only
$1:1$ resonances are permitted and the Lenard-Balescu equation (\ref{ilb3})
reduces to 
\begin{equation}
\frac{\partial f_a}{\partial t}=2\pi^2  \frac{\partial}{\partial J}
\sum_b\sum_{k}\int dJ' \, |k|\,
|A^d_{kk}(J,J',k \Omega)|^2\delta(\Omega-\Omega') \left
(m_b f'_b \frac{\partial f_a}{\partial J}-m_a f_a \frac{\partial
f'_b}{\partial J'}\right ).
\label{ilb31d}
\end{equation}
This equation is similar to the kinetic equation describing the evolution of an
axisymmetric
or unidirectional distribution of 2D point vortices (this analogy was
mentioned early in \cite{kin2007,unified}). Therefore, the discussion is
similar to the one given in \cite{Kvortex2023}. Below, we study the general
properties of the kinetic equation (\ref{ilb31d})  when the frequency profile is
monotonic.

\subsection{Multi-species systems}
\label{sec_mscq}

Introducing the notation  $\chi(J,J',\Omega(J))=2\pi^2\sum_{k} |k|\,
|A^d_{kk}(J,J',k \Omega)|^2$ we can rewrite Eq. (\ref{ilb31d}) as
\begin{equation}
\frac{\partial f_a}{\partial t}= \frac{\partial}{\partial J}
\sum_b\int dJ' \, \chi(J,J',\Omega(J))\delta(\Omega-\Omega') \left
(m_b f'_b \frac{\partial f_a}{\partial J}-m_a f_a \frac{\partial
f'_b}{\partial J'}\right ).
\label{aham105}
\end{equation}
If the frequency profile $\Omega(J)$ is
monotonic, using the identity 
\begin{eqnarray}
\delta(\Omega(J')-\Omega(J))=\frac{1}{|\Omega'(J)|}\delta(J-J'),
\label{gei8}
\end{eqnarray}
the
kinetic equation (\ref{aham105}) becomes\footnote{It is possible that,
initially, the frequency profile is
non-monotonic 
but that it becomes monotonic during the evolution. In that case, the evolution
of the DF is first described by Eq. (\ref{aham105}), then by Eq.
(\ref{aham105bg}).}
\begin{equation}
\frac{\partial f_a}{\partial t}= \frac{\partial}{\partial J}
\sum_b 
\frac{\chi(J,J,\Omega(J))}{|\Omega'(J)|} \left
(m_b f_b \frac{\partial f_a}{\partial J}-m_a f_a \frac{\partial
f_b}{\partial J}\right ).
\label{aham105bg}
\end{equation}
It can be written as a Fokker-Planck equation of the form of Eq.
(\ref{ilb10}) with a
diffusion coefficient 
\begin{eqnarray}
D=\sum_b  \frac{\chi(J,J,\Omega(J))}{|\Omega'(J)|}m_b f_b=
\frac{\chi(J,J,\Omega(J))}{|\Omega'(J)|} f_2
\label{aham94b}
\end{eqnarray}
and a friction by polarization
\begin{eqnarray}
F_{\rm pol}^{(a)}=m_a\sum_b   \frac{\chi(J,J,\Omega(J))}{|\Omega'(J)|}
\frac{\partial f_b}{\partial J}=m_a   \frac{\chi(J,J,\Omega(J))}{|\Omega'(J)|}
\frac{\partial f}{\partial J}.
\label{aham95b}
\end{eqnarray}
Equation
(\ref{aham105bg}) conserves the energy, the impulse and the particle number of
each species, and
satisfies an $H$-theorem for the Boltzmann entropy. At equilibrium, the
currents ${\cal J}_a$ defined by $\partial f_a/\partial t=-\partial
{\cal J}_a/\partial J$
vanish (for each species), and  we have the relation
\begin{eqnarray}
f_a^{\rm eq}(J)=C_{ab} [f_b^{\rm eq}(J)]^{m_a/m_b},
\label{aham109b}
\end{eqnarray}
where $C_{ab}$ is a constant (independent of $J$).  This relation is similar to
Eq. (\ref{ilb5})
which was obtained in the case where the equilibrium DF is given by the
Boltzmann distribution (\ref{ilb4}). However, it is important to stress
that
Eq.
(\ref{aham105bg})  does {\it not} relax towards the Boltzmann distribution (see
below). Therefore, the equilibrium DF $f_a^{\rm eq}(J)$
in Eq. (\ref{aham109b}) is generally not given by Eq. (\ref{ilb4}).

The equation for the total DF  reduces to
\begin{eqnarray}
\frac{\partial f}{\partial t}=0.
\label{aham106b}
\end{eqnarray}
This equation is closed and shows that the total DF profile does not
change (the total current vanishes). This kinetic blocking is discussed in more
detail in the following section. We note, by contrast, that the DF
$f_b$ of the different species evolves in time. Because of the friction term the
particles with large mass tend to concentrate around the maxima of the total DF.
The hierarchy of equations
for the moments of the DF reads
\begin{eqnarray}
\frac{\partial f_n}{\partial t}=\frac{\partial}{\partial J}\left\lbrack 
\frac{\chi(J,J,\Omega(J))}{|\Omega'(J)|}\left (f_{2} \frac{\partial
f_n}{\partial J}-f_{n+1}\frac{\partial f}{\partial J}\right
)\right\rbrack.
\label{aham107b}
\end{eqnarray}
This hierarchy of moments   is not closed (for $n\ge 2$) since the equation for
$f_{n}$ depends on $f_{n+1}$.

\subsection{Single-species systems}
\label{sec_ssc}

For a single species system of particles, the Lenard-Balescu equation 
(\ref{aham105})  reduces to
\begin{equation}
\frac{\partial f}{\partial t}=m\frac{\partial}{\partial J}
\int dJ' \, \chi(J,J',\Omega(J))\delta(\Omega-\Omega') \left
(f' \frac{\partial f}{\partial J}-f \frac{\partial
f'}{\partial J'}\right ).
\label{ham106}
\end{equation}
If the frequency profile is
monotonic we find, using identity
(\ref{gei8}), that
\begin{eqnarray}
\frac{\partial f}{\partial t}=m\frac{\partial}{\partial J} \int dJ'\,
\chi(J,J',\Omega(J))   \frac{1}{|\Omega'(J)|}\delta(J'-J)\left (f'
\frac{\partial f}{\partial J}-f\frac{\partial f'}{\partial J'}\right
)=0.
\label{ham107}
\end{eqnarray}
We recall that the Lenard-Balescu equation (\ref{ham106})  is 
valid at the order $1/N$ so it describes the evolution of the average DF
on a timescale $N t_D$ under the effect of two-body correlations. Equation
(\ref{ham107}) shows that the DF does not change on this timescale (the
Lenard-Balescu current vanishes at the order $1/N$). This is a situation of
kinetic
blocking due to the absence of resonance at the order $1/N$.\footnote{As long as
the frequency profile is nonmonotonic there are resonances leading to a
nonzero current (${\cal J}\neq
0$). However, the relaxation stops (${\cal J}=0$) when the frequency profile
becomes monotonic even if the system has not reached the Boltzmann distribution
of statistical equilibrium. This ``kinetic blocking'' is
illustrated in \cite{fbc}.} The DF may evolve on a longer timescale due
to higher order correlations  between the particles. For example,
three-body correlations are of order $1/N^2$ and induce an evolution of the DF
on a timescale $N^2\, t_D$. A kinetic equation valid at the order $1/N^2$ has
been derived by Fouvry \cite{fouvry} when collective effects are neglected. It
generalizes the results of \cite{fbcn2,fcpn2} for homogeneous 1D system with
long-range interactions. For generic potentials (see \cite{jbr} for more
details), this equation does not display kinetic blocking and relaxes
towards the Boltzmann distribution on a timescale
\begin{equation}
t_{\rm R}\sim {N}^2 t_D.
\label{rwholef}
\end{equation}
By contrast, for inhomogeneous systems with long-range interactions different
from those considered in the present section,  the kinetic equation is given by
Eq. (\ref{ilb3}) and the collision
term does not vanish even in the single species case.\footnote{This is
because there are more resonances at the order $1/N$.}  In that
case, the system relaxes towards the
Boltzmann distribution on a timescale $Nt_D$ \cite{physicaA}.

{\it Remark:} The same situation of 
kinetic blocking at the order $1/N$ occurs for 1D homogeneous systems of
material particles with long-range interactions  \cite{epjp,unified} such as
1D plasmas \cite{ef,dawson,kp} and the HMF model above the critical energy
\cite{cvb,bd}, and for
2D vortices with a monotonic profile of angular velocity in the axisymmetic
case or with a monotonic profile of velocity in the unidirectional case
\cite{pre,dubin2,cl,Kvortex2023}.

\subsection{Out-of-equilibrium bath}
\label{sec_ooeb}

The kinetic equation governing the evolution of a test particle of mass
$m$ in a 
bath of field particles with masses $\lbrace m_b\rbrace$ and
DFs $\lbrace f_b\rbrace$ is given by
\begin{equation}
\frac{\partial P}{\partial t}= \frac{\partial}{\partial J}
\sum_b\int dJ' \, \chi(J,J',\Omega(J))\delta(\Omega-\Omega') \left
(m_b f'_b \frac{\partial P}{\partial J}-m P \frac{\partial
f'_b}{\partial J'}\right ).
\label{pagny}
\end{equation}
If the frequency profile is monotonic, using identity
(\ref{gei8}), this Fokker-Planck equation becomes
\begin{eqnarray}
\frac{\partial P}{\partial t}=\frac{\partial}{\partial J}\sum_b
\frac{\chi(J,J,\Omega(J))}{|\Omega'(J)|}\left (m_b f_b \frac{\partial
P}{\partial
J}-m P \frac{df_b}{d J}\right ).
\label{tb1b}
\end{eqnarray}
It can be written as Eq. (\ref{ilb15}) with
\begin{eqnarray}
D=\sum_b
\frac{\chi(J,J,\Omega(J))}{|\Omega'(J)|}m_b f_b=\frac{\chi(J,J,\Omega(J))}{
|\Omega'(J)| } f_2,
\label{tb3b}
\end{eqnarray}
and
\begin{eqnarray}
F_{\rm pol}=m\sum_b \frac{\chi(J,J,\Omega(J))}{|\Omega'(J)|}
\frac{df_b}{dJ}=m \frac{\chi(J,J,\Omega(J))}{|\Omega'(J)|} \frac{df}{dJ}.
\label{tb4b}
\end{eqnarray}
As explained in Sec.
\ref{sec_tbaw}, this approach is
self-consistent in the
multispecies case only if the field particles are at statistical equilibrium
with
the Boltzmann distribution from Eq. (\ref{ilb4}) otherwise their
distribution would
evolve under
the effect of collisions (see Sec.
\ref{sec_mscq}). However, if the
field particles have the same mass, their distribution does not change on
a timescale $N\, t_D$ (see Sec. \ref{sec_ssc}). They are in a blocked state. In
that case, the bath approximation is justified (on this timescale) for an
arbitrary DF $f_b$, not only for the Boltzmann
distribution. The Fokker-Planck equation (\ref{tb1b}) can then be rewritten as
\begin{eqnarray}
\frac{\partial P}{\partial t}=\frac{\partial}{\partial J}\left\lbrack D(J)\left
(\frac{\partial P}{\partial
J}-\frac{m}{m_b}P\frac{d\ln f_b}{dJ}\right )\right\rbrack
\label{ham98}
\end{eqnarray}
with
\begin{eqnarray}
D=\frac{\chi(J,J,\Omega(J))}{|\Omega'(J)|}m_b f_b.
\label{tb3c}
\end{eqnarray}
The friction by polarization satisfies the generalized Einstein relation
\begin{eqnarray}
F_{\rm pol}=\frac{m}{m_b}D\frac{d\ln f_b}{dJ}.
\label{tb4bh}
\end{eqnarray}
The
distribution of the test particle relaxes towards the equilibrium distribution
\begin{eqnarray}
P_{\rm eq}(J)=A f_b^{m/m_b}
\end{eqnarray}
on a relaxation time
\begin{eqnarray}
t_R^{\rm bath}\sim Nt_D.
\end{eqnarray}
In the thermal bath approximation, using Eq. (\ref{ilb4}), the
Fokker-Planck
equation (\ref{ham98}) reduces to 
\begin{eqnarray}
\frac{\partial P}{\partial t}=\frac{\partial}{\partial J}\left \lbrack D\left
(\frac{\partial P}{\partial J}
+\beta m P  \Omega \right )\right\rbrack,
\label{ham103}
\end{eqnarray}
in agreement with Eq. (\ref{ilb19}). In that case, we recover the usual
Einstein relation $F_{\rm pol}=-D\beta m\Omega$.

{\it Remark:} The Fokker-Planck equation (\ref{ham98}) can be obtained from an
out-of-equilibium fluctuation-dissipation theorem as discussed in Appendix H of
\cite{Kvortex2023}.

\section{Brownian particles with long-range interactions}
\label{sec_tdbv}

In the previous sections,  we have considered an
isolated system of particles with long-range interactions
described by the $N$-body Hamiltonian equations
\begin{equation}
\frac{d{\bf r}_i}{dt}={\bf v}_i,\qquad \frac{d{\bf v}_i}{dt}=-
\sum_j m_j \nabla u(|{\bf r}_i-{\bf r}_j|).
\label{ham}
\end{equation}
This model is associated with
the
microcanonical ensemble where the energy $E$ of the system is conserved.  When
$N\rightarrow +\infty$ with $m\sim 1/N$, the collisions between the particles
are negligible and the
evolution of the mean DF is described by the Vlasov equation [see Eqs. 
(\ref{n4b}) and (\ref{n2bzero})]. This equation
generically experiences a
process of
violent relaxation towards a QSS in a few dynamical
times
$t_D$. On a
longer timescale $\sim N t_D$, the evolution of
the mean DF is governed by the Lenard-Balescu equation (\ref{ilb3}),
which is valid at the order $1/N$. This equation takes into account
the collisions between the particles. It relaxes towards the
mean field  Boltzmann
distribution (\ref{ilb4}) with a temperature $T(E)$ on a timescale $t_R\sim N\,
t_D$. For  $1D$ systems where only $1:1$ resonances are
permitted, and when the frequency profile is monotonic, the Lenard-Balescu
equation (for a single
species system of particles) reduces to
$\partial f/\partial
t=0$. This leads to a situation of kinetic blocking \cite{fbc}. In that
case, the relaxation towards the
Boltzmann distribution
(\ref{ilb4}) is described  by a kinetic equation valid at the
order $1/N^2$ \cite{fbcn2,fcpn2,fouvry}. For generic potentials of
interaction, this equation relaxes towards the Boltzmann
distribution on a time scale $t_R\sim N^2 t_D$.

In Ref. \cite{hb5} 
we have considered a system of Brownian particles with long-range
interactions  described by the $N$-body stochastic Langevin
equations\footnote{We
have specifically studied a system of self-gravitating
Brownian particles in Ref. \cite{crs}, a system of Brownian particles with a
cosine interaction -- called the Brownian mean field (BMF) model -- in Ref.
\cite{bmf}, and a system of 2D Brownian vortices in Ref. \cite{bv}. The
self-gravitating Brownian gas model could describe the evolution of
planetesimals in the solar nebula in the context of
planet formation. In that case, the particles experience a friction with the
gas and a stochastic force due to Brownian motion or turbulence
\cite{aaplanetes}. If the gas of particles is sufficiently dense, self-gravity
can become important leading to gravitational collapse and planet formation.}
\begin{equation}
\frac{d{\bf r}_i}{dt}={\bf v}_i,\qquad \frac{d{\bf v}_i}{dt}=-
\sum_j m_j \nabla u(|{\bf r}_i-{\bf r}_j|)-\xi_i {\bf v}_i+\sqrt{2D}{\bf
R}_i(t),
\label{brow1}
\end{equation}
where
${\bf R}_i(t)$ is a Gaussian
white noise satisfying $\langle {\bf R}_i(t)\rangle ={\bf 0}$ and $\langle
{R}^{\alpha}_i(t){R}^{\beta}_j(t')\rangle=\delta_{ij}\delta_{\alpha\beta}
\delta(t-t')$. Here, $i=1,...,N$ label the particles and $\alpha=1,...,d$ the
coordinates of space. As compared to the Hamiltonian model (first term),
this Brownian model includes a friction force (second term) and a random
force (third term). The diffusion coefficient and the friction coefficient
satisfy the Einstein relation $D=\xi_i k_B T/m_i$ in order to correctly
reproduce the Gibbs distribution at statistical equilibrium. This
Brownian gas model is associated with the canonical
ensemble where the
temperature $T$ is
fixed. 
When
$N\rightarrow +\infty$  with $m\sim 1/N$,  the collisions between
the particles
are negligible and the
evolution of the mean DF is described by a mean field Fokker-Planck
equation which has the form of a Vlasov-Kramers
equation\footnote{In the present section, we
write the equations for a single species system, but these
equations can be
straightforwardly generalized to a multi-species system (see e.g.
Ref. \cite{multisopik} and Appendix \ref{sec_sim}).} 
\begin{eqnarray}
\label{brow2}
\frac{\partial f}{\partial t}+{\bf v}\cdot \frac{\partial
f}{\partial {\bf r}}-
\nabla\Phi\cdot
\frac{\partial f}{\partial {\bf v}}=\frac{\partial}{\partial {\bf
v}}\cdot\left\lbrack D
\left (\frac{\partial f}{\partial {\bf v}}+\beta m f {\bf v}\right
)\right\rbrack,
\end{eqnarray}
\begin{eqnarray}
\Phi({\bf r},t)=\int u(|{\bf r}-{\bf r}'|)\rho({\bf
r}',t)\, d{\bf
r}'.
\label{brow3}
\end{eqnarray}
This equation relaxes towards the mean field Boltzmann distribution
$f=Ae^{-\beta m(v^2/2+\Phi)}$ with a
temperature $T$ on a frictional timescale $t_{K}\sim 1/\xi$. When
$\xi\rightarrow 0$, the Vlasov-Kramers equation
first
experiences  a process of violent relaxation towards a QSS  in a few dynamical
times
$t_D$ before slowly relaxing towards
the  Boltzmann distribution, as discussed in \cite{bco} in the
context of the BMF model.\footnote{When $N$ is finite and 
$N\ll 1/(\xi D)$, the
system  may first achieve a QSS
on a timescale $t_D$, followed by a microcanonical equilibrium state at
energy $E$ on a timescale $t_R\sim N\, t_D$, itself followed by a
canonical
equilibrium state at temperature
$T$ on a timescale $t_K\sim 1/\xi$ (see Ref. \cite{bco}). When the fluctuations
are
taken into account, the phenomenology is even richer as discussed at the end of
this section and in Sec. \ref{sec_spv}.} In that case, after the phase of
violent relaxation, the DF is a function
 of the actions only and the deterministic equation governing the evolution of
the 
mean DF $f=f({\bf J},t)$ reads (see
Appendix \ref{sec_oake}) 
\begin{equation}
\frac{\partial f}{\partial t}=\frac{\partial}{\partial {\bf J}}\cdot
\left
\lbrack D_K\left (\frac{\partial f}{\partial {\bf J}}+\beta m f {\bf
\Omega}\right )\right\rbrack,
\label{brow4}
\end{equation}
where $D_K({\bf J},t)$ is the diffusion coefficient in action space and ${\bf
\Omega}({\bf J},t)=\partial H/\partial {\bf J}$ is the pulsation of an orbit of
action ${\bf J}$ determined by the system itself. Eq.
(\ref{brow4}) will be called the inhomogeneous mean
field Kramers
equation. It relaxes towards the
mean field  Boltzmann
distribution (\ref{ilb4}) with temperature $T$ on a timescale $t_K\sim 1/\xi$.

The self-consistent Kramers equation (\ref{brow4})
associated
with the canonical ensemble is structurally
very different from the Lenard-Balescu equation (\ref{ilb3})
associated
with
the microcanonical ensemble. It is also
very
different from the Kramers equation (\ref{ilb19})
describing the
evolution 
of the mean DF of a system of noninteracting test particles in a
thermal bath of field particles at statistical equilibrium.  In that case,
the pulsation ${\bf \Omega}({\bf J})$ is determined by the equilibrium
distribution (\ref{fdi3voit}) of
the field particles whereas in Eq. (\ref{brow4}) the pulsation
${\bf\Omega}({\bf J},t)$  is self-consistently produced by the DF
$f({\bf J},t)$ of the system itself. Furthermore, the  self-consistent
Kramers
equation
(\ref{brow4}) is valid when $N\rightarrow
+\infty$ while the Lenard-Balescu equation (\ref{ilb3}) and the Kramers
equation (\ref{ilb19}) are valid at the
order $1/N$.

The mean field Boltzmann equation
may have several solutions. The Vlasov-Kramers
equation
describes
the evolution of the mean DF towards one of these equilibrium
states which is stable in the canonical ensemble (minimum of free energy at
fixed mass). Following
\cite{hb5}, if we
consider  the evolution of Brownian particles on a mesoscopic scale, we
have to add a stochastic term in the kinetic equation. This noise term arises
from finite $N$
effects and takes fluctuations into account (note that we are still neglecting
the collisions between the particles). This leads to the stochastic
Vlasov-Kramers equation \cite{hb5}\footnote{The evolution of the
discrete (exact) DF $f_d({\bf r},{\bf v},t)=\sum_{i=1}^{N} m_i \delta({\bf
r}-{\bf
r}_i(t))\delta({\bf
v}-{\bf
v}_i(t))$ is
also determined  by a stochastic Vlasov-Kramers equation
 of the form of Eqs. (\ref{brow5}) and (\ref{brow6}) with $\overline{f}$ and
$\overline{\Phi}$  replaced by $f_d$ and $\Phi_d$ \cite{hb5}. These
equations (sometimes called the Dean \cite{dean} equations) are exact in the
sense that they contain the same information as
the $N$-body Langevin equations (\ref{brow1}). They are the counterpart of
the Klimontovich equations for Hamiltonian systems. Taking the
ensemble average of
these equations and making a mean field approximation ($N\rightarrow +\infty$
with $m\sim 1/N$),
we obtain Eqs. (\ref{brow2}) and (\ref{brow3}). Keeping track of the
fluctuations at
a
mesoscopic scale, we obtain  Eqs. (\ref{brow5}) and (\ref{brow6}).}
\begin{eqnarray}
\label{brow5}
\frac{\partial \overline{f}}{\partial t}+{\bf v}\cdot \frac{\partial
\overline{f}}{\partial {\bf r}}-
\nabla\overline{\Phi}\cdot
\frac{\partial \overline{f}}{\partial {\bf v}}=\frac{\partial}{\partial {\bf
v}}\cdot\left\lbrack D
\left (\frac{\partial \overline{f}}{\partial {\bf v}}+\beta m \overline{f} {\bf
v}\right
)\right\rbrack+\frac{\partial}{\partial{\bf v}}\cdot \left
(\sqrt{2Dm\overline{f}}\, {\bf Q}({\bf r},{\bf v},t)\right ),
\end{eqnarray}
\begin{eqnarray}
\overline{\Phi}({\bf r},t)=\int u(|{\bf r}-{\bf
r}'|)\overline{\rho}({\bf
r}',t)\, d{\bf
r}',
\label{brow6}
\end{eqnarray}
where ${\bf Q}({\bf r},{\bf v},t)$ is a Gaussian white noise satisfying $\langle
{\bf
Q}({\bf r},{\bf v},t)\rangle={\bf 0}$ and  $\langle Q_\alpha({\bf r},{\bf
v},t)Q_\beta({\bf
r}',{\bf v}',t')\rangle=\delta_{\alpha\beta}\delta({\bf r}-{\bf r}')\delta({\bf
v}-{\bf v}')\delta(t-t')$. In the limit $\xi\rightarrow 0$, we obtain
a stochastic partial differential equation in angle-action variables of the form
\begin{equation}
\frac{\partial \overline{f}}{\partial t}=\frac{\partial}{\partial {\bf J}}\cdot
\left
\lbrack D_K\left (\frac{\partial \overline{f}}{\partial {\bf J}}+\beta m
\overline{f} {\bf \Omega}\right )\right\rbrack+\frac{\partial}{\partial{\bf
J}}\cdot \left
(\sqrt{2D_K m\overline{f}}\, {\bf Q}({\bf J},t)\right ),
\label{brow7}
\end{equation}
where ${\bf Q}({\bf J},t)$ is a Gaussian white noise satisfying $\langle
{\bf Q}({\bf J},t)\rangle ={0}$ and $\langle {Q}_\alpha({\bf J},t){Q}_\beta({\bf
J}',t')\rangle=\delta_{\alpha\beta}\delta({\bf J}-{\bf J}')\delta(t-t')$. 
We will call it the stochastic inhomogeneous
Kramers equation. When the
 mean field Boltzmann equation admits several
equilibrium states,  Eqs. (\ref{brow5}) and
(\ref{brow7}) can be used to study random
transitions from one state to the other (see Ref. \cite{random} in a similar
context). The probability of transition is given
by the Kramers formula, which can be established from the 
instanton theory associated with the Onsager-Machlup functional
\cite{random,gsse}.

In the preceding discussion, we have considered the low friction limit
$\xi\rightarrow 0$ where the system is dominated by inertial
effects. Inversely, in the high friction limit $\xi\rightarrow +\infty$
considered in \cite{hb5,crs,bmf}, we can neglect
inertial effects at leading order. In that case, the DF $f({\bf r},{\bf v},t)$
is close to a Maxwellian distribution at each time $t$ and the evolution of the
mean
spatial density
$\rho({\bf r},t)$ is governed by the mean field Smoluchowski equation 
\begin{eqnarray}
\frac{\partial\rho}{\partial t}=\nabla\cdot
\left\lbrack
D_*(\nabla\rho+\beta m \rho\nabla\Phi)\right\rbrack
\label{brow8}
\end{eqnarray}
with Eq. (\ref{brow3}) where $D_*=k_B T/\xi m$ is the
diffusion
coefficient in configuration space given by the ordinary Einstein relation.
The mean field Smoluchowski equation 
relaxes toward the mean field Boltzmann distribution $\rho=A' e^{-\beta
m\Phi}$ on a diffusive timescale $t_S\sim R^2/D_*$ where $R$ is the system's
size. If we consider a mesoscopic level of
description and take fluctuations into account we obtain the stochastic
Smoluchowski equation\footnote{In the high friction limit, the evolution of the
discrete (exact) density $\rho_d({\bf r},t)=\sum_{i=1}^{N} m_i
\delta({\bf
r}-{\bf
r}_i(t))$ is
also determined  by a stochastic Smoluchowski equation
 of the form of Eqs. (\ref{brow6}) and (\ref{brow9}) with $\overline{\rho}$ and
$\overline{\Phi}$  replaced by $\rho_d$ and $\Phi_d$ \cite{hb5}. These
equations are exact \cite{dean}. Taking the
ensemble average of
these equations and making a mean field approximation ($N\rightarrow +\infty$
with $m\sim 1/N$),
we obtain Eqs. (\ref{brow3}) and (\ref{brow8}). Keeping track of the
fluctuations at
a
mesoscopic scale, we obtain  Eqs. (\ref{brow6}) and (\ref{brow9}).}
\begin{eqnarray}
\frac{\partial\rho}{\partial t}=\nabla\cdot
\left\lbrack
D_*(\nabla\rho+\beta m \rho\nabla\Phi)\right\rbrack+\nabla\cdot
\left \lbrack \sqrt{2D_* m\rho}\, {\bf R}({\bf r},t)\right \rbrack
\label{brow9}
\end{eqnarray}
with Eq. (\ref{brow6}), where ${\bf R}({\bf r},t)$ is a Gaussian white noise
satisfying $\langle {\bf
R}({\bf r},t)\rangle={\bf 0}$ and  $\langle R_\alpha({\bf r},t)R_\beta({\bf
r}',t)\rangle=\delta_{\alpha\beta}\delta({\bf r}-{\bf r}')\delta(t-t')$. As
explained
above, when the deterministic
equation (\ref{brow8}) admits several equilibrium states, the noise term 
in Eq. (\ref{brow9}) allows the system to switch from one equilibrium state to
another one through
random transitions (see, e.g., 
\cite{random,gsse,nardini2,bouchetsimmonet,rbs,brs} in
different contexts).

{\it Remark:} Equations (\ref{brow5})-(\ref{brow7}) take
into account finite $N$ effects ($m\sim 1/N$) which
are responsible for the noise term (fluctuations) but they ignore the collisions
between the particles that would lead to the Lenard-Balescu collision term, as
well as the
correlations
arising from the noise term. The correlations arising from the noise term induce
an additional nonlinear diffusion of the mean DF on a timescale $Nt_D$
which is discussed in Sec. \ref{sec_spv}.

\section{Secular dressed diffusion equation}
\label{sec_sdd}

In this section, we consider the
case where an inhomogeneous system of particles with long-range interactions is
submitted to an external stochastic  potential $\Phi_e$ [see
Eq. (\ref{n1})]. We study a rather
general situation where the
external forcing is not necessarily induced by a discrete collection of field
particles. We consider the limit $N\rightarrow +\infty$ with $m\sim 1/N$
where the collisions between the particles of the system can be neglected. We
show that, under the
effect of the external forcing, the
evolution of the system satisfies a SDD equation. We
derive the SDD equation from the Klimontovich equation and from the
Fokker-Planck equation and analyze its main properties.

\subsection{From the Klimontovich equation}
\label{sec_sddk}

The basic equations
governing the evolution of
the mean DF
$f({\bf J},t)$ of a spatially inhomogeneous system of particles 
with long-range interactions\footnote{This DF which is a steady
state of the Vlasov equation assumed to be dynamically stable may
be imposed initially or results from a process
of violent relaxation.}   forced by an external
perturbation are given by Eqs. (\ref{i56}) and (\ref{i57}).
Introducing the total fluctuating potential $\delta\Phi_{\rm
tot}=\delta\Phi+\Phi_e$,  we can rewrite Eq. (\ref{i56}) as
\begin{eqnarray}
\frac{\partial f}{\partial t}=\frac{\partial}{\partial {\bf J}}\cdot
\left\langle\delta f
\frac{\partial}{\partial {\bf
w}}\delta\Phi_{\rm tot}\right\rangle.
\end{eqnarray}
Introducing the
Fourier transforms  of the fluctuations of
DF and potential, we get 
\begin{equation}
\frac{\partial f}{\partial t}=-i\sum_{\bf
k}\int
\frac{d\omega}{2\pi}\sum_{\bf k'}\int\frac{d\omega'}{2\pi} \,
\left ({\bf k}'\cdot \frac{\partial}{\partial {\bf J}}\right ) e^{i({\bf k}\cdot
{\bf w}-\omega t)}e^{-i({\bf k}'\cdot {\bf w}-\omega'
t)} \langle \delta\hat f({\bf k},{\bf
J},\omega)\delta\hat\Phi_{\rm tot}^*({\bf
k}',{\bf J},\omega')\rangle.
\label{i73}
\end{equation}
Equation (\ref{i57}) can be written in Fourier space as [see Eq.
(\ref{i60})]
\begin{equation}
\delta\hat f ({\bf k},{\bf J},\omega)=\frac{{\bf k}\cdot \frac{\partial
f}{\partial {\bf J}}}{{\bf k}\cdot {\bf
\Omega}-\omega}\delta\hat\Phi_{\rm tot}({\bf
k},{\bf J},\omega).
\label{i60k}
\end{equation}
Substituting Eq. (\ref{i60k}) into Eq. (\ref{i73}), we obtain
\begin{equation}
\frac{\partial f}{\partial t}=-i\sum_{\bf
k}\int
\frac{d\omega}{2\pi}\sum_{\bf k'}\int\frac{d\omega'}{2\pi} \,
\left ({\bf k}'\cdot \frac{\partial}{\partial {\bf J}}\right ) e^{i({\bf k}-{\bf
k}')\cdot {\bf w}}e^{-i(\omega-\omega')t}
\frac{{\bf k}\cdot \frac{\partial
f}{\partial {\bf J}}}{{\bf k}\cdot {\bf
\Omega}-\omega}\langle \delta\hat\Phi_{\rm tot}({\bf
k},{\bf J},\omega) \delta\hat\Phi_{\rm tot}^*({\bf
k}',{\bf J},\omega')\rangle.
\label{i74}
\end{equation}
Since the mean DF $f$ depends only on the action ${\bf J}$, we
can average the
collision term over the angle ${\bf w}$ without loss of
information. This
brings a Kronecker factor
$\delta_{{\bf k},{\bf k}'}$ which amounts to taking ${\bf
k}'={\bf k}$. Therefore,
\begin{equation}
\frac{\partial f}{\partial t}=-i\sum_{\bf
k}\int
\frac{d\omega}{2\pi}\int\frac{d\omega'}{2\pi} \,
\left ({\bf k}\cdot \frac{\partial}{\partial {\bf J}}\right )
e^{-i(\omega-\omega')t}
\frac{{\bf k}\cdot \frac{\partial
f}{\partial {\bf J}}}{{\bf k}\cdot {\bf
\Omega}-\omega}\langle \delta\hat\Phi_{\rm tot}({\bf
k},{\bf J},\omega) \delta\hat\Phi_{\rm tot}^*({\bf
k},{\bf J},\omega')\rangle.
\label{i74b}
\end{equation}
Introducing the power spectrum of the total potential fluctuations from Eq.
(\ref{psaab}), and
performing the
integration over $\omega'$, we get
\begin{equation}
\frac{\partial f}{\partial t}=-i\sum_{\bf
k}\int
\frac{d\omega}{2\pi} \,
\left ({\bf k}\cdot \frac{\partial}{\partial {\bf J}}\right )
\frac{{\bf k}\cdot \frac{\partial
f}{\partial {\bf J}}}{{\bf k}\cdot {\bf
\Omega}-\omega}P({\bf
k},{\bf J},\omega).
\label{i84}
\end{equation}
Recalling the Landau prescription $\omega\rightarrow \omega+i0^+$ and using the
Sokhotski-Plemelj
\cite{sokhotski,plemelj}  formula (\ref{n25}), we can replace $1/({\bf k}\cdot
{\bf
\Omega}-\omega-i0^+)$ by $+i\pi\delta({\bf k}\cdot {\bf
\Omega}-\omega)$. 
Accordingly, 
\begin{eqnarray}
\frac{\partial f}{\partial t}=\pi\sum_{\bf
k}\int
\frac{d\omega}{2\pi} \,
\left ({\bf k}\cdot \frac{\partial}{\partial {\bf J}}\right )
\delta({\bf k}\cdot {\bf \Omega}-\omega)P({\bf
k},{\bf J},\omega)
\left ({\bf k}\cdot \frac{\partial
f}{\partial {\bf J}}\right ).
\label{i85}
\end{eqnarray}
Integrating over the $\delta$-function (resonance), we get
\begin{eqnarray}
\frac{\partial f}{\partial t}=\frac{1}{2}\sum_{\bf
k}\,
\left ({\bf k}\cdot \frac{\partial}{\partial {\bf J}}\right )
P({\bf
k},{\bf J},{\bf k}\cdot {\bf \Omega})
\left ({\bf k}\cdot \frac{\partial
f}{\partial {\bf J}}\right ).
\label{i86}
\end{eqnarray}
Therefore, for spatially inhomogeneous
systems with long-range
interactions, the
secular
evolution of
the mean DF $f({\bf J},t)$ of the particles sourced by an
external stochastic force is governed by a nonlinear diffusion equation of the
form
\begin{equation}
\frac{\partial f}{\partial t}=\frac{\partial}{\partial J_i}\left (D_{ij}[f,{\bf
J}]\frac{\partial f}{\partial J_j}\right ),
\label{i87b}
\end{equation}
with an anisotropic diffusion tensor 
\begin{equation}
D_{ij}[f,{\bf J}]=\frac{1}{2}\sum_{\bf
k}\, k_ik_j
P({\bf
k},{\bf J},{\bf k}\cdot {\bf \Omega}).
\label{i88}
\end{equation}

Using Eq. (\ref{i83}), we can express the diffusion tensor in terms of
the correlation tensor of the external perturbation as
\begin{equation}
D_{ij}[f,{\bf
J}]=\frac{1}{2}\sum_{\bf k}\sum_{
\alpha\alpha'} \, k_i k_j \hat\Phi_{\alpha}({\bf
k},{\bf J})\left \lbrack\epsilon^{-1} {\hat
C} (\epsilon^{\dagger})^{-1}\right
\rbrack_{\alpha\alpha'}({\bf k}\cdot {\bf\Omega}({\bf J}))
\hat\Phi_{\alpha'}({\bf
k},{\bf J})^*.
\label{i88bc}
\end{equation}
As detailed in Sec. \ref{sec_diffco}, the diffusion
tensor  from Eq. (\ref{i88bc}) is a functional of the
DF $f({\bf J},t)$ itself. As a result, Eq. (\ref{i87b})  with  Eq.
(\ref{i88bc})
is
a complicated integrodifferential equation called the inhomogeneous SDD
equation \cite{weinberg,pa,fpp,fp15,fbp,fb,sdduniverse}. This is the counterpart
of
the SDD equation derived in Ref. \cite{epjp} for homogeneous systems of
material particles with
long-range
interactions forced by an external stochastic perturbation.\footnote{A similar,
but different, equation is derived in Refs. \cite{nardini,nardini2}. See
Ref. \cite{epjp2} for a
comparison between these two approaches.}

When
collective
effects
are neglected, the
diffusion 
tensor reduces to
\begin{equation}
D_{ij}^{\rm bare}=\frac{1}{2}\sum_{\bf k} \, k_i k_j {\hat
C}({\bf k},{\bf J},{\bf k}\cdot {\bf\Omega}({\bf J})).
\label{i88bzb}
\end{equation}
Equation
(\ref{i87b}) with Eq. (\ref{i88bzb}) is called the inhomogeneous SBD
equation \cite{bl,mab,ba,bf,sdduniverse}.

{\it Remark:} The SDD equation describes the diffusive evolution
of a near-equilibrium system caused by scattering of orbits by fluctuations
in the potential due to the external
perturbation. For large but finite $N$, the
evolution
timescale is intermediate between
the violent collisionless relaxation timescale $\sim t_D$ and the collisional
relaxation
timescale $\sim N t_D$ (or $\sim N^2 t_D$ in the situation
discussed in Sec. \ref{sec_mono}).

\subsection{From the Fokker-Planck equation}
\label{sec_sddfp}

The SDD equation (\ref{i87b}) can also be derived from the Fokker-Planck
equation
(\ref{fp8}). In the
limit $N\rightarrow +\infty$ with $m\sim 1/N$ where the
collisions between
the particles
are negligible, the friction by polarization, which is proportional to $m$
(see Sec. \ref{sec_ifpol}),
vanishes
\begin{equation}
\label{ham119}
{\bf F}_{\rm pol}={\bf 0}.
\end{equation}
Indeed, the perturbation on the system caused by the test particle is
negligible. In that case, the Fokker-Planck equation (\ref{fp8}) reduces to
\begin{equation}
\frac{\partial f}{\partial t}=\frac{\partial}{\partial J_i}\left
(D_{ij}\frac{\partial f}{\partial J_j}\right ).
\label{i87wq}
\end{equation}
The diffusion coefficient can be calculated as in Sec. \ref{sec_diffco}
returning the expression from Eqs. (\ref{i88}) and (\ref{i88bc}) obtained from
the Klimontovich approach. Therefore, the Klimontovich approach
 and the Fokker-Planck approach coincide.

{\it Remark:} We note that, in Eq. (\ref{i87wq}), the diffusion tensor is
``sandwiched''
between
the two derivatives $\partial/\partial J$ in agreement with Eq.
(\ref{i87b}). As
explained in Sec. \ref{sec_fp}, this is not the usual form of the Fokker-Planck
equation which is given by Eq. (\ref{fp5}). Therefore, the test particle
experiences a friction [see Eq. (\ref{fp9})] 
\begin{equation}
\label{ham120}
F^{\rm tot}_i=\frac{\partial D_{ij}}{\partial J_j}, 
\end{equation}
arising from the inhomogeneity of the diffusion
tensor.\footnote{This
formula is established by a direct calculation in Sec. 3.4 of \cite{physicaA}.}
Using Eqs.
(\ref{fp6}) and (\ref{fp7}), the relation (\ref{ham120}) can be written as 
\begin{equation}
\label{ham120y}
\frac{\langle\Delta J_i\rangle}{\Delta t}=\frac{1}{2}\frac{\partial}{\partial
J_j}\frac{\langle \Delta J_i \Delta J_j\rangle}{\Delta t},
\end{equation}
returning the result of \cite{bl,ba}.

\subsection{Properties of the SDD equation}

Some general properties of the inhomogeneous SDD equation
(\ref{i87b}) can be
given. First
of all, the total mass $M=\int f\, d{\bf J}$ of the system is conserved since
the right hand side of
Eq. (\ref{i87b})
is the divergence of a current in action space. By contrast, the energy of the
system is not
conserved, contrary to the case of the  inhomogeneous Lenard-Balescu equation
\cite{heyvaerts,physicaA},
since the system
is forced by an external medium. Taking the time
derivative
of the energy
\begin{equation}
E=\int f H({\bf J})\, d{\bf v},
\label{in30}
\end{equation}
using Eq. (\ref{i87b}), and integrating by parts, we get
\begin{equation}
\dot E=-\int D_{ij}[f,{\bf J}]\frac{\partial f}{\partial J_j} \Omega_i \,
d{\bf J}.
\label{in31}
\end{equation}
In general, $\dot E$ has not a definite sign. However, in the usual situation
where $f=f(H)$ with $f'(H)\le 0$, we get $\dot E=-\int f'(H)
D_{ij}\Omega_i\Omega_j \,
d{\bf J}\ge 0$ showing that energy is injected in the system. Finally,
introducing
the $H$-functions \cite{thlb}
\begin{equation}
S=-\int C(f) \, d{\bf J},
\label{in32}
\end{equation}
where $C(f)$ is any convex function, we get  \cite{bl}
\begin{equation}
\dot S=\int C''(f) \frac{\partial f}{\partial J_i} D_{ij}[f,{\bf
J}]\frac{\partial f}{\partial J_j} \, d{\bf J}.
\label{in33}
\end{equation}
Because of
the convexity condition $C''\ge 0$ and the fact that the
quadratic form $x_iD_{ij}x_j$ is definite positive (see Sec.
\ref{sec_sddk}),
we find that $\dot S\ge 0$. Therefore, all the $H$-functions increase
monotonically with time through the SDD equation. This is
different from the case of the inhomogeneous
Lenard-Balescu equation where only the Boltzmann entropy
increases
monotonically with time \cite{heyvaerts,physicaA}.

\subsection{Connection between the SDD equation and the multi-species
Lenard-Balescu equation}
\label{sec_conn}

In this section, we discuss the connection between the
SDD equation (\ref{i87b})
with Eq. (\ref{i88bc}) and the
multi-species Lenard-Balescu equation (\ref{ilb3}). 
The Lenard-Balescu 
equation governs the mean evolution of the DF $f_a({\bf J},t)$ of
particles of species $a$
under the effects of ``collisions'' with particles of all species ``$b$''
(including the particles of
species $a$) with DF $f_b({\bf J}',t)$. The dielectric matrix is
given by Eq. (\ref{i70}) where $f({\bf J},t)$ denotes the
total DF $\sum_b f_b({\bf J},t)$ and ${\bf\Omega}({\bf J},t)$ is the
corresponding pulsation. The set of equations (\ref{ilb3}), in which all
the DFs $f_a({\bf J},t)$ evolve in a self-consistent manner, is
closed.

We now make the following approximations to simplify
these
equations. The particles of mass $m_a$ with DF
$f_a({\bf J},t)$ form our
system. They will
be called the test particles. The particles of masses $\lbrace
m_b\rbrace_{b\neq a}$ with DFs
$\lbrace f_b({\bf J},t)\rbrace_{b\neq a}$ form 
the external -- background -- medium. They will be called the field particles.
We take
into account the
collisions induced by the field particles of mass $\lbrace m_b\rbrace_{b\neq a}$
on the test particles
but we neglect the
collisions induced by the test particles on the field particles and on
themselves. This approximation is valid for very light test particles  $m_a\ll
m_b$ or, more precisely, in the limit $N_a\rightarrow +\infty$ with $m_a\sim
1/N_a$ (see Sec. \ref{sec_purediff}).  Finally, the DFs
$\lbrace f_b({\bf J},t)\rbrace_{b\neq a}$ of the field particles are
either assumed to be fixed
(bath) or evolve according to their own dynamics (i.e. following equations that
we do not write explicitly). Under these
conditions,  
the inhomogeneous Lenard-Balescu equation (\ref{ilb3}) reduces to 
\begin{equation}
\frac{\partial f_a}{\partial t}=\pi (2\pi)^d 
\frac{\partial}{\partial {\bf
J}}\cdot \sum_{b\neq a}\sum_{{\bf k},{\bf k}'}\int d{\bf J}' \, {\bf k}\,
|A^d_{{\bf k},{\bf k}'}({\bf J},{\bf J}',{\bf k}\cdot {\bf
\Omega})|^2\delta({\bf k}\cdot {\bf \Omega}-{\bf k}'\cdot {\bf \Omega}') \left
(m_b f'_b {\bf k}\cdot \frac{\partial f_a}{\partial {\bf J}}\right
).
\label{can2}
\end{equation}
This equation can be
interpreted as a nonlinear diffusion equation. The diffusion arises from the
discrete
distribution of the field particles which creates a fluctuating potential
(Poisson shot noise) acting on the test particles.
For that reason, the diffusion tensor of the test particles is proportional
to the masses $\lbrace m_b\rbrace_{b\neq a}$ of the field particles. The
diffusion tensor has no contribution from particles
of species $a$. The
condition $m_a\ll m_b$ justifies neglecting the fluctuations induced  by the
test particles on
themselves.  On the
other hand, the friction by polarization vanishes (${\bf F}_{\rm
pol}={\bf 0}$). Indeed, since the mass $m_a$ of the  test particles
is
small,
the test particles do not significantly perturb the DF of the
medium, so there
is no friction by polarization (no retroaction).  As a
result, the mass $m_a\rightarrow 0$ of the particles  of species $a$
does not
appear in the kinetic equation (\ref{can2}). This returns the diffusion equation
(\ref{ilb22}) with Eq. (\ref{ilb16}).

Equation (\ref{can2}) can be written as an inhomogeneous
SDD equation
\begin{equation}
\frac{\partial f}{\partial t}=\frac{\partial}{\partial J_i}\left
(D_{ij}[{\bf J},f]\frac{\partial f}{\partial {J}_j}\right
)
\label{can3}
\end{equation}
with a
diffusion tensor 
\begin{equation}
D_{ij}[{\bf J},f]=\pi (2\pi)^d \sum_{b}\sum_{{\bf
k},{\bf k}'}\int d{\bf J}' \, {k}_i k_j\,
|A^d_{{\bf k},{\bf k}'}({\bf J},{\bf J}',{\bf k}\cdot {\bf
\Omega})|^2\delta({\bf k}\cdot {\bf \Omega}-{\bf k}'\cdot {\bf \Omega}') m_b
f'_b, 
\label{can4}
\end{equation}
where we have droped the subscript $a$ for clarity. The expression
(\ref{can4}) of the diffusion tensor is consistent with Eq.
(\ref{i88bc}) where ${\hat C}_{\alpha\alpha'}(\omega)$ is the
bare correlation function of the potential created by a discrete collection of
field particles of masses
$\lbrace m_b\rbrace$ given by Eq. (\ref{ci10}). To exactly recover the
SDD equation (\ref{i87b}) with Eq. (\ref{i88bc}), we
have to replace
$f({\bf J},t)+\sum_{b} f_b({\bf J})$ by  $f({\bf J},t)$ in the dielectric
matrix. This
assumes that the external medium -- field particles -- is not polarizable (i.e.
collective effects
can be neglected) while our system -- test particles -- is
polarizable (i.e.
collective effects must be taken into account).\footnote{In the present case,
Eq.
(\ref{can3}) with Eq. (\ref{can4})
is more accurate than the SDD equation (\ref{i87b}) with Eq.
(\ref{i88bc}) 
because it takes into account the
polarizability of the external medium.} As we
have already mentioned,
the SDD equation (\ref{can3}) is a nonlinear
diffusion equation involving a
diffusion tensor which depends on the DF of the system
$f({\bf J},t)$ itself. This is
therefore an integrodifferential equation.

\subsection{SDD equation with damping}
\label{sec_sddwd}

Let us add a small linear damping term $-\nu f$ on
the right hand side of the SDD equation (\ref{i87b}) in order to
account for a
possible dissipation. This yields
\begin{equation}
\label{eqsddg}
\frac{\partial f}{\partial t}=\frac{\partial}{\partial J_i}\left (D_{ij}[f,{\bf
J}]\frac{\partial f}{\partial J_j}\right )-\nu f.
\end{equation}
The general behavior of this nonlinear equation is
difficult to
predict because it depends on the correlation function of the external
potential. Furthermore, since the diffusion tensor $D_{ij}[f]$ is a
functional of $f$, the SDD equation (\ref{eqsddg})  is a very nonlinear equation
 which presents a rich and complex behavior. It may relax towards a
non-Boltzmannian steady state determined by the equation
\begin{equation}
\frac{\partial}{\partial J_i}\left (D_{ij}[f,{\bf
J}]\frac{\partial f}{\partial J_j}\right )-\nu f=0.
\label{eqsdd}
\end{equation}
or exhibit a
complicated (e.g. periodic) dynamics.

\subsection{SDD equation with friction}

By analogy with the inhomogeneous Kramers equation (\ref{ilb19}) we can
heuristically add
a friction term on
the right hand side of the SDD equation
(\ref{i87b}).
This leads to an equation of the form
\begin{equation}
\frac{\partial f}{\partial t}=\frac{\partial}{\partial J_i}\left (D_{ij}[f,{\bf
J}]\frac{\partial f}{\partial J_j}+\xi f \Omega_i({\bf
J})\right ),
\label{wq}
\end{equation}
where ${\bf \Omega}({\bf J})$ is
either a given external ``force'' or
the pulsation generated by the DF $f({\bf J},t)$ itself in a
self-consistent
manner. We recall
that
${\bf
\Omega}({\bf J})$ can be written as a gradient ${\bf \Omega}({\bf
J})=\partial H/\partial {\bf J}$ where $H$ plays the role of the potential. The
stationary solutions of Eq. (\ref{wq}) satisfy
\begin{equation}
D_{ij}[f,{\bf J}]\frac{\partial f}{\partial J_j}+\xi f \Omega_i({\bf J})=0,
\label{in34st}
\end{equation}
As in Sec. \ref{sec_sddwd} this equation admits nontrivial solutions.

\subsection{Stochastic SDD equation}

The SDD equation (\ref{wq}), which is a deterministic
partial differential
equation, describes the evolution of the mean DF $f({\bf J},t)$. If we
take fluctuations into account, by
analogy with the results presented in \cite{hb5}, we expect that the
mesoscopic DF $\overline{f}({\bf J},t)$ will
satisfy a stochastic partial differential equation of the form
\begin{equation}
\frac{\partial \overline{f}}{\partial t}=\frac{\partial}{\partial J_i}\left
(D_{ij}[\overline{f},{\bf
J}]\frac{\partial \overline{f}}{\partial J_j}+\xi \overline{f} \Omega_i({\bf
J})\right )+\zeta({\bf
J},t),
\label{gin34}
\end{equation}
where $\zeta({\bf J},t)$ is a noise term with zero mean that generally depends
on
$\overline{f}({\bf J},t)$. When $D_{ij}=D\delta_{ij}$ is constant and
isotropic and when the
fluctuation-dissipation theorem is fulfilled so that $\xi=D\beta m$, as in the
case of Brownian particles with long-range interactions (see Sec.
\ref{sec_tdbv}), the
noise term is given by \cite{hb5}
\begin{equation}
\zeta({\bf J},t)=\frac{\partial}{\partial{\bf J}}\cdot \left
(\sqrt{2Dm\overline{f}}\, {\bf Q}({\bf J},t)\right ),
\end{equation}
where ${\bf Q}({\bf J},t)$ is a Gaussian white noise satisfying $\langle
{\bf Q}({\bf J},t)\rangle ={0}$ and $\langle {Q}_\alpha({\bf J},t){Q}_\beta({\bf
J}',t')\rangle=\delta_{\alpha\beta}\delta({\bf J}-{\bf J}')\delta(t-t')$. This
expression
can be obtained from an adaptation of the theory of
fluctuating hydrodynamics \cite{hb5}. When $D[\overline{f}]$ is a
functional of
$\overline{f}$, the noise term may be more complicated.\footnote{A
general
approach to obtain the noise term and the corresponding action is to use
the theory of large deviations \cite{bouchetld}.} When the deterministic
equation (\ref{wq}) admits several equilibrium states, the noise term in Eq.
(\ref{gin34}) can trigger random transitions from one state to the
other (see,
e.g.,
\cite{random,gsse,nardini2,bouchetsimmonet,rbs,brs} in various contexts).

{\it Remark:}  Similarly, we can introduce a stochastic
term (noise) in Eq.
(\ref{eqsddg}) to describe the evolution of the system on a mesoscopic
scale.
This leads to a stochastic SDD equation of the form   
\begin{equation}
\frac{\partial \overline{f}}{\partial t}=\frac{\partial}{\partial J_i}\left
(D_{ij}[\overline{f},{\bf
J}]\frac{\partial \overline{f}}{\partial J_j}\right )-\nu
\overline{f}+\zeta({\bf
J},t),
\label{gin34h}
\end{equation}
The previous comments also apply to this stochastic partial
differential equation.

\section{Stochastic damped Vlasov equation}
\label{sec_fd}

In this section, we consider the stochastic damped Vlasov
equation 
\begin{equation}
\frac{\partial f_c}{\partial t}+{\bf v}\cdot\frac{\partial f_c}{\partial {\bf
r}}-\nabla\Phi_c\cdot \frac{\partial f_c}{\partial {\bf
v}}=-\nu f_c+\sqrt{2 m \nu f_c}\, Q({\bf r},{\bf v},t),
\label{fh1}
\end{equation}
where $f_c({\bf r},{\bf v},t)$ is a continuous DF, ${Q}({\bf r},{\bf v},t)$
is a Gaussian white noise
satisfying $\langle {Q}({\bf
r},{\bf v},t)\rangle =0$ and $\langle {Q}({\bf r},{\bf v},t){Q}({\bf
r}',{\bf v}',t')\rangle=\delta({\bf
r}-{\bf r}')\delta({\bf
v}-{\bf v}')\delta(t-t')$, $\nu$ is a small damping coefficient, and $m$ has
the dimension of a mass (we will
ultimately take the limit $\nu\rightarrow 0$). We introduce
this equation in
an {\it ad hoc} manner but we will show below that the stochastic term generates
a power spectrum of the potential fluctuations that coincides with the power
spectrum produced by an isolated
distribution of particles of mass $m\sim 1/N\ll 1$. This
approach therefore provides another manner to determine
the power
spectrum of an inhomogeneous system of particles with long-range interactions. 
This is an additional
motivation
to study Eq. (\ref{fh1}) in detail. The calculations of this section are
inspired by similar
calculations on
fluctuating hydrodynamics performed in \cite{hb5}.

When $\nu\rightarrow 0$, the mean DF $f({\bf
r},{\bf v},t)=\langle f_c({\bf r},{\bf v},t)\rangle$ rapidly reaches a QSS
which is a steady state of the Vlasov equation. 
This process of
violent relaxation takes place on a few dynamical times. If the evolution of
the system is
ergodic, the QSS can be determined
by the Lynden-Bell statistical 
theory \cite{lb}. The kinetic theory of violent collisionless relaxation
is discussed in \cite{kinvr,ewart}. On a longer timescale, the mean DF
evolves through
a sequence of QSSs sourced by the noise. 
Adapting the procedure of
Sec. \ref{sec_inhos} to the present context, 
we obtain the quasilinear equations
\begin{eqnarray}
\label{nfh1l}
\frac{\partial f}{\partial t}=-\nu f+\frac{\partial}{\partial
{\bf J}}\cdot \left\langle \delta
f \frac{\partial\delta\Phi}{\partial {\bf w}} \right\rangle,
\end{eqnarray}
\begin{equation}
\frac{\partial \delta f}{\partial t}+{\bf \Omega}\cdot \frac{\partial
\delta f}{\partial {\bf
w}}-\frac{\partial\delta\Phi}{\partial {\bf w}}\cdot \frac{\partial
f}{\partial {\bf J}}=-\nu \delta f+\sqrt{2m\nu f({\bf J})}\,
{Q}({\bf w},{\bf J},t),
\label{nfh1}
\end{equation}
where ${Q}({\bf w},{\bf J},t)$ is a Gaussian white noise satisfying $\langle
{Q}({\bf w},{\bf J},t)\rangle =0$ and $\langle {Q}({\bf w},{\bf J},t){Q}({\bf
w}',{\bf J}',t')\rangle=\delta({\bf w}-{\bf w}')\delta({\bf J}-{\bf
J}')\delta(t-t')$. 
Introducing the Fourier transforms of the
fluctuations of DF and potential, we can rewrite these
equations as  
\begin{equation}
\frac{\partial f}{\partial t}=-\nu f-i\sum_{\bf
k}\int
\frac{d\omega}{2\pi}\sum_{\bf k'}\int\frac{d\omega'}{2\pi} \,
\left ({\bf k}'\cdot \frac{\partial}{\partial {\bf J}}\right ) e^{i({\bf k}\cdot
{\bf w}-\omega t)}e^{-i({\bf k}'\cdot {\bf w}-\omega'
t)} \langle \delta\hat f({\bf k},{\bf
J},\omega)\delta\hat\Phi^*({\bf
k}',{\bf J},\omega')\rangle,
\label{fh4}
\end{equation}
\begin{equation}
\delta{\hat f}({\bf k},{\bf J},\omega)=\frac{{\bf k}\cdot\frac{\partial
f}{\partial {\bf J}}}{{\bf k}\cdot{\bf\Omega}-\omega-i\nu}\delta{\hat\Phi}({\bf
k},{\bf J},\omega)-i\sqrt{2m\nu f({\bf J})}\frac{{\hat Q}({\bf k},{\bf
J},\omega)}{{\bf k}\cdot{\bf\Omega}-\omega-i\nu},
\label{fh4b}
\end{equation}
where ${\hat Q}({\bf k},{\bf J},\omega)$ is a Gaussian white noise satisfying
$\langle {\hat
Q}({\bf k},{\bf J},\omega)\rangle=0$ and $\langle {\hat Q}({\bf
k},{\bf J},\omega){\hat Q}({\bf
k}',{\bf J}',\omega')^*\rangle=(2\pi)^{1-d}\delta_{{\bf k},{\bf k}'}\delta({\bf
J}-{\bf J}
')\delta(\omega-\omega')$. 
Substituting Eq.
(\ref{fh4b}) into Eq. (\ref{fh4}), we get
\begin{equation}
\frac{\partial f}{\partial t}=-\nu f-i\sum_{\bf
k}\int
\frac{d\omega}{2\pi}\sum_{\bf k'}\int\frac{d\omega'}{2\pi} \,
\left ({\bf k}'\cdot \frac{\partial}{\partial {\bf J}}\right ) e^{i({\bf k}\cdot
{\bf w}-\omega t)}e^{-i({\bf k}'\cdot {\bf w}-\omega'
t)}\frac{{\bf k}\cdot\frac{\partial
f}{\partial {\bf J}}}{{\bf k}\cdot{\bf\Omega}-\omega-i\nu} \langle \delta\hat
\Phi({\bf k},{\bf
J},\omega)\delta\hat\Phi^*({\bf
k}',{\bf J},\omega')\rangle.
\label{ham110w}
\end{equation}
In writing Eq. (\ref{ham110w}) we have only considered
the contribution of the
term $\langle \delta\hat
\Phi({\bf k},{\bf
J},\omega)\delta\hat\Phi^*({\bf
k}',{\bf J},\omega')\rangle$ which leads to a diffusive evolution. The
other terms will be investigated elsewhere \cite{prep1}.

\subsection{Calculation of the power spectrum using the biorthogonal basis}

Let us study Eq. (\ref{nfh1}) for the fluctuations and determine the
power
spectrum $P({\bf k},{\bf J},\omega)$ of the fluctuating potential defined
by
\begin{equation}
\label{psaabann}
\langle \delta\hat\Phi({\bf
k},{\bf J},\omega) \delta\hat\Phi({\bf
k},{\bf J},\omega')^*\rangle=2\pi \delta(\omega-\omega')P({\bf
k},{\bf J},\omega).
\end{equation}
Substituting Eq.
(\ref{fh4b}) into Eq. (\ref{i67}) and using Eq. (\ref{i62}) we
get
\begin{eqnarray}
{\hat A}_{\alpha}(\omega)=-(2\pi)^d \int d{\bf J}\sum_{\bf k} \frac{{\bf
k}\cdot\frac{\partial
f}{\partial {\bf J}}}{{\bf
k}\cdot{\bf\Omega}-\omega-i\nu}\sum_{\alpha'} {\hat
A}_{\alpha'}(\omega) {\hat\Phi}_{\alpha'}({\bf k},{\bf J}) 
{\hat\Phi}_{\alpha}^*({\bf k},{\bf J})\nonumber\\
+(2\pi)^d \int d{\bf
J}\sum_{\bf k} i\sqrt{2m\nu f({\bf J})}\frac{{\hat Q}({\bf k},{\bf
J},\omega)}{{\bf k}\cdot{\bf\Omega}-\omega-i\nu}{\hat\Phi}_{\alpha}^*({\bf
k},{\bf J})
\label{nfh3}
\end{eqnarray}
or, equivalently,
\begin{eqnarray}
\sum_{\alpha'} {\hat A}_{\alpha'}(\omega)\left\lbrack
\delta_{\alpha\alpha'}+(2\pi)^d \int d{\bf J}\sum_{\bf k} \frac{{\bf
k}\cdot\frac{\partial
f}{\partial {\bf J}}}{{\bf
k}\cdot{\bf\Omega}-\omega-i\nu} {\hat\Phi}_{\alpha'}({\bf
k},{\bf J}) 
{\hat\Phi}_{\alpha}^*({\bf k},{\bf J})\right\rbrack\nonumber\\
=(2\pi)^d \int d{\bf
J}\sum_{\bf k} i\sqrt{2m\nu f({\bf J})}\frac{{\hat Q}({\bf k},{\bf
J},\omega)}{{\bf k}\cdot{\bf\Omega}-\omega-i\nu}{\hat\Phi}_{\alpha}^*({\bf
k},{\bf J}).
\label{nfh4}
\end{eqnarray}
Introducing the dielectric matrix from Eq. (\ref{i70}), the
foregoing equation can be rewritten as
\begin{eqnarray}
\sum_{\alpha'} \epsilon_{\alpha\alpha'}(\omega){\hat
A}_{\alpha'}(\omega)=(2\pi)^d \int d{\bf
J}\sum_{\bf k} i\sqrt{2m\nu f({\bf J})}\frac{{\hat Q}({\bf k},{\bf
J},\omega)}{{\bf k}\cdot{\bf\Omega}-\omega-i\nu}{\hat\Phi}_{\alpha}^*({\bf
k},{\bf J}).
\label{nfh5}
\end{eqnarray}
The Fourier transform of the amplitude of the fluctuating potential is therefore
given by\footnote{In matrix form, Eq. (\ref{nfh5}) can be written as $\epsilon
A=B$ hence
$A=(\epsilon^{-1})B$.}
\begin{eqnarray}
{\hat
A}_{\alpha}(\omega)=\sum_{\alpha'}
(\epsilon^{-1})_{\alpha\alpha'}(\omega) (2\pi)^d \int d{\bf
J}\sum_{\bf k} i\sqrt{2m\nu f({\bf J})}\frac{{\hat Q}({\bf k},{\bf
J},\omega)}{{\bf k}\cdot{\bf\Omega}-\omega-i\nu}{\hat\Phi}_{\alpha'}^*({\bf
k},{\bf J})
\label{nfh6}
\end{eqnarray}
and its correlation function is
\begin{eqnarray}
\langle {\hat
A}_{\alpha}(\omega){\hat
A}_{\alpha'}(\omega')^*\rangle=\sum_{\alpha''\alpha'''}
(\epsilon^{-1})_{\alpha\alpha''}(\omega)
(\epsilon^{-1})_{\alpha'\alpha'''}^*(\omega') \nonumber\\
\times (2\pi)^{2d}\int d{\bf
J}d{\bf J}'\sum_{{\bf k}{\bf k}'} 2m\nu \sqrt{f({\bf J})f({\bf
J}')}\frac{\langle {\hat Q}({\bf k},{\bf
J},\omega){\hat Q}({\bf k}',{\bf
J}',\omega')^*\rangle}{({\bf
k}\cdot{\bf\Omega}-\omega-i\nu)({\bf
k}'\cdot{\bf\Omega}'-\omega'+i\nu)}{\hat\Phi}_{\alpha''}^*({\bf
k},{\bf J}){\hat\Phi}_{\alpha'''}({\bf
k}',{\bf J}').
\label{nfh7}
\end{eqnarray}
For a Gaussian white noise, we get
\begin{eqnarray}
\langle {\hat
A}_{\alpha}(\omega){\hat
A}_{\alpha'}(\omega')^*\rangle=\sum_{\alpha''\alpha'''}
(\epsilon^{-1})_{\alpha\alpha''}(\omega)
(\epsilon^{-1})_{\alpha'\alpha'''}^*(\omega)\nonumber\\
\times (2\pi)^{d+1}\int d{\bf
J}\sum_{{\bf k}} 2m\nu f({\bf J})\delta(\omega-\omega')\frac{1}{({\bf
k}\cdot{\bf\Omega}-\omega)^2+\nu^2}{\hat\Phi}_{\alpha''}^*({\bf
k},{\bf J}){\hat\Phi}_{\alpha'''}({\bf
k},{\bf J}).
\label{nfh8}
\end{eqnarray}
Taking the limit $\nu\rightarrow 0$ and using the identity
\begin{equation}
\lim_{\epsilon\rightarrow 0}\frac{\epsilon}{x^2+\epsilon^2}=\pi\, \delta(x),
\label{fh6}
\end{equation}
we obtain 
\begin{eqnarray}
\langle {\hat
A}_{\alpha}(\omega){\hat
A}_{\alpha'}(\omega')^*\rangle=\sum_{\alpha''\alpha'''}
(\epsilon^{-1})_{\alpha\alpha''}(\omega)
(\epsilon^{-1})_{\alpha'\alpha'''}^*(\omega)\nonumber\\
\times (2\pi)^{d+2}\int d{\bf
J}\sum_{{\bf k}} m f({\bf J})\delta(\omega-\omega')\delta({\bf
k}\cdot{\bf\Omega}-\omega){\hat\Phi}_{\alpha''}^*({\bf
k},{\bf J}){\hat\Phi}_{\alpha'''}({\bf
k},{\bf J}).
\label{nfh9}
\end{eqnarray}
From Eqs. (\ref{i62})  and (\ref{nfh9}) we find that
\begin{eqnarray}
\langle \delta\hat\Phi({\bf k},{\bf J},\omega) \delta\hat\Phi^*({\bf k},{\bf
J},\omega')\rangle=\sum_{\alpha\alpha'\alpha''\alpha'''}
(\epsilon^{-1})_{\alpha\alpha''}(\omega)
(\epsilon^{-1})_{\alpha'\alpha'''}^*(\omega)\nonumber\\
\times (2\pi)^{d+2}\int d{\bf
J}'\sum_{{\bf k}'} m f({\bf J}')\delta(\omega-\omega')\delta({\bf
k}'\cdot{\bf\Omega}'-\omega){\hat\Phi}_{\alpha''}^*({\bf
k}',{\bf J}'){\hat\Phi}_{\alpha'''}({\bf
k}',{\bf J}'){\hat\Phi}_{\alpha}({\bf
k},{\bf J}){\hat\Phi}_{\alpha'}({\bf
k},{\bf J})^*.
\label{nfh10}
\end{eqnarray}
Introducing the dressed potential of interaction from Eq. (\ref{can1b}) we
can rewrite the foregoing equation as 
\begin{eqnarray}
\langle \delta\hat\Phi({\bf k},{\bf J},\omega) \delta\hat\Phi^*({\bf k},{\bf
J},\omega')\rangle=(2\pi)^{d+2}\int d{\bf
J}'\sum_{{\bf k}'} |A^d_{{\bf k}{\bf k}'}({\bf J},{\bf J}',\omega)|^2 m
f({\bf J}')\delta(\omega-\omega')\delta({\bf
k}'\cdot{\bf\Omega}'-\omega).
\label{nfh11}
\end{eqnarray}
Finally, using Eq. (\ref{psaabann}), we obtain the expression of the power
spectrum
\begin{eqnarray}
P({\bf k},{\bf J},\omega)=(2\pi)^{d+1}\int d{\bf
J}'\sum_{{\bf k}'} |A^d_{{\bf k}{\bf k}'}({\bf J},{\bf J}',\omega)|^2 m
f({\bf J}')\delta({\bf
k}'\cdot{\bf\Omega}'-\omega).
\label{nfh12}
\end{eqnarray}
This returns the
power
spectrum (\ref{pi4}) produced by a
random distribution of particles.

\subsection{Calculation of the power spectrum without using the biorthogonal
basis}

We can also calculate the power spectrum without using the biorthogonal
basis. Substituting Eq. (\ref{fh4b}) into Eq. (\ref{i60g}) we get
\begin{eqnarray}
\delta{\hat\Phi}({\bf k},{\bf J},\omega)=(2\pi)^d\sum_{{\bf k}'}\int d{\bf J}'
A_{{\bf k},{\bf k}'}({\bf J},{\bf J}')\frac{{\bf k}'\cdot\frac{\partial
f'}{\partial {\bf J}'}}{{\bf
k}'\cdot{\bf\Omega}'-\omega-i\nu}\delta{\hat\Phi}({\bf
k}',{\bf J}',\omega)\nonumber\\
-(2\pi)^d\sum_{{\bf k}'}\int d{\bf J}'
A_{{\bf k},{\bf k}'}({\bf J},{\bf J}')i\sqrt{2m\nu f({\bf J}')}\frac{{\hat
Q}({\bf k}',{\bf
J}',\omega)}{{\bf k}'\cdot{\bf\Omega}'-\omega-i\nu}.
\end{eqnarray}
Proceeding like in Appendix \ref{sec_dnwout}, we can write the solution of this
integral equation as
\begin{eqnarray}
\delta{\hat\Phi}({\bf k},{\bf J},\omega)=
-(2\pi)^d\sum_{{\bf k}'}\int d{\bf J}'
A^d_{{\bf k},{\bf k}'}({\bf J},{\bf J}',\omega)i\sqrt{2m\nu f({\bf
J}')}\frac{{\hat
Q}({\bf k}',{\bf
J}',\omega)}{{\bf k}'\cdot{\bf\Omega}'-\omega-i\nu}.
\end{eqnarray}
We then obtain
\begin{eqnarray}
\langle \delta{\hat\Phi}({\bf k},{\bf J},\omega)
\delta{\hat\Phi}({\bf k},{\bf J},\omega')^*\rangle=
(2\pi)^{2d}\sum_{{\bf k}'{\bf k}''}\int d{\bf J}'d{\bf J}''
A^d_{{\bf k},{\bf k}'}({\bf J},{\bf J}',\omega)A^d_{{\bf k},{\bf k}''}({\bf
J},{\bf J}'',\omega')^* \nonumber\\
\times 2m\nu\sqrt{f({\bf
J}')f({\bf
J}'')}\frac{\langle {\hat
Q}({\bf k}',{\bf
J}',\omega){\hat
Q}({\bf k}'',{\bf
J}'',\omega')^*\rangle}{({\bf k}'\cdot{\bf\Omega}'-\omega-i\nu)({\bf
k}''\cdot{\bf\Omega}''-\omega'+i\nu)}.
\end{eqnarray}
For a Gaussian noise, we get
\begin{eqnarray}
\langle \delta{\hat\Phi}({\bf k},{\bf J},\omega)
\delta{\hat\Phi}({\bf k},{\bf J},\omega')^*\rangle=
(2\pi)^{d+1}\sum_{{\bf k}'}\int d{\bf J}'
|A^d_{{\bf k},{\bf k}'}({\bf J},{\bf J}',\omega)|^2 
\delta(\omega-\omega')\frac{2m\nu f({\bf
J}')}{({\bf
k}'\cdot{\bf\Omega}'-\omega)^2+\nu^2}.
\end{eqnarray}
Taking the limit $\nu\rightarrow 0$ and using the identity (\ref{fh6}) we
recover Eqs. (\ref{nfh11}) and (\ref{nfh12}).

\subsection{The nonlinear
diffusion equation}

Substituting Eqs. (\ref{psaabann}) and (\ref{nfh12})
into Eq. (\ref{ham110w}) and
repeating the calculations of Sec. \ref{sec_sdd}, we obtain the
nonlinear
diffusion equation
\begin{equation}
\frac{\partial f}{\partial t}=-\nu f+\frac{\partial}{\partial J_i}\left
(D_{ij}[f,{\bf
J}]\frac{\partial f}{\partial J_j}\right )
\label{i87bG}
\end{equation}
with a diffusion coefficient
\begin{equation}
D_{ij}[f,{\bf J}]=\frac{1}{2}\sum_{\bf
k}\, k_ik_j
P({\bf
k},{\bf J},{\bf k}\cdot {\bf \Omega})=\pi(2\pi)^{d}
m\sum_{{\bf k},{\bf k}'}\int d{\bf
J}'\, k_ik_j
|A^d_{{\bf k},{\bf k}'}({\bf J},{\bf
J'},{\bf k}\cdot {\bf\Omega})|^2 \delta({\bf
k}\cdot {\bf\Omega}-{\bf k}'\cdot
{\bf\Omega}')f({\bf J}')
\label{i88g}
\end{equation}
coinciding with the diffusion coefficient produced by a random distribution of
particles (see
Sec. \ref{sec_diffco}). The term $-\nu f$ describes the damping
of the system on a timescale $1/\nu$ and
the   diffusion tensor $D_{ij}[f,{\bf J}]\sim 1/N$ describes its
evolution on a timescale $N t_D$.

On a mesoscopic scale, we can keep track of the
fluctuations in the
evolution of the DF and write  
\begin{eqnarray}
\frac{\partial \overline{f}}{\partial t}=-\nu
\overline{f}+\frac{\partial}{\partial J_i}\left
(D_{ij}[\overline{f},{\bf
J}]\frac{\partial \overline{f}}{\partial J_j}\right )+\sqrt{2 m \nu
\overline{f}}\, Q({\bf J},t),
\label{fh17b}
\end{eqnarray}
where ${Q}({\bf J},t)$ is a Gaussian white noise satisfying $\langle
{Q}({\bf J},t)\rangle =0$ and $\langle
{Q}({\bf J},t){Q}({\bf J}',t')\rangle=\delta({\bf J}-{\bf J}')\delta(t-t')$.
Equation (\ref{fh17b})
without the noise term may have several equilibrium
states. The noise
term 
allows the system to switch from one equilibrium state to another one through
random transitions (see, e.g., 
\cite{random,gsse,nardini2,bouchetsimmonet,rbs,brs}  in various
contexts).

{\it Remark:} In conclusion, the power spectrum of an inhomogeneous system of
particles with long-range interactions can be obtained in
different manners. It can be obtained from the linearized
Klimontovich equation by solving
an initial value problem 
(see Eq. (43) of Ref. \cite{physicaA} and Eq. (\ref{int3}) of Appendix
\ref{sec_j}).  It can also be obtained by considering
the
dressing of the bare correlation function of the potential created by a
random distribution of
particles viewed as an external perturbation (see Eq.
(\ref{pi4})
of Sec. \ref{sec_inhoscf}). Finally, in this section,  we have
determined the
power spectrum (\ref{nfh12}) directly from  the stochastic
damped
Vlasov equation (\ref{fh1}) in the spirit of fluctuating
hydrodynamics \cite{hb5}.

\section{Stochastically forced Brownian particles with long-range interactions}
\label{sec_spv}

In this section, we consider a stochastic model of particles with long-range
interactions described
by the $N$
coupled Langevin equations
\begin{equation}
\frac{d{\bf r}_i}{dt}={\bf v}_i,\qquad \frac{d{\bf v}_i}{dt}=-
\sum_j m_j \nabla u(|{\bf r}_i-{\bf r}_j|)+\sqrt{2D}{\bf
R}_i(t),
\label{gbrow1}
\end{equation}
where $i=1,...,N$ label the particles and ${\bf R}_i(t)$ is a Gaussian
white noise satisfying $\langle {\bf R}_i(t)\rangle ={\bf 0}$ and $\langle
{R}^{\alpha}_i(t){R}^{\beta}_j(t')\rangle=\delta_{ij}\delta_{\alpha\beta}
\delta(t-t')$. The constant $D$ is the bare diffusion
coefficient  (we will ultimately take the limit
$D\rightarrow 0$). This stochastic model of particles corresponds to the model
of Brownian particles with long-range interactions introduced in \cite{hb5} with
$\xi=0$, i.e., without friction (see Sec. \ref{sec_tdbv}). 

The exact equation satisfied by the discrete DF $f_d({\bf r},{\bf
v},t)=\sum_{i=1}^{N} m_i \delta({\bf
r}-{\bf
r}_i(t))\delta({\bf
v}-{\bf
v}_i(t))$ is
\cite{hb5}
\begin{eqnarray}
\label{gbrow5}
\frac{\partial f_d}{\partial t}+{\bf v}\cdot \frac{\partial
f_d}{\partial {\bf r}}-
\nabla\Phi_d\cdot
\frac{\partial f_d}{\partial {\bf v}}=\frac{\partial}{\partial {\bf
v}}\cdot\left ( D \frac{\partial f_d}{\partial {\bf v}}\right
)+\frac{\partial}{\partial{\bf v}}\cdot \left
(\sqrt{2Dmf_d}\, {\bf Q}({\bf r},{\bf v},t)\right ),
\end{eqnarray}
\begin{eqnarray}
{\Phi}_d({\bf r},t)=\int u(|{\bf r}-{\bf
r}'|){\rho}_d({\bf
r}',t)\, d{\bf
r}',
\label{gbrow6p}
\end{eqnarray}
where ${\bf Q}({\bf r},{\bf v},t)$ is a Gaussian white noise satisfying $\langle
{\bf
Q}({\bf r},{\bf v},t)\rangle={\bf 0}$ and  $\langle Q_\alpha({\bf r},{\bf
v},t)Q_\beta({\bf
r}',{\bf v}',t')\rangle=\delta_{\alpha\beta}\delta({\bf r}-{\bf r}')\delta({\bf
v}-{\bf v}')\delta(t-t')$. 
For
simplicity, we consider a single species gas of particles but the generalization
to multiple species of particles is
straightforward.

In this section, we neglect the collisions  between the
particles (finite $N$ effects) that would lead to the Lenard-Balescu collision
term (see
Secs. \ref{sec_avp} and \ref{sec_mono}) and focus on the effect of the noise. In
that case, the mean DF $f({\bf r},{\bf v},t)=\langle f_{d}({\bf
r},{\bf v},t)\rangle$ satisfies the equation
\begin{eqnarray}
\label{gbrow5j}
\frac{\partial f}{\partial t}+{\bf v}\cdot \frac{\partial
f}{\partial {\bf r}}-
\nabla\Phi\cdot
\frac{\partial f}{\partial {\bf v}}=\frac{\partial}{\partial {\bf
v}}\cdot\left ( D \frac{\partial f}{\partial {\bf v}}\right
)
\end{eqnarray}
with Eq. (\ref{brow3}) and the  mesoscopic DF satisfies the equation
\begin{eqnarray}
\label{gbrow6}
\frac{\partial \overline{f}}{\partial t}+{\bf v}\cdot \frac{\partial
\overline{f}}{\partial {\bf r}}-
\nabla\overline{\Phi}\cdot
\frac{\partial \overline{f}}{\partial {\bf v}}=\frac{\partial}{\partial {\bf
v}}\cdot\left ( D \frac{\partial \overline{f}}{\partial {\bf v}}\right
)+\frac{\partial}{\partial{\bf v}}\cdot \left
(\sqrt{2Dm\overline{f}}\, {\bf Q}({\bf r},{\bf v},t)\right )
\end{eqnarray}
with Eq. (\ref{brow6}).

When $D\rightarrow 0$ the mean DF $f({\bf
r},{\bf v},t)$ rapidly reaches a QSS
which is a stable steady state of the Vlasov equations (violent relaxation). On
a longer
timescale, the mean DF evolves through
a sequence of QSSs sourced by the noise. This is similar to the problem
discussed in Sec.  \ref{sec_fd}. Adapting the procedure of
Sec. \ref{sec_inhos} to the present context (see also Sec.  \ref{sec_tdbv}),
we obtain the quasilinear equations 
\begin{eqnarray}
\label{gk1}
\frac{\partial f}{\partial t}=\frac{\partial}{\partial {\bf
J}}\cdot\left ( D_K \frac{\partial f}{\partial {\bf J}}\right
)+\frac{\partial}{\partial
{\bf J}}\cdot \left\langle \delta
f \frac{\partial\delta\Phi}{\partial {\bf w}} \right\rangle,
\end{eqnarray}
\begin{equation}
\frac{\partial \delta f}{\partial t}+{\bf \Omega}\cdot \frac{\partial
\delta f}{\partial {\bf
w}}-\frac{\partial\delta\Phi}{\partial {\bf w}}\cdot \frac{\partial
f}{\partial {\bf J}}=\frac{\partial}{\partial {\bf
w}}\cdot\left ( D_K \frac{\partial \delta f}{\partial {\bf w}}\right
)+\frac{\partial}{\partial{\bf w}}\cdot \left
(\sqrt{2D_Km f({\bf J})}\, {\bf Q}({\bf w},{\bf J},t)\right ),
\label{gk2}
\end{equation}
where ${\bf Q}({\bf w},{\bf J},t)$ is a Gaussian white noise satisfying $\langle
{\bf Q}({\bf w},{\bf J},t)\rangle =0$ and $\langle {Q}_{\alpha}({\bf w},{\bf
J},t){Q}_{\beta}({\bf
w}',{\bf J}',t')\rangle=\delta_{\alpha\beta}\delta({\bf w}-{\bf w}')\delta({\bf
J}-{\bf
J}')\delta(t-t')$.

Repeating
the calculations of Sec. \ref{sec_fd}  with only
minor modifications, we obtain
the nonlinear
diffusion equation 
\begin{eqnarray}
\label{gk3}
\frac{\partial f}{\partial t}=\frac{\partial}{\partial {\bf
J}}\cdot\left ( D_K \frac{\partial f}{\partial {\bf J}}\right
)+\frac{\partial}{\partial 
J_i}\cdot\left ( D_{ij}[f,{\bf J}] \frac{\partial f}{\partial J_j}\right
)
\end{eqnarray}
with the diffusion tensor $D_{ij}[f,{\bf J}]$ from Eq. (\ref{i88g})
which
coincides with the Lenard-Balescu
diffusion tensor of a gas of particles with long-range interactions due to
finite $N$ effects (see
Sec. \ref{sec_diffco}). The diffusion coefficient $D_K$
describes the
evolution
of the system on a diffusive timescale $J_{\rm max}^2/D_K$
(where $J_{\rm max}$
is the maximum value of the action) and the   diffusion tensor $D_{ij}[f,{\bf
J}]\sim 1/N$
describes its
evolution
on a timescale $N t_D$. We should also take into account
the long-range collisions between the particles, leading to the Lenard-Balescu
current from Eq. (\ref{ilb3}),
which develop on the same timescale. 

At a mesoscopic level,\footnote{We stress that there are several levels
of description. The mesoscopic description leading to Eq. (\ref{gk4}) takes
place at a higher scale than the mesoscopic description leading to Eq.
(\ref{gbrow6}).} we can keep track of the
forcing (fluctuations) in the evolution of the DF and write
\begin{eqnarray}
\label{gk4}
\frac{\partial f}{\partial t}=\frac{\partial}{\partial {\bf
J}}\cdot\left ( D_K \frac{\partial f}{\partial {\bf J}}\right
)+\frac{\partial}{\partial J_i}\cdot\left ( D_{ij}[f,{\bf J}] \frac{\partial
f}{\partial {J}_j}\right
)+\frac{\partial}{\partial{\bf J}}\cdot \left
(\sqrt{2D_K m f({\bf J})}\, {\bf Q}({\bf J},t)\right ),
\end{eqnarray}
where ${\bf Q}({\bf J},t)$ is a Gaussian white noise satisfying $\langle
{\bf Q}({\bf J},t)\rangle =0$ and $\langle {Q}_{\alpha}({\bf
J},t){Q}_{\beta}({\bf
J}',t')\rangle=\delta_{\alpha\beta}\delta({\bf J}-{\bf
J}')\delta(t-t')$.

\section{Summary of the different kinetic equations}
\label{sec_diff}

In this section, we recapitulate the different kinetic equations discussed in
our paper and their different derivations.

\subsection{Lenard-Balescu equation}

The Lenard-Balescu equation (\ref{ilb3}) governs the
mean evolution of an isolated  system of particles with long-range interactions
due to discreteness
effects (``collisions'').

It can be derived from the Klimontovich formalism by taking $\Phi_e=0$ in Eq.
(\ref{n1}) and considering an initial
value problem as explained in Sec. 2 of \cite{physicaA}. In that case, Eq.
(\ref{i57}) with $\Phi_e=0$ has to
be solved by using a Fourier transform in angles and a Laplace transform in
time. This introduces in Eq. (\ref{i60}) a term related to the initial
condition [see Eq. (21) in \cite{physicaA}] instead of the term related to
$\Phi_e$.
We can then compute the
collision term in Eq. (\ref{i56}) with
$\Phi_e=0$ as in Sec. 2 of \cite{physicaA} and obtain the
Lenard-Balescu equation (\ref{ilb3}).

Another approach is to start from the Fokker-Planck equation (\ref{fp5}) or
(\ref{fp8}) and compute the
diffusion and friction coefficients individually.

(i) To compute the diffusion coefficient from Eq. (\ref{fp6}) leading to
Eq.
(\ref{gei3}), we have to
evaluate
the power spectrum of the fluctuations of the force created by a random
distribution of
field particles. This can be done
in two manners. 

(a) The first possibility is to take  $\Phi_e=0$ in Eq. (\ref{n1}) and
solve Eq. Eq. (\ref{i57}) with
$\Phi_e=0$ by using a Fourier transform in angle and a Laplace
transform in time as mentioned above. This leads to the expression (21) of
\cite{physicaA} for the fluctuations, which involves the initial
condition. The
power spectrum is then given by Eq. (43) of \cite{physicaA} and the diffusion
coefficient by Eq. (79) of \cite{physicaA}.

(b) Another possibility is to introduce a stochastic perturbation $\Phi_e$
in Eq. (\ref{n1}) and 
solve Eq. (\ref{i57}) by introducing Fourier transforms in angle and time.
This leads to Eq. (\ref{i60}) for the fluctuations, which involves the
external perturbation. The power
spectrum is then given by Eq. (\ref{i83}). If we assume that the external
perturbation is due to a random distribution of field particles, we can use  Eq.
(\ref{ci10}) to obtain the expression (\ref{pi4}) of the power spectrum. The
diffusion tensor is then given by Eq. (\ref{di1}).

(ii) We can compute  the friction in two manners.

(a) The first possibility is to
compute the total friction (\ref{fp7}) arising in Eq. (\ref{fp5}) by proceeding
like in Secs. 3.3 and 3.4 of \cite{physicaA}. This leads to Eq. (112) of
\cite{physicaA}.
We then find that the total friction splits in two terms: a term 
interpreted as a ``friction by polarization'' (see  Sec. 3.3 of
\cite{physicaA}) and another term related to the
gradient of the diffusion tensor (see  Sec. 3.4 of
\cite{physicaA}). Substituting the diffusion tensor (Eq. (79) of
\cite{physicaA})
and the total friction (Eq. (112) of \cite{physicaA}) in the ordinary expression
(\ref{fp5}) of the Fokker-Planck
equation, and using an integration by parts, we obtain the Lenard-Balescu
equation (\ref{ilb3}). This
is the
approach followed in Sec. 3.5 of \cite{physicaA}.

(b) Alternatively, we can directly compute
the friction by polarization arising in Eq. (\ref{fp8}) by considering the
response
of the system to the
perturbation created by the test particle (see Sec. \ref{sec_ifpol}). This
leads to Eq. (\ref{gei4zz}). This
calculation shows that ${\bf F}_{\rm pol}$ can be truly interpreted as a
friction
by polarization. Substituting the diffusion tensor (\ref{di1}) and the
friction by
polarization (\ref{gei4zz}) in the expression (\ref{fp8}) of the Fokker-Planck
equation, we obtain the
Lenard-Balescu equation (\ref{ilb3}). This derivation is simpler
(less
technical) and
more physical that the one given in \cite{physicaA}.

\subsection{SDD equation}

The SDD equation (\ref{i87b}) with Eqs. (\ref{i88}) and (\ref{i88bc})
governs the mean evolution of a system of particles with long-range interactions
submitted to an external
stochastic perturbation in the limit where the collisions between the
particles are negligible, i.e., in the limit $N\rightarrow +\infty$ with
$m\sim 1/N\rightarrow 0$. In that case, we can solve Eq. (\ref{i57})
by using Fourier transforms in angle and time. This leads to the
expression (\ref{i83}) of the power spectrum. The SDD equation
can be
derived from the Klimontovich formalism (see Sec.
\ref{sec_sddk}) or
from
the Fokker-Planck formalism (see Sec.
\ref{sec_sddfp}).
Since
there is
no friction by polarization ($m\rightarrow 0$), the Fokker-Planck equation
(\ref{fp8}) reduces to Eq. (\ref{i87wq}). Using Eq.
(\ref{i83}), the
diffusion coefficient
(\ref{gei3}) can be written as in Eq. (\ref{i88b}). Substituting
Eqs. (\ref{gei3}) and (\ref{i88b}) into Eq. (\ref{i87wq}) we
obtain the SDD equation (\ref{i87b}) with Eqs. (\ref{i88}) and
(\ref{i88bc}). If the external perturbation is due to a random distribution of
field particles, the SDD equation takes the form of Eq. (\ref{can2}) in
connection to the Lenard-Balescu equation.

\subsection{General kinetic equation}
\label{sec_gke}

We now present a kinetic equation that generalizes the Lenard-Balescu equation 
(\ref{ilb3}) and the SDD equation  (\ref{i87b})-(\ref{i88bc}). We consider a
collection of particles of
mass $m$
submitted to a stochastic perturbation that can be internal or external to the
system (or both). The test particles experience a diffusion
due to the
stochastic
perturbation and a friction by polarization due to the retroaction (response) of
the background medium to
the
deterministic perturbation that they induce. The
evolution of their density (mean DF) is
thus governed by
a general Fokker-Planck equation of the form of Eq. (\ref{fp8}) where $D_{ij}$
is given by Eq.
(\ref{gei3})
and ${\bf F}_{\rm pol}$ is given by Eq. (\ref{ci1bnh}). Explicitly,
\begin{equation}
\label{hybrid1}
\frac{\partial f}{\partial t}=\frac{\partial}{\partial J_{i}} \sum_{{\bf
k}}\left
\lbrace\frac{1}{2} k_i k_j  P({\bf k},{\bf J},{\bf
k}\cdot {\bf\Omega}({\bf J}))\frac{\partial f}{\partial
J_{j}}+ m f   k_i\,  {\rm
Im}\left\lbrack A^d_{{\bf k}{\bf k}}({\bf J},{\bf J},{\bf k}\cdot {\bf
\Omega}) \right\rbrack\right \rbrace.
\end{equation}
Alternative expressions of this kinetic equation can be obtained by using Eqs.
(\ref{i88b}) and (\ref{gei4zz}) instead of Eqs.  (\ref{gei3}) and 
(\ref{ci1bnh}). This kinetic equation is more general than the
Lenard-Balescu
equation (\ref{ilb3}) because the noise is not necessarily due to
a discrete
collection of field particles. It is
also more general than the SDD equation (\ref{i87b})-(\ref{i88bc})
because it takes into account the friction by polarization of the test
particles. If we neglect the friction
by polarization (i.e. if we take $m\rightarrow 0$) we recover the SDD equation.
If we
assume that the external perturbation is only due to
field particles and use Eqs. (\ref{di1}) and (\ref{gei4zzb}), we
recover the
Lenard-Balescu equation.  If we assume that a part of the stochastic
perturbation is due to field particles and another part is due to an external
noise, we get an hybrid 
(mixed) kinetic equation with a Lenard-Balescu term and a SDD term (see Sec.
\ref{sec_hybrid}).

\subsection{Hybrid kinetic equation and its subcases}
\label{sec_hybrid}

In order to be as general as possible, we consider a
system of test particles
of mass $m$ in ``collision'' with field particles of masses
$\lbrace m_b\rbrace$ and submitted in addition to an
external noise of a general nature. In that case, the
evolution of the mean DF is governed by a mixed kinetic equation of the
form
\begin{eqnarray}
\frac{\partial f}{\partial t}=C_{\rm LB}+C_{\rm SDD},
\label{hybrid2}
\end{eqnarray}
with a Lenard-Balescu collision term $C_{\rm LB}$ [see Eq. (\ref{ilb3})] due to
the collisions between the particles (finite $N$ effects) and a collision
term $C_{\rm SDD}$ [see Eqs. (\ref{i87b})-(\ref{i88bc})] due to the external
noise. This corresponds to Eq.
(\ref{hybrid1}) with a power spectrum $P=P_{\rm LB}+P_{\rm SDD}$.

Let us consider particular cases of this equation:

(i) When $m\rightarrow 0$, we can neglect the friction by polarization
and we get a diffusion equation with two terms of diffusion $D_{\rm LB}$ and
$D_{\rm SDD}$.

(i-a)  In the absence of external noise, we recover the diffusion equation
(\ref{ilb22}) with Eq. (\ref{ilb16}). 

(i-b) When $m_b\rightarrow 0$ we can neglect the diffusion induced by the
field particles and we recover the SDD equation (\ref{i87b})
with Eqs. (\ref{i88}) and (\ref{i88bc}).

(ii) When $m_b\rightarrow 0$, we can neglect the diffusion induced by
the field particles and we get a Fokker-Planck equation of the form of Eq.
(\ref{hybrid1}) with a diffusion term
$D_{\rm SDD}$ due to the external noise  and a friction by
polarization.

(ii-a) In the absence of external noise, we recover the advection
equation (\ref{ilb24}) with Eq. (\ref{ilb17}).

(ii-b) When $m\rightarrow 0$, we
can neglect the friction by polarization and 
we recover the SDD equation  (\ref{i87b})
with Eqs. (\ref{i88}) and (\ref{i88bc}).

(ii-c) In the case where the field particles are at
statistical equilibrium, we can simplify the friction by polarization and we
get an equation of the form of Eq. (\ref{wq}).

(iii) In the absence of external noise, we recover the
Lenard-Balescu equation (\ref{ilb3}).

(iii-a) When $m\rightarrow 0$, we can neglect the friction by polarization
and  we recover the diffusion equation (\ref{ilb22}) with Eq. (\ref{ilb16}). 

(iii-b) When $m_b\rightarrow 0$, we can neglect the diffusion induced by 
the field vortices and we recover the advection
equation (\ref{ilb24}) with Eq. (\ref{ilb17}).

(iii-c) In the case where the field particles are at
statistical equilibrium we can simplify the friction by polarization and we
recover the Kramers equation (\ref{ilb19}).

\section{Conclusion}
\label{sec_conc}

In this paper, we have complemented the kinetic theory of inhomogeneous systems
with long-range interactions initiated in \cite{heyvaerts,physicaA}. We 
used an approach inspired by the one developed by Hubbard \cite{hubbard1} in the
homogeneous case. We computed the power spectrum of the total fluctuating
potential acting on a test particle and the corresponding diffusion tensor. We
also
computed the friction by polarization induced by the wake produced by the moving
particle. When the diffusion tensor and the friction by polarization are
substituted into the Fokker-Planck equation  in ``sandwiched''
form, we obtain the inhomogeneous Lenard-Balescu equation which reduces to the
inhomogeneous Landau kinetic equation when collective effects are neglected.
This allowed us to derive these equations in a more physical
manner than in the traditional approaches based on the BBGKY
hierarchy \cite{heyvaerts} and on the Klimontovich
formalism \cite{physicaA}.\footnote{This procedure of derivation from the
Fokker-Planck equation was initiated
in Sec. 3 of \cite{physicaA} and in \cite{hfcp}.}

We also derived the proper form of fluctuation-dissipation theorem for
inhomogeneous systems with long-range interactions at statistical equilibrium
(thermal bath). This theorem, which relates
the response function to the correlation function, is of fundamental
significance in statistical physics. Using
the fluctuation-dissipation theorem, we showed that the
friction  by polarization experienced by a test particle is proportional to the
pulsation of the orbit of the particle with action ${\bf J}$ and that the 
friction tensor satisfies the Einstein relation.  In that case, the
Fokker-Planck equation takes the form of a Kramers
equation in action space. We also showed that the friction tensor can be
written in the form of a Green-Kubo formula expressing the
fluctuation-dissipation theorem.

We considered the case of particles with different masses. The diffusion tensor
$D_{ij}$ of the test particle is due to the fluctuations
of the force produced by the field particles. It is therefore proportional to
the mass $m_b$ of the field particles. The friction by polarization ${\bf
F}_{\rm pol}$ arises from the retroaction (response) of the mean DF to the
perturbation caused by the test particle, like in a polarization process.
It can be obtained from a linear response theory. It is therefore proportional
to the mass $m$ of the test particle. For a thermal bath, the friction by
polarization is related to the diffusion tensor by the Einstein relation
involving the mass $m$ of the test particle.

We considered the case of 1D inhomogeneous systems for which only $1:1$
resonances are permitted. These systems can present a situation of kinetic
blocking when the frequency profile is monotonic. We emphasized the analogy
with the kinetic theory of 2D point vortices
\cite{Kvortex2023}.

We also discussed the inhomogeneous SDD
equation which describes the collisionless evolution of a system with
long-range
interactions submitted to an external stochastic perturbation. This equation
has the form of a nonlinear (integrodifferential) diffusion equation.

We presented a kinetic equation [see Eq. (\ref{hybrid1})] that generalizes the
Lenard-Balescu equation and the SDD equation. This equation includes a diffusion
term of arbitrary origin (instead of being produced by a collection of $N$ field
particles) and a friction by polarization taking into account the retroaction of
the
medium to the perturbation caused by the test particle.

Finally, we compared the inhomogeneous Lenard-Balescu
equation describing the collisional evolution of an isolated Hamiltonian system
of particles with long-range interactions undergoing
perturbations arising from its own discreteness (finite $N$ effects) to the
inhomogeneous mean field Kramers equation describing the collisionless
evolution of
a dissipative Brownian system of particles with long-range interactions
experiencing a friction force and a stochastic force due to an inert medium.
The Lenard-Balescu
equation is valid at the order  $1/N$ while the mean field Kramers equation  is
valid in the limit $N\rightarrow +\infty$. These deterministic equations
describe the
evolution of the mean DF. Following our previous
works \cite{hb5,random,entropy,gsse}, we discussed
stochastic kinetic equations which take fluctuations into account.
We explicitly considered the stochastic Kramers and Smoluchowski
equations which are associated with the canonical ensemble. Similarly, one could
consider the stochastic Landau and Lenard-Balescu equations which are
associated with the microcanonical ensemble \cite{bouchetld}. These stochastic
equations could be used to describe random transitions between different
equilibrium states \cite{random,gsse,nardini2,bouchetsimmonet,rbs,brs}.

The Lenard-Balescu equation has a limited domain of validity. Since it assumes
that the damped modes decay rapidly (i.e. the system must be strongly stable),
it is
valid only sufficiently far from the critical point where the system becomes
unstable. Close to the critical point, the fluctuations
become very important \cite{hb5}. The dielectric matrix $\epsilon(\omega)$
vanishes for some real pulsation $\omega$ while the
response
function  and the amplitude of the correlation function (power
spectrum) diverge. This is similar to  the phenomenon of critical opalescence
in
the theory of liquids \cite{sdduniverse}. As a result,
the diffusion coefficient $D\propto 1/|\epsilon({\bf k}\cdot {\bf\Omega})|^2$
can become very large and the relaxation time decreases, i.e., the relaxation
accelerates (see, e.g., Refs. \cite{weinberglb,hb5,aa,magorrian} and Eq. (56) of
\cite{epjp3}). In parallel, the decorrelation time increases and the
damped modes
decay slowly \cite{hb1,hb2,hb3,hb4,hb5}. Therefore, close to the critical point,
many assumptions made to
derive the Lenard-Balescu break down: the Bogoliubov ansatz is not valid
anymore, the evolution is non-markovian, nonlinear terms have to be taken into
account etc. On the other hand, since the fluctuations become large close to the
critical point it becomes crucial to use stochastic
kinetic equations (as discussed above) rather than deterministic kinetic
equations. These are important effects to consider in future works.

Recently, Hamilton and Heinemann \cite{hh} have suggested that the linear
response
of stellar systems does not diverge at marginal stability, at
least for sufficiently (realistic) short times.\footnote{Similar
results were obtained in \cite{linres}  although they were not explicitly
stated. See in particular the convergence of the response function
$J(k,t)$ in Eq. (54) when
$\omega_0(k)\rightarrow 0$ at fixed $t$ and the results of Sec. 5 for the
Cauchy distribution.} As a result, according to these authors,
the Lenard-Balescu
equation is not valid close to the critical point, not because of nonlinear
effects, but because the contribution of the damped modes has not been taken
into account. Actually, it is possible that certain quantities like the
correlation
functions (power spectrum) and the diffusion coefficient diverge at the critical
point (if we ignore nonlinear effects) while other quantities like the wake
potential and the
friction by polarization remain finite. This is an interesting problem to
investigate further.

\appendix

\section{Orbit-averaged Kramers equation}
\label{sec_oake}

We consider a system of Brownian particles with long-range interactions (see
Sec. \ref{sec_tdbv}). For $N\rightarrow +\infty$, the evolution of their mean
DF $f({\bf
r},{\bf v},t)$ is governed by the mean field Kramers (or Vlasov-Kramers)
equation
\begin{eqnarray}
\label{oake1}
\frac{\partial f}{\partial t}+{\bf v}\cdot \frac{\partial
f}{\partial {\bf r}}-
\nabla\Phi\cdot
\frac{\partial f}{\partial {\bf v}}=\frac{\partial}{\partial {\bf
v}}\cdot\left\lbrack D
\left (\frac{\partial f}{\partial {\bf v}}+\beta m f {\bf v}\right
)\right\rbrack.
\end{eqnarray}
We consider the low friction
limit $\xi=D\beta m\rightarrow 0$ with $\beta,m\sim O(1)$. In that case, the
Kramers ``collision'' term
[r.h.s of Eq. (\ref{oake1})] is a small perturbation to the Vlasov equation
[l.h.s of Eq. (\ref{oake1})]. We assume that,
after a process of collisionless violent relaxation \cite{lb}, the system has
reached a QSS (i.e. a stable steady state of the
Vlasov equation) in which the DF is a function of the energy alone: $f\simeq
f(\epsilon)$ with $\epsilon=v^2/2+\Phi({\bf r})$. This is a
particular case of the Jeans theorem \cite{jeans}. On a secular timescale, the
DF
slowly changes with time because of the Kramers ``collision'' term taking into
account the effect of the stochastic force and friction force
in the Langevin equations (\ref{brow1}).\footnote{In the
case of Hamiltonian systems with long-range interactions for which $\xi=D=0$,
the DF slowly changes with time because of finite $N$ effects
representing ``collisions'' between the particles. These encounters are
usually modelled by the Landau operator which is of order
$1/N\ll 1$
while for Brownian particles with long-range interactions
the Kramers operator (obtained for $N\rightarrow +\infty$) is of order
$\xi t_{D}\ll 1$.} Since
$\xi\rightarrow 0$, this change is slow so the perturbations cause only a
small variation of the energy. We
shall therefore consider that the DF $f({\bf r},{\bf v},t)\simeq
f(\epsilon,t)$ remains a function of the energy alone that slowly
evolves in time. We want
to derive the kinetic equation satisfied by $f(\epsilon,t)$ by averaging the
Kramers equation over the orbits of the particles. This will lead to the
orbit-averaged Kramers equation for Brownian systems with long-range
interactions which is the counterpart of the orbit-averaged Landau equation for
Hamiltonian systems with long-range interactions.\footnote{For self-gravitating
systems, this procedure leads to the orbit-averaged Fokker-Planck
equation (\ref{oak14})-(\ref{oak16}) which was originally derived by Kuzmin
\cite{kuzmin} and H\'enon \cite{henonbinary}.}

To that purpose, we introduce the phase space hypersurface with
energy less than $\epsilon$:
\begin{eqnarray}
\label{oake2}
J(\epsilon)=\int_{v^2/2+\Phi\le
\epsilon} d{\bf r}d{\bf v}.
\end{eqnarray}
Using the fact that, for given ${\bf r}$, the
integral over ${\bf v}$ is just the volume of a sphere of ``radius''
$v=\left\lbrack 2(\epsilon-\Phi({\bf r},t))\right\rbrack^{1/2}$ in $d$
dimensions, we get
\begin{eqnarray}
\label{oake3}
J(\epsilon)=\frac{S_d}{d}\int
\left\lbrack 2(\epsilon-\Phi)\right\rbrack^{d/2}\, d{\bf r}=\frac{S_d}{d}\int
v^d\, d{\bf r}.
\end{eqnarray}
In $d=1$, this quantity represents the action $J(\epsilon)=\oint
v\, d{x}$ of a particle with energy $\epsilon$.

The density of state is
\begin{eqnarray}
\label{oake4}
g(\epsilon)=\frac{dJ}{d\epsilon}.
\end{eqnarray}
The quantity $g(\epsilon)d\epsilon$ represents the phase space hypersurface
 with energy between $\epsilon$ and $\epsilon+d\epsilon$. Taking the derivative
of Eq. (\ref{oake3}) with respect to $\epsilon$, we get
\begin{eqnarray}
\label{oake5}
g(\epsilon)= S_d\int \left\lbrack 2(\epsilon-\Phi)\right\rbrack^{d/2-1}\, d{\bf
r}=S_d\int  v^{d-2}\, d{\bf r}.
\end{eqnarray}
In $d=1$, the density of state $g(\epsilon)=\oint
d{x}/v$ is equal to the inverse of
the
pulsation $\Omega=d\epsilon/dJ$ of the orbit of energy $\epsilon$. We have
$g(\epsilon)={dJ}/{d\epsilon}={1}/{\Omega(\epsilon)}$.

The density of particles on the hypersurface between $\epsilon$ and
$\epsilon+d\epsilon$ is uniform since their DF depends  only on
the
energy. We shall therefore average the kinetic equation (\ref{oake1}) on each
``ring'' of isoenergy $\epsilon$, using 
\begin{eqnarray}
\label{oake6} 
\langle X\rangle=\frac{\int X v^{d-2}\, d{\bf
r}}{\int v^{d-2} \, d{\bf r}}
\end{eqnarray}
for any function $X(\epsilon,v,t)$. Here, for brevity, we shall assume that the
temporal variations of $\Phi({\bf
r},t)$, $J(\epsilon,t)$ and $g(\epsilon,t)$ can be
neglected.\footnote{See \cite{virial2} for a more precise
treatment. This changes $g(\epsilon)\frac{\partial f}{\partial t}$ by
$\frac{\partial J}{\partial \epsilon}\frac{\partial f}{\partial
t}-\frac{\partial J}{\partial t}\frac{\partial f}{\partial \epsilon}$ in Eq.
(\ref{oake8}).} Using $f=f(\epsilon,t)$ and $\partial\epsilon/\partial v=v$, we
obtain
\begin{eqnarray}
\label{oake7} 
\frac{\partial f}{\partial t}&=&\left\langle  \frac{\partial}{\partial
{\bf
v}}\cdot\left\lbrack D
\left (\frac{\partial f}{\partial {\bf v}}+\beta m f {\bf v}\right
)\right\rbrack \right\rangle=\left\langle \frac{1}{v^{d-1}}
\frac{\partial}{\partial
{v}}\left\lbrack v^{d-1} D
\left (\frac{\partial f}{\partial {v}}+\beta m f {v}\right
)\right\rbrack \right\rangle\nonumber\\
&=&\left\langle \frac{1}{v^{d-2}}
\frac{\partial}{\partial\epsilon}\left\lbrack v^{d} D
\left (\frac{\partial f}{\partial \epsilon}+\beta m f \right
)\right\rbrack \right\rangle=\frac{1}{g(\epsilon)}S_d \int
\frac{\partial}{\partial\epsilon} \left\lbrack v^{d} D
\left (\frac{\partial f}{\partial \epsilon}+\beta m f \right
)\right\rbrack\, d{\bf r}.
\end{eqnarray}
Then, using Eq. (\ref{oake3}), we get
\begin{eqnarray}
\label{oake8} 
\frac{\partial f}{\partial
t}=\frac{1}{g(\epsilon)}\frac{\partial}{\partial\epsilon}
\left\lbrack d D J
\left (\frac{\partial f}{\partial \epsilon}+\beta m f \right
)\right\rbrack.
\end{eqnarray}
This is the orbit-averaged Kramers equation. It was established under this
form in Sec. II.F of \cite{virial2} in the case of self-gravitating Brownian
particles. In $d=1$  it can be rewritten in terms of the action as
\begin{eqnarray}
\label{oake9} 
\frac{\partial f}{\partial t}=\frac{\partial}{\partial
J}\left\lbrack D J
\left (\frac{\partial f}{\partial \epsilon}+\beta m f
\right ) \right\rbrack
\end{eqnarray}
or, equivalently, as
\begin{eqnarray}
\label{oake10} 
\frac{\partial f}{\partial t}=\frac{\partial}{\partial
J}\left\lbrack D \frac{J}{\Omega}
\left (\frac{\partial f}{\partial J}+\beta m f\Omega
\right ) \right\rbrack.
\end{eqnarray}
The diffusion coefficient in action space is
$D_K(J)=DJ/\Omega(J)$.  This returns the
results of  Kramers \cite{kramers}.
Introducing the free energy
$F=\int f
\epsilon\, dJ+k_B T\int \frac{f}{m}\ln \frac{f}{m}\, dJ$, we get $\dot
F=-\int D_K\frac{k_B T}{mf}(\frac{\partial
f}{\partial J}+\beta m f \Omega)^2\, dJ\le 0$. The free energy is monotonically
decreasing ($H$-theorem). At equilibrium, we obtain the Boltzmann distribution 
$f_{\rm eq}=Ae^{-\beta m\epsilon}$.

Following \cite{lifetime}, we can write the orbit-averaged Kramers equation  in
a different form that makes a clear
connection with thermodynamics. We introduce the quantity 
\begin{eqnarray}
\label{oake11} 
N(\epsilon,t)=g(\epsilon,t)\frac{f}{m}(\epsilon,t),
\end{eqnarray}
so that $N(\epsilon,t) d\epsilon$ gives the number of particles of energy
between $\epsilon$ and
$\epsilon+d\epsilon$. Neglecting as before the temporal variations of
$g(\epsilon,t)$ and using Eq. (\ref{oake8}), we find that the Fokker-Planck
equation satisfied by
$N(\epsilon,t)$ reads 
\begin{eqnarray}
\label{oake12} 
\frac{\partial N}{\partial t}=\frac{\partial}{\partial
\epsilon}\left\lbrack D(\epsilon)
\left (\frac{\partial N}{\partial \epsilon}+\beta N\frac{\partial
F}{\partial\epsilon}
\right ) \right\rbrack,
\end{eqnarray}
where
\begin{eqnarray}
\label{oake13} 
D(\epsilon)=dD\frac{J(\epsilon)}{g(\epsilon)}
\end{eqnarray}
is the diffusion coefficient in energy and $F(\epsilon)=m\epsilon-T S(\epsilon)$
is the
free energy defined in terms of the microcanonical entropy $S(\epsilon)=k_B \ln
g(\epsilon)$. We note that $P(\epsilon,t)=N(\epsilon,t)/N$ is
the probability density that a particle has the energy $\epsilon$ at time $t$.
Eq. (\ref{oake12}) is the correct Fokker-Planck equation for the energy
distribution. The friction force can be written as $F_{\rm
friction}=D\beta\frac{\partial F}{\partial \epsilon}=Dm(\beta-\beta_{\rm MCE})$
where $k_B\beta_{\rm MCE}(\epsilon)=\partial S/\partial (m\epsilon)$ is
the inverse microcanonical  temperature. This Fokker-Planck equation was first
established in
\cite{lifetime} at a general level. Introducing the
free energy $F=\int N
F(\epsilon)\, d\epsilon+k_B T\int N\ln N\, d\epsilon$, we get $\dot
F=-\int D\frac{k_B T}{N}(\frac{\partial N}{\partial\epsilon}+\beta
N\frac{\partial F}{\partial\epsilon})^2\, d\epsilon\le 0$. The free energy is
monotonically decreasing ($H$-theorem). At equilibrium, we obtain the Boltzmann
distribution $N_{\rm eq}=Ae^{-\beta F(\epsilon)}=Ag(\epsilon)e^{-\beta
m\epsilon}$.

{\it Remark:} We can compare these results with those obtained in Sec. 7 of
Ref. \cite{kin2007} in the context of the kinetic theory of 1D inhomogeneous
systems with long-range interactions in the thermal bath approximation.

\section{Bare and dressed potentials of interaction}
\label{sec_bdpi}

\subsection{Bare potential of interaction}
\label{sec_bdpia}

Following Refs. \cite{fpc,hfcp,fbc},
we show that the Fourier transform of the
bare potential of interaction $A_{{\bf k}{\bf
k}'}({\bf J},{\bf J}')$ in
angle-action
variables can be
expressed in terms of the elements of the biorthogonal basis.

The Fourier transform in
angles of the bare potential of interaction is 
\begin{eqnarray}
A_{{\bf k}{\bf k}'}({\bf J},{\bf J}')=\frac{1}{(2\pi)^{2d}}\int d{\bf w}d{\bf
w}' \, u[{\bf r}({\bf w},{\bf J})-{\bf r}({\bf w}',{\bf J}')]e^{-i({\bf k}\cdot
{\bf w}-{\bf k}'\cdot {\bf w}')}.
\label{pi31}
\end{eqnarray}
Inversely, we have
\begin{eqnarray}
u[{\bf r}({\bf w},{\bf J})-{\bf r}({\bf w}',{\bf J}')]=\sum_{{\bf k}{\bf k}'}
e^{i({\bf k}\cdot {\bf w}-{\bf k}'\cdot {\bf w}')}
A_{{\bf k}{\bf k}'}({\bf J},{\bf J}').
\label{pi32}
\end{eqnarray}

The potential of interaction $u({\bf r}-{\bf r}')$, which is a function of
${\bf r}$ for fixed ${\bf r}'$, can be expanded on the
biorthogonal basis according to
\begin{eqnarray}
u({\bf r}-{\bf r}')=\sum_{\alpha}U_\alpha({\bf r}')\Phi_{\alpha}({\bf r}).
\label{pi33}
\end{eqnarray}
Using Eq. (\ref{i63}), the coefficients $U_\alpha({\bf r}')$ are given by
\begin{eqnarray}
U_\alpha({\bf r}')=-\int u({\bf r}-{\bf r}')\rho_\alpha({\bf r})^*\, d{\bf
r}=-\Phi_{\alpha}({\bf r}')^*.
\label{pi34}
\end{eqnarray}
Therefore,
\begin{eqnarray}
u({\bf r}-{\bf r}')=-\sum_{\alpha}\Phi_{\alpha}({\bf r})\Phi_\alpha({\bf
r}')^*.
\label{pi34h}
\end{eqnarray}
Substituting this expression into Eq. (\ref{pi31}), we obtain
\begin{eqnarray}
A_{{\bf k}{\bf k}'}({\bf J},{\bf J}')=-\frac{1}{(2\pi)^{2d}}\sum_{\alpha}\int
d{\bf w}d{\bf w}' \, \Phi_{\alpha}({\bf r}({\bf w},{\bf J}))\Phi_\alpha({\bf
r}({\bf w}',{\bf J}'))^*e^{-i({\bf k}\cdot {\bf w}-{\bf k}'\cdot {\bf w}')}.
\label{pi35}
\end{eqnarray}
Introducing the Fourier transform in angles of the elements of the
biorthogonal
basis
\begin{eqnarray}
{\hat \Phi}_{\alpha}({\bf k},{\bf J})=\frac{1}{(2\pi)^d}\int d{\bf w}\,e^{-i{\bf
k}\cdot {\bf w}} \Phi_{\alpha}({\bf r}({\bf w},{\bf J})),
\label{pi36}
\end{eqnarray}
we find that
\begin{eqnarray}
A_{{\bf k}{\bf k}'}({\bf J},{\bf J}')=-\sum_{\alpha} {\hat \Phi}_{\alpha}({\bf
k},{\bf J}){\hat \Phi}_\alpha({\bf k}',{\bf J}')^*.
\label{pi37}
\end{eqnarray}
When ${\bf J}'={\bf J}$ and ${\bf k}'={\bf k}$, we get
\begin{eqnarray}
A_{{\bf k}{\bf k}}({\bf J},{\bf
J})=-\sum_{\alpha}{\hat\Phi}_{\alpha}({\bf k},{\bf
J}){\hat\Phi}^*_{\alpha}({\bf k},{\bf
J})=-\sum_{\alpha} |{\hat\Phi}_{\alpha}({\bf k},{\bf
J})|^2.
\label{fdi13}
\end{eqnarray}
We note that $A_{{\bf k}{\bf k}}({\bf J},{\bf
J})$ is a real negative quantity.  Finally, we note the identities
\begin{eqnarray}
A_{-{\bf k}-{\bf k}'}({\bf J},{\bf J}')=A_{{\bf k}{\bf k}'}({\bf
J},{\bf
J}')^*
\label{dn14bare}
\end{eqnarray}
and
\begin{eqnarray}
A_{{\bf k}'{\bf k}}({\bf J}',{\bf J})=-\sum_{\alpha} {\hat \Phi}_{\alpha}({\bf
k}',{\bf J}'){\hat \Phi}_\alpha({\bf k},{\bf J})^*=A_{{\bf k}{\bf k}'}({\bf
J},{\bf J}')^*.
\label{pi37tax}
\end{eqnarray}

\subsection{Dressed potential of interaction}

The Fourier transform of the
dressed potential of interaction in angle-action variables is given
by\footnote{It is written as $-1/D_{{\bf k},{\bf
k}'}({\bf J},{\bf J}',\omega)$ in Refs. \cite{heyvaerts,physicaA}.}
\begin{eqnarray}
A^d_{{\bf k},{\bf k}'}({\bf J},{\bf
J'},\omega)=-\sum_{\alpha\alpha'}{\hat\Phi}_{\alpha}({\bf k},{\bf
J})(\epsilon^{-1})_{\alpha\alpha'}(\omega){\hat\Phi}^*_{\alpha'}({\bf k}',{\bf
J}').
\label{can1b}
\end{eqnarray}
This is the proper generalization of the expression of the dressed potential of
interaction  ${\hat u}_{\rm dressed}={\hat u}(k)/|\epsilon({\bf k},{\bf
k}\cdot {\bf v})|$ in the
homogeneous case \cite{epjp,epjp2} (see Appendix \ref{sec_bona}). Using the
result of footnote 16, we
have the property
\begin{eqnarray}
A^d_{-{\bf k}-{\bf k}'}({\bf J},{\bf J}',-\omega)=A^d_{{\bf k}{\bf k}'}({\bf
J},{\bf
J}',\omega)^*.
\label{dn14}
\end{eqnarray}
On the other hand, taking
${\bf J}'={\bf J}$ and ${\bf k}'={\bf k}$ in Eq. (\ref{can1b}), we obtain
\begin{eqnarray}
A^d_{{\bf k},{\bf k}}({\bf J},{\bf
J},\omega)=-\sum_{\alpha\alpha'}{\hat\Phi}_{\alpha}({\bf k},{\bf
J})(\epsilon^{-1})_{\alpha\alpha'}(\omega){\hat\Phi}^*_{\alpha'}({\bf k},{\bf
J}).
\end{eqnarray}
If we
neglect collective effects by taking $\epsilon=1$, we recover the 
Fourier transform of the bare potential of interaction in angle-action
variables from Eq. (\ref{pi37}). Therefore, neglecting collective effects
amounts to
replacing the dressed potential (\ref{can1b}) by the bare potential
(\ref{pi37}).

\section{Relation between the potential and the DF in
angle-action variables}
\label{sec_bona}

\subsection{Without collective effects}
\label{sec_ts}

The
potential
is related to the density
by a product of convolution [see Eq. (\ref{n2bzero})]
\begin{eqnarray}
\Phi({\bf r},t)&=&\int u({\bf r}-{\bf r}')\rho({\bf r}',t)\, d{\bf
r}'\nonumber\\
&=&\int u({\bf r}-{\bf r}') f({\bf r}',{\bf v}',t)\, d{\bf r}'d{\bf v}'.
\label{ciw1}
\end{eqnarray}
Introducing angle-action variables and using the fact that the
transformation $({\bf r},{\bf
v})\rightarrow ({\bf w},{\bf J})$ is canonical we get
\begin{eqnarray}
\Phi({\bf w},{\bf J},t)=\int u[{\bf r}({\bf w},{\bf J})-{\bf r}({\bf
w}',{\bf J}')]f({\bf w}',{\bf J}',t)\, d{\bf w}'d{\bf J}'.
\label{ciw2}
\end{eqnarray}
Its Fourier transform is therefore
\begin{eqnarray}
{\hat \Phi}({\bf k},{\bf J},\omega)&=&\int \frac{d{\bf w}}{(2\pi)^d}
\int_{-\infty}^{+\infty} dt \int d{\bf
w}'\int d{\bf J}'\, u[{\bf r}({\bf w},{\bf J})-{\bf r}({\bf w}',{\bf
J}')]{f}({\bf w}',{\bf J}',t)e^{-i({\bf k}\cdot {\bf
w}-\omega t)}\nonumber\\
&=&\sum_{{\bf k}'}\int\frac{d\omega'}{2\pi}\int \frac{d{\bf w}}{(2\pi)^d}
\int_{-\infty}^{+\infty} dt \int d{\bf
w}'\int d{\bf J}'\, u[{\bf r}({\bf w},{\bf J})-{\bf r}({\bf w}',{\bf
J}')]e^{-i({\bf k}\cdot {\bf
w}-\omega t)}{\hat f}({\bf k}',{\bf J}',\omega')e^{i({\bf k}'\cdot {\bf
w}'-\omega' t)}\nonumber\\
&=&\sum_{{\bf k}'}\int \frac{d{\bf w}}{(2\pi)^d} \int d{\bf w}'\int d{\bf J}'\,
u[{\bf
r}({\bf w},{\bf J})-{\bf r}({\bf w}',{\bf J}')]
 {\hat f}({\bf k}',{\bf J}',\omega)e^{-i({\bf k}\cdot {\bf w}-{\bf k}'\cdot
{\bf w}')},
\label{bof3}
\end{eqnarray}
where we have used Eq. (\ref{i59}) to get the second line and Eq. (\ref{deltac})
to get the third line. Introducing the bare
potential of interaction from Eq. (\ref{pi31}), we obtain 
\begin{eqnarray}
{\hat \Phi}({\bf k},{\bf J},\omega)=(2\pi)^d\sum_{{\bf k}'} \int d{\bf J}'\, 
A_{{\bf k}{\bf k}'}({\bf J},{\bf
J}')
 {\hat f}({\bf k}',{\bf J}',\omega).
\label{bof3b}
\end{eqnarray}
This relation is general. It relates the Fourier transform in angle-action
variables of the potential $\Phi$ to the corresponding  DF
$f$. For example, it determines 
$\delta{\hat \Phi}$ as a
function of $\delta {\hat f}$ (see Eq. (A5) in
\cite{physicaA}):
\begin{eqnarray}
\delta{\hat \Phi}({\bf k},{\bf J},\omega)=(2\pi)^d\sum_{{\bf k}'} \int d{\bf
J}'\,  A_{{\bf k}{\bf k}'}({\bf J},{\bf J}')
 \delta{\hat f}({\bf k}',{\bf J}',\omega),
\label{edi}
\end{eqnarray}
and ${\hat \Phi}_e$ as a
function of ${\hat f}_e$:
\begin{eqnarray}
{\hat \Phi_e}({\bf k},{\bf J},\omega)=(2\pi)^d\sum_{{\bf k}'} \int d{\bf
J}'\,  A_{{\bf k}{\bf k}'}({\bf J},{\bf J}')
 {\hat f_e}({\bf k}',{\bf J}',\omega).
\label{gh4h}
\end{eqnarray}
If the potential
is produced by a
collection of $N$ particles, using Eq. (\ref{ci3}), the foregoing
equation becomes 
\begin{eqnarray}
{\hat \Phi}_e({\bf k},{\bf J},\omega)=2\pi \sum_i m_i \sum_{{\bf k}'} A_{{\bf
k}{\bf k}'}({\bf J},{\bf J}_i) e^{-i{\bf k}'\cdot {\bf w}_i}\delta(\omega-{\bf
k}'\cdot{\bf \Omega}_i).
\label{ciw3}
\end{eqnarray}

\subsection{With collective effects by using a biorthogonal basis}
\label{sec_dngh}

According to Eq.
(\ref{i62}), the total fluctuating potential in angle-action variables
experienced by a test particle can be
written as 
\begin{eqnarray}
\delta\Phi_{\rm tot}({\bf r}({\bf w},{\bf J}),t)= \sum_{\alpha}A_{\alpha}^{\rm
tot}(t)\Phi_{\alpha}({\bf r}({\bf w},{\bf J})).
\label{dn1}
\end{eqnarray}
Its Fourier transform in angle and time is
\begin{eqnarray}
\delta{\hat \Phi}_{\rm tot}({\bf k},{\bf J},\omega)=
\sum_{\alpha}{\hat A}_{\alpha}^{\rm tot}(\omega){\hat \Phi}_{\alpha}({\bf
k},{\bf
J}).
\label{dn2}
\end{eqnarray}
Using Eqs. (\ref{i72}) and (\ref{ci1b}), we  successively obtain
\begin{eqnarray}
\delta{\hat \Phi}_{\rm tot}({\bf k},{\bf
J},\omega)=\sum_{\alpha\alpha'}(\epsilon^{-1})_{\alpha\alpha'}(\omega){\hat
A}^e_{\alpha'}(\omega){\hat \Phi}_{\alpha}({\bf k},{\bf J})
\label{ci1bnj}
\end{eqnarray}
and
\begin{eqnarray}
\delta{\hat \Phi}_{\rm tot}({\bf k},{\bf
J},\omega)=-\sum_{\alpha\alpha'}\int
d{\bf w}'d{\bf J}'\, (\epsilon^{-1})_{\alpha\alpha'}(\omega) {\hat
\Phi}_{\alpha}({\bf k},{\bf J}){\hat f}_e({\bf
w}',{\bf J}',\omega) {\Phi}_{\alpha'}^*({\bf w}',{\bf J}').
\label{dn6re}
\end{eqnarray}
Introducing the Fourier transform in angles of the DF, we obtain
\begin{eqnarray}
\delta{\hat \Phi}_{\rm tot}({\bf k},{\bf
J},\omega)&=&-\sum_{\alpha\alpha'}\sum_{{\bf k}'}\int
d{\bf w}'d{\bf J}'\, e^{i{\bf k}'\cdot
{\bf w}'}(\epsilon^{-1})_{\alpha\alpha'}(\omega) {\hat
\Phi}_{\alpha}({\bf k},{\bf J}){\hat f}_e({\bf
k}',{\bf J}',\omega) {\Phi}_{\alpha'}^*({\bf w}',{\bf J}')\nonumber\\
&=&-(2\pi)^d\sum_{\alpha\alpha'}\sum_{{\bf k}'}\int
d{\bf J}'\, (\epsilon^{-1})_{\alpha\alpha'}(\omega)
{\hat
\Phi}_{\alpha}({\bf k},{\bf J}){\hat f}_e({\bf
k}',{\bf J}',\omega) {\hat \Phi}_{\alpha'}^*({\bf k}',{\bf J}').
\label{dn6a}
\end{eqnarray}
Introducing the dressed potential of interaction from Eq. (\ref{can1b}),
we can rewrite the foregoing
equation as 
\begin{eqnarray}
\delta{\hat \Phi}_{\rm tot}({\bf k},{\bf
J},\omega)=(2\pi)^d\sum_{{\bf k}'}\int
d{\bf J}'\, A^d_{{\bf k}{\bf k}'}({\bf J},{\bf
J}',\omega) {\hat f}_e({\bf
k}',{\bf J}',\omega). 
\label{dn6b}
\end{eqnarray}
This equation generalizes Eq. (\ref{bof3b}) by taking into account collective
effects. It justifies the definition of the dressed potential of
interaction from Eq. (\ref{can1b}). If the potential is produced by a collection
of $N$ particles, using  Eq. (\ref{ci3}), the foregoing equation becomes
\begin{eqnarray}
\delta{\hat \Phi}_{\rm tot}({\bf k},{\bf J},\omega)
=2\pi \sum_i\sum_{{\bf k}'}m_i A^d_{{\bf k}{\bf k}'}({\bf J},{\bf
J}_i,\omega) e^{-i{\bf k}'\cdot {\bf w}_i}\delta(\omega-{\bf k}'\cdot
{\bf\Omega}_i).
\label{dn7}
\end{eqnarray}
This equation generalizes Eq. (\ref{ciw3}) by taking into account
collective
effects.

\subsection{With collective effects without using the biorthogonal basis}
\label{sec_dnwout}

It is possible to derive Eq. (\ref{dn6b}) without using the
biorthogonal
basis. According to Eq. (\ref{i60}) we have
\begin{equation}
\delta\hat f ({\bf k},{\bf J},\omega)=\frac{{\bf k}\cdot \frac{\partial
f}{\partial {\bf J}}}{{\bf k}\cdot {\bf
\Omega}-\omega}\delta\hat\Phi_{\rm tot}({\bf
k},{\bf J},\omega).
\label{ni60app}
\end{equation} 
On the other hand, according to Eq. (\ref{bof3b}) we have
\begin{eqnarray}
\delta{\hat \Phi}_{\rm tot}({\bf k},{\bf J},\omega)=(2\pi)^d\sum_{{\bf k}'} \int
d{\bf J}'\, 
A_{{\bf k}{\bf k}'}({\bf J},{\bf
J}')\left\lbrack {\hat f}_e({\bf k}',{\bf J}',\omega)+\delta{\hat f}({\bf
k}',{\bf J}',\omega)\right\rbrack.
\end{eqnarray}
Combining these two equations we get
\begin{eqnarray}
\delta{\hat \Phi}_{\rm tot}({\bf k},{\bf J},\omega)-(2\pi)^d\sum_{{\bf k}'} \int
d{\bf J}'\, 
A_{{\bf k}{\bf k}'}({\bf J},{\bf
J}')\frac{{\bf k}'\cdot \frac{\partial
f'}{\partial {\bf J}'}}{{\bf k}'\cdot {\bf
\Omega}'-\omega}\delta\hat\Phi_{\rm tot}({\bf
k}',{\bf J}',\omega)=(2\pi)^d\sum_{{\bf k}'} \int
d{\bf J}'\, 
A_{{\bf k}{\bf k}'}({\bf J},{\bf
J}'){\hat f}_e({\bf k}',{\bf J}',\omega).\qquad
\label{fred}
\end{eqnarray}
This is a Fredholm integral equation (see Appendix A of
\cite{physicaA}) which relates the Fourier transform of the fluctuations of
the
potential to the Fourier transform of the fluctuations of the
external DF. The formal solution of this equation is
\begin{eqnarray}
\delta{\hat \Phi}_{\rm tot}({\bf k},{\bf J},\omega)=(2\pi)^d\sum_{{\bf k}'} \int
d{\bf J}'\, 
A^d_{{\bf k}{\bf k}'}({\bf J},{\bf
J}',\omega){\hat f}_e({\bf k}',{\bf J}',\omega),
\label{fem}
\end{eqnarray}
where $A^d_{{\bf k}{\bf k}'}({\bf J},{\bf J}',\omega)$ is the
dressed potential
of interaction defined by (see Appendix A of
\cite{physicaA})
\begin{eqnarray}
\label{nmat3}
A^d_{{\bf k}{\bf k}'}({\bf J},{\bf J}',\omega)-(2\pi)^d\sum_{{\bf
k}''} \int d{\bf
J}''\, 
A_{{\bf k}{\bf k}''}({\bf J},{\bf
J}'') \frac{{\bf k}''\cdot \frac{\partial
f''}{\partial
{\bf J}''}}{{\bf
k}''\cdot {\bf
\Omega}''-\omega}A^d_{{\bf k}''{\bf k}'}({\bf J}'',{\bf J}',\omega)
=A_{{\bf k}{\bf k}'}({\bf J},{\bf
J}'). 
\end{eqnarray}
It can be interpreted  as a Green
function.\footnote{This is the
counterpart of the Green function $G(k,y,y',\sigma)$ in the case of 2D point
vortices \cite{Kvortex2023}.} Indeed, from Eq. (\ref{nmat3}), we get
\begin{eqnarray}
\label{nmat4}
(2\pi)^d\sum_{{\bf k}'}\int d{\bf
J}' A^d_{{\bf k}{\bf k}'}({\bf J},{\bf J}',\omega) {\hat f}_e({\bf k}',{\bf
J}',\omega)\nonumber\\
=(2\pi)^d\sum_{{\bf k}'}\int d{\bf
J}'  {\hat f}_e({\bf k}',{\bf J}',\omega)
(2\pi)^d\sum_{{\bf
k}''} \int d{\bf
J}''\, 
A_{{\bf k}{\bf k}''}({\bf J},{\bf
J}'') \frac{{\bf k}''\cdot \frac{\partial
f''}{\partial
{\bf J}''}}{{\bf
k}''\cdot {\bf
\Omega}''-\omega}A^d_{{\bf k}''{\bf k}'}({\bf J}'',{\bf
J}',\omega)\nonumber\\
+(2\pi)^d\sum_{{\bf k}'}\int d{\bf
J}' A_{{\bf k}{\bf k}'}({\bf J},{\bf J}') {\hat f}_e({\bf k}',{\bf J}',\omega).
\end{eqnarray}
Interchanging $({\bf k}',{\bf J}')$ and $({\bf k}'',{\bf J}'')$ in the second
line we obtain
\begin{eqnarray}
\label{nmat5}
(2\pi)^d\sum_{{\bf k}'}\int d{\bf
J}' A^d_{{\bf k}{\bf k}'}({\bf J},{\bf J}',\omega){\hat f}_e({\bf k}',{\bf
J}',\omega) \nonumber\\
=(2\pi)^d\sum_{{\bf k}''}\int d{\bf
J}''  {\hat f}_e({\bf k}'',{\bf J}'',\omega)
(2\pi)^d\sum_{{\bf
k}'} \int d{\bf
J}'\, 
A_{{\bf k}{\bf k}'}({\bf J},{\bf
J}') \frac{{\bf k}'\cdot \frac{\partial
f'}{\partial
{\bf J}'}}{{\bf
k}'\cdot {\bf
\Omega}'-\omega}A^d_{{\bf k}'{\bf k}''}({\bf J}',{\bf
J}'',\omega)\nonumber\\
+(2\pi)^d\sum_{{\bf k}'}\int d{\bf
J}' A_{{\bf k}{\bf k}'}({\bf J},{\bf J}') {\hat f}_e({\bf k}',{\bf J}',\omega),
\end{eqnarray}
which is identical to Eq. (\ref{fred}) if we use Eq. (\ref{fem}). Eq.
(\ref{nmat3}) can be used as a definition of $A^d_{{\bf k}{\bf k}'}({\bf J},{\bf
J}',\omega)$ which does not require to introduce a biorthogonal basis (see
Appendix A of \cite{physicaA}). The two manners to define $A^d_{{\bf k}{\bf
k}'}({\bf J},{\bf
J}',\omega)$, either by Eq. (\ref{can1b}) or by Eq. (\ref{nmat3}),  are
equivalent.

\section{Alternative derivations of the diffusion tensor}

In this Appendix, we provide alternative derivations of the diffusion tensor to
the one given in Sec. \ref{sec_diffco}.

\subsection{General expression of the diffusion tensor using Fourier transforms
in angles and time}
\label{sec_eug}

The change in action of a test particle due to the total fluctuating potential
is given by [see Eq.
(\ref{fp1b})] 
\begin{eqnarray}
\frac{d{\bf J}}{dt}=-\frac{\partial \delta\Phi_{\rm tot}}{\partial {\bf w}}({\bf
r}({\bf w},{\bf J}),t).
\label{hama1}
\end{eqnarray}
Integrating this equation between $0$ and $t$, we obtain
\begin{eqnarray}
\Delta{\bf J}&=&-\int_0^t \frac{\partial \delta\Phi_{\rm tot}}{\partial {\bf w}}({\bf w}(t'),{\bf J}(t'),t')\, dt'\nonumber\\
&=&-\int_0^t \frac{\partial \delta\Phi_{\rm tot}}{\partial {\bf w}}({\bf
w}+{\bf\Omega}t',{\bf J},t')\, dt',
\label{hama2}
\end{eqnarray}
where we have used the unperturbed equations of motion (\ref{unp}) in the second
equation (this accounts for the fact that the test particle follows the mean
field
trajectory at leading order).
Decomposing the potential in Fourier modes, we get
\begin{eqnarray}
\Delta{\bf J}&=&-\int_0^t dt'\, \frac{\partial}{\partial {\bf w}}\sum_{\bf k}\int \frac{d\omega}{2\pi}e^{i{\bf k}\cdot ({\bf w}+{\bf\Omega}t')}e^{-i\omega t'}\delta{\hat \Phi}_{\rm tot}({\bf k},{\bf J},\omega)\nonumber\\
&=&-\sum_{\bf k}\int \frac{d\omega}{2\pi} i {\bf k}  e^{i{\bf k}\cdot {\bf w}}\delta{\hat \Phi}_{\rm tot}({\bf k},{\bf J},\omega)\int_0^t e^{i({\bf k}\cdot {\bf\Omega}-\omega)t'}\, dt'\nonumber\\
&=&-\sum_{\bf k}\int \frac{d\omega}{2\pi} i {\bf k}  e^{i{\bf k}\cdot {\bf
w}}\delta{\hat \Phi}_{\rm tot}({\bf k},{\bf J},\omega)\frac{e^{i({\bf k}\cdot
{\bf\Omega}-\omega)t}-1}{i({\bf k}\cdot {\bf\Omega}-\omega)}.
\label{hama3}
\end{eqnarray}
The diffusion tensor is defined by [see Eq. (\ref{fp6})]
\begin{eqnarray}
D_{ij}=\lim_{t\rightarrow +\infty}\frac{\langle \Delta J_i\Delta
J_j\rangle}{2t}.
\label{hama5}
\end{eqnarray}
Substituting Eq. (\ref{hama3}) into Eq. (\ref{hama5}) and averaging over ${\bf
w}$, we obtain
\begin{equation}
D_{ij}=\lim_{t\rightarrow +\infty}\frac{1}{2t}\int \frac{d{\bf
w}}{(2\pi)^d}\sum_{{\bf k}{\bf k}'}\int \frac{d\omega}{2\pi} \int
\frac{d\omega'}{2\pi}  k_i k'_j  e^{i({\bf k}-{\bf k}')\cdot {\bf w}}\langle
\delta{\hat \Phi}_{\rm tot}({\bf k},{\bf J},\omega)\delta{\hat \Phi}_{\rm
tot}({\bf k}',{\bf J},\omega')^*\rangle \frac{e^{i({\bf k}\cdot
{\bf\Omega}-\omega)t}-1}{i({\bf k}\cdot {\bf\Omega}-\omega)}
\frac{e^{-i({\bf k}'\cdot {\bf\Omega}-\omega')t}-1}{-i({\bf k}'\cdot
{\bf\Omega}-\omega')}.
\label{hama6}
\end{equation}
Performing the integral over ${\bf w}$ and the sum over ${\bf k}'$, we get
\begin{equation}
D_{ij}=\lim_{t\rightarrow +\infty}\frac{1}{2t}\sum_{{\bf k}}\int
\frac{d\omega}{2\pi} 
\int \frac{d\omega'}{2\pi}  k_i k_j  \langle \delta{\hat \Phi}_{\rm tot}({\bf
k},{\bf J},\omega)\delta{\hat \Phi}_{\rm tot}({\bf k},{\bf J},\omega')^*\rangle
\frac{e^{i({\bf k}\cdot {\bf\Omega}-\omega)t}-1}{i({\bf k}\cdot
{\bf\Omega}-\omega)}
\frac{e^{-i({\bf k}\cdot {\bf\Omega}-\omega')t}-1}{-i({\bf k}\cdot
{\bf\Omega}-\omega')}.
\label{hama7}
\end{equation}
Introducing the power spectrum from Eq. (\ref{psaab}), the foregoing equations
can be
rewritten as 
\begin{equation}
D_{ij}=\lim_{t\rightarrow +\infty}\frac{1}{2t}\sum_{{\bf k}}\int
\frac{d\omega}{2\pi}  k_i k_j  P({\bf k},{\bf J},\omega) \frac{|e^{i({\bf
k}\cdot {\bf\Omega}-\omega)t}-1|^2}{({\bf k}\cdot {\bf\Omega}-\omega)^2}.
\label{hama8}
\end{equation}
Using the identity 
\begin{equation}
\lim_{t\rightarrow +\infty}\frac{|e^{ixt}-1|^2}{x^2t}=2\pi\delta(x),
\label{dn16}
\end{equation}
we find that
\begin{equation}
D_{ij}=\pi\sum_{{\bf k}}\int \frac{d\omega}{2\pi}  k_i k_j  P({\bf k},{\bf
J},\omega)\delta({\bf k}\cdot {\bf\Omega}-\omega)
\label{hama9}
\end{equation}
Integrating over the $\delta$-function (resonance), we get
\begin{equation}
D_{ij}=\frac{1}{2}\sum_{{\bf k}}  k_i k_j  P({\bf k},{\bf J},{\bf k}\cdot
{\bf\Omega}),
\label{hama10}
\end{equation}
which returns Eq. (\ref{gei3}). Then, using Eq. (\ref{i83}), we
recover Eq. (\ref{i88b}).

{\it Remark:} If we do not take the limit $t\rightarrow +\infty$ in Eq.
(\ref{hama8}), we
obtain a
time-dependent diffusion coefficient of the form
\begin{equation}
D_{ij}(t)=\pi\sum_{{\bf k}}\int
\frac{d\omega}{2\pi}  k_i k_j  P({\bf k},{\bf J},\omega) \Delta({\bf k}\cdot
{\bf\Omega}-\omega,t)
\label{hama8del}
\end{equation}
with the regularized function
\begin{equation}
\Delta(x,t)=\frac{1}{2\pi t}\frac{|e^{ixt}-1|^2}{x^2}=\frac{1-\cos(xt)}{\pi t
x^2}.
\label{hama8def}
\end{equation}
When $t\rightarrow +\infty$, we can make the replacement
$\Delta(x,t)\rightarrow \delta(x)$ corresponding to the diffusive regime. When
$t\rightarrow 0$, we have $\Delta(x,t)\sim t/2\pi$ corresponding to the
ballistic regime.

\subsection{General expression of the diffusion tensor using
a Fourier transform in angles}
\label{sec_ele}

We can make the calculations of the previous Appendix in a slightly different
manner. In Eq. (\ref{hama2}) we decompose the total fluctuating potential in
Fourier
modes in angle but not in time. In that case, we get
\begin{eqnarray}
\Delta{\bf J}&=&-\int_0^t dt'\, \frac{\partial}{\partial {\bf w}}\sum_{\bf k}
e^{i{\bf k}\cdot ({\bf w}+{\bf\Omega}t')}\delta{\hat \Phi}_{\rm tot}({\bf
k},{\bf J},t')\nonumber\\
&=&-\int_0^t dt'\, \sum_{\bf k} i{\bf k} e^{i{\bf k}\cdot ({\bf
w}+{\bf\Omega}t')}\delta{\hat \Phi}_{\rm tot}({\bf k},{\bf J},t').
\label{an9}
\end{eqnarray}
Substituting Eq. (\ref{an9}) into Eq. (\ref{hama5}), we obtain
\begin{eqnarray}
D_{ij}=\lim_{t\rightarrow +\infty}  \frac{1}{2t}\int_0^t dt'\int_0^t dt'' \,
\sum_{{\bf k}{\bf k}'} k_i k'_j e^{i{\bf k}\cdot ({\bf w}+{\bf \Omega}t')}
e^{-i{\bf k}'\cdot ({\bf w}+{\bf \Omega}t'')} \left\langle
\delta{\hat \Phi}_{\rm tot}({\bf k},{\bf J},t')\delta{\hat \Phi}_{\rm tot}({\bf
k}',{\bf J},t'')^*\right\rangle. 
\label{an10}
\end{eqnarray}
Since the mean DF $f$ depends only on the action ${\bf J}$,
we can average $D_{ij}$ over the angle ${\bf w}$. This amounts to taking ${\bf
k}'={\bf k}$, yielding
\begin{eqnarray}
D_{ij}=\lim_{t\rightarrow +\infty} \frac{1}{2t}\int_0^t dt'\int_0^t dt'' \,
\sum_{{\bf k}} k_i k_j  e^{i{\bf k}\cdot {\bf\Omega}(t'-t'')} \left\langle
\delta{\hat \Phi}_{\rm tot}({\bf k},{\bf J},t')\delta{\hat \Phi}_{\rm tot}({\bf
k},{\bf J},t'')^*\right\rangle. 
\label{an11}
\end{eqnarray}
Using the relation
\begin{eqnarray}
\left\langle
\delta{\hat \Phi}_{\rm tot}({\bf k},{\bf J},t)\delta{\hat \Phi}_{\rm tot}({\bf
k},{\bf J},t')^*\right\rangle={\cal P}({\bf k},{\bf J},t-t'),
\label{an11b}
\end{eqnarray}
where ${\cal P}({\bf k},{\bf J},t)$ is the temporal inverse  Fourier transform
of $P({\bf k},{\bf J},\omega)$, we can rewrite the foregoing equation as
\begin{eqnarray}
D_{ij}
=\lim_{t\rightarrow +\infty} \frac{1}{2t}\int_0^t dt'\int_0^t dt'' \,
\sum_{{\bf k}} k_i k_j  e^{i{\bf k}\cdot {\bf\Omega}(t'-t'')} {\cal P}({\bf
k},{\bf
J},t'-t'').
\label{an12}
\end{eqnarray}
Using the identity from Eq. (\ref{diff4}), we get
\begin{eqnarray}
D_{ij}=\lim_{t\rightarrow +\infty} \frac{1}{t}\int_0^t ds\, (t-s) \sum_{{\bf
k}} k_i k_j  e^{i{\bf k}\cdot {\bf\Omega}s}  {\cal P}({\bf k},{\bf
J},s).
\label{an13i}
\end{eqnarray}
Assuming that ${\cal P}({\bf k},{\bf
J},s)$ decreases
more rapidly than $s^{-1}$, we obtain
\begin{eqnarray}
D_{ij}
=\int_0^{+\infty} ds\, \sum_{{\bf k}} k_i k_j  e^{i{\bf k}\cdot
{\bf\Omega}s} 
{\cal P}({\bf k},{\bf J},s).
\label{an13ii}
\end{eqnarray}
Making the change of
variables $s\rightarrow -s$ and ${\bf k}\rightarrow -{\bf k}$, and using the
fact that ${\cal P}(-{\bf k},{\bf J},-s)={\cal P}({\bf k},{\bf J},s)$, we see
that we can replace
$\int_{0}^{+\infty}ds$ by $(1/2)\int_{-\infty}^{+\infty}ds$. Therefore,
\begin{eqnarray}
D_{ij}
=\frac{1}{2}\int_{-\infty}^{+\infty} ds\, \sum_{{\bf k}} k_i k_j  e^{i{\bf
k}\cdot {\bf\Omega}s} 
{\cal P}({\bf k},{\bf J},s).
\label{an13iii}
\end{eqnarray}
Finally, taking the inverse Fourier transform of ${\cal P}({\bf k},{\bf J},s)$
we find that
\begin{eqnarray}
D_{ij}=\frac{1}{2} \sum_{{\bf k}} k_i k_j  {P}({\bf k},{\bf J},{\bf k}\cdot
{\bf\Omega}),
\label{an13iv}
\end{eqnarray}
which returns Eq. (\ref{gei3}). Then, using Eq. (\ref{i83}), we
recover Eq. (\ref{i88b}).

{\it Remark:} If we introduce the temporal Fourier transform of  ${\cal P}({\bf
k},{\bf J},t)$ in Eq. (\ref{an12}), we get
\begin{eqnarray}
D_{ij}=\lim_{t\rightarrow +\infty} \frac{1}{2t}\int_0^t dt'\int_0^t dt'' \,
\sum_{\bf k}\int \frac{d\omega}{2\pi}\, k_i k_j  e^{i{\bf k}\cdot {\bf
\Omega}(t'-t'')} e^{-i\omega(t'-t'')} {P}({\bf
k},{\bf J},\omega),
\label{an8}
\end{eqnarray}
which is equivalent to Eq. (\ref{diff1}) with Eq.
(\ref{diff3}). If we
integrate over $t'$ and $t''$, we recover Eq. (\ref{hama8}).

\subsection{Diffusion tensor created by $N$ particles}
\label{sec_dn}

The total fluctuating potential in angle-action
variables produced by an ensemble of $N$ particles is given by Eq.
(\ref{dn7})].
Substituting this expression into Eq. (\ref{hama3}) 
and
integrating over $\omega$, we obtain
\begin{eqnarray}
\Delta{\bf J}=-\sum_i m_i\sum_{{\bf k}{\bf k}'} i {\bf k}  e^{i{\bf k}\cdot {\bf
w}}\frac{e^{i({\bf k}\cdot {\bf\Omega}-{\bf k}'\cdot {\bf\Omega}_i)t}-1}{i({\bf
k}\cdot {\bf\Omega}-{\bf k}'\cdot {\bf\Omega}_i)}
A^d_{{\bf k}{\bf k}'}({\bf J},{\bf J}_i,{\bf k}'\cdot
{\bf\Omega}_i) e^{-i{\bf k}'\cdot {\bf w}_i}.
\label{dn8}
\end{eqnarray}
The diffusion coefficient from Eq. (\ref{hama5})
is then given by
\begin{eqnarray}
D_{ij}=-\lim_{t\rightarrow +\infty}\frac{1}{2t}\Biggl\langle
\sum_{IJ}m_I m_J\sum_{{\bf k}{\bf k}'{\bf K}{\bf K}'} k_i K_j  e^{i{\bf
k}\cdot {\bf w}}e^{i{\bf K}\cdot {\bf w}}
\frac{e^{i({\bf k}\cdot {\bf\Omega}-{\bf k}'\cdot {\bf\Omega}_I)t}-1}{i({\bf
k}\cdot {\bf\Omega}-{\bf k}'\cdot {\bf\Omega}_I)}
\frac{e^{i({\bf K}\cdot {\bf\Omega}-{\bf K}'\cdot {\bf\Omega}_J)t}-1}{i({\bf
K}\cdot {\bf\Omega}-{\bf K}'\cdot {\bf\Omega}_J)}\nonumber\\
 A^d_{{\bf k}{\bf k}'}({\bf J},{\bf J}_I,{\bf k}'\cdot {\bf\Omega}_I)
A^d_{{\bf K}{\bf K}'}({\bf J},{\bf J}_J,{\bf K}'\cdot {\bf\Omega}_J)
e^{-i{\bf k}'\cdot {\bf w}_I}e^{-i{\bf K}'\cdot {\bf w}_J}
\Biggr\rangle.
\label{dn10}
\end{eqnarray}
Since the particles are initially uncorrelated and since the particles of the
same
species are identical,  we get (see the similar steps detailed after Eq.
(\ref{ci6}) in Sec. \ref{sec_inhoscf})
\begin{eqnarray}
D_{ij}=-\lim_{t\rightarrow +\infty}\frac{1}{2t}\sum_b\int d{\bf
w}'d{\bf J}' \sum_{{\bf k}{\bf k}'{\bf K}{\bf K}'} k_i K_j 
e^{i({\bf k}+{\bf K})\cdot {\bf w}}
\frac{e^{i({\bf k}\cdot {\bf\Omega}-{\bf k}'\cdot {\bf\Omega}')t}-1}{i({\bf k}\cdot {\bf\Omega}-{\bf k}'\cdot {\bf\Omega}')}
\frac{e^{i({\bf K}\cdot {\bf\Omega}-{\bf K}'\cdot {\bf\Omega}')t}-1}{i({\bf K}\cdot {\bf\Omega}-{\bf K}'\cdot {\bf\Omega}')}\nonumber\\
\times A^d_{{\bf k}{\bf k}'}({\bf J},{\bf J}',{\bf k}'\cdot
{\bf\Omega}')
A^d_{{\bf K}{\bf K}'}({\bf J},{\bf J}',{\bf K}'\cdot {\bf\Omega}')
e^{-i({\bf k}'+{\bf K}')\cdot {\bf w}'} m_b f_b({\bf J}').
\label{dn11}
\end{eqnarray}
Integrating over ${\bf w}'$ and averaging over ${\bf w}$ (which amounts to
making ${\bf K}=-{\bf k}$), we obtain
\begin{eqnarray}
D_{ij}=\lim_{t\rightarrow +\infty}\frac{1}{2t}\sum_b  (2\pi)^d\int d{\bf J}'
\sum_{{\bf k}{\bf k}'} k_i k_j 
\frac{e^{i({\bf k}\cdot {\bf\Omega}-{\bf k}'\cdot {\bf\Omega}')t}-1}{i({\bf k}\cdot {\bf\Omega}-{\bf k}'\cdot {\bf\Omega}')}
\frac{e^{-i({\bf k}\cdot {\bf\Omega}-{\bf k}'\cdot {\bf\Omega}')t}-1}{-i({\bf k}\cdot {\bf\Omega}-{\bf k}'\cdot {\bf\Omega}')}\nonumber\\
\times A^d_{{\bf k}{\bf k}'}({\bf J},{\bf J}',{\bf k}'\cdot
{\bf\Omega}')
A^d_{-{\bf k}-{\bf k}'}({\bf J},{\bf J}',-{\bf k}'\cdot
{\bf\Omega}') m_b f_b({\bf J}').
\label{dn12}
\end{eqnarray}
Using the identity (\ref{dn14}) we can rewrite the foregoing equation as
\begin{eqnarray}
D_{ij}=\lim_{t\rightarrow +\infty}\frac{1}{2t}\sum_b  (2\pi)^d\int d{\bf J}'
\sum_{{\bf k}{\bf k}'} k_i k_j 
\frac{|e^{i({\bf k}\cdot {\bf\Omega}-{\bf k}'\cdot {\bf\Omega}')t}-1|^2}{({\bf k}\cdot {\bf\Omega}-{\bf k}'\cdot {\bf\Omega}')^2}
 |A^d_{{\bf k}{\bf k}'}({\bf J},{\bf J}',{\bf k}'\cdot
{\bf\Omega}')|^2 m_b f_b({\bf J}').
\label{dn15}
\end{eqnarray}
Finally, using Eq. (\ref{dn16}), we get
\begin{eqnarray}
D_{ij}=\frac{1}{2}\sum_b (2\pi)^{d+1}\int d{\bf J}' \sum_{{\bf k}{\bf k}'}
k_i k_j 
 \delta({\bf k}\cdot {\bf\Omega}-{\bf k}'\cdot {\bf\Omega}') |A^d_{{\bf
k}{\bf k}'}({\bf J},{\bf J}',{\bf k}\cdot {\bf\Omega})|^2 m_b f_b({\bf J}'),
\label{dn17}
\end{eqnarray}
which returns Eq. (\ref{di1}).

{\it Remark:} It we do not take the limit $t\rightarrow +\infty$ in the
foregoing equations, we obtain a
time-dependent diffusion coefficient
\begin{eqnarray}
D_{ij}(t)=\frac{1}{2}\sum_b (2\pi)^{d+1}\int d{\bf J}' \sum_{{\bf k}{\bf k}'}
k_i k_j 
 \Delta({\bf k}\cdot {\bf\Omega}-{\bf k}'\cdot {\bf\Omega}',t) |A^d_{{\bf
k}{\bf k}'}({\bf J},{\bf J}',{\bf k}\cdot {\bf\Omega})|^2 m_b f_b({\bf J}'),
\label{dn17w}
\end{eqnarray}
where the regularized function $\Delta(x,t)$ is defined in Eq. (\ref{hama8def}).

\section{Force auto-correlation function and
diffusion coefficient with collective effects}
\label{sec_geic}

The force experienced by the test particle is 
\begin{eqnarray}
{\bf F}=-\frac{\partial\delta\Phi_{\rm tot}}{\partial {\bf w}},
\label{ham178}
\end{eqnarray}
where $\delta\Phi_{\rm tot}({\bf w},{\bf J},t)$ is the total fluctuating
potential acting on the test particle. Introducing the Fourier transform of the
total fluctuating potential, the force
auto-correlation function of the test particle accounting for collective effects
can be written as
\begin{equation}
\left\langle F_i({\bf w},{\bf J},0)F_j({\bf w}+{\bf\Omega}t,{\bf J},t)\right
\rangle=\sum_{{\bf k}{\bf k}'}\int\frac{d\omega}{2\pi}\frac{d\omega'}{2\pi}k_i
k'_j e^{i {\bf k}\cdot {\bf w}} e^{-i{\bf k}'\cdot ({\bf w}+{\bf\Omega}t)}
e^{i\omega' t}\left\langle
\delta\hat\Phi_{\rm tot}({\bf k},{\bf J},\omega)  \delta\hat\Phi_{\rm
tot}({\bf k}',{\bf J},\omega')^*\right\rangle.
 \label{ham179}
\end{equation}
To define the correlation function, we have used a Lagrangian point of view
and we have used the fact that the test particle follows the mean field
trajectory
from Eq. (\ref{unp})  at leading order. Since the correlation function should
not depend on ${\bf w}$ we can take ${\bf k}'={\bf k}$ and we get
\begin{equation}
\left\langle F_i({\bf w},{\bf J},0)F_j({\bf w}+{\bf\Omega}t,{\bf J},t)\right
\rangle=\sum_{{\bf k}}\int\frac{d\omega}{2\pi}\frac{d\omega'}{2\pi}k_i
k_j e^{-i{\bf k}\cdot {\bf\Omega}t}
e^{i\omega' t}\left\langle
\delta\hat\Phi_{\rm tot}({\bf k},{\bf J},\omega)  \delta\hat\Phi_{\rm
tot}({\bf k},{\bf J},\omega')^*\right\rangle.
 \label{ham179b}
\end{equation}
Using the expression (\ref{psaab})
of the correlation function of the
total fluctuating potential (power spectrum), we obtain
\begin{eqnarray}
\left\langle F_i({\bf w},{\bf J},0)F_j({\bf w}+{\bf\Omega}t,{\bf J},t)\right
\rangle=\sum_{{\bf k}}\int\frac{d\omega}{2\pi}k_i
k_j e^{i(\omega-{\bf k}\cdot {\bf\Omega}) t}P({\bf k},{\bf J},\omega).
\label{ham180}
\end{eqnarray}
The power spectrum can be related to the correlation function of the external
perturbation in the general case by using Eq. (\ref{i83}) or Eq.
(\ref{ciw4tay}). 
If the external potential is created by a random distribution of $N$ particles
then, using Eq.
(\ref{pi4}), the foregoing equation becomes 
\begin{eqnarray}
\left\langle F_i({\bf w},{\bf J},0)F_j({\bf w}+{\bf\Omega}t,{\bf J},t)\right
\rangle&=&(2\pi)^{d+1}\sum_b m_b\sum_{{\bf k}{\bf
k}'}\int d{\bf
J}'\,\int\frac{d\omega}{2\pi}k_i
k_j e^{i(\omega-{\bf k}\cdot {\bf\Omega}) t}
|A^d_{{\bf k},{\bf k}'}({\bf J},{\bf
J'},\omega)|^2 \delta(\omega-{\bf k}'\cdot {\bf\Omega}')f_b({\bf J}')\nonumber\\
&=&(2\pi)^{d}\sum_b m_b\sum_{{\bf k}{\bf
k}'}\int d{\bf
J}'\,k_i
k_j e^{i({\bf k}'\cdot {\bf\Omega}'-{\bf k}\cdot {\bf\Omega}) t}
|A^d_{{\bf k},{\bf k}'}({\bf J},{\bf
J'},{\bf k}'\cdot {\bf\Omega}')|^2 f_b({\bf J}').
\label{ham182}
\end{eqnarray}
Using
Eq. (\ref{gei1}), we
find that the diffusion tensor of the test particle is
\begin{eqnarray}
D_{ij}=\pi(2\pi)^{d}\sum_b m_b\sum_{{\bf k}{\bf
k}'}\int d{\bf
J}'\,k_i
k_j \delta({\bf k}'\cdot {\bf\Omega}'-{\bf k}\cdot {\bf\Omega})
|A^d_{{\bf k},{\bf k}'}({\bf J},{\bf
J'},{\bf k}'\cdot {\bf\Omega}')|^2 f_b({\bf J}').
\label{ham183}
\end{eqnarray}
This returns the result from Eq. (\ref{di1}).

\section{Polarization cloud}
\label{sec_pc}

The variation of the DF of an inhomogeneous system of particles with long-range
interactions resulting from an external perturbation is [see Eq. (\ref{i60})] 
\begin{equation}
\delta\hat f ({\bf k},{\bf J},\omega)=\frac{{\bf k}\cdot \frac{\partial
f}{\partial {\bf J}}}{{\bf k}\cdot {\bf
\Omega}-\omega}\delta\hat\Phi_{\rm tot}({\bf
k},{\bf J},\omega).
\label{pc1}
\end{equation}
The Fourier transform of the total fluctuating potential (including collective
effects) created by a test particle is given by Eq. (\ref{arm1}). Combining
Eqs. (\ref{arm1}) and (\ref{pc1}), we obtain 
\begin{equation}
\delta\hat f ({\bf k},{\bf J},\omega)=2\pi m\frac{{\bf k}\cdot \frac{\partial
f}{\partial {\bf J}}}{{\bf k}\cdot {\bf
\Omega}-\omega} \sum_{{\bf k}'} A^d_{{\bf
k}{\bf k}'}({\bf J},{\bf J}_0,\omega) e^{-i{\bf k}'\cdot {\bf
w}_0}\delta(\omega-{\bf
k}'\cdot{\bf \Omega}_0).
\label{pc2}
\end{equation}
Returning to physical space, we get
\begin{equation}
\delta f ({\bf w},{\bf J},t)=\sum_{\bf k}\int \frac{d\omega}{2\pi} e^{i({\bf
k}\cdot {\bf w}-\omega t)} 2\pi m\frac{{\bf k}\cdot \frac{\partial
f}{\partial {\bf J}}}{{\bf k}\cdot {\bf
\Omega}-\omega} \sum_{{\bf k}'} A^d_{{\bf
k}{\bf k}'}({\bf J},{\bf J}_0,\omega) e^{-i{\bf k}'\cdot {\bf
w}_0}\delta(\omega-{\bf
k}'\cdot{\bf \Omega}_0),
\label{pc3}
\end{equation}
giving
\begin{equation}
\delta f ({\bf w},{\bf J},t)=m\sum_{{\bf k}{\bf k}'}
e^{i{\bf k}\cdot {\bf w}} 
\frac{{\bf k}\cdot
\frac{\partial
f}{\partial {\bf J}}}{{\bf k}\cdot {\bf
\Omega}-{\bf k}'\cdot {\bf\Omega}_0} A^d_{{\bf
k}{\bf k}'}({\bf J},{\bf J}_0,{\bf
k}'\cdot{\bf \Omega}_0) e^{-i{\bf k}'\cdot ({\bf w}_0+{\bf\Omega}_0 t)}.
\label{pc4}
\end{equation}

{\it Remark:} If we neglect collective effects, we just have to replace 
$A^d_{{\bf k}{\bf k}'}({\bf J},{\bf J}_0,\omega)$ by $A_{{\bf k}{\bf k}'}({\bf
J},{\bf J}_0)$ in Eq. (\ref{pc2}), returning Eq. (\ref{gh2nj}).

\section{Simplified kinetic equations}
\label{sec_sim}

In this Appendix, we propose a simplified kinetic
equation that may
approximately describe the dynamical evolution of an inhomogeneous Hamiltonian
system of particles with long-range interactins in certain cases. To obtain this
equation, we make the thermal
bath approximation in the Lenard-Balescu equation 
(\ref{ilb3}), leading to Eq.
(\ref{ilb19}), but we assume that ${\bf \Omega}({\bf J},t)=\partial
H/\partial {\bf J}$ evolves
self-consistently
with time, being
determined by the total DF $f({\bf J},t)=\sum_a f_a({\bf J},t)$ through the
Hamiltonian $H$ instead of being prescribed as in
Sec.
\ref{sec_tbaw}.\footnote{In Sec.
\ref{sec_tbaw}, ${\bf \Omega}({\bf J})$ is determined by the DF
$\sum_b f_b({\bf J})$ of
the field particles assumed to be independent of time.} This gives
\begin{eqnarray}
\frac{\partial f_a}{\partial t}=\frac{\partial}{\partial J_i}\left \lbrack
D_{ij}\left (\frac{\partial f_a}{\partial J_j}+\beta m_a f_a \Omega_j
\right )\right\rbrack,
\label{sim1}
\end{eqnarray}
where $D_{ij}$ is given by Eq. (\ref{ilb16}). This equation
does not conserve the
energy
contrary to the Lenard-Balescu equation (\ref{ilb3}). However, we can enforce
the
energy conservation by allowing $\beta$ to depend on time in such
a way that $\dot E=\int H \frac{\partial f}{\partial
t}\, d{\bf J}=0$. This
yields
\begin{eqnarray}
\beta(t)=-\frac{\int D_{ij} \frac{\partial f}{\partial
J_j}\Omega_i\, d{\bf J}}{\int D_{ij} f_2 \Omega_i\Omega_j\, d{\bf J}}.
\label{sim3}
\end{eqnarray}
Equation (\ref{sim1}) with Eq. (\ref{sim3})
conserves the
mass of each species of particles, the total energy, and increases the entropy
($H$-theorem).
It relaxes towards the Boltzmann
distribution (\ref{fdi3voit}) of statistical
equilibrium on a timescale $Nt_D$. This equation is well-posed mathematically
and interesting in
its own right. It can be seen as a heuristic approximation of the
Lenard-Balescu equation (\ref{ilb3})
providing a simplified kinetic equation for an inhomogeneous Hamiltonian system
of particles with long-range interactions. However, since the approximation
leading to Eq. (\ref{sim1}) is
uncontrolled, the solution of this equation may substantially differ from the
solution of the Lenard-Balescu equation (\ref{ilb3}). The relevance of Eq.
(\ref{sim1}) with Eq. (\ref{sim3}) should be
determined
case by case by
solving this equation numerically and comparing its solution with the solution
of the Lenard-Balescu equation (\ref{ilb3}) or with direct
numerical
simulations of the $N$-body system.

In principle, the diffusion tensor $D_{ij}$ is a functional of
$f_a$ but we
shall assume, for simplicity, that $D_{ij}=D\delta_{ij}$ is isotropic and
constant. We also take $\beta={\rm cst}$ in Eq. (\ref{sim1}) like in
the case of Brownian particles with long-range interactions described by the
canonical
ensemble. In that case, Eq. (\ref{sim1})
conserves the mass of the different species of particles and
decreases the free energy
$F=E-TS$. With this setting, we can take fluctuations due to finite
$N$
effects into account
by adding a noise term in the kinetic equation like in Sec. \ref{sec_tdbv}.
This leads
to a stochastic partial differential equation of the form\footnote{It is also
possible to take into account
fluctuations in the more general equations (\ref{sim1})-(\ref{sim3}) and in the
Lenard-Balescu equation (\ref{ilb3})
but the expression of the noise is more
complicated \cite{bouchetld}.} 
\begin{eqnarray}
\frac{\partial f_a}{\partial t}=\frac{\partial}{\partial {\bf J}}\cdot \left
\lbrack
D\left (\frac{\partial f_a}{\partial {\bf J}}+\beta m_a f_a {\bf \Omega}({\bf
J},t) \right
)\right\rbrack+\frac{\partial}{\partial {\bf J}}\cdot \left\lbrack
\sqrt{2D m_a f_a} {\bf R}({\bf J},t)\right\rbrack.
\label{sim4}
\end{eqnarray}
This equation could be used to describe random transitions
between different equilibrium states as discussed in Sec. \ref{sec_tdbv}.

{\it Remark:} We can also perform this procedure of simplification directly on
the kinetic equations written in terms of the original variables $({\bf r},{\bf
v})$. In
that case, we obtain a deterministic equation of the form
\begin{eqnarray}
\label{sim6}
\frac{\partial f_a}{\partial t}+{\bf v}\cdot \frac{\partial
f_a}{\partial {\bf r}}-
\nabla\Phi\cdot
\frac{\partial f_a}{\partial {\bf v}}=\frac{\partial}{\partial {\bf
v}}\cdot\left\lbrack D
\left (\frac{\partial f_a}{\partial {\bf v}}+\beta m_a f_a {\bf v}\right
)\right\rbrack
\end{eqnarray}
coupled to Eq. (\ref{brow3}). We can enforce the conservation of energy by
allowing $\beta$ to depend on time in such
a way that $\dot E=\int (v^2/2+\Phi) \frac{\partial f}{\partial
t}\, d{\bf r}d{\bf v}=0$. This
yields\footnote{Equations (\ref{sim6}) and (\ref{sim7}) can be obtained
from a maximum
entropy production principle
(MEPP) as discussed in \cite{csr} in the context of the theory of
violent collisionless relaxation (see also \cite{gen,kinvr} for
generalizations).}
\begin{eqnarray}
\beta(t)=-\frac{\int D  \frac{\partial f}{\partial {\bf v}}\cdot {\bf v}\,
d{\bf r}d{\bf v}}{\int D f_2 v^2\, d{\bf r}d{\bf v}}.
\label{sim7}
\end{eqnarray}
Taking fluctuations into account, we obtain a stochastic
partial differential equation of the form
\begin{eqnarray}
\label{sim8}
\frac{\partial f_a}{\partial t}+{\bf v}\cdot \frac{\partial
f_a}{\partial {\bf r}}-
\nabla\Phi\cdot
\frac{\partial f_a}{\partial {\bf v}}=\frac{\partial}{\partial {\bf
v}}\cdot\left\lbrack D
\left (\frac{\partial f_a}{\partial {\bf v}}+\beta m_a f_a {\bf v}\right
)\right\rbrack+\frac{\partial}{\partial{\bf v}}\cdot
\left
(\sqrt{2Dm_a f_a}\, {\bf Q}({\bf r},{\bf v},t)\right ),
\end{eqnarray}
coupled to Eq. (\ref{brow6}), where ${\bf Q}({\bf r},{\bf v},t)$ is a Gaussian
white noise satisfying $\langle
{\bf
Q}({\bf r},{\bf v},t)\rangle={\bf 0}$ and  $\langle Q_\alpha({\bf r},{\bf
v},t)R_\beta({\bf
r}',{\bf v}',t')\rangle=\delta_{\alpha\beta}\delta({\bf r}-{\bf r}')\delta({\bf
v}-{\bf v}')\delta(t-t')$. In the low friction limit $\xi\rightarrow 0$, the
system first settles on a QSS resulting from a process of violent relaxation.
The DF $f({\bf J},t)$
then evolves secularly according to Eqs. (\ref{sim1}) and (\ref{sim4}).

{\it Remark:} As mentioned above, the simplified kinetic equations of this
section may not always give a good agreement with the Lenard-Balescu
equation. For example, for 1D homogeneous systems where there is no
resonance, Eq. (\ref{sim6}) gives a non-vanishing flux 
($\partial f/\partial t\neq 0$) driving the system towards the Boltzmann
distribution on a timescale $Nt_D$ while the Lenard-Balescu flux
vanishes identically ($\partial f_{\rm LB}/\partial t=0$) and the Boltzmann
distribution is reached on a longer timescale
$N^2t_D$.\footnote{This
timescale discrepency could be corrected by empirically changing the value of
$D$ in Eq. (\ref{sim6}) to make it of order $1/N^2$.} This is an obvious case
of strong discrepency. However, such a situation does not occur for spatially
inhomogeneous systems for which there are always resonances and the
Lenard-Balescu equation relaxes towards the Boltzmann distribution on a
timescale $N t_D$.

\section{Kinetic equations for inhomogeneous systems with long-range
interactions}
\label{sec_j}

\subsection{Inhomogeneous Lenard-Balescu equation}
\label{sec_jbl}

We consider an isolated system of $N$ particles with identical mass
$m$. We treat the case of a possibly spatially inhomogeneous system. We want to
derive the Lenard-Balescu equation governing the evolution of the DF at the
order $1/N$ by using the Klimontovich approach. The
derivation follows the one given
in Ref. \cite{physicaA} (see also \cite{klim,Kvortex2023} for the similar
problem of 2D
point
vortices) and it can be easily adapted to the case of homogeneous systems
\cite{epjp,prep1}. We start from the quasilinear equations
(\ref{i56}) and (\ref{i57}) without the external
potential ($\Phi_e=0$) that we rewrite as
\begin{eqnarray}
\label{j1}
\frac{\partial f}{\partial t}=\frac{\partial}{\partial
{\bf J}}\cdot \left\langle \delta
f \frac{\partial\delta\Phi}{\partial {\bf w}}\right\rangle,
\end{eqnarray}
\begin{eqnarray}
\label{j2}
\frac{\partial \delta f}{\partial t}+{\bf \Omega}\cdot \frac{\partial\delta
f}{\partial {\bf w}}- \frac{\partial\delta\Phi}{\partial {\bf w}}
\cdot \frac{\partial
f}{\partial {\bf J}}=0.
\end{eqnarray}
We make the Bogoliubov ansatz which treats $f$ as
time independent. Taking the Fourier-Laplace transform of Eq.
(\ref{j2}), we obtain
\begin{eqnarray}
\label{j3}
\delta{\tilde f}({\bf k},{\bf
J},\omega)=\frac{{\bf k}\cdot \frac{\partial
f}{\partial
{\bf J}}}{{\bf
k}\cdot {\bf
\Omega}-\omega}\delta{\tilde\Phi}({\bf k},{\bf
J},\omega)+\frac{\delta{\hat f} ({\bf k},{\bf J},0)}{i(
{\bf k}\cdot {\bf \Omega}-\omega)},
\end{eqnarray}
where $\delta{\hat f}({\bf k},{\bf J},t=0)$ is the Fourier transform of the
initial value of the perturbed DF due to finite $N$ effects. According to Eq.
(\ref{bof3b}),
we
have
\begin{eqnarray}
\delta{\tilde \Phi}({\bf k},{\bf J},\omega)=(2\pi)^d\sum_{{\bf k}'} \int d{\bf
J}'\, 
A_{{\bf k}{\bf k}'}({\bf J},{\bf
J}') \delta{\tilde f}({\bf k}',{\bf J}',\omega),
\label{bof3bann}
\end{eqnarray}
where $A_{{\bf k}{\bf k}'}({\bf J},{\bf J}')$ is the Fourier transform of
the bare potential of
interaction in angle-action variables. Substituting Eq. (\ref{j3}) into Eq.
(\ref{bof3bann}) we get
\begin{eqnarray}
\delta{\tilde \Phi}({\bf k},{\bf J},\omega)&=&(2\pi)^d\sum_{{\bf k}'} \int d{\bf
J}'\, 
A_{{\bf k}{\bf k}'}({\bf J},{\bf
J}') \frac{{\bf k}'\cdot \frac{\partial
f'}{\partial
{\bf J}'}}{{\bf
k}'\cdot {\bf
\Omega}'-\omega}\delta{\tilde\Phi}({\bf k}',{\bf
J}',\omega)
\nonumber\\
&+&(2\pi)^d\sum_{{\bf k}'} \int d{\bf
J}'\, 
A_{{\bf k}{\bf k}'}({\bf J},{\bf
J}') \frac{\delta{\hat f} ({\bf k}',{\bf J}',0)}{i(
{\bf k}'\cdot {\bf \Omega}'-\omega)}.
\label{mat1}
\end{eqnarray}
This is a Fredholm integral equation (see Appendix A of
\cite{physicaA}) which relates the Fourier-Laplace transform of the fluctuations
of
the
potential to the Fourier transform of the initial fluctuations of the
DF. The formal solution of this equation is (see
Appendix \ref{sec_dnwout})
\begin{eqnarray}
\label{j5}
\delta{\tilde\Phi}({\bf k},{\bf J},\omega)=(2\pi)^d\sum_{{\bf k}'}\int d{\bf
J}' A^d_{{\bf k}{\bf k}'}({\bf J},{\bf J}',\omega) \frac{\delta{\hat f}({\bf
k}',{\bf J}',0)}{i({\bf k}'\cdot {\bf\Omega}'-\omega)},
\end{eqnarray}
where $A^d_{{\bf k}{\bf k}'}({\bf J},{\bf J}',\omega)$ is the Fourier
transform of the dressed potential
of interaction defined by Eq. (\ref{nmat3}). 
Equation (\ref{nmat3}) can
be used as a definition of $A^d_{{\bf k}{\bf k}'}({\bf J},{\bf
J}',\omega)$ which does not require to introduce a biorthogonal basis (see
Appendix A of \cite{physicaA}). However, it is shown in Sec. 2 of
\cite{physicaA} that
Eq. (\ref{j5}) can also be
obtained from the matrix method where $A^d_{{\bf k}{\bf k}'}({\bf J},{\bf
J}',\omega)$ is defined by Eq. (\ref{can1b}). This
shows that the two manners to define $A^d_{{\bf k}{\bf k}'}({\bf J},{\bf
J}',\omega)$ are equivalent.

Taking the inverse
Laplace transform of Eq. (\ref{j5}), using the Cauchy residue theorem, and
neglecting the contribution of the damped
modes for sufficiently
late times,\footnote{We only consider the contribution of the
pole $\omega-{\bf k}'\cdot {\bf \Omega}'$ in Eq. (\ref{j5}) and ignore the
contribution of the
proper modes of the system which are the solutions of the dispersion relation 
(\ref{disrel}). This assumes that the system is sufficiently far from the
threshold of instability (see the Conclusion). We refer to \cite{linres} for
general considerations about the linear response theory
of
systems with long-range interactions.} we obtain
\begin{eqnarray}
\label{j6}
\delta{\hat\Phi}({\bf k},{\bf J},t)=(2\pi)^d\sum_{{\bf k}'}\int d{\bf
J}' A^d_{{\bf k}{\bf k}'}({\bf J},{\bf J}',{\bf k}'\cdot {\bf \Omega}')
\delta{\hat f}({\bf
k}',{\bf J}',0)e^{-i {\bf k}'\cdot {\bf\Omega}'t}.
\end{eqnarray}
On the other hand, taking the Fourier transform of Eq. (\ref{j2}), we find
that
\begin{eqnarray}
\label{j7}
\frac{\partial \delta{\hat f}}{\partial t}+i{\bf k}\cdot{\bf\Omega}\delta{\hat
f}=i{\bf k}\cdot \frac{\partial f}{\partial {\bf J}}\delta{\hat\Phi}.
\end{eqnarray}
This first order differential equation in time can be solved with the method of
the variation of the constant, giving
\begin{eqnarray}
\label{j8}
\delta{\hat f}({\bf k},{\bf J},t)=\delta{\hat
f}({\bf
k},{\bf J},0)e^{-i {\bf k}\cdot
{\bf\Omega} t}+i{\bf k}\cdot
\frac{\partial f}{\partial {\bf J}}\int_0^t dt'\, \delta{\hat \Phi}({\bf
k},{\bf J},t')e^{i{\bf k}\cdot{\bf\Omega}(t'-t)}.
\end{eqnarray}
Substituting Eq. (\ref{j6}) into Eq. (\ref{j8}), we obtain
\begin{equation}
\label{j9}
\delta{\hat f}({\bf k},{\bf J},t)=\delta{\hat
f}({\bf
k},{\bf J},0)e^{-i {\bf k}\cdot
{\bf\Omega} t}+i{\bf k}\cdot
\frac{\partial f}{\partial {\bf J}} (2\pi)^d\sum_{{\bf k}'}\int d{\bf
J}' A^d_{{\bf k}{\bf k}'}({\bf J},{\bf J}',{\bf k}'\cdot {\bf \Omega}')
\delta{\hat f}({\bf
k}',{\bf J}',0)e^{-i{\bf
k}\cdot{\bf\Omega}t}\int_0^t dt'\, e^{i({\bf
k}\cdot{\bf\Omega}-{\bf k}'\cdot {\bf\Omega}')t'}.
\end{equation}
Eqs.
(\ref{j6}) and
(\ref{j9}) relate $\delta{\hat\Phi}({\bf k},{\bf J},t)$ and
$\delta{\hat f}({\bf k},{\bf J},t)$ to
the
initial fluctuation $\delta{\hat f}({\bf k},{\bf J},0)$. For sufficiently large
times, we can write
\begin{equation}
\label{j9b}
\delta{\hat f}({\bf k},{\bf J},t)=\delta{\hat
f}({\bf
k},{\bf J},0)e^{-i {\bf k}\cdot
{\bf\Omega} t}+{\bf k}\cdot
\frac{\partial f}{\partial {\bf J}} (2\pi)^d\sum_{{\bf k}'}\int d{\bf
J}' A^d_{{\bf k}{\bf k}'}({\bf J},{\bf J}',{\bf k}'\cdot {\bf \Omega}')
\delta{\hat f}({\bf
k}',{\bf J}',0)\frac{e^{-i{\bf k}'\cdot {\bf\Omega}'t}}{{\bf
k}\cdot{\bf\Omega}-{\bf k}'\cdot {\bf\Omega}'}.
\end{equation}
Eqs. (\ref{j6}) and (\ref{j9b}) [or Eq. (\ref{j9})] give the
asymptotic expressions of the fluctuations $\delta{\hat\Phi}({\bf k},{\bf J},t)$
and $\delta{\hat f}({\bf k},{\bf J},t)$ for $t\rightarrow +\infty$ that can be 
used to calculate the correlation functions.

Let us compute the flux
\begin{eqnarray}
\label{j10}
\left\langle \delta f\frac{\partial\delta\Phi}{\partial {\bf
w}}\right\rangle=\sum_{{\bf k}{\bf k}'} i{\bf k}' e^{i{\bf k}\cdot {\bf
w}}e^{i{\bf k}'\cdot {\bf w}}\langle \delta{\hat f}({\bf k},{\bf
J},t)\delta{\hat \Phi}({\bf k}',{\bf J},t)\rangle.
\end{eqnarray}
From Eqs. (\ref{j6}) and (\ref{j9}) we get
\begin{eqnarray}
\label{j11}
\langle \delta{\hat f}({\bf k},{\bf
J},t)\delta{\hat \Phi}({\bf k}',{\bf J},t)\rangle=(2\pi)^d\sum_{{\bf k}''}\int
d{\bf
J}' A^d_{{\bf k}'{\bf k}''}({\bf J},{\bf J}',{\bf k}''\cdot {\bf \Omega}')
e^{-i {\bf k}''\cdot {\bf\Omega}'t}e^{-i {\bf k}\cdot
{\bf\Omega} t}\langle \delta{\hat f}({\bf
k}'',{\bf J}',0)\delta{\hat
f}({\bf
k},{\bf J},0)\rangle\nonumber\\
+(2\pi)^d\sum_{{\bf k}''}\sum_{{\bf k}'''}\int d{\bf
J}' \int d{\bf
J}''\, A^d_{{\bf k}'{\bf k}''}({\bf J},{\bf J}',{\bf k}''\cdot {\bf \Omega}')
\langle\delta{\hat f}({\bf
k}'',{\bf J}',0)\delta{\hat f}({\bf
k}''',{\bf J}'',0)\rangle e^{-i {\bf k}''\cdot {\bf\Omega}'t}\nonumber\\
\times i{\bf k}\cdot
\frac{\partial f}{\partial {\bf J}} (2\pi)^d A^d_{{\bf k}{\bf k}'''}({\bf
J},{\bf J}'',{\bf k}'''\cdot {\bf \Omega}'')
e^{-i{\bf
k}\cdot{\bf\Omega}t}\int_0^t dt'\, e^{i({\bf
k}\cdot{\bf\Omega}-{\bf k}'''\cdot {\bf\Omega}'')t'}.
\end{eqnarray}
We assume that at $t=0$ there are no correlations among the particles.
Therefore, we have (see Appendix C of \cite{physicaA}) 
\begin{eqnarray}
\label{j12}
\langle \delta{\hat f}({\bf
k},{\bf J},0)\delta{\hat
f}({\bf
k}',{\bf J}',0)\rangle=\frac{1}{(2\pi)^d}\delta_{{\bf k},-{\bf k}'}\delta({\bf
J}-{\bf J}')mf({\bf J}).
\end{eqnarray}
Eq. (\ref{j11}) then reduces to
\begin{eqnarray}
\label{j13}
&&\langle \delta{\hat f}({\bf k},{\bf
J},t)\delta{\hat \Phi}({\bf k}',{\bf J},t)\rangle=A^d_{{\bf
k}',-{\bf k}}({\bf J},{\bf J},-{\bf k}\cdot {\bf \Omega})mf({\bf J})\nonumber\\
&&+\sum_{{\bf k}''}\int d{\bf
J}' \, A^d_{{\bf k}'{\bf k}''}({\bf J},{\bf J}',{\bf k}''\cdot {\bf
\Omega}') m f({\bf J}') i{\bf k}\cdot
\frac{\partial f}{\partial {\bf J}} (2\pi)^d A^d_{{\bf k},-{\bf k}''}({\bf
J},{\bf J}',-{\bf k}''\cdot {\bf \Omega}')
\int_0^t ds\, e^{-i({\bf
k}\cdot{\bf\Omega}+{\bf k}''\cdot {\bf\Omega}')s}.
\end{eqnarray}
where we have set $s=t-t'$. Substituting this
relation
into Eq. (\ref{j10}), we
obtain
\begin{eqnarray}
\label{j14a}
&&\left\langle \delta f\frac{\partial\delta\Phi}{\partial {\bf
w}}\right\rangle=\sum_{{\bf k}{\bf k}'} i{\bf k}' e^{i{\bf k}\cdot {\bf
w}}e^{i{\bf k}'\cdot {\bf w}}A^d_{{\bf
k}',-{\bf k}}({\bf J},{\bf J},-{\bf k}\cdot {\bf \Omega})mf({\bf J})\nonumber\\
&&+\sum_{{\bf k}{\bf k}'} i{\bf k}' e^{i{\bf k}\cdot {\bf
w}}e^{i{\bf k}'\cdot {\bf w}}\sum_{{\bf k}''}\int d{\bf
J}' \, A^d_{{\bf k}'{\bf k}''}({\bf J},{\bf J}',{\bf k}''\cdot {\bf
\Omega}') m f({\bf J}') i{\bf k}\cdot
\frac{\partial f}{\partial {\bf J}} (2\pi)^d A^d_{{\bf k},-{\bf k}''}({\bf
J},{\bf J}',-{\bf k}''\cdot {\bf \Omega}')\int_0^t ds\, e^{-i({\bf
k}\cdot{\bf\Omega}+{\bf k}''\cdot {\bf\Omega}')s}.\nonumber\\
\end{eqnarray}
Averaging this expression over ${\bf w}$, which amounts to replacing ${\bf k}'$
by $-{\bf k}$,\footnote{Recall that the average flux should not depend on ${\bf
w}$.}
we find that
\begin{eqnarray}
\label{j14b}
&&\left\langle \delta f\frac{\partial\delta\Phi}{\partial {\bf
w}}\right\rangle=-\sum_{{\bf k}} i{\bf k} A^d_{-{\bf
k},-{\bf k}}({\bf J},{\bf J},-{\bf k}\cdot {\bf \Omega})mf({\bf J})\nonumber\\
&&-\sum_{{\bf k}} i{\bf k}\sum_{{\bf k}''}\int d{\bf
J}' \, A^d_{-{\bf k},{\bf k}''}({\bf J},{\bf J}',{\bf k}''\cdot {\bf
\Omega}') m f({\bf J}') i{\bf k}\cdot
\frac{\partial f}{\partial {\bf J}} (2\pi)^d A^d_{{\bf k},-{\bf k}''}({\bf
J},{\bf J}',-{\bf k}''\cdot {\bf \Omega}')\int_0^t ds\, e^{-i({\bf
k}\cdot{\bf\Omega}+{\bf k}''\cdot {\bf\Omega}')s}.\nonumber\\
\end{eqnarray}
If we make the change of notation ${\bf k}''=-{\bf k}'$ and let
$t\rightarrow +\infty$ in the second
term, we can rewrite the foregoing equation as
\begin{eqnarray}
\label{j14c}
&&\left\langle \delta f\frac{\partial\delta\Phi}{\partial {\bf
w}}\right\rangle=-\sum_{{\bf k}} i{\bf k} A^d_{-{\bf
k},-{\bf k}}({\bf J},{\bf J},-{\bf k}\cdot {\bf \Omega})mf({\bf J})\nonumber\\
&&-\sum_{{\bf k}} i{\bf k}\sum_{{\bf k}'}\int d{\bf
J}' \, A^d_{-{\bf k},-{\bf k}'}({\bf J},{\bf J}',-{\bf k}'\cdot {\bf
\Omega}') m f({\bf J}') i{\bf k}\cdot
\frac{\partial f}{\partial {\bf J}} (2\pi)^d A^d_{{\bf k},{\bf k}'}({\bf
J},{\bf J}',{\bf k}'\cdot {\bf \Omega}')\int_0^{+\infty} ds\, e^{-i({\bf
k}\cdot{\bf\Omega}-{\bf k}'\cdot {\bf\Omega}')s}.\nonumber\\
\end{eqnarray}
Making the transformations
$s\rightarrow -s$, ${\bf
k}\rightarrow -{\bf k}$ and 
${\bf k}'\rightarrow -{\bf k}'$ we see that
we can replace $\int_0^{+\infty} ds$ by $\frac{1}{2}\int_{-\infty}^{+\infty}
ds$. We then get 
\begin{eqnarray}
\label{j15}
&&\left\langle \delta f\frac{\partial\delta\Phi}{\partial {\bf
w}}\right\rangle=-\sum_{{\bf k}} i{\bf k} A^d_{-{\bf
k},-{\bf k}}({\bf J},{\bf J},-{\bf k}\cdot {\bf \Omega})mf({\bf J})\nonumber\\
&&-\sum_{{\bf k}} i{\bf k}\sum_{{\bf k}'}\int d{\bf
J}' \, A^d_{-{\bf k},-{\bf k}'}({\bf J},{\bf J}',-{\bf k}'\cdot {\bf
\Omega}') m f({\bf J}') i{\bf k}\cdot
\frac{\partial f}{\partial {\bf J}} (2\pi)^d A^d_{{\bf k},{\bf k}'}({\bf
J},{\bf J}',{\bf k}'\cdot {\bf \Omega}')\frac{1}{2}\int_{-\infty}^{+\infty}
ds\, e^{-i({\bf
k}\cdot{\bf\Omega}-{\bf k}'\cdot {\bf\Omega}')s}.\nonumber\\
\end{eqnarray}
Using the identity (\ref{deltac}), we obtain
\begin{eqnarray}
\label{j17}
&&\left\langle \delta f\frac{\partial\delta\Phi}{\partial {\bf
w}}\right\rangle=-\sum_{{\bf k}} i{\bf k} A^d_{-{\bf
k},-{\bf k}}({\bf J},{\bf J},-{\bf k}\cdot {\bf \Omega})mf({\bf J})\nonumber\\
&&-\sum_{{\bf k}} i{\bf k}\sum_{{\bf k}'}\int d{\bf
J}' \, A^d_{-{\bf k},-{\bf k}'}({\bf J},{\bf J}',-{\bf k}'\cdot {\bf
\Omega}') m f({\bf J}') i{\bf k}\cdot
\frac{\partial f}{\partial {\bf J}} (2\pi)^d A^d_{{\bf k},{\bf k}'}({\bf
J},{\bf J}',{\bf k}'\cdot {\bf \Omega}') \pi \delta({\bf
k}\cdot{\bf\Omega}-{\bf k}'\cdot {\bf\Omega}').
\end{eqnarray}
Finally, using Eq. (\ref{dn14}), we obtain the Lenard-Balescu flux
\begin{eqnarray}
\label{j18}
&&\left\langle \delta f\frac{\partial\delta\Phi}{\partial {\bf
w}}\right\rangle=-\sum_{{\bf k}} i{\bf k} A^d_{{\bf
k},{\bf k}}({\bf J},{\bf J},{\bf k}\cdot {\bf \Omega})^* mf({\bf J})\nonumber\\
&&+\pi (2\pi)^d \sum_{{\bf k}{\bf k}'} {\bf k}\int d{\bf
J}' \,   {\bf k}\cdot
\frac{\partial f}{\partial {\bf J}} |A^d_{{\bf k},{\bf k}'}({\bf
J},{\bf J}',{\bf k}'\cdot {\bf \Omega}')|^2
\delta({\bf
k}\cdot{\bf\Omega}-{\bf k}'\cdot {\bf\Omega}')mf({\bf J}').
\end{eqnarray}
The first term is the friction by polarization term and the second term is the
diffusion term.
Using the identity (\ref{fdi2a}) and substituting the flux from Eq.
(\ref{j18}) into Eq. (\ref{j1}), we obtain
the inhomogeneous Lenard-Balescu equation \cite{heyvaerts,physicaA}
\begin{equation}
\frac{\partial f}{\partial t}=\pi (2\pi)^d m \frac{\partial}{\partial {\bf
J}}\cdot \sum_{{\bf k},{\bf k}'}\int d{\bf J}' \, {\bf k}\,
|A^d_{{\bf k},{\bf k}'}({\bf J},{\bf J}',{\bf k}\cdot {\bf
\Omega})|^2\delta({\bf k}\cdot {\bf \Omega}-{\bf k}'\cdot {\bf \Omega}') \left
(f' {\bf k}\cdot \frac{\partial f}{\partial {\bf J}}-f {\bf
k}'\cdot \frac{\partial f'}{\partial {\bf J}'}\right ).
\label{j19}
\end{equation}
We can easily extend the derivation of the Lenard-Balescu equation to the
multispecies case, leading to Eq. (\ref{ilb3}).

{\it Remark:} From Eq. (\ref{j6}) we get
\begin{eqnarray}
\label{int1}
\langle \delta{\hat\Phi}({\bf k},{\bf J},t)\delta{\hat\Phi}({\bf k},{\bf
J},t')^*\rangle=(2\pi)^{2d}\sum_{{\bf k}'{\bf k}''}\int d{\bf
J}'d{\bf J}''\, A^d_{{\bf k}{\bf k}'}({\bf J},{\bf J}',{\bf k}'\cdot {\bf
\Omega}') A^d_{{\bf k}{\bf k}''}({\bf J},{\bf J}'',{\bf k}''\cdot {\bf
\Omega}'')^* \nonumber\\
\times\langle
\delta{\hat f}({\bf
k}',{\bf J}',0)\delta{\hat f}({\bf
k}'',{\bf J}'',0)^*\rangle e^{-i {\bf k}'\cdot {\bf\Omega}'t} e^{i {\bf
k}''\cdot {\bf\Omega}''t'}.
\end{eqnarray}
Using Eq. (\ref{j12}) we obtain
\begin{eqnarray}
\label{int2}
\langle \delta{\hat\Phi}({\bf k},{\bf J},t)\delta{\hat\Phi}({\bf k},{\bf
J},t')^*\rangle=(2\pi)^{d}\sum_{{\bf k}'}\int d{\bf
J}'\, |A^d_{{\bf k}{\bf k}'}({\bf J},{\bf J}',{\bf k}'\cdot {\bf
\Omega}')|^2   m f({\bf J}') e^{-i {\bf k}'\cdot {\bf\Omega}'(t-t')}.
\end{eqnarray}
This is the temporal correlation function  ${\cal P}({\bf k},{\bf J},t-t')$ of
the potential fluctuations [see Eq.
(\ref{an11bh})]. Taking its Fourier transform, we obtain the power spectrum
produced by a random distribution of particles
\begin{eqnarray}
P({\bf
k},{\bf J},\omega)=(2\pi)^{d+1} m\sum_{{\bf k}'}\int d{\bf
J}'\,
|A^d_{{\bf k},{\bf k}'}({\bf J},{\bf
J'},\omega)|^2 \delta(\omega-{\bf k}'\cdot {\bf\Omega}')f({\bf J}').
\label{int3}
\end{eqnarray}
This returns Eq. (43) of \cite{physicaA} [see also Eq.
(\ref{pi4}) of the present paper]. On the other hand, if we compare
Eqs. (\ref{fem}) and (\ref{j5}) we find the correspondance
\begin{eqnarray}
\frac{\delta{\hat f}({\bf
k}',{\bf J}',0)}{i({\bf k}'\cdot {\bf\Omega}'-\omega)}\leftrightarrow {\hat
f}_e({\bf k}',{\bf J}',\omega).
\end{eqnarray}
If we use this correspondance in Eqs. (39) and (40) of \cite{physicaA} together
with Eq. (\ref{ciw5}) we recover Eqs. (43) and (44) of \cite{physicaA}.

\subsection{Inhomogeneous Landau equation}

If we ignore collective effects, we would be tempted to replace $A^d$ by $A$ in
Eq. (\ref{j18}). But, in that case, the friction by polarization would vanish
since
$A$ is
real. Therefore, we must first compute ${\rm Im}(A^d)$ by using Eq.
(\ref{fdi2a}),
{\it then} replace $A^d$
by $A$. To directly derive the Landau equation from the quasilinear equations
(\ref{j1}) and (\ref{j2}), we can proceed as
follows.\footnote{This approach is related to the iterative method used in
\cite{hb4} to derive the Landau equation in physical space.}

According to Eq. (\ref{bof3b})  we have
\begin{eqnarray}
\label{p27}
\delta{\hat\Phi}({\bf k},{\bf J},t)=(2\pi)^d\sum_{{\bf k}'}\int d{\bf
J}' A_{{\bf k}{\bf k}'}({\bf J},{\bf J}')
\delta{\hat f}({\bf
k}',{\bf J}',t).
\end{eqnarray}
Substituting Eq. (\ref{j8}) into Eq. (\ref{p27}) we obtain
\begin{eqnarray}
\label{p28}
\delta{\hat\Phi}({\bf k},{\bf J},t)=(2\pi)^d\sum_{{\bf k}'}\int d{\bf
J}' A_{{\bf k}{\bf k}'}({\bf J},{\bf J}')
\delta{\hat f}({\bf
k}',{\bf J}',0)e^{-i{\bf k}'\cdot {\bf\Omega}' t}\nonumber\\
+(2\pi)^d\sum_{{\bf k}'}\int d{\bf
J}' A_{{\bf k}{\bf k}'}({\bf J},{\bf J}')i{\bf k}'\cdot
\frac{\partial f'}{\partial {\bf J}'}\int_0^t dt'\, \delta{\hat \Phi}({\bf
k}',{\bf J}',t')e^{i{\bf k}'\cdot{\bf\Omega}'(t'-t)}.
\end{eqnarray}
This equation is an exact  Fredholm integral equation equivalent to Eq.
(\ref{mat1}). In order to relate
$\delta{\hat\Phi}({\bf k},{\bf J},t)$ to $\delta{\hat f}({\bf
k}',{\bf J}',0)$, instead of using Eq. (\ref{j6}), we shall use an
iterative
method. We assume that
we can replace $\delta{\hat\Phi}$ in the second line of Eq. (\ref{p28}) by its
expression
obtained by keeping only the contribution from the first line. This gives
\begin{eqnarray}
\label{p29}
\delta{\hat\Phi}({\bf k},{\bf J},t)=(2\pi)^d\sum_{{\bf k}'}\int d{\bf
J}' A_{{\bf k}{\bf k}'}({\bf J},{\bf J}')
\delta{\hat f}({\bf
k}',{\bf J}',0)e^{-i{\bf k}'\cdot {\bf\Omega}' t}\nonumber\\
+(2\pi)^d\sum_{{\bf k}'}\int d{\bf
J}' A_{{\bf k}{\bf k}'}({\bf J},{\bf J}')i{\bf k}'\cdot
\frac{\partial f'}{\partial {\bf J}'}\int_0^t dt'\, e^{i{\bf
k}'\cdot{\bf\Omega}'(t'-t)}(2\pi)^d\sum_{{\bf k}''}\int d{\bf
J}'' A_{{\bf k}'{\bf k}''}({\bf J}',{\bf J}'')
\delta{\hat f}({\bf
k}'',{\bf J}'',0)e^{-i{\bf k}''\cdot {\bf\Omega}'' t'}.
\end{eqnarray}
We use the same strategy in Eq. (\ref{j8}), thereby obtaining
\begin{eqnarray}
\label{p30}
\delta{\hat f}({\bf k},{\bf J},t)=\delta{\hat
f}({\bf
k},{\bf J},0)e^{-i {\bf k}\cdot
{\bf\Omega} t}+i{\bf k}\cdot
\frac{\partial f}{\partial {\bf J}}\int_0^t dt'\, e^{i{\bf
k}\cdot{\bf\Omega}(t'-t)}(2\pi)^d\sum_{{\bf k}'}\int d{\bf
J}' A_{{\bf k}{\bf k}'}({\bf J},{\bf J}')
\delta{\hat f}({\bf
k}',{\bf J}',0)e^{-i{\bf k}'\cdot {\bf\Omega}' t'}.
\end{eqnarray}
We see that Eqs. (\ref{p29}) and (\ref{p30}) are similar to Eqs.
(\ref{j6}) and (\ref{j9}) except that $A^d$ is replaced by $A$ and there
is a ``new'' term
in Eq.   (\ref{p29}) in addition to the ``old'' one. This new term
accounts for the friction by polarization. It substitutes itself to the term
proportional to ${\rm Im}(A^d)$ which vanishes when
$A^d$ is replaced by $A$ (as noted above). On the other hand, the old term with
$A^d$ replaced by $A$ accounts for the diffusion. Therefore, repeating
the calculations of Appendix \ref{sec_jbl}, or directly using Eq. (\ref{j18})
with $A^d$
replaced by $A$, we get
\begin{eqnarray}
\label{p26}
\left\langle \delta f\frac{\partial\delta\Phi}{\partial {\bf
w}}\right\rangle_{\rm diffusion}=
\pi (2\pi)^d \sum_{{\bf k}{\bf k}'} {\bf k}\int d{\bf
J}' \,   {\bf k}\cdot
\frac{\partial f}{\partial {\bf J}} |A_{{\bf k},{\bf k}'}({\bf
J},{\bf J}')|^2
\delta({\bf
k}\cdot{\bf\Omega}-{\bf k}'\cdot {\bf\Omega}')mf({\bf J}').
\end{eqnarray}
This quantity corresponds to the product of the first term in Eq. (\ref{p29})
with the two terms in Eq. (\ref{p30}). Let us now compute the friction term.
To that purpose, we have
to compute
\begin{eqnarray}
\label{aaa3}
\langle \delta{\hat f}({\bf k},{\bf
J},t)\delta{\hat \Phi}({\bf k}',{\bf J},t)\rangle_{\rm
friction}=(2\pi)^d\sum_{{\bf k}''}\int
d{\bf
J}' A_{{\bf k}'{\bf k}''}({\bf J},{\bf J}')i{\bf k}''\cdot
\frac{\partial f'}{\partial {\bf J}'}\int_0^t dt'\, e^{i{\bf
k}''\cdot{\bf\Omega}'(t'-t)}\nonumber\\
\times(2\pi)^d\sum_{{\bf k}'''}\int d{\bf
J}'' A_{{\bf k}''{\bf k}'''}({\bf J}',{\bf J}'')
e^{-i{\bf k}'''\cdot {\bf\Omega}'' t'}e^{-i {\bf k}\cdot
{\bf\Omega} t}
\langle \delta{\hat
f}({\bf
k},{\bf J},0)\delta{\hat f}({\bf
k}''',{\bf J}'',0)\rangle. 
\end{eqnarray}
This quantity corresponds to the product of the second term in Eq. (\ref{p29})
with the first term in Eq. (\ref{p30}). In line with our perturbative approach
we neglect the product of the second terms in Eqs. (\ref{p29}) and (\ref{p30}).
Using Eq. (\ref{j12}) we get
\begin{eqnarray}
\label{p28b}
\langle \delta{\hat f}({\bf k},{\bf
J},t)\delta{\hat \Phi}({\bf k}',{\bf J},t)\rangle_{\rm
friction}=(2\pi)^d\sum_{{\bf k}''}\int
d{\bf
J}' A_{{\bf k}'{\bf k}''}({\bf J},{\bf J}')i{\bf k}''\cdot
\frac{\partial f'}{\partial {\bf J}'}\int_0^t ds\, e^{-i({\bf
k}\cdot{\bf\Omega}+{\bf k}''\cdot{\bf\Omega}')s}A_{{\bf k}'',-{\bf k}}({\bf
J}',{\bf J}) m f({\bf J}),\nonumber\\
\end{eqnarray}
where we have set $s=t-t'$. Substituting this relation into Eq. (\ref{j10}), we
obtain
\begin{eqnarray}
\label{aaa1}
\left\langle \delta f\frac{\partial\delta\Phi}{\partial {\bf
w}}\right\rangle_{\rm friction}=\sum_{{\bf k}{\bf k}'} i{\bf k}' e^{i{\bf
k}\cdot {\bf
w}}e^{i{\bf k}'\cdot {\bf w}}(2\pi)^d\sum_{{\bf k}''}\int
d{\bf
J}' A_{{\bf k}'{\bf k}''}({\bf J},{\bf J}')i{\bf k}''\cdot
\frac{\partial f'}{\partial {\bf J}'}\int_0^t ds\, e^{-i({\bf
k}\cdot{\bf\Omega}+{\bf k}''\cdot{\bf\Omega}')s}
A_{{\bf k}'',-{\bf k}}({\bf J}',{\bf J}) m f({\bf
J}).\nonumber\\
\end{eqnarray}
Averaging this expression over ${\bf w}$, which
amounts to replacing ${\bf k}'$
by $-{\bf k}$, we find that
\begin{eqnarray}
\label{aaa2}
\left\langle \delta f\frac{\partial\delta\Phi}{\partial {\bf
w}}\right\rangle_{\rm friction}=-\sum_{{\bf k}} i{\bf k}(2\pi)^d\sum_{{\bf
k}''}\int
d{\bf
J}' A_{-{\bf k},{\bf k}''}({\bf J},{\bf J}')i{\bf k}''\cdot
\frac{\partial f'}{\partial {\bf J}'}\int_0^t ds\, 
e^{-i({\bf
k}\cdot{\bf\Omega}+{\bf k}''\cdot{\bf\Omega}')s}
A_{{\bf k}'',-{\bf
k}}({\bf J}',{\bf J}) m f({\bf J}).
\end{eqnarray}
If we make the change of notation ${\bf k}''=-{\bf k}'$ and let $t\rightarrow
+\infty$, we can rewrite the foregoing equation as
\begin{eqnarray}
\label{p31}
\left\langle \delta f\frac{\partial\delta\Phi}{\partial {\bf
w}}\right\rangle_{\rm friction}=\sum_{{\bf k}} i{\bf k}(2\pi)^d\sum_{{\bf
k}'}\int
d{\bf
J}' A_{-{\bf k},-{\bf k}'}({\bf J},{\bf J}')i{\bf k}'\cdot
\frac{\partial f'}{\partial {\bf J}'} A_{-{\bf k}',-{\bf k}}({\bf J}',{\bf J})
 m f({\bf J})
\int_0^{+\infty} ds\,
e^{-i({\bf k}\cdot {\bf\Omega}-{\bf
k}'\cdot{\bf\Omega}')s}.\nonumber\\
\end{eqnarray}
Making the transformations
$s\rightarrow -s$, ${\bf
k}\rightarrow -{\bf k}$ and 
${\bf k}'\rightarrow -{\bf k}'$ we see that
we can replace $\int_0^{+\infty} ds$ by $\frac{1}{2}\int_{-\infty}^{+\infty}
ds$. We then get 
\begin{eqnarray}
\label{p32}
\left\langle \delta f\frac{\partial\delta\Phi}{\partial {\bf
w}}\right\rangle_{\rm friction}=\sum_{{\bf k}} i{\bf k}(2\pi)^d\sum_{{\bf
k}'}\int
d{\bf
J}' A_{-{\bf k},-{\bf k}'}({\bf J},{\bf J}')i{\bf k}'\cdot
\frac{\partial f'}{\partial {\bf J}'} A_{-{\bf k}',-{\bf k}}({\bf J}',{\bf J})
 m f({\bf J})
\frac{1}{2}\int_{-\infty}^{+\infty} ds\,
e^{-i({\bf k}\cdot {\bf\Omega}-{\bf
k}'\cdot{\bf\Omega}')s}.\nonumber\\
\end{eqnarray}
Using the identity (\ref{deltac}) we obtain
\begin{eqnarray}
\label{p34}
\left\langle \delta f\frac{\partial\delta\Phi}{\partial {\bf
w}}\right\rangle_{\rm friction}=-\pi(2\pi)^d\sum_{{\bf k}} {\bf k}\sum_{{\bf
k}'}\int
d{\bf
J}' A_{-{\bf k},-{\bf k}'}({\bf J},{\bf J}'){\bf k}'\cdot
\frac{\partial f'}{\partial {\bf J}'} A_{-{\bf k}',-{\bf k}}({\bf J}',{\bf J})
\delta({\bf k}\cdot {\bf\Omega}-{\bf
k}'\cdot{\bf\Omega}')m f({\bf J}).
\end{eqnarray}
Finally, using the identities (\ref{dn14bare}) and (\ref{pi37tax}), we arrive
at the expression
\begin{eqnarray}
\label{p35}
\left\langle \delta f\frac{\partial\delta\Phi}{\partial {\bf
w}}\right\rangle_{\rm friction}=-\pi(2\pi)^d\sum_{{\bf k}} {\bf k}\sum_{{\bf
k}'}\int
d{\bf
J}' |A_{{\bf k},{\bf k}'}({\bf J},{\bf J}')|^2 {\bf k}'\cdot
\frac{\partial f'}{\partial {\bf J}'} 
\delta({\bf k}\cdot {\bf\Omega}-{\bf
k}'\cdot{\bf\Omega}')m f({\bf J}).
\end{eqnarray}
Adding Eqs. (\ref{p26}) and (\ref{p35}), we obtain the Landau flux
\begin{eqnarray}
\label{p36}
\left\langle \delta f\frac{\partial\delta\Phi}{\partial {\bf
w}}\right\rangle=
\pi (2\pi)^d \sum_{{\bf k}{\bf k}'} {\bf k}\int d{\bf
J}' \,   {\bf k}\cdot
\frac{\partial f}{\partial {\bf J}} |A_{{\bf k},{\bf k}'}({\bf
J},{\bf J}')|^2
\delta({\bf
k}\cdot{\bf\Omega}-{\bf k}'\cdot {\bf\Omega}')mf({\bf J}')\nonumber\\
-\pi(2\pi)^d\sum_{{\bf k}} {\bf k}\sum_{{\bf
k}'}\int
d{\bf
J}' |A_{{\bf k},{\bf k}'}({\bf J},{\bf J}')|^2 {\bf k}'\cdot
\frac{\partial f'}{\partial {\bf J}'} 
\delta({\bf k}\cdot {\bf\Omega}-{\bf
k}'\cdot{\bf\Omega}')m f({\bf J}).
\end{eqnarray}
Substituting this flux into Eq. (\ref{j1}), we obtain
the inhomogeneous Landau equation \cite{aa}
\begin{equation}
\frac{\partial f}{\partial t}=\pi (2\pi)^d m \frac{\partial}{\partial {\bf
J}}\cdot \sum_{{\bf k},{\bf k}'}\int d{\bf J}' \, {\bf k}\,
|A_{{\bf k},{\bf k}'}({\bf J},{\bf J}')|^2\delta({\bf k}\cdot
{\bf \Omega}-{\bf k}'\cdot {\bf \Omega}') \left
(f' {\bf k}\cdot \frac{\partial f}{\partial {\bf J}}-f {\bf
k}'\cdot \frac{\partial f'}{\partial {\bf J}'}\right ).
\label{j19land}
\end{equation}
We can easily extend the derivation of the Landau equation to the
multispecies case.

\end{document}